\documentclass[11pt]{book}

\pdfoutput=1

\usepackage[latin1]{inputenc}
\usepackage{amsmath}    
\usepackage{graphicx}   
\usepackage{verbatim}   
\usepackage{color}      
\usepackage{subfigure}  
\usepackage{hyperref}   

\usepackage{slashed}

\usepackage{amssymb}
\usepackage{bm}

\usepackage[Bjornstrup]{fncychap}  

\newcommand{\im}{\text{i}}

\newcommand{\be}{\begin{equation}}
\newcommand{\eeq}{\end{equation}}
\newcommand{\bea}{\begin{eqnarray}}
\newcommand{\eea}{\end{eqnarray}}
\newcommand{\ba}{\begin{array}}
\newcommand{\ea}{\end{array}}

\newcommand{\eq}[1]{Eq.~(\ref{#1})}
\newcommand{\comm}[2]{\left[#1,#2\right]}
\newcommand{\ii}{\mathrm{i}}
\newcommand{\ee}{\mathrm{e}}

\newcommand{\Tr}{\mathrm{Tr}\,}
\newcommand{\tr}{\mathrm{tr}\,}
\newcommand{\diag}{\mathrm{diag}\,}

\newcommand{\cD}{{\mathcal{D}}}

\newcommand{\cE}{{\mathcal{E}}}
\newcommand{\cF}{{\mathcal{F}}}

\newcommand{\cM}{{\mathcal{M}}}
\newcommand{\cN}{{\mathcal{N}}}
\newcommand{\cP}{{\mathcal{P}}}
\newcommand{\cQ}{{\mathcal{Q}}}
\newcommand{\cR}{{\mathcal{R}}}

\newcommand{\cZ}{{\mathcal{Z}}}

\newcommand{\one}{{\rm 1\kern -.9mm l}}

\newcommand{\Pf}{\mathrm{Pf}}
\newcommand{\ve}[1]{{\vec e}_{#1}}
\newdimen\tableauside\tableauside=1.0ex
\newdimen\tableaurule\tableaurule=0.4pt
\newdimen\tableaustep

\def\phantomhrule#1{\hbox{\vbox to0pt{\hrule height\tableaurule
width#1\vss}}}
\def\phantomvrule#1{\vbox{\hbox to0pt{\vrule width\tableaurule
height#1\hss}}}
\def\sqr{\vbox{%
 \phantomhrule\tableaustep
\hbox{\phantomvrule\tableaustep\kern\tableaustep\phantomvrule\tableaustep}%
 \hbox{\vbox{\phantomhrule\tableauside}\kern-\tableaurule}}}
\def\squares#1{\hbox{\count0=#1\noindent\loop\sqr
 \advance\count0 by-1 \ifnum\count0>0\repeat}}
\def\tableau#1{\vcenter{\offinterlineskip
 \tableaustep=\tableauside\advance\tableaustep by-\tableaurule
 \kern\normallineskip\hbox
   {\kern\normallineskip\vbox
     {\gettableau#1 0 }%
    \kern\normallineskip\kern\tableaurule}%
 \kern\normallineskip\kern\tableaurule}}
\def\gettableau#1 {\ifnum#1=0\let\next=\null\else
 \squares{#1}\let\next=\gettableau\fi\next}
\newcommand{\Yfund}{\tableau{1}}
\newcommand{\Ysymm}{\tableau{2}}
\newcommand{\Yasymm}{\tableau{1 1}}

\usepackage{textcase}
\usepackage{xspace}

\begin{document}


\newcommand{\myTitle}{D-branes and Non-Perturbative Quantum Field Theory: Stringy Instantons and Strongly Coupled Spintronics\xspace}

\newcommand{\mySubtitle}{Subtitle or whatever \xspace}
\newcommand{\myName}{{\bfseries Daniele Musso}\xspace}
\newcommand{\myProf}{Prof. Alberto Lerda \xspace}
\newcommand{\rel}{{\bf Relatore} \xspace}
\newcommand{\corel}{{\bf Co-Relatore} \xspace}
\newcommand{\myaldo}{{Dr. Aldo Cotrone} \xspace}
\newcommand{\com}{{\bf Commissione}\xspace}
\newcommand{\comdue}{Prof. Silvia Penati\xspace}
\newcommand{\comtre}{Prof. Matteo Bertolini\xspace}
\newcommand{\comuno}{Prof. Marco Bill\'o\xspace}

\newcommand{\mySupervisora}{Prof. Nick Evans\xspace}
\newcommand{\mySupervisorb}{Prof. Riccardo Argurio\xspace}

\newcommand{\contrel}{{\bf Controrelatori}\xspace}

\newcommand{\myFaculty}{Facolt\`a di Scienze Matematiche, Fisiche e Naturali \xspace}
\newcommand{\myDepartment}{Dipartimento di Fisica Teorica \xspace}
\newcommand{\myUni}{\protect{\bf Universit\`a degli studi di Torino}\xspace}

\newcommand{\myTime}{March 16th, 2012}

\begin{titlepage}
\pagestyle{empty}

\begin{center}
\begin{large}
{\bf Scuola di Dottorato in Scienza ed Alta Tecnologia} \\
{Indirizzo in Fisica ed Astrofisica} \\
\end{large}

\hrulefill
        \large  

        \hfill

        \vfill

        \begingroup
           \huge{\scshape\myTitle} \\ \bigskip
        \endgroup
\bigskip
        {\scshape \myName}

        \vfill

        \includegraphics[width=4cm]{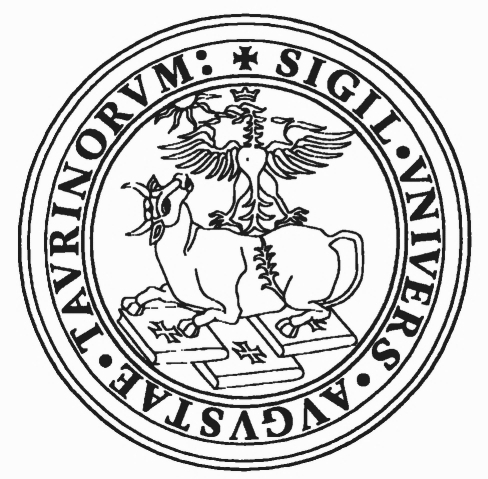} \hspace{4cm}
        \includegraphics[width=4cm]{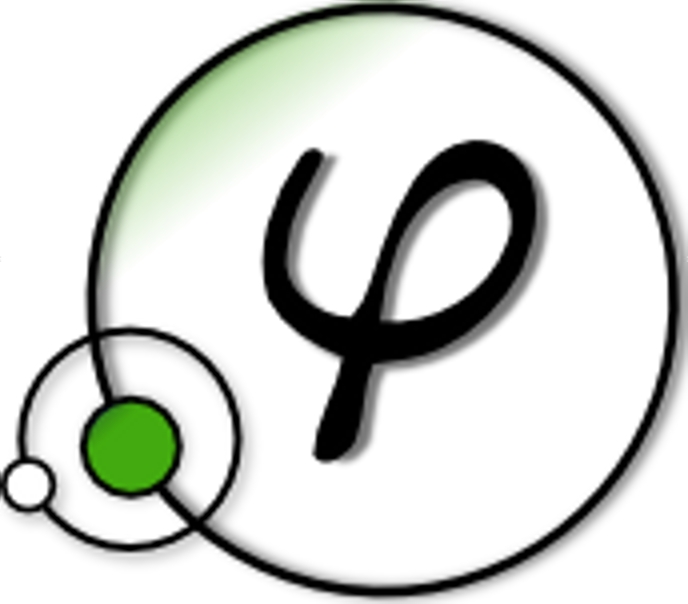}
        \bigskip \\
	{\bfseries Universit\'a degli Studi di Torino} \hspace{3.25cm} {\bfseries Dipartimento di Fisica Teorica}
\bigskip

        \vfill

\bigskip
\bigskip

\noindent{\rel \hfill \com}\\
\noindent{\myProf \hfill \comtre}\\
\noindent{\corel \hfill \comuno}\\
\noindent{\myaldo \hfill \comdue}\\
\noindent{\contrel \hfill {\mbox{}}}\\
\noindent{\mySupervisorb \hfill {\mbox{}}}\\
\noindent{\mySupervisora \hfill {\mbox{}}}\\
\bigskip
        \myDepartment \\                            
        \myFaculty \\
        \myUni 
\vfill

        \myTime

    \end{center}  

\newpage 

\begin{center}
 {\color{white}aaaaaa\\ aaaaaa\\ aaaaaa\\ aaaaaa\\ aaaaaa\\ aaaaaa\\ aaaaaa\\ aaaaaa\\ aaaaaa\\ aaaaaa\\ aaaaaa\\ aaaaaa\\ aaaaaa\\ aaaaaa\\}
 To my parents, my grandmother and Claudio 
\end{center}

\end{titlepage}   


\pagenumbering{roman}

\tableofcontents

\makeatletter\@openrightfalse

\setcounter{chapter}{-1}
\chapter{Abstract}


\pagenumbering{arabic}
\setcounter{page}{1}

The non-perturbative dynamics of quantum field theories is studied
using theoretical tools inspired by string formalism.
Two main lines are developed: the analysis of stringy instantons in a class of four-dimensional ${\cal N}=2$ gauge theories
and the holographic study of the minimal model for a strongly coupled unbalanced superconductor.

The field theory instanton calculus admits a natural and efficient description in terms of D-brane models.
In addition, the string viewpoint offers the possibility of generalizing the ordinary instanton configurations.
Even though such generalized, or stringy, instantons would be absent in a purely
field-theoretical, low-energy treatment, we demonstrate that they do alter the
IR effective description of the brane dynamics by
introducing contributions related to the string scale $\alpha'$.
In the first part of this thesis we compute explicitly the stringy instanton corrections to the effective prepotential
in a class of quiver gauge theories.

In the second part of the thesis, we present a detailed analysis of the minimal holographic setup yielding an effective description
of a superconductor with two Abelian currents. 
The model contains a scalar field whose condensation produces a spontaneous symmetry breaking
which describes the transition to a superfluid phase.
This system has important applications in both
QCD and condensed matter physics;
moreover, it allows us to study mixed electric-spin transport properties (i.e. spintronics)
at strong coupling.

\chapter{Preamble}
The subject of the present thesis consists in the study of the non-perturbative dynamics of quantum field theory 
using string-inspired theoretical tools.

The non-perturbative dynamics of quantum field theory is relevant for an extremely wide scenario of physical contexts.
Indeed, it plays a crucial role in various sectors of physics ranging from the dynamics of fundamental 
constituents and interactions to the condensed matter panorama.
The string approach offers a versatile and powerful framework in which many distinct non-perturbative aspects of quantum field theory
can be accommodated and studied.
As the range of application is wide, also the ensemble of possible techniques is very wealthy; our treatment concentrates especially
on two different lines, namely the {\bfseries D-brane instanton calculus} and the {\bfseries $\bm{AdS}$/CFT correspondence}. 
They both originate from a common string environment.

In the string formalism, the dynamics is described in terms of the evolution and the interactions
of some fundamental objects: the membranes.
In contrast to particles and as the name intuitively suggests, membranes can have extended directions.
If we agree on defining a generic $n$-brane as an object that is extended in $n$ space-time directions,
we can think of strings as $1$-branes.
In addition to the strings, the set of basic objects in the string formalism comprehends also the D$p$-branes that are 
particular $p$-dimensional membranes to which the open string endpoints can be attached.
Focusing the attention on the concept of membrane rather than just on strings, sets a democratic viewpoint
encompassing all the fundamental constituents of the formalism, which, not so democratically, we
will still indicate as ``string formalism''.

The D$p$-branes, or D-branes for short, are the pivotal ingredient that allows us to explore the non-perturbative 
aspects of quantum field theory of interest here. 
An essential feature of D-branes is their relation and interactions with strings. 
Historically they have been introduced as surfaces on which the open string endpoints can lie;
they actually define Dirichlet (from which the ``D'' of D-brane) boundary conditions for the strings\footnote{More precisely, 
we have Dirichlet boundary conditions (i.e. constrained) for the directions that are transverse to the D-brane
and Neumann (i.e. free) boundary condition along the D-brane itself.}.
It is possible to argue that the open strings attached to a brane offer a description of the dynamics of the brane itself, hence D-branes
and open strings are closely related.
Open strings are objects with tension; an open string connected to a D-brane can be naively thought of as a quantum ``tension fluctuation''
or excitation of the D-brane itself.

In the low-energy spectrum of closed strings there is a massless mode which can be identified with the graviton, i.e. the spin $2$ quantum mediating
gravitational interactions.
Since the D-branes are objects possessing a rest intrinsic energy, they interact gravitationally
and it is therefore straightforward to expect that the D-branes can source and absorb closed strings.

The relation with both open and closed strings puts the D-branes in a central position for the developments we are going to analyze
throughout the thesis.
Depicting a naive image which will be clarified and made precise in the following treatment, the open strings attached to a D-brane
are described at low-energy by means of a gauge theory whose base manifold coincides with the $p+1$-dimensional hyper-volume
spanned by the D-brane in its time-evolution\footnote{Henceforth referred to as the $p+1$ dimensional \emph{world-volume} of the D$p$-brane.}.
Instead, the closed strings propagating at low-energy in the whole ambient space-time containing the D-branes are described 
at an effective level by supergravity or even classical gravity models with extended solutions.

If we concentrate on the physics of closed strings in the proximity of a D-brane, we can wonder if the closed-string/gravitational behavior
could account as well for the whole dynamics of the brane itself. 
At a sketchy and low-energy level, we could hope that looking at the local space-time deformations
induced by the presence of a D-brane we could recover full information on the D-brane dynamics.
Since, as we have just mentioned, the brane dynamics is already encoded in the physics of the open string modes attached to it, 
we are here speculating about an open/closed string connection relying crucially on the D-brane physics.
Such idea lies at the heart of the $AdS$/CFT correspondence and holographic models in general.

We can also consider a different aspect of D-brane dynamics which leads us to the stringy instanton calculus. 
Since, in appropriate low-energy regimes, the open strings ``living on a brane'' are well-described
by a quantum supersymmetric gauge field theory, it is natural to ask whether such effective field theory emerges
with all its perturbative and non-perturbative content. The proper answer is that from the analysis of D-brane models we do not only recover all the standard 
perturbative and non-perturbative features of the low-energy quantum field theory but, in addition, the string framework makes it
possible to study some significant generalized configurations.
Such generalizations are outside of the reach of an IR (i.e. low-energy) purely field theoretical approach and
we henceforth refer to them as \emph{stringy} or \emph{exotic}. 
In particular, our focus is on the stringy instantonic configurations and the modifications that they induce in the couplings of the low-energy effective field theory.
As we will see, although the origin of the exotic effects is intrinsically related to the stringy nature of the model, they do produce 
important modifications to the low-energy quantum field theory.

\section{Historical and ``philosophical'' note}

Historically, the string formalism has been firstly introduced to describe strong interactions, specifically
as a model for meson scattering\footnote{Two synthetic historical accounts about the early steps of string models
can be found in the introductory chapters of \cite{green1988superstring} and \cite{Ferro:2009gv}.}.
The string description of meson scattering is particularly suitable to account for the $t$-$s$ duality of meson scattering
where the two Mandelstam channels presenting the same amplitude value are naturally encoded in a single open string amplitude. 
Moreover, also the proliferation of mesons is describable in terms of spinning string states possessing a precise interconnection 
between the rest energy $s$ and the spin $J$,
\begin{equation}\label{slop}
 J(s) = \alpha_0 + \alpha' s \ ,
\end{equation}
where $\alpha_0$ represents a constant shift while $\alpha'$ is the celebrated \emph{Regge slope}.
From relation \eqref{slop} we can observe that $\alpha'$ is inversely related to the string tension;
indeed, keeping fixed the energy $s$ we have greater angular momentum $J$ if $\alpha'$ is increased.
We can then expect that, increasing $\alpha'$, we increase the length of the spinning string and then, at fixed energy, 
this is equivalent to reducing its tension\footnote{The tension corresponds to the energy density on the world-sheet spanned by the string:
as such it is related to the ``linear energy density'' along the string.
Here ``increase the length of the string'' means ``let the length increase''; since we keep the energy fixed the increase in length
does not correspond to a mechanical stretch of the string (i.e. there is no ``work'' done on the string).}.

In the context of the strong interaction, the later introduction of a renormalizable quantum field theory, namely the QCD, overcame the string description, 
at least in the high-energy perturbative regime.
Before long, the recognition in the closed string spectrum of a spin $2$ graviton-like particle 
interacting democratically\footnote{See Appendix \ref{demogravi} for a brief argument.} 
with all the objects possessing mass gave a tremendous new momentum to the theoretical research in this field.
Indeed, this discovery opened the doors to a completely new description of gravity hopefully admitting a 
consistent quantum treatment.

Another greatly interesting feature of the string formalism is the possibility of comprehending various,
and maybe all, kinds of fundamental interactions in one single theory.
This aspect of unification of all fundamental interactions produced widespread interest and even radical
enthusiasm leading some people to call string theory the ``theory of everything''.
Without spending much time to discuss this point, it is generally possible to assume a different and 
more moderate perspective.
As the frequent use of the name ``string formalism'' instead of ``string theory'' suggests,
we want to adopt an instrumental attitude towards the stringy mathematical tools and techniques; 
these are extremely fertile and insightful in describing many important physical systems.
In the part of the present thesis regarding the holographic approaches to study strongly coupled systems,
in some cases we will even adopt a phenomenological effective approach, meaning that the string inspired models we study describe macroscopic
physical properties without the pretension of accounting precisely for the microscopic features of the system under analysis.
Already at the bottom-up level we will observe how insightful a string inspired model can be in shedding light on the strongly coupled
dynamics of quantum field theories.
The moderate attitude does not deny the conceptual and philosophical fascination of aiming to a unified
theory of everything, it simply focuses on the operative purpose of studying and exploiting
the string formalism as deeply as possible before, or even independently, of how the dispute on
whether we deal with the theory of everything or not will be settled.

\section{Purpose and Original Content}
This thesis is aimed at producing a text which could result as clear as possible
also to a partial or even a ``localized'' reading.
The body of the text is indeed divided in many paragraphs containing various footnotes and references 
both to the numerous appendices and to papers in the literature;
this fragmentation is deliberate considering that an exiting Ph.D. student
feels the moral duty to be as useful as possible to the doctorate students following him.

The treated subjects require quite massive introductions. At the outset we underline that the original content of the thesis is contained mainly in
the chapters:

\begin{itemize}
 \item Stringy Instantons (Chapter \ref{mio1})
 \item Holographic superconductors with two fermion species and spintronics (Chapter \ref{mio2})
\end{itemize}

\section{Disclaimer}
\label{disclaimer}
The thesis was publicly defended on March the 16th 2012.
All the content and in particular the bibliographic references are referred to that date.
In the meantime, there have been further developments in the field which do not appear here.

\chapter{Introduction}

\section[Strings, Branes and SUSY Gauge Theories]{Strings, Branes and Gauge Theories}
When regarding the string formalism as a candidate description of the fundamental interactions, various important questions arise.
One among the most significant points consists in how the physics that we have already experimentally tested could admit a stringy description,
or rather, how can it be embedded in a string model.
Specifically, we are interested in the way in which General Relativity and the quantum field theory describing the electro-weak and
strong interactions (i.e. the Standard Model) can appear in the string context.

Let us remind ourselves that the characteristic energy scale of strings is Planck's scale ($\sim 10^{19}\,\text{GeV}$)
which is much higher than all the scales directly probed in human particle physics experiments so far%
\footnote{This is strictly true when the string model under consideration does not include compactified directions.
The compactification scale can ``lower'' the \emph{string scale} (i.e. the scale at which stringy effects have to be taken into account)
many orders of magnitude below Plank's scale; as it is natural to expect thinking of Kaluza-Klein modes, the larger the compact directions, the lower the string scale. 
For details see for instance \cite{Antoniadis:1998ig}.}.
This can be naively thought of as a consequence of the fact that the string degrees of freedom appear as 
a quantum description (among other things) of gravity.
The string scale has therefore a close relationship with the scale at which gravity becomes sensitive to quantum corrections,
i.e. Planck's scale. 
The physical theories in which we are confident, i.e. the theories that passed many experimental tests,
must therefore emerge in the low-energy (with respect to Planck's scale) regime of string theory.
The string formalism aims to furnish the unifying UV completion of General Relativity and the Standard Model, hence
the low-energy limit of string theory is required
to reproduce them, both at the perturbative and the non-perturbative levels.

Studying the low-energy limit of a string model is equivalent to analyzing it at an energy level
which is small with respect to the characteristic energy needed to excite the strings.
The string excitation energy is measured by the tension, therefore the low-energy limit can be
taken considering the infinite tension limit.
Indeed, the string tension $T$ appears as an overall factor which scales the string action;
in a would-be string-field-theory\footnote{With
string-field-theory is usually meant the second-quantized formulation of a string model.
Note that the existence of such a formulation is still an open question in all the 
string models studied so far, including the bosonic string.} in which the exponential
of minus the action $e^{-S}$ weights the probability amplitude of a possible evolution (i.e. the amplitude associated to a path in a path integral formulation),
the string excitation ``cost'' scales as the action and then according to the tension $T$.

As already mentioned, the string tension $T$ is expressed in terms of the dimensionful constant $\alpha'$; 
in natural units $\hbar = c = 1$, we have:
\begin{equation}
T = \frac{1}{2\pi \alpha'} \ . 
\end{equation}
Moreover, it is straightforward to define a characteristic string length $l$.
Actually, the world-sheet $\Sigma$ is the two-dimensional surface spanned by the string in its evolution and then
the tension $T$ has the dimensions of an inverse area, that is an inverse squared length.
In natural units, we define:
\begin{equation}\label{car_length}
 l = \frac{1}{\sqrt{\pi T}} = \sqrt{2 \alpha'}\ .
\end{equation}
Notice that $l$ is not to be thought of as the string length \emph{tout court}.
In fact, a world-sheet amplitude can be regarded as describing different
classical string propagations. 
Consider for instance a rectangular plane world-sheet much longer than wide.
It can either represent the propagation of a long string on a short path or
the propagation of a short string on a long path. 
Since, as we will see shortly, the action measures the world-sheet proper area, in this simple rectangular case, 
the characteristic length we defined is related to the geometric mean
of the two sides of the rectangular world-sheet.

The low-energy regime of a string model as the infinite string-tension limit implies,
through \eqref{car_length}, that the original extended strings whose tension
diverges become effectively point-like, $l \rightarrow 0$.
It is pretty reasonable that the physics of relativistic point-like objects can be described with quantum field theory;
this is indeed the case and the naive expectation can be precisely tested.

In the perturbative analysis of string dynamics one considers the scattering amplitudes of the lowest lying string vibrational modes, namely
the massless ones. Actually, the massive modes correspond to excited and so more energetic vibrational modes of the strings which are then strongly suppressed in the
low-energy limit.
Within the string formalism, the low-energy limit is achieved performing the computations at finite $\alpha'$ and then 
considering the results in the infinite tension limit $\alpha'\rightarrow 0$.
The results obtained in this fashion are to be compared to the scattering amplitudes computed for the corresponding\footnote{The correspondence
is set by the identification of particles and string vibrational modes presenting the same quantum numbers.} massless
particles in quantum field theory. 
The string computations yield, in general, the field theoretical results with additional corrections; these corrections
are weighted by positive powers of the string constant $\alpha'$. 
Such corrections vanish in the $\alpha'\rightarrow 0$ limit.
At the perturbative level, the field theory for the set of massless particles
corresponding to the massless string modes, can be regarded as an effective low-energy 
description of the corresponding string model.

Let us show an explicit example, namely the scattering amplitude of three non-Abelian massless vectors\footnote{I.e.
massless vectors corresponding to gauge bosons of a non-Abelian gauge theory.}.
As shown in detailed in e.g. \cite{polchinski2001string}, the string computation of this
amplitude returns:
\begin{equation}\label{tering}
 \begin{split}
 {\cal A}(\bm{k}_1,\bm{\epsilon}_1;\bm{k}_2,\bm{\epsilon}_2;\bm{k}_3,\bm{\epsilon}_3) \propto &\, \delta(\sum_i \bm{k}_i)
 \left[(\bm{\epsilon}_1\cdot \bm{k}_{23})(\bm{\epsilon}_2\cdot \bm{\epsilon}_3)
 +(\bm{\epsilon}_2\cdot \bm{k}_{31})(\bm{\epsilon}_3\cdot \bm{\epsilon}_1)\right.\\
 &\left.+(\bm{\epsilon}_3\cdot \bm{k}_{12})(\bm{\epsilon}_1\cdot \bm{\epsilon}_2)
 + \frac{\alpha'}{2} (\bm{\epsilon}_1 \cdot \bm{k}_{23})(\bm{\epsilon}_2 \cdot \bm{k}_{31})(\bm{\epsilon}_3 \cdot \bm{k}_{12})\right]
 \end{split}
\end{equation}
where $\bm{k}_i$ and $\bm{\epsilon}_i$ are respectively the momenta and polarization vectors of the three
vector modes labeled by the index $i=1,2,3$.
For the sake of compactness, we have dropped the color structure and the corresponding factor from the amplitude \eqref{tering}; 
we also defined $\bm{k_{ij}}=\bm{k}_i-\bm{k}_j$.
Equation \eqref{tering} coincides with the Yang-Mills theory result for three gauge bosons up to first order in the momenta.
It is evident that in the infinite-tension limit $\alpha'\rightarrow 0$ the additional stringy contribution vanishes and
we recover precisely the field theoretical result.

The complete perturbative test of the consistency of the low-energy 
field theory description for a string model is a wide topic.
Furthermore, one can start considering many different string models and ask whether some features of field theories such
as masses or potentials can arise in the low-energy regime of appropriate string configurations.
Leaving this important aspect somewhat aside, our interest would be instead directed towards the non-perturbative side of the story.

Gauge theories and their supersymmetric versions have a rich vacuum structure.
We can study perturbatively these theories around different classical field configurations that, at least locally, minimize the action.
The perturbative agreement between string formalism
and its low-energy field theory description holds also around a non-trivial vacuum.
Of course, this can be tested explicitly and, even more importantly, we have to understand
how the non-trivial background itself could arise from a string model.

This relevant question has been first addressed as soon as the D-branes were discovered.
Actually, D-branes can be seen as non-perturbative objects in string theory,
meaning that they have a solitonic nature in the string context.
A nice feature of these extended objects is that, even though they have a non-perturbative origin,
their dynamics is describable in terms of perturbative modes as we will see accurately in Section \ref{pert_brane}.
One can expect this observing, for instance, that it is perfectly legitimate to consider
small oscillations, i.e. fluctuations, of a field around a non-trivial vacuum in which the field itself assumes a big VEV.
The perturbative picture of low-energy D-brane dynamics is realized by open strings whose
endpoints live on the D-branes and by closed string modes emitted and absorbed by the D-branes themselves.

One could ask why the membranes such D-branes do not shrink to point-like
objects themselves in the $\alpha'\rightarrow 0$ limit. The answer is related to topology because D-branes are solitonic objects
representing topologically non-trivial ground states of the string theory. The low-energy limit of a topologically-non trivial sector
is the topologically non-trivial vacuum; this vacuum configuration can contain extended objects or extended field configurations. 
In the low-energy picture, the strings describe the small fluctuations around the topologically non-trivial vacuum represented by
the D-branes at rest.
Hence we can guess that their ``solitonic character'' would not be spoiled by the $\alpha'\rightarrow 0$ limit,
and ``naively'' the expectation is true.
In other terms, the D-branes are topologically protected.
To be slightly more precise, we must say that D-brane models, containing the appropriate kinds and
number of D-branes, give indeed rise to setups whose low-energy dynamics could be encoded in
a supersymmetric gauge theory with all its perturbative and non-perturbative features.

\section{Perturbative description of the D-branes}
\label{pert_brane}
In the present section we indulge on how supergravity and gauge theories emerge in the low-energy description of the closed and open string sector respectively
in the presence of D-branes.
As a necessary introductory step, we must concentrate first on the quantum description of string dynamics.
Later we will focus on the low-energy limit.

\subsection{Quantum treatment of strings}

 \begin{figure}
  \centering
  \includegraphics[width=\textwidth]{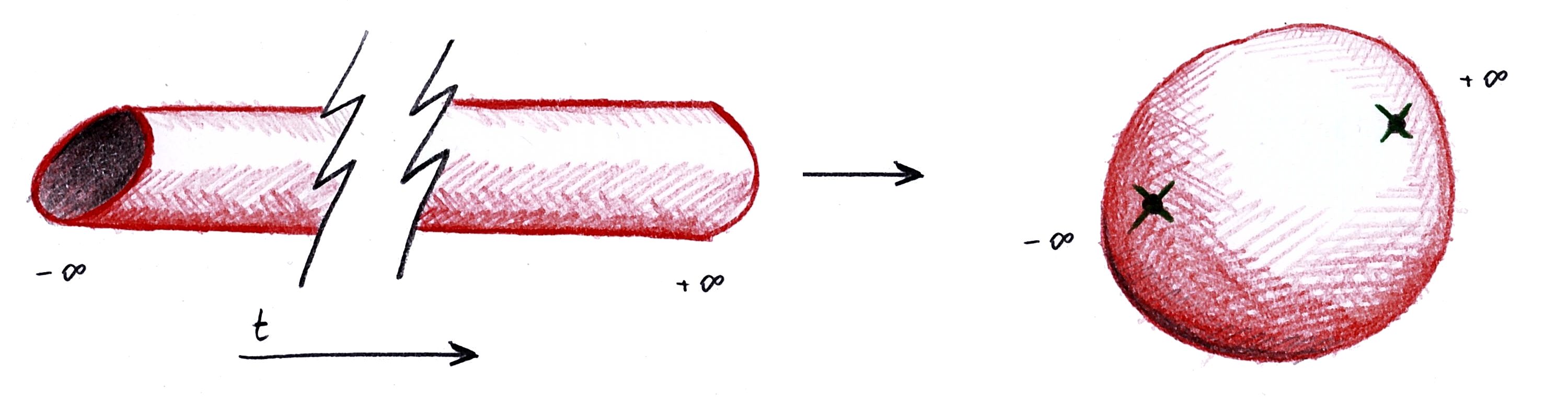}
  \caption{Closed string free propagation: the conformal invariance of the action allows us to map it to a punctured sphere.}
 \end{figure}

\begin{figure}
  \centering
  \includegraphics[width=\textwidth]{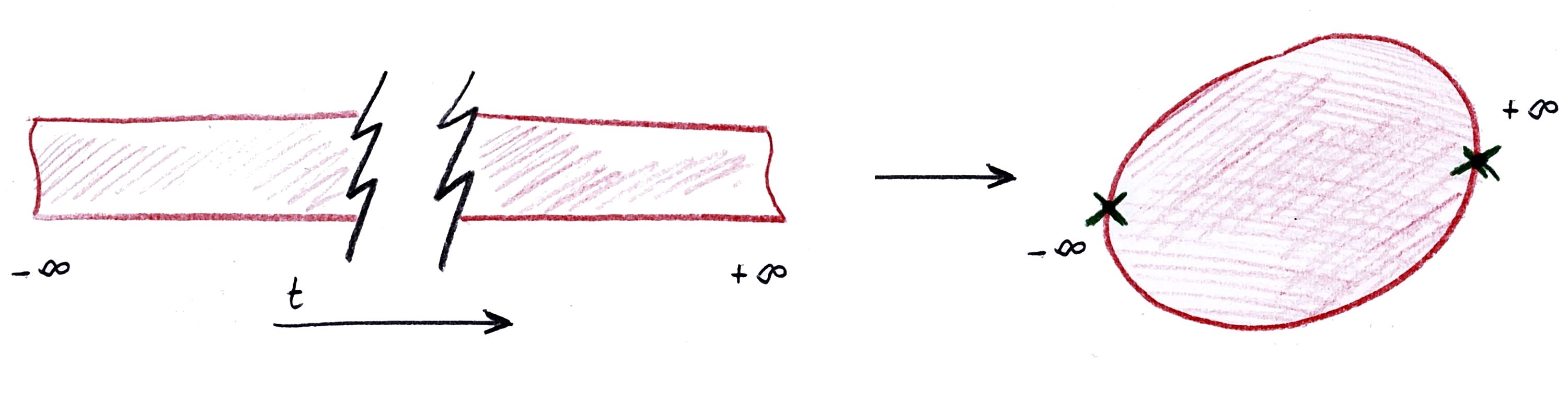}
  \caption{Open string free propagation: the conformal invariance of the action allows us to map it to a disk.}
 \end{figure}

From a classical point of view, a propagating string sweeps the two-dimensional world-sheet embedded into space-time.
In order to associate an action to a particular string evolution, we generalize what is standard
in particle dynamics. 
Actually, considering a relativistic particle, we associate to its propagation an action that measures the proper length
of the world-line representing the particle evolution in space-time.
The proper length is invariant with respect to reparametrizations
of the world-line, in accordance with the relativistic coordinate invariance requirement.
Another important characteristic of the proper length is its additivity: namely, the action of the composition of two word-lines sharing an endpoint
is given by the sum of the values of the action associated to the component paths.

Inspired by the classical relativistic particle, we are straightforwardly led to think that the classical evolution of the string is encoded in
the world-sheet with minimal proper area, being this expressed by the following action 
\begin{equation}\label{proper_area}
 S_{\text{NG}} = T \int_{\Sigma} d^2 \sigma\ \sqrt{g_{\alpha\beta}}
=T \int_{\Sigma} d^2 \sigma\ \left|G_{MN}\frac{\partial X^M}{\partial \sigma^\alpha}
\frac{\partial X^N}{\partial \sigma^\beta}\right|^{1/2}\ .
\end{equation}
This functional is usually referred to as the Nambu-Goto action.
As usual, we indicated the tension with $T$, the world-sheet with $\Sigma$, the two world-sheet coordinates with
$\sigma^\alpha$ where $\alpha=1,2$, the space-time coordinates with $X^M$, the space-time metric with $G_{MN}$ and
the corresponding induced world-sheet metric with $g_{\alpha\beta}$.
Notice that the variational study of \eqref{proper_area} has to be performed choosing suitable boundary conditions.

The relativistic string is endowed with a new crucial feature with respect to the relativistic particle: the
presence of ``internal'' freedom. 
Actually, a particle is just a point-like object without internal characteristics
whereas a string can oscillate. 
Since the string length is usually of the order of Plank's length,
\begin{equation}
 l_P = \sqrt{\frac{\hbar G}{c^3}} \sim 1.61 \cdot 10^{-35} \text{m}\ ,
\end{equation}
its oscillations are clearly a quantum effect.
The study of the internal modes of the string brings us to the question of the string quantization.

The detail of string quantization is far beyond the purpose of this introductory part;
for a thorough treatment of the topic we refer to the literature (see for instance \cite{green1988superstring,polchinski2001string}).
The Nambu-Goto action \eqref{proper_area} is not suitable for a quantum treatment because the presence of the square root in the integrand
makes the quantization process cumbersome. 
Indeed, for the sake of treating the string at the quantum level one actually considers the action
\begin{equation}\label{sheet}
  S = \frac{T}{2} \int_\Sigma d^2 \sigma \ \sqrt{-\gamma} \, \gamma_{\alpha\beta} \, \partial^\alpha X^M \partial^\beta X^N G_{MN}\ ,
\end{equation}
that is classically equivalent to the Nambu-Goto action where $\gamma$ is the metric on the world-sheet; notice that $\gamma$
is here regarded as a dynamical field.
The action \eqref{sheet} is usually referred to as the Brink-Di Vecchia-Howe-Deser-Zumino-Polyakov action and
putting $\gamma$ on-shell we recover \eqref{proper_area} (see for instance \cite{polchinski2001string}).
Choosing appropriate boundary conditions, one studies the equations of motion descending from the variational analysis of the action.
The oscillatory modes of such string solutions describe the profile in space-time of the
string itself.
More precisely, any point of the string is mapped into space-time by the so-called 
embedding functions $X^M(\sigma^1,\sigma^2)$. 
Notice that we are actually embedding the world-sheet spanned by the two $\sigma$'s into space-time .
These embedding functions can be regarded as a collection of scalar fields living on the world-sheet whose Fourier modes are promoted
to operators in a world-sheet-Fock space.
The space-time quantum dynamics of the string is encoded in the quantum field theory defined on the world-sheet.
This crucial point is both natural and surprising.
Its naturalness descends from the fact that we are actually generalizing straightforwardly
the approach which is standard for relativistic particles;
its novelty originates from the fact that scattering amplitudes for string processes in space-time
are obtained computing matrix elements of the quantum field theory living on the world-sheet.

\subsection{Supersymmetry and superstrings}

So far we have considered only bosonic string modes. 
A crucial ingredient in developing string theory and defining the D-branes is represented by \emph{supersymmetry}.
It relates bosonic and fermionic degrees of freedom and it can be thought of as an extension
of the standard Poincar\'e invariance\footnote{Supersymmetry
can be expressed as a generalized Poincar\'e invariance on an extended space-time comprehending also fermionic (i.e. Grassmann) directions.}.
In a supersymmetric theory any bosonic mode has a corresponding fermionic partner.
To promote the bosonic string model \eqref{sheet} to a superstring (i.e. supersymmetric string) model one possibility is to introduce
supersymmetry in the world-sheet theory.
At the level of the action we add to \eqref{sheet} the fermion term
\begin{equation}
 S_{\text{ferm}} = -\im \frac{T}{2} \int_\Sigma d^2\sigma \ G_{MN}\ \overline{\psi}^M \Gamma^\alpha \partial_\alpha \psi^N \ ,
\end{equation}
where the $\psi^M$ are a collection of $d$ Majorana spinors where $d$ is the ambient space-time dimensionality.
The matrices $\Gamma^\alpha$ satisfy the bi-dimensional world-sheet Clifford algebra.

Once that the world-sheet model is supersymmetric, in order to obtain a supersymmetric string theory also from the ambient space-time
viewpoint a careful analysis is required. At first, anomaly cancellation implies that a superstring theory can only be consistent 
for $d=10$. In addition, Gliozzi, Sherk and Olive found a way of projecting the superstring spectrum to render it actually supersymmetric\footnote{The GSO projection
can be regarded as a consequence of one-loop modular invariance requirement, see for instance \cite{green1987superstring}.}.
Depending on the relative chirality choice between left and right fermion modes on the closed strings%
\footnote{Strings can be open ore closed. Closed-string vibrations admit both ``clockwise'' and ``counter-clockwise'' 
running wave solutions around the string.}, we have two possible GSO
projections leading to two consistent string theories usually referred to as Type IIA and Type IIB.

\subsection{String scattering amplitudes and vertex operators}

The building blocks of the perturbative analysis are the string scattering amplitudes.
These are classified in accordance with the topology of the world-sheet $\Sigma$;
indeed, the number of ``handles'' of a world-sheet topology generalizes
the number of loops of a Feynman diagram in particle theories.

The asymptotic string states participating in a scattering process,
are encoded in localized operators (the so-called \emph{vertex operators}) defined on the world-sheet.
Any external or asymptotic string state can be associated to a puncture on the world-sheet; the latter
is then a bi-dimensional punctured Riemann surface.
The punctures (or \emph{vertices}), where we insert the vertex operators, correspond to the external legs
of particle diagrams.

Since in a string diagram the punctures are associated to the emission/absorption of a state in the string spectrum,
it is quite natural to expect that, in accordance with the viewpoint of the second-quantized field theory living on the world-sheet,
they are represented by operators.
Without entering into further detail, this is indeed the case\footnote{We recommend to look at \cite{green1988superstring} for a deeper analysis.}.
The vertex operators carry the quantum numbers of the string state they create/annihilate.
Let us underline that the second-quantized treatment of the world-sheet field theory corresponds to a first-quantized
picture of string in space-time.
More precisely, in studying string scattering at the first-quantized level, the fundamental object is the world-sheet, i.e. the string trajectory, 
which is specified \emph{a priori}. In this sense, we consider string fluctuations propagating on a given
world-sheet ``background''. 
In a second-quantized picture for the strings, the world-sheet would be dynamically determined
and the fundamental object would be the string-Fock space.

A scattering amplitude represents a matrix element between asymptotic states which, by definition,
involve a string propagation for infinite time. 
To be neat, think about the $2$-point (i.e. propagator) amplitude for a closed string.
It is obviously described by a cylindrical world-sheet extending from negative to positive infinite time.
Exploiting the invariances of the theory\footnote{Namely the reparametrization and Weyl freedom, see for instance \cite{green1988superstring}.}
it is possible to map this infinite cylinder to a sphere with two punctures. 
The topology of the cylinder and the topology of the sphere with two punctures are the same.
In a similar fashion, it is possible to map any ``$n$-loop'' and ``$N$-particle'' scattering diagram on a sphere or on an $n$-torus with the appropriate number $n$ of handles
and the appropriate number $N$ of punctures.

Proceeding in analogy with the closed string case, the free open string $2$-point function is associated to a rectangular world-sheet
extending from minus to plus infinite time; this presents the same topology of a disk. 
The asymptotic open string states are again represented by vertex operators, in this case they are localized on points belonging to the disk boundary.

Given a certain topology for the world-sheet $\Sigma$, we can compute the associated $N$-point string amplitude
evaluating the vacuum expectation value of the $N$ vertex operators 
$V_{\phi_1}, ... , V_{\phi_N}$ in the framework of the conformal field theory\footnote{The 
tracelessness of the energy-momentum tensor together with the fact that the base manifold (i.e. the world-sheet) is two-dimensional
implies that the Poincar\'e and Weyl invariance of the world-sheet field theory is promoted to full \emph{conformal invariance}, 
look for instance \cite{polchinski2001string}.}
living on the world-sheet,
\begin{equation}
 {\cal A}_N = \int_{\Sigma} \left< V_{\phi_1} ... V_{\phi_N} \right>\ .
\end{equation}
The integral ${\cal A}_N$ could be in general quite complicated.
As it will be useful in what follows, a vertex operator associated to a generic mode $\phi$ can be split 
in its operator part and the polarization part (again indicated with $\phi$), namely
\begin{equation}
 V_{\phi} = \phi\ {\cal V}_{\phi}\ .
\end{equation}

\subsection{Disk and sphere diagrams in the presence of D-branes}
\label{tadpole}

 \begin{figure}
  \centering
  \includegraphics[width=\textwidth]{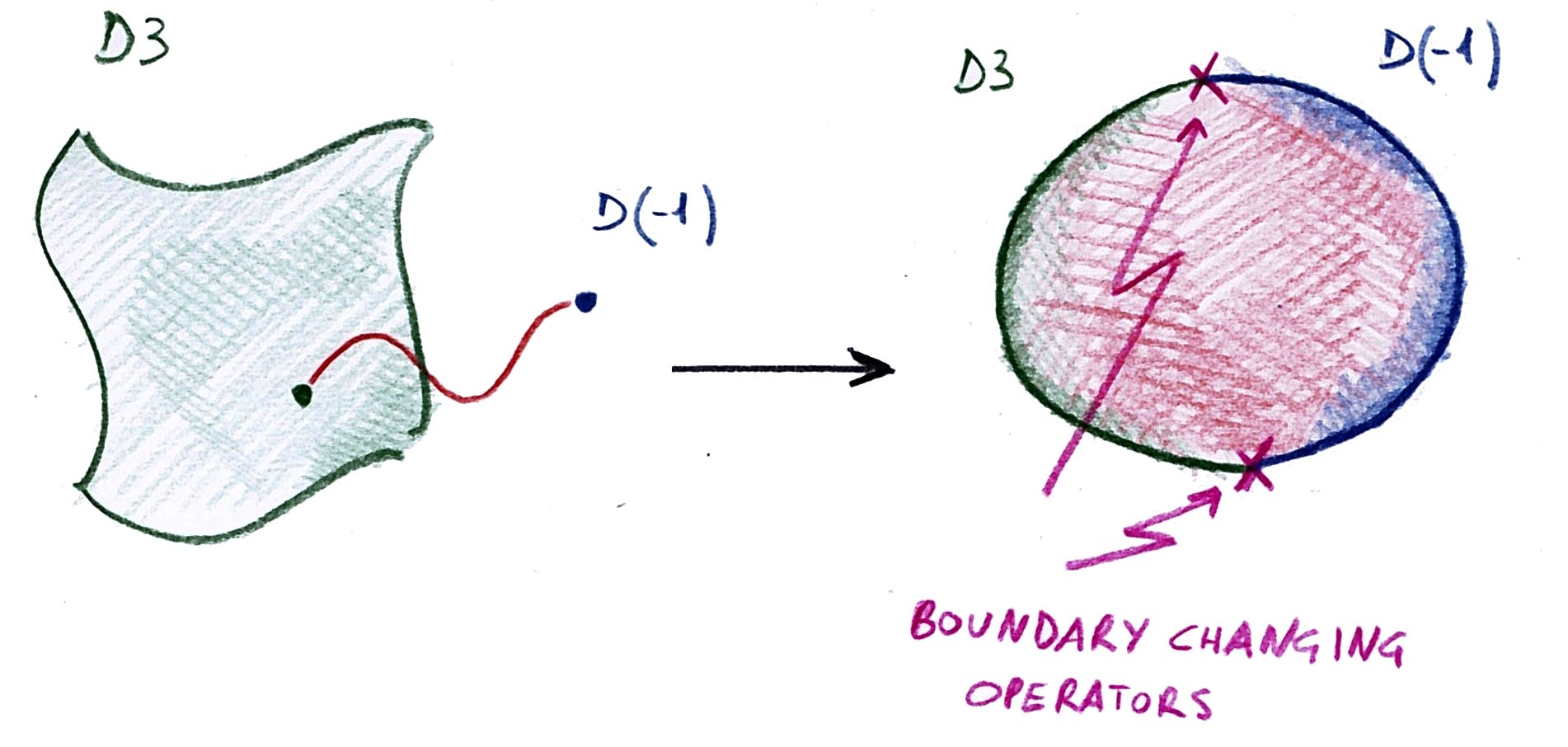}
  \caption{String stretching between different kinds of branes mapped to a ``mixed'' disk with boundary changing operators.}
  \label{mis}
 \end{figure}

The tree level propagation of closed and open strings is encoded by world-sheets having the sphere and disk
topology respectively. 
They in fact correspond to Feynman diagrams with no handles, i.e. no loops.

Computing explicitly the vacuum expectation value of a generic closed string 
vertex operator $\phi$ on the sphere we obtain zero,
\begin{equation}
 \left<{\cal V}_{\phi}\right>_{\text{sphere}} = 0\ ,
\end{equation}
meaning that there is no tadpole amplitude associated to any closed-string mode $\phi$.
Analogously, for the generic open string mode $\psi$, we can compute directly by means of conformal theory methods the
vacuum expectation value of a single vertex operator on a disk.
Again we obtain zero,
\begin{equation}
 \left<{\cal V}_{\psi}\right>_{\text{disk}} = 0\ .
\end{equation}
We interpret these zero results as the absence of tadpoles for both open and closed string modes.
This picture matches the idea of a trivial vacuum in which all the fields have vanishing VEV.
To rephrase the point, we obtained that in a model possessing just open and closed string,
the lowest scattering topologies describe the perturbative physics around the trivial vacuum.

The next step consists in introducing new actors on the stage.
The new objects we consider are the D-branes.
Assuming the viewpoint of the strings, the presence of the D-branes implies the possibility of having
world-sheets with new characteristics. 
In addition to the sphere and the disk amplitudes already considered without the D-branes,
we can now have diagrams with different boundary conditions.
For the sake of simplicity let us start introducing a single D-brane.
Due to the presence of the brane, it is possible to consider the insertion of a boundary into the closed string sphere diagram
(think about a soap bubble on the surface of the sink);
in other terms closed string diagrams with the topology of the disks. 
Such a topology accounts for the possibility of having an emission/absorption of closed strings by the D-brane.
On the boundary just introduced the brane induces boundary identifications between right and left-moving
closed string modes, \cite{Polchinski:1996na}. As a consequence, we can have non-null expectations even for
a diagram with a single close vertex insertion,
\begin{equation}
 \left< V_{\phi}\right>_{\text{D-brane}} \neq 0\ . 
\end{equation}

The open string counterpart consists instead in the possibility of having different boundary conditions
at the two endpoints of the strings. 
In a configuration with two different D-branes (for instance a D$p$ and a D$q$ with $p\neq q$) we can have strings stretching
between them.
The propagation of such a string stretching between two different kinds of D-branes 
is computationally described introducing appropriate operators on the boundary 
of the disk diagrams. 
These operators are called \emph{boundary changing operators}. 

The expectation of a single vertex operator on a disk containing also boundary changing operators
can be non-vanishing,
\begin{equation}
 \left< {\cal V}_{\psi} \right>_{\text{D}p/\text{D}q} \neq 0\ .
\end{equation}
For a pictorial explicit example corresponding to the D$3/$D$(-1)$ case see Figure\ref{mis}.

The string scattering computations are performed in the framework of the world-sheet conformal field theory;
indeed scattering amplitudes are obtained studying expectation values on the world-sheet;
to have more details on the scattering computations in the presence of D-branes see \cite{DiVecchia:1999rh,DiVecchia:1999fx,Billo:2002hm}
and references therein.
A self-contained account of the actual techniques and conformal computations is beyond the purpose of the present treatment;
we refer the reader to the review \cite{Kostelecky:1986xg} for a throughout analysis.

\subsection{Effective supersymmetric gauge theory on the D-brane world-volume}
\label{DBIsub}

We already faced the question of defining an action on the world-sheet which is a two-dimensional surface embedded into ten-dimensional space-time.
In order to specify an action for a D$p$-brane we generalize the same approach to $p+1$-dimensional hyper-surfaces.
These represent the world-volume of the branes and on them we have $10$ bosonic fields each associated to a space-time coordinate;
we can always choose (at least locally) a convenient coordinatization (usually said ``well adapted'') in which the first $p+1$
space-time coordinates parametrize the D-brane world-volume. The remaining space-time coordinates span the so called transverse space.

Intuitively the fields associated to the transverse coordinates represent the oscillations of the brane itself in the surrounding space-time, while
the longitudinal field can be organized in a $p+1$-dimensional array $A_\mu$ with $\mu=0,...,p$. Notice that a $p+1$ brane breaks the entire ten-dimensional
Poincar\'e invariance preserving a Poincar\'e sub-invariance corresponding to its world-volume. 
The array $A_\mu$ behaves indeed as a vector of the preserved Poincar\'e invariance and, being a massless mode, it is straightforwardly
interpreted as a U$(1)$ gauge field.

Having observed the presence of a U$(1)$ gauge vector we can naturally accept that the bosonic part of the D-brane action 
assumes the following form:
\begin{equation}\label{DBI}
 S_{\text{DBI}} \propto \int_{W_{p+1}} d^{p+1}\xi \ e^{-\phi(\bm{X})} \sqrt{-\text{det}\left[g_{\mu\nu}(\bm{X})+ (2\pi\alpha')F_{\mu\nu}(\bm{X})\right]}
\end{equation}
where $i,j=0,...,p$. 
This is the renown Dirac-Born-Infeld action%
\footnote{Historically the Dirac-Born-Infeld action was studied in the attempt of clarifying the problem of
infinite Coulomb energy of point-like particles like electrons. 
In the DBI theory there is an infinite tail of non-linear terms in $F$ generalizing the usual Maxwell electrodynamics.
As a result, the point-like particles presents finite field-strength and finite total energy;
the fields are however not smooth.} (DBI). 
The metric $g_{\mu\nu}$ is induced on the D$p$-brane world-volume by the ambient $G_{MN}$ metric, while $F_{\mu\nu}$
is the field-strength associated to $A_\mu$. The field $\phi$, called \emph{dilaton}, is a scalar mode emerging in the closed string spectrum
and it will be introduced in the Subsection \ref{effebu}.

Let us observe that in the absence of the field-strength, \eqref{DBI} returns simply the Nambu-Goto action \eqref{proper_area} generalized to
the $p$-dimensional brane. 
The introduction of the DBI dependence on $F$ is naturally understandable in the framework of T-duality arguments (see for instance \cite{polchinski2005string}).
In the low-energy theory we can expand the DBI action in powers of the field-strength and retain only
the lowest order, namely the one quadratic in $F$; in this way we obtain the pure Maxwell Lagrangian. 
Similarly, the supersymmetric extension of the DBI action, at quadratic order in the derivatives, returns the Super-Maxwell gauge field theory Lagrangian.
We remind the reader that the expansion of the DBI action contains kinetic and interaction terms that can be computed from the study of string scattering.

 \begin{figure}
  \centering
  \includegraphics[scale=.6]{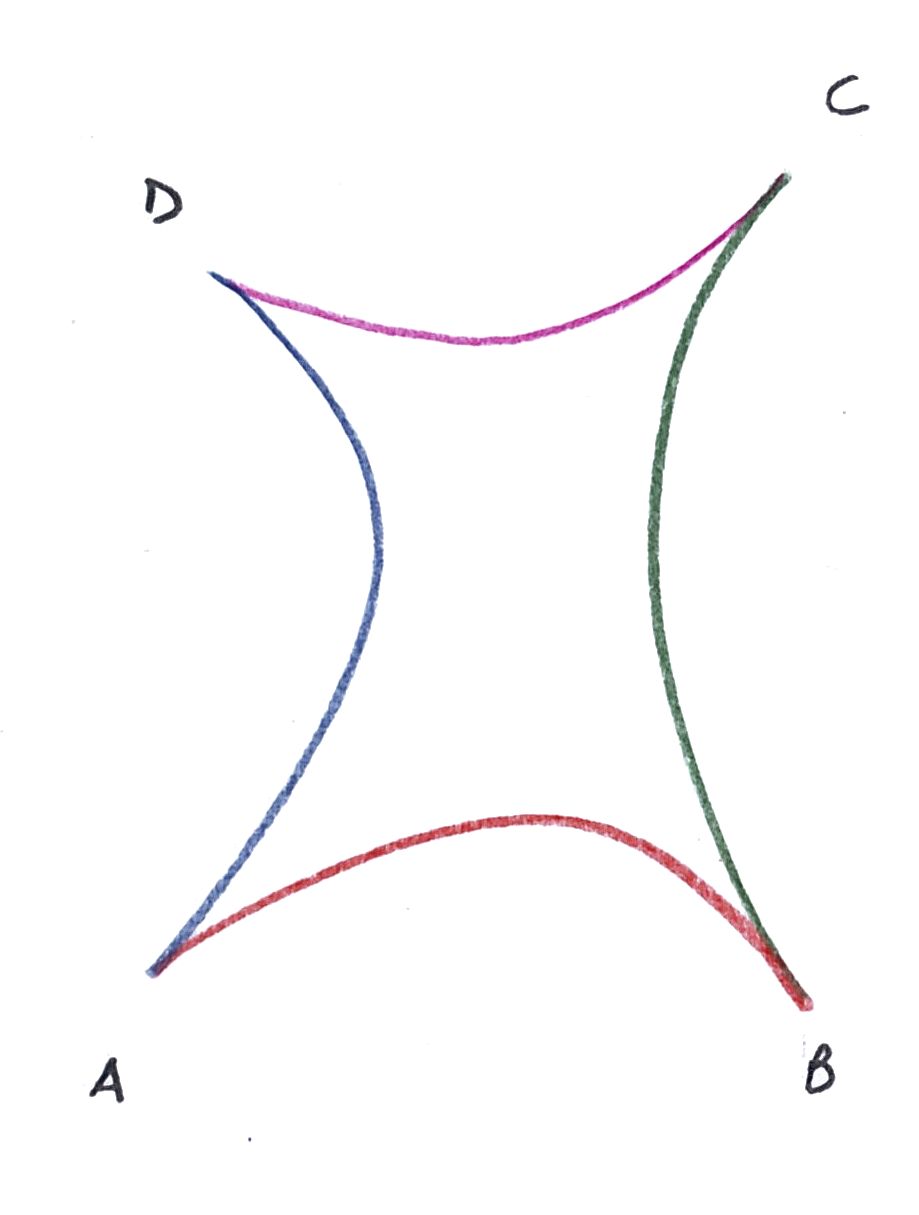}
  \caption{Four-string diagram: along the boundary between two vertices the Chan-Paton label is preserved.}
  \label{ChaPa}
 \end{figure}

Let us concentrate on more than a single brane at a time and, more specifically, let us consider 
a stack of $N$ coinciding (i.e. on top of each others) D$3$-branes all of the same kind\footnote{The
branes in the stack have the same geometrical arrangement and symmetry properties with respect to background 
operators as orbifolds or orientifolds (see Section \ref{OrbiOrie}).}.
In such a setup each open string can start and end on any brane in the stack. 
To account for this possibility, we can associate a label to each D-brane so that any open string state
is characterized by a couple of labels expressing the additional information concerning the branes to which it is attached.
These are called Chan-Paton indexes.

As observed in \cite{Polchinski:1996na}, it is possible to consider additional non-dynamical degrees of freedom
to the endpoints of open strings.
Indeed, such an addition respects the symmetry of the theory, namely the space-time Poincar\'e invariance of the D$3$ world-volume
and the world-sheet conformal invariance.
Given their non-dynamical character, the Chan-Paton indexes can be regarded as mere labels which are preserved in free string evolution.

The Chan-Paton labels running over the values $1,..,N$
have to be introduced in the space of asymptotic string states.
Any state has therefore the following form: 
\begin{equation}\label{CP_fock}
 | \bm{k}; i, j \rangle \ ,
\end{equation}
where $\bm{k}$ represents the center of mass momentum of the string.
Note that, before the introduction of the Chan-Paton labels, the momentum furnished a complete set of quantum numbers
for defining an open string state.

We can use a different representation and consider a basis $\lambda^a_{ij}$
for the space of $N\times N$ Chan-Paton matrices in $i,j$ indexes; $a$ runs over $1,...,N^2$.
The relation with the old basis is:
\begin{equation}
 | k; a \rangle = \sum_{i,j = 1}^N \lambda^a_{i,j}| k; i, j \rangle \ .
\end{equation}

For the sake of clarity, let us consider an explicit example: the $4$ strings scattering amplitude.
The corresponding diagram has four external ``legs'' that are associated
to vertex operators. 
Since we have introduced
the Chan-Paton indexes to label the states, also the vertex operators have to carry the Chan-Paton
structure. They therefore contain as a factor a matrix expressible in the $\lambda^a$ basis.
The Chan-Paton factors are non-dynamical and as a consequence they have to be conserved along the
world-sheet boundary comprehended between two vertex operators.

The Chan-Paton factor is going to describe the gauge structure of the low-energy effective model; let us
consider this point carefully.
In general, our interest is concentrated on gauge-invariant quantities, i.e. objects whose ``gauge indexes'' are saturated.
Let us consider string amplitudes summed over all the possible Chan-Paton configurations.
We indicate with $\lambda^{A,B,C,D}_{ij}$ the Chan-Paton factors corresponding to the four asymptotic states in Figure \ref{ChaPa}.
The amplitude we want to compute results from summing all the adjacent Chan-Paton indexes, leading to an overall factor
\begin{equation}\label{CP_trace}
 \lambda^A_{ij}\lambda^B_{jk}\lambda^C_{kl}\lambda^D_{li} = \text{tr}\left(\lambda^A\lambda^B\lambda^C\lambda^D\right)\ .
\end{equation}
Notice the important fact that the factor \eqref{CP_trace} is manifestly invariant with respect to the following transformation:
\begin{equation}
 \lambda^a \rightarrow M \lambda^a M^{-1}\ ,
\end{equation}
where $M\in\text{GL}(N)$ .
We have to remember that a matrix $M_{ij}$ transforms a quantum state labeled with $i$
into a quantum state labeled with $j$; as a quantum transformation, it is required to be unitary and we have to limit our attention
to U$(N)\subset\text{GL}(N)$,
\begin{equation}
 \lambda^a \rightarrow U \lambda^a U^\dagger\ .
\end{equation}
It is natural to interpret the two indexes of a Chan-Paton matrix $\lambda^a_{ij}$ as belonging respectively
to the fundamental and anti-fundamental representation of U$(N)$.
The lambda's transform then in the $N\times\overline{N}$ representation that is actually the adjoint representation of U$(N)$.
One of the low-energy open string excitation modes is a massless vector that, as just stated, transforms in the adjoint
representation of U$(N)$. 
In the low-energy effective field theory, this mode plays the role of the gauge field.
To realize this, one has to analyze the string scattering and check that at low-energy the 
string amplitudes involving gauge vectors coincide with the amplitudes obtained by the standard effective field Lagrangian $\propto F^2$,
being $F$ the non-Abelian field-strength.

For a stack of $N$ coinciding branes the generalized version of \eqref{DBI} is not know in a closed form.
An expansion for the non-Abelian generalization of DBI action can be in principle obtained 
(with such an effort that usually only the first terms are computable) following specific requirements or prescriptions such as off-shell supersymmetry,
\cite{Collinucci:2002ac}.
The choice is not unique and the literature presents several possibilities which, however, lead all to classically equivalent results (i.e. upon using the equations of
motion). 
Our particular preference for the off-shell supersymmetry requirement has a profound motivation in relation to instanton solutions;
indeed, in this framework, the instantons are believed to represent solutions of the complete quantum theory and not only of its first-terms approximation.

Up to quadratic order, all the non-Abelian versions of the DBI expansion coincide with SYM theory and,
specifically for the case of D$3$-branes, we have four-dimensional ${\cal N}=4$ SYM theory.
The gauge group is U$(N)$ but the U$(1)$ part (associated to the trace) constitutes an infrared-free Abelian sub-sector;
at low-energy scales this Abelian part decouples from the remaining SU$(N)$ part because the former becomes
negligible with respect to the running non-Abelian coupling. Henceforth we will simply understand this caveat,
and indicate the D-brane stack as simply supporting an SU$(N)$ gauge theory.
Let us write the explicit ${\cal N}=4$ SYM action:
\begin{equation}\label{afga}
 \begin{split}
 S_{{\cal N}=4} = \frac{1}{g_{YM}^2} \int d^4x\ \text{tr} &\left\{ \frac{1}{2}F^2_{\mu\nu}
- 2 \overline{\Lambda}_{\dot{\alpha}A} \slashed{\cD}^{\dot{\alpha}\beta} \Lambda_\beta^{\ A}
+ (\cD_\mu \phi_a)^2 - \frac{1}{2}[\phi_a,\phi_b]^2 \right.\\
&\left.-\im (\Sigma^a)^{AB} \overline{\Lambda}_{\dot{\alpha} A}[\phi_a,\overline{\Lambda}^{\dot{\alpha}}_{\ B}]
-\im (\overline{\Sigma}^a)_{AB} \Lambda^{\alpha A} [\phi_a,\Lambda_\alpha^{\ B}]\right\}\ .
 \end{split}
\end{equation}
The index $a$ labels the scalars and, in the stringy picture, is associated to the internal space directions;
the index $A$ is the spinorial counterpart of $a$ so it runs on the internal space spinor components.
The action \eqref{afga} can be recovered from a systematic study of the low-energy dynamics as performed in detail in \cite{Billo:2002hm}.

\subsection{Effective supergravity in the bulk}
\label{effebu}

Studying the low-energy, closed superstring spectrum we find a set of massless modes including a complex scalar $\phi$ called \emph{dilaton},
the already mentioned spin $2$ \emph{graviton}, and some totally anti-symmetric fields $A_{M_1...M_n}$ referred to as \emph{Ramond-Ramond forms}\footnote{We
do not enter into much detail here since the subject is described in any introductory string theory book. Let us mention
that we have a different set of Ramond-Ramond forms depending on the kind of superstring model we are considering, namely Type IIA or Type IIB.
We will consider Type IIB whose spectrum contains all the $A_{M_1...M_n}$ forms where $n$ is even.
We will then have $A_{(4)}$ (i.e. with $4$ indexes) which naturally couples with the D$3$-branes that are the central object for the subjects presented in 
the thesis.}.
Repeating somehow the approach we followed with open strings, we can study the scattering of low-energy closed strings and 
account for their propagation and interactions by means of an effective field theory. 
Such an effective field theory for closed-string, massless modes is called \emph{supergravity}.

In general superstring models and their effective supergravity descriptions admit extended solutions.
Among these we find the D$p$-branes which are $p$-dimensional spatial surfaces.
The world-volume of a $p$-brane has $p+1$ dimensions including time and then it naturally couples to the Ramond-Ramond field with $p+1$ indices.
Indeed, we can regard the branes as generalizing the relativistic particle electro-dynamics; there we have a zero-dimensional object, i.e. the charged particle,
spanning in its evolution a one-dimensional manifold, the world-line. A vector field with one space-time index couples with the particle because
the world-line has a one-dimensional tangent space in any of its points. The electromagnetic coupling of a charged particle is given by
the following term in the action:
\begin{equation}\label{parta}
 q_1 \int_{W_1} d\xi\ \frac{d X^M}{d \xi} A_M = q_1 \int_{W_1} A_M dX^M\ , 
\end{equation}
where $W_1$ is the world-line, $q_1$ is the charge and $d X^M/d \xi$ is the tangent vector to the world-line.
Let us generalize this to a D$p$-brane, for example for $p=3$,
\begin{equation}
 q_4  \int_{W_4} A_{MNOP}\ dX^M\wedge dX^N \wedge dX^O \wedge dX^P \ .
\end{equation}
The integration is performed on the four-dimensional D$3$-brane world-volume.
The D$3$-brane then can emit and absorb $A_{(4)}$ quanta. In a supergravity picture, the presence of such a D$3$-brane
translates to the possibility of having a source for the $A_{(4)}$ generalized gauge field. From now on, the coupling constant $q_4$
will be fixed to $1$. 

Without entering into details, let us give the relevant terms in the supergravity action (in the string frame) that are needed to describe at low-energy
the dynamics of the ambient space-time in the presence of a stack of coinciding D$3$-branes:
\begin{equation}\label{SUGRA3}
 S_{SUGRA} = \frac{1}{(2\pi)^7 (\alpha')^4} \int d^{10} x \sqrt{-g} \left[ e^{-2\phi} \left(R + 4 (\nabla\phi)^2\right) - \frac{1}{2\cdot 5!} F^2_5 \right] + ... \ ,
\end{equation} 
where $g$ is the space-time metric and $R$ the associated Ricci scalar, $\phi$ is the dilaton
and $F_{(5)} = d A_{(4)}$ is the self-dual part of the field-strength associated to the Ramond-Ramond potential with 
respect to which the D$3$-branes are charged. The $...$ indicate
the omission of the other Ramond-Ramond forms and of the $B$ form and also of the fermionic terms%
\footnote{It is appropriate to remind ourselves that supergravity models generalize 
supersymmetric ones making the supersymmetry local. 
In supergravity we then have again a fermionic partner for any boson of the theory.}.

It is possible to show that the equations of motion deriving from \eqref{SUGRA3} admit the following solution:
\begin{eqnarray}\label{sugabrane}
 d^2s     &=& H^{-1/2}(r)\, \sum_{\mu=0}^{3} (dX^\mu)^2 + H^{1/2}(r)\, \sum_{j=4}^{9} (dY^j)^2\\
 A_{MNOP} &=& \epsilon_{MNOP}\ H(r)\\
 e^{\phi} &=& g_s 
\end{eqnarray}
where the coordinates $X^\mu$ with $\mu=0,..,3$ are longitudinal while the $Y^j$ with $j=4,...,9$ are transverse with respect to the stack of D$3$-branes.
The coordinate $r^2=\sum_{j} (Y^j)^2$ is the hyper-spherical radius in the transverse space where the solution \eqref{sugabrane} is spherically
symmetric. Finally, $g_s$ is the string coupling constant
and $H(r)$ is the following harmonic function:
\begin{equation}\label{armo}
 H(r) = 1 + \frac{g_s N \alpha'^2}{r^4}\ ,
\end{equation}
being $N$ the number of branes in the stack. 
This solution describes a stack of $N$ parallel and coinciding $3$-branes\footnote{The case of parallel but non coinciding branes is described by the following
harmonic function:
\begin{equation}\label{Hstack}
 H(\vec{r}) = 1 + \sum_{I=1}^{N} \frac{d_3 g_s l^4}{|\vec{r}-\vec{r}_I|^4}\ \ \ \text{with}\ \ \ \ d_3= 4 \pi\, \Gamma(2)\ ,
\end{equation}
where $\vec{r_I}$ indicates the position of the $I$-th brane in the space spanned by the $Y$ coordinates. Any brane in \eqref{Hstack}
carries one unit of Ramond-Ramond charge.}. Notice that the dilaton solution is a constant, this feature is peculiar of D$3$-brane solutions.

From \eqref{sugabrane} and \eqref{armo} it is possible to see that computing the flux of the R-R field through a hyper-sphere 
containing the branes (i.e. a hyper-sphere in the transverse space) we obtain
\begin{equation}
 \int_{S^5} F_5 = N\ ,
\end{equation}
that is the number of the branes (remember that we have fixed the brane coupling constant $q_4=1$).

The SYM theory living on the D$3$-branes is coupled with the dynamics of the bulk fields living in the ambient space-time.
The coupling is related to the string constant $\alpha'$; once we take $\alpha'\rightarrow 0$, it is possible to consider the following expansion
\begin{equation}
 \frac{1}{g_s} \int d^4 x F^2_{\mu\nu} + \frac{1}{\alpha'^4} \int d^{10} x \sqrt{g} R e^{-2 \phi} + ...
\end{equation}
and effectively regard the two theories on the world-volume and in the bulk as independent; this is usually referred as \emph{decoupling limit}.
As explained for instance in \cite{Zaffaroni:2000vh}, the decoupling limit should be more precisely defined in order to maintain the physical 
interesting
quantities finite (e.g. the ``Higgs mass'' of a string stretching between two separated branes, namely $\Delta Y/\alpha'$). 
Specifically, the decoupling or Maldacena limit is given by
\begin{equation} \label{deco}
  \alpha' \rightarrow 0,\ \ \ \
  g_s \ \ \ \text{fixed},\ \ \ \
  N \ \ \ \ \text{fixed},\ \ \ \
  \phi^j = \frac{Y^j}{\alpha'} \ \ \ \ \ \text{fixed}\ .
\end{equation}

Let us rewrite the metric \eqref{sugabrane} using hyper-spherical coordinates in the space orthogonal to the D-branes,
\begin{equation}
 d^2s = H^{-1/2}(r)\, \sum_{\mu=0}^{3} (dX^\mu)^2 + H^{1/2}(r)\, (dr^2 + r^2 d\Omega_5)\ ,
\end{equation}
where $d\Omega_5$ represents the elemental solid angle; in the decoupling limit \eqref{deco} and using \eqref{armo} we have
\begin{equation}\label{lame}
 ds^2 \sim \sqrt{\frac{r^4}{g_s N\alpha'^2}}\ dX_\mu dX^\mu + \sqrt{\frac{g_s N\alpha'^2}{r^4}} \left(dr^2 + r^2 d\Omega_5 \right)\ .
\end{equation}
We can define $L^2 \doteq \sqrt{g_sN}\alpha'$ and substitute it into \eqref{lame} obtaining:
\begin{equation} \label{Mallim}
 ds^2 \sim \frac{r^2}{L^2} dX_\mu dX^\mu + \frac{L^2}{r^2} \left(dr^2 + r^2 d\Omega_5 \right) 
      \sim \alpha'^2 \frac{\phi^2}{L^2} dX_\mu dX^\mu + \frac{L^2}{\phi^2} \left(d\phi^2 + \phi^2 d\Omega_5 \right)
\end{equation}
where we have used \eqref{deco} and $\phi^2= \sum_j (\phi^j)^2$.
Notice that $L^2$ contains $\alpha'$ and therefore vanishes in the Maldacena limit.
We ought to define $\hat{L}^2 \doteq L^2/\alpha' = \sqrt{g_{YM}^2 N}$. 
Substituting in the metric \eqref{Mallim} resulting from the Maldacena (or decoupling) limit, we obtain
\begin{equation}
 \begin{split}
 ds^2 &\sim\alpha' \bigg\{\frac{\phi^2}{\hat{L}^2} dX_\mu dX^\mu + \frac{\hat{L}^2}{\phi^2} \left(d\phi^2 + \phi^2 d\Omega_5 \right)\bigg\}\ ;
 \end{split}
\end{equation}
this metric is manifestly the product of two Einstein spaces of constant curvature: $AdS_5 \times S^5$.
Notice that both spaces are characterized by the same curvature radius, namely $\hat{L}$ (in string units).
In addition, we should note that the decoupling limit works for any value of $g_s$ and $N$.

\section{Holography and the \texorpdfstring{$AdS$}{}/CFT correspondence}
The $AdS$/CFT correspondence is a conjectured duality between a specific gravity theory defined on an Anti-de Sitter space ($AdS$) 
and a corresponding
conformal field theory (CFT) that can be thought of as living on the ``conformal boundary'' (see Section \ref{colf}) of the $AdS$ space-time.

The word \emph{holography} comes from the combination of the two ancient Greek terms: \emph{holos} meaning ``whole''
and \emph{graf\'e} meaning ``writing'' or ``painting''; it is usually referred to optical techniques
which are able to reconstruct the \emph{whole} three-dimensional information of an image by means of a
two-dimensional support. 
The term has been adopted in the gauge/gravity correspondences framework because the gauge and gravity theories related by the correspondence 
are defined on space-times with different dimensionality. 
In its stronger sense, a duality is a map between two theories describing the same physics and therefore
``containing'' the same information; the content of the theory is just written in different terms, i.e. using different degrees of freedom.
In this sense, the gauge/gravity dualities are \emph{holographic} because they show that the content of the theory living in 
a higher dimensional space-time is encoded \emph{holographically} in the lower dimensional theory.

The first holographic hint in theoretical physics was suggested by black hole thermodynamics%
\footnote{The seminal papers in which the holographic principle has been proposed are \cite{'tHooft:1993gx} and
\cite{Susskind:1994vu}.}.
The entropy of a black hole, which is related to the number of quantum states contained in the black hole volume,
is proportional to the surface of the black hole horizon \cite{Hawking:1971tu,Bekenstein:1972tm}.
A property of the bulk volume of a black hole is indeed related to the surface or boundary containing the same volume.
Let us note that the holographic hint given by black holes comes from a context in which
gravity needs to be treated at the quantum level.

At the outset of describing holography and AdS/CFT it should be understood that the topic is very wide and
we must often refer to the numerous reviews present in the literature. 
In particular, for an introductory treatment of some fundamental ingredients such as conformal field theory and $AdS$ gravity we refer especially
to \cite{Aharony:1999ti,Zaffaroni:2000vh} and references therein.

\subsection{String/field connections}
\label{stringVSfield}

As we have mentioned, the dynamics of D-brane models can be described in the low-energy regime with appropriate gauge field theories. 
In the low-energy and infinite tension (or zero length limit) for the strings, the open strings themselves become
effectively point-like objects accountable for within a quantum field theory living on the world-volume of the branes.
However, this direct connection is not the only link between field and string theory.
Indeed, in hindsight, we can observe that string theory was originally developed in the context of strongly coupled
hadronic interactions, i.e. a context which should be also describable with a strongly coupled quantum field theory in the 
confining regime\footnote{Notice that sometimes this early string models are called \emph{dual models}.
Here the term ``dual'' refers to a property of hadronic scattering consisting in the equality
of hadronic scatterings in the $s$ and $t$-channels for small values of $s$ and $t$
($s$ and $t$ are Mandelstam variables), \cite{green1988superstring}.
Note that the $s\leftrightarrow t$ world-sheet crossing symmetry can be seen as a first hint of open/closed string duality.}.
Actually there are many string-like objects involved in this context like \emph{flux tubes} and \emph{Wilson lines}.
Flux tubes give an effective description of the interaction between two quarks and their behavior 
resembles the dynamics of strings.
Think for instance of the bag-like potential for quarks that is related to the area spanned by the flux tube 
in space-time evolution of the quark pair, \cite{Creutz:1984mg};
this is analogous to Nambu-Goto action for a string propagation where the action actually measures the proper area of the world-sheet.

Even though, in the context of strong interactions, string models have been superseded by QCD,
we are generally not able to employ analytical tools for the analysis of its strongly coupled and confining regime.
Before the employment of $AdS$/CFT inspired techniques, the main theoretical instrument to investigate the strongly coupled regime
of QCD and more in general non-Abelian gauge theories was provided by numerical simulations on the lattice
and effective models (such as NJL, $\sigma-$models,...).

The intimate relation between field and string theory has been significantly boosted in the 90's after
the proposal of the $AdS$/CFT Maldacena's conjecture \cite{Maldacena:1997re}.
In its strongest version, the conjecture claims a complete equivalence between Type IIB string theory compactified on 
an asymptotically $AdS_5\times S^5$ background and ${\cal N}=4$ SYM (Super Yang Mills) theory living in four-dimensional
flat space-time.
With $AdS_5\times S^5$ we indicate the manifold obtained from the Cartesian product of five-dimensional Anti-de Sitter
space-time and the $5$-dimensional hypersphere.
Notice that the duality relates a theory containing quantum gravity (among other interactions) to a gauge field theory without gravity.
$AdS$/CFT has stimulated much interest on the possibility of having gauge/gravity dualities and much effort has been tributed to
this field in the last fifteen years.

One crucial point to be highlighted at the outset is that $AdS$/CFT is a strong/weak duality, meaning that it relates
the strongly coupled regime of one theory with the weakly regime of the other and vice versa.
This feature, which renders extremely ambitious the task of finding a direct proof of the conjecture\footnote{To find
a succinct account of $AdS$/CFT tests look at \cite{Aharony:1999ti}.}, is its most interesting
practical characteristics. 
In fact, because of its strong/weak character the $AdS$/CFT provides a powerful tool to obtain analytical results at strong coupling
in field theory by means of string low-energy and perturbative calculations.
Such a possibility is particularly interesting because the theoretical methods to perform analytical computation
at strong coupling in field theory are generically quite poor and, a part from numerical simulations on the lattice,
the strong coupling regime has been quite often theoretically unaccessible.

\subsection{'t Hooft's large \texorpdfstring{$N$}{} limit}

A very suggestive relation between non-Abelian gauge theory and string theory follows from an observation proposed by 't Hooft in 1974.
He noticed how a U$(N)$ Yang-Mills theory in the large $N$, i.e. large number of colors, admits
a classification of Feynman's diagrams according to their topological properties,
\cite{'tHooft:1973jz}.

Let us look to the large $N$ limit of U$(N)$ Yang-Mills theory in more detail.
Apparently the $N\rightarrow\infty$ limit seems to lead to ill-defined quantities;
actually, instead of performing just the large $N$ limit, we have to act on the coupling constant
$g_{YM}$ as well.
Consider for instance a self-energy diagram for the gauge field that belongs to the adjoint
representation of the gauge group;
the gauge field $A$ is then a Hermitian matrix expressible on a basis of $N^2$ independent gluons.
It is possible to show, \cite{Zaffaroni:2000vh}, that the self-energy for a gluon scales as $N$
and, in the large $N$, limit it is then of order ${\cal O}(N)$.
Nevertheless, if we include in the analysis the coupling $g_{YM}$ we notice that actually the self-energy behavior is
${\cal O}(g_{YM}^2N)$. It is then natural to consider the so called 't Hooft limit, namely
\begin{equation}\label{dou_sca}
 N\rightarrow\infty\ , \ \ \ \ g_{YM}\rightarrow 0 \ , \ \ \ \ \text{with}\ \ \lambda=g_{YM}^2N\ \ \text{fixed}\ .
\end{equation}
In this limit all the diagrams either remain finite or vanish.

In pure Yang-Mills theory the only dimensional parameter is the QCD scale $\Lambda_{QCD}$.
Let us observe that in 't Hooft's limit the QCD scale remains constant, indeed the $\beta$-function
equation (which defines\footnote{We can define $\Lambda_{QCD}$ as the scale at which $g_{YM}$ runs to
an infinite value. However, we ought to note that the $\beta$-function equation (or renormalization equation) 
\eqref{ren_eq} is derived in perturbation theory and then it is reliable only for small values of the coupling.
When $g_{YM}$ becomes large the perturbation scheme ceases to be justified.
Keeping clear mind about this caveat, we can nevertheless retain the formal definition of the QCD scale
$\Lambda_{QCD}$, \cite{Nason:1997zu}.}$\Lambda_{QCD}$) for pure SU$(N)$ YM theory is given by:
\begin{equation}\label{ren_eq}
 \mu \frac{d g_{YM}}{d\mu} = -\frac{11}{3} N \frac{g^3_{YM}}{16\pi^2} + {\cal O}(g^5_{YM})\ ,
\end{equation}
where $\mu$ is a reference renormalization mass scale.
The two sides of Eq.\eqref{ren_eq} scale in the same way in the 't Hooft limit \eqref{dou_sca}.

Following \cite{Zaffaroni:2000vh}, in a non-Abelian Yang-Mills theory it is possible to express
the adjoint field with a double-line notation essentially treating the adjoint representation
in line with its bi-fundamental nature. 
Adopting such double-line notation, it is possible to show that any Feynman's
diagram can be accommodated on a Riemann surface whose genus $g$ 
(i.e. the number of holes) is
related to the features of the diagram itself, namely
\begin{equation}
 2 - 2g = F - E + V \ ,
\end{equation}
where $F$ corresponds to the number of loops (faces), $V$ is the number of vertices and $E$
is the number of propagators (edges).

Let us focus on the pure U$(N)$ gauge theory\footnote{We are concentrating just on the adjoint fields;
the diagrams containing also fundamental fields result suppressed with respect to the leading contribution. 
An observation which is nevertheless interesting is that,
in the presence of fundamental fields, the topological classifications of the diagrams has to involve also
varieties with boundaries, \cite{Zaffaroni:2000vh}.}.
Its Lagrangian density is
\begin{equation}
 {\cal L} = \frac{1}{g_{YM}^2} \text{tr} F^2 = \frac{N}{\lambda} \text{tr} F^2 \ ,
\end{equation}
where we have put in evidence the Yang-Mills coupling constant.
Any propagator is accompanied with a factor of $\lambda/N$ and any interaction vertex is instead 
accompanied with a factor $N/\lambda$.
Furthermore, any loop contributes is accompanied by a factor of $N$;
the reason is that, since we are adopting the double-line notation,
we are considering loops associated to the fundamental representation which has indeed $N$ components.
Collecting these observations in one formula, we have that the generic diagram scales as
\begin{equation}
 \lambda^{E-V}\ N^{F-E+V} = {\cal O}(N^{2-2g})\ .
\end{equation}
Any amplitude in the field theory can be expanded accordingly to the topology of the contributing Feynman's
diagrams,
\begin{equation}
 {\cal A} = \sum_{g=0}^{\infty} N^{2-2g} f_g(\lambda)\ ,
\end{equation}
where the functions $f_g$ are polynomials in $\lambda$.
In the 't Hooft limit, the expansion is clearly dominated by the low-genus configurations, 
and in particular by the planar $g=0$ graphs possessing the topology of a sphere.
This expansion resembles precisely the topological expansion of perturbative multi-loop string diagrams.

Further detail on the large $N$ limit can be found in \cite{Aharony:1999ti,Zaffaroni:2000vh}

\part{Stringy Instanton Calculus}
\label{part1}

\chapter{Instanton Preliminaries}
In this part of the thesis we devote keen attention to non-perturbative effects and particularly
to \emph{instantons} both in SUSY gauge field theories and in superstring theories.

The term ``non-perturbative'' refers to configurations whose action $S$ is proportional to a negative
power of the coupling constant.
In the partition function, any configuration is weighted by $e^{-S}$, therefore the non-perturbative 
effects are exponentially suppressed at small coupling.
In this regime they are generally negligible with respect to any perturbative contribution
whose weight vanishes instead as a positive power of the coupling constant.
The non-perturbative physics is relevant either when we consider a strongly coupled regime
or whenever the competing perturbative effects are absent.
In relation to the latter case, some perturbative contributions can be forbidden, for instance, 
by non-renormalization theorems induced by the supersymmetry of the theory.

In the $1970$s the theoretical physics community started to study systematically the
non-trivial solutions of the classical field equations of motion of many field theories comprehending
Yang-Mills theory and its supersymmetric generalizations%
\footnote{As an aside curiosity, it is interesting to recall that far before the systematic study of non-trivial
field solutions (or \emph{solitons}) and even before the modern atomic theory was proposed, 
Kelvin suggested a model based on vortexes (that are a particular kind of solitons) in a fluid to represent atoms.
In Kelvin's picture, the chemical variety of atoms was explained in terms
of different topological arrangements (i.e. different topological charges);
the stability of atoms corresponded to the topological stability of solitons with respect to small fluctuations.
As we will see, topological features are an essential property of solitons.}.
In the quantum field theory framework, a classical solution of the equations of motion represents a
background around which the quantum fluctuations are studied.
Notably, the non-trivial solutions usually mingle global and localized features.
On the one side they are related to topological characteristics corresponding to global properties
of the classical field configuration as a whole, on the other their energy density is non-vanishing
on a finite support%
\footnote{It is intuitive to expect that a field configuration whose potential (i.e. static) energy density is non-vanishing
everywhere has a diverging action. 
In a path-integral (or partition function) formulation, a diverging action is translated into the total suppression of any amplitude
involving such configuration.}.
Because of their localized character, the non-trivial solutions of the equations of motion 
are usually referred to as particle-like configurations or \emph{pseudo-particles}%
\footnote{For an ample panoramic view on the topic of solitons
and their particle-like behavior (e.g. in scattering phenomena), consult \cite{manton2007topological}. }.
They are nevertheless distinguished from the fundamental particle excitations arising from the perturbative quantization of the fields.
In fact, as opposed to solitons, the perturbative quantum fluctuations around classical configurations emerge
from the quantization of continuous deformations of
the background field profile. As such they cannot, by definition, change the topology of the background itself.
Indeed, the topological sectors in the field configuration space are closed (i.e. not connected with each other)
with respect to continuous deformations of the fields.

Instantons constitute a prototypical example of totally localized non-perturbative field configurations
of Yang-Mills theory; the name is formed by the prefix ``instant-'' suggesting
localization also in the time direction, and the suffix ``-on'', usually attributed to particles.
The first analysis of instantons dates back to $1975$ and was performed by Belavin, Polyakov, Schwarz and Tyupkin 
in \cite{Belavin:1975fg}.
Notice that the localized nature also in the time direction makes it impossible to think of instantons as
stable propagating particles.

\section{Topological charge}
Instantons are non-trivial classical solutions of the equations of motion of pure Yang-Mills theory
defined on four-dimensional Euclidean space-time. 
They have finite action and enjoy the property of self-duality, i.e.
\begin{equation}
 F= {}^*F\ ,
\end{equation}
where $F$ is the standard non-Abelian field-strength
\begin{equation}
 F_{\mu\nu} = \partial_\mu A_\nu - \partial_\nu A_\mu + [A_\mu,A_\nu]\ ,
\end{equation}
and ${}^*F$ is its Hodge dual,
\begin{equation}
 {\,}^*F_{\mu\nu} = \frac{1}{2} \epsilon_{\mu\nu\rho\sigma} F_{\rho\sigma}\ .
\end{equation}
In Yang-Mills theory the Hodge duality constitutes the non-Abelian generalization of the electro-magnetic duality.
We define the non-Abelian ``electric'' and ``magnetic'' fields as follows:
\begin{eqnarray}\label{physfield}
   & E^{ia} &= F^a_{0i} \\
   & B^{ia} &= -\frac{1}{2} \epsilon^{ijk} F^a_{jk}
\end{eqnarray}
where $i,j,k$ are spatial indexes and $a$ represents the index associated to the gauge group generators $T^a$
which are traceless anti-Hermitian matrices satisfying
\begin{eqnarray}
 & [T^a, T^b] &= f^{abc} T^c\\
 & \text{tr}\left(T^aT^b\right) &= -\frac{1}{2} \delta^{ab}
 \label{tracia}
\end{eqnarray}
where $f^{abc}$ are real structure constants.
Remember that we are adopting an Euclidean metric; the upper or lower position of space-time indexes is unimportant
and the Hodge duality squares to $1$. Euclidean electro-magnetic duality transforms
\begin{eqnarray}
 &\bm{E} &\rightarrow \bm{B} \\
 &\bm{B} &\rightarrow \bm{E}\ .
 \label{econd}
\end{eqnarray}
This is in contrast with Minkowskian electro-magnetic duality which introduces a minus in \eqref{econd}.
Indeed, in Minkowski space-time the Hodge dual squares to $-1$ and it is impossible to define (non-trivial)
self-dual or anti-self-dual configurations.

Instantons correspond to Euclidean classical configurations locally minimizing the action 
and are therefore stable against field fluctuations.
Instantons possess a finite but non-vanishing value for the action and are characterized by an integer number $k\in\mathbb{Z}$
called \emph{topological charge} or \emph{Pontryagin number} or also \emph{winding number}.
It corresponds to the integral
\begin{equation}\label{top_charge}
 k = - \frac{1}{16\, \pi^2} \int d^4 x\ \text{tr} \left( F_{\mu\nu} \,^*F_{\mu\nu}\right)\ ,
\end{equation}
and its meaning will be clarified shortly.
The Euclidean action of Yang-Mills gauge theory is
\begin{equation}\label{actiongau}
 S = -\frac{1}{2g^2_{YM}} \int d^4x\ \text{tr} F_{\mu\nu} F_{\mu\nu}\ ;
\end{equation}
this, in the absence of sources, leads to the Yang-Mills equations of motion,
\begin{equation}
 D_\mu F_{\mu\nu} = 0 \ .
\end{equation}
A field configuration that, like an instanton, has finite action must then correspond to a field-strength $F$ 
tending to zero faster than $1/r^2$ for large values of the four-dimensional Euclidean space-time radius 
\begin{equation}
 r = \sum_{\mu=0}^3 (x^i)^2 \ .
\end{equation}
Explicitly, a finite value for the action requires
\begin{equation}\label{vani}
 F_{\mu\nu} \overset{r\rightarrow\infty}{\longrightarrow}\ {\cal O}(1/r^{2+\epsilon})\ ,
\end{equation}
with $\epsilon>0$. 
In order to present such an asymptotically vanishing field-strength $F_{\mu\nu}$,
the gauge field $A_\mu$ has to tend for large $r$ to a \emph{pure gauge} plus terms vanishing faster than
$1/r$. A pure gauge configuration is a field configuration that can be obtained applying a gauge transformation to the trivial vacuum.
Remember in fact that the field-strength is a gauge invariant quantity and its value on the trivial vacuum is zero.
Mathematically, we then have:
\begin{equation}\label{pure}
 A_\mu \overset{r\rightarrow\infty}{\longrightarrow}\ U^{-1} \partial_\mu U + {\cal O}(1/r^{1+\epsilon'})\ ,
\end{equation}
being $\epsilon'$ a positive quantity
and $U(x)$ the matrix field representing a gauge transformation.

The integrand $F_{\mu\nu}{\,}^*F_{\mu\nu}$ in the definition of the topological charge \eqref{top_charge}
can be expressed as a total derivative,
\begin{equation}\label{deri_tot}
 \begin{split}
  \text{tr}\ F_{\mu\nu}{\,}^*F_{\mu\nu} 
&= 2\, \epsilon_{\mu\nu\rho\sigma}\, \text{tr} \left( \partial_\mu A_\nu \partial_\rho A_\sigma 
                                      + 2 \partial_\mu A_\nu A_\rho A_\sigma
                                      + A_\mu A_\nu A_\rho A_\sigma \right)\\
&= 2\, \epsilon_{\mu\nu\rho\sigma}\, \text{tr}\ \partial_\mu \left( A_\nu \partial_\rho A_\sigma
                                      + \frac{2}{3} A_\nu A_\rho A_\sigma \right)\\
&= \epsilon_{\mu\nu\rho\sigma}\, \text{tr}\ \partial_\mu \left( A_\nu F_{\rho\sigma}
                                      - \frac{2}{3} A_\nu A_\rho A_\sigma \right)\ ,
 \end{split}
\end{equation}
where we have used the cyclic property of the trace to discard the $AAAA$ term%
\footnote{Note that the tensor $\epsilon_{\mu\nu\rho\sigma}$ acquires a minus upon a cyclic permutation of its indexes.} 
and the symmetry of $\partial_\mu \partial_\rho$.
We apply Stoke's theorem and compute $k$ via an integral on the ``boundary'' at infinite radius%
\footnote{Assuming that any field tends to a constant for $r\rightarrow\infty$, with \emph{boundary}
we mean a spherical shell with asymptotic radius.}.
From the asymptotical behavior of the field-strength \eqref{vani}, we have that the term in \eqref{deri_tot} containing $F_{\rho\sigma}$ is neglectable.
Therefore the topological charge is given by
\begin{equation}\label{boun_surf}
 k = \frac{1}{24\pi^2} \int_{\overset{S_3}{r\rightarrow \infty}} d\Sigma_\mu \epsilon_{\mu\nu\rho\sigma}\ \text{tr}
\left[ \left(U^{-1}\partial_\nu U\right) \left(U^{-1}\partial_\rho U\right) \left(U^{-1}\partial_\sigma U\right) \right]
\end{equation}
where $d\Sigma_\mu$ is the radial hyper-surface element of the asymptotic three-sphere.
The function $U(x)$ considered on the asymptotic $S_3$ defines a map from the boundary itself
to the gauge group manifold.
It is possible to show\footnote{Look at the appendices of \cite{Belitsky:2000ws} to have an explicit example in the case
of SU$(2)$ gauge group.} that the integer $k$ counts how many times the asymptotic $S_3$ ``winds'' around
the gauge group manifold according to the map $U$. 
This justifies the name ``winding number'' for $k$.

Given the topological nature of the winding number $k$, it cannot be affected by continuous deformations of the gauge field
configuration.
The possibility of obtaining a field configuration $\tilde{A}$ by continuously deforming a configuration $A$
can be regarded as an equivalence relation between $\tilde{A}$ and $A$.
In this framework, the gauge field configuration space splits into distinct equivalence classes (usually called \emph{topological sectors})
associated to different values of the topological charge $k$. 
Within a generic topological sector the action of any element of the sector 
has a value satisfying the inequality
\begin{equation}\label{BPSbound}
 S \geq \frac{8\pi^2}{g_{YM}^2} |k|\ .
\end{equation}
This is called BPS bound from the names of Bogomol'nyi, Prasad and Sommerfield
who studied it for the first time.
The BPS inequality \eqref{BPSbound} can be proven rewriting the action \eqref{actiongau} as follows:
\begin{equation}\label{BPS}
 \begin{split}
S &= -\frac{1}{2g_{YM}} \int d^4x\, \text{tr}\, F^2\\
  &= -\frac{1}{4g_{YM}} \int d^4x\, \text{tr}\, \left(F\pm {\,}^*F\right)^2 
\pm \frac{1}{2g^2_{YM}} \int d^4x\, \text{tr}\, F{\,}^*F\\
  &\geq \pm \frac{1}{2g^2_{YM}} \int d^4x\, \text{tr}\, F{\,}^*F\\
  &= \mp \frac{8\pi^2}{g^2_{YM}} k\ .
 \end{split}
\end{equation}
Recalling \eqref{tracia}, in the third passage of \eqref{BPS} we have discarded a positive quantity.
Notice that in \eqref{BPSbound} the equality holds if and only if the field configuration corresponds to a field-strength
that is either self-dual or anti-self-dual, namely
\begin{equation}\label{dualadual}
 {\,}^*F = \pm F\ .
\end{equation}
Moreover, the BPS argument implies that the configurations satisfying the self-duality condition 
\eqref{dualadual} minimize the action within the topological sector to which they belong.
Conventionally we refer to the self-dual configurations as instantons and to the 
anti-self-dual configurations as anti-instantons%
\footnote{In the Mathematical community the opposite definition is usually considered.}. 
From the definition of the topological charge \eqref{top_charge} and \eqref{tracia},
we have that instantons and anti-instantons have positive and negative $k$ respectively.

\section{Vacua and tunneling amplitudes}
\label{vacua}

Instantons can be interpreted as tunneling processes interpolating
between different vacua of the Minkowskian formulation of the YM model, \cite{weinberg1996quantum}.
In order to illustrate this crucial point and before moving from the Euclidean to the Minkowskian formulation,
it is necessary to consider the temporal gauge, namely
\begin{equation}\label{temp_gau}
 A_0^a(t,\bm{x}) = 0\ .
\end{equation}
In the temporal gauge (sometimes also referred to as Weyl gauge)
it is possible to canonically quantize the theory.
We have the following (Euclidean) Lagrangian and Hamiltonian densities:
\begin{eqnarray}\label{eu}
 &{\cal L}^{(Eu)} &= \frac{1}{2} \left(\bm{E}^a\cdot \bm{E}^a + \bm{B}^a\cdot \bm{B}^a \right)\\
 &{\cal H}^{(Eu)} &= \frac{1}{2} \left(\bm{E}^a\cdot \bm{E}^a - \bm{B}^a\cdot \bm{B}^a \right)\ .
\end{eqnarray}
Note that the relative signs are opposite to the usual Minkowskian expectations, indeed
the terms in $\bm{E}$ represent the kinetic part of the densities ($\bm{E}^a=-\partial_t\bm{A}^a$)
and the terms in $\bm{B}$ constitute the potential energy density.
\footnote{The Euclidean formulation of YM theory can be regarded as the Wick-rotated (i.e. imaginary time)
version of the Minkowskian version.}

A general feature of field theories defined on a non-compact base manifold
is that the topological considerations are strictly related to the asymptotic (i.e. at large radius)
behavior of the fields.
To rephrase \eqref{deri_tot} and \eqref{boun_surf}, the topological charge is given by the flux integral of the Chern current
\begin{equation}\label{tempchern}
 J_\mu =  \epsilon_{\mu\nu\rho\sigma}\, \text{tr}\ \left( A_\nu F_{\rho\sigma}
                                      - \frac{2}{3} A_\nu A_\rho A_\sigma \right)
       \underset{r\rightarrow \infty}{\sim} - \frac{2}{3} \epsilon_{\mu\nu\rho\sigma}\, \text{tr}\ A_\nu A_\rho A_\sigma
\end{equation}
through an asymptotic hyper-surface.
From the temporal gauge condition \eqref{temp_gau} we have that $J_\mu$ is directed in the $0$ direction for large $r$.
Since we deal with a local theory, we add the general assumption that the fields vanish at spatial infinity, namely
\begin{equation}\label{spatialdrop}
 A_i(\bm{x},t) \overset{\rho\rightarrow \infty}{\longrightarrow} 0\ \ \text{with} \ \ \rho=\sum_{i=1}^3 (x^i)^2\ .
\end{equation}
Equations \eqref{tempchern} and \eqref{spatialdrop} combined together
mean that the total flux $\Phi[J]$ of $J_\mu$ at ``infinity'' receives contributions only from the asymptotic regions 
corresponding to $t=\pm\infty$, that is
\begin{equation}
 \Phi[J] = \phi[J]_{+\infty} - \phi[J]_{-\infty}\ .
\end{equation}
More precisely, $\phi[J]_{+\infty}$ and $\phi[J]_{-\infty}$ represent the fluxes of $J$ ``through'' 
the spatial three-dimensional manifolds corresponding to positive and negative temporal infinity respectively.
We can repeat the argument connecting \eqref{deri_tot} to \eqref{boun_surf} for the three-dimensional configurations
at asymptotic time; namely, we can associate to the two $t=\pm\infty$ configurations a (spatial) topological charge:
\begin{eqnarray}\label{topspa}
 k_+ &=& \frac{1}{24\pi^2} \int_{t=+\infty} d\Sigma_0^{(+)} \epsilon_{0\nu\rho\sigma}\ \text{tr}
\left[ \left(U^{-1}\partial_\nu U\right) \left(U^{-1}\partial_\rho U\right) \left(U^{-1}\partial_\sigma U\right) \right]\\
 k_- &=& \frac{1}{24\pi^2} \int_{t=-\infty} d\Sigma_0^{(-)} \epsilon_{0\nu\rho\sigma}\ \text{tr}
\left[ \left(U^{-1}\partial_\nu U\right) \left(U^{-1}\partial_\rho U\right) \left(U^{-1}\partial_\sigma U\right) \right]\ ,
 \label{topspameno}
\end{eqnarray}
where $d\Sigma_0^{(\pm)}$ represent the three-dimensional hyper-surface element (i.e. the volume element) oriented along the time direction.
Note that the four-dimensional overall topological charge $k$ is given by
\begin{equation}
 k = k_+ - k_- \ .
\end{equation}

The formal similarity between \eqref{boun_surf} and \eqref{topspa}, \eqref{topspameno} is evident, however a doubt could arise.
Indeed, while \eqref{boun_surf} is defined on the asymptotic $S_3$ of Euclidean $\mathbb{R}^4$ space,
\eqref{topspa} and \eqref{topspameno} are defined on the $\mathbb{R}^3$ spatial manifold. 
The former is a compact space while the latter is not. 
The homotopy argument that led us to interpret $k$ as the winding number seems to be impossible
for $k_\pm$ because it apparently lacks one of the essential ingredients: the compactness of the manifold on which the integral is considered.
A subtle observation comes to our help.
Note that we assumed in \eqref{spatialdrop} that the gauge potential vanishes at spatial infinity.
For configurations related to the trivial vacuum by a gauge transformation $U$,
\begin{equation}\label{gaueq}
 A_i = U^{-1} \partial_i U\ , \ \ A_0 = U^{-1} \partial_0 U = 0 \ ,
\end{equation}
we have that at spatial infinity $U$ tends to a constant value that can be fixed to be the identity%
\footnote{We can discard rigid gauge rotations from our analysis without spoiling its generality.},
\begin{equation}
 U(\bm{x},t) \overset{\rho\rightarrow \infty}{\longrightarrow} \one
\end{equation}
The gauge fields and transformations assume a fixed value in the limit $\rho\rightarrow\infty$ independently
of the particular direction along which we move towards spatial infinity.
In this sense we can add ``the point at infinity'' assigning
\begin{equation}
 A_\mu^{(\infty)} = 0, \ \ U^{(\infty)} = \one \ ,
\end{equation}
``completing'' the spatial manifold $\mathbb{R}^3$ to a compact $S_3$.
In this sense, the integrals \eqref{topspa} expressing $k_\pm$ can be regarded as properly defined spatial winding numbers.

As a consequence of the preceding arguments, it is natural to interpret the exponentiated action of the instanton as the transition amplitude between
the two configurations at temporal infinity. 
This transition connects field configurations presenting different spatial winding numbers $k_\pm$.
Let us remark the fact that the instanton amplitude is given by its classical action
\begin{equation}
 A(k_-\rightarrow k_+) = e^{-\frac{8\pi^2}{g^2_{YM}}|k|}\ ,
\end{equation}
where the coupling constant appears at the denominator of the exponent.
This non-perturbative feature reminds us the semi-classical WKB tunneling amplitudes. 
Indeed, we are interpreting the instanton as the transition amplitude through the barrier
dividing distinct topological sectors.
The semi-classical character of the present analysis arises from the fact that we are considering just the instanton amplitude
with lowest action, i.e. only the minimal classical path in a quantum path integral.

If we consider the gauge configurations \eqref{gaueq} which are obtained by applying a constant-time gauge transformation%
\footnote{The constant-time gauge transformations constitute a residual gauge symmetry of the temporal gauge \eqref{temp_gau}.} to the vacuum,
we have that the corresponding spatial part $F_{ij}$ of the field-strength vanishes everywhere.
From equation \eqref{physfield} we have then $\bm{B}=0$ and since the potential energy density in \eqref{eu} is 
\begin{equation}
 V[A] = \frac{1}{2} \bm{B}^{a}\cdot\bm{B}^a = 0 \ ,
\end{equation}
it vanishes too. 
The configurations with zero potential energy are degenerate with the vacuum and we henceforth refer to them as the \emph{vacua} of the theory.
Let us argue that an instanton solution describes a semi-classical transition amplitude connecting two such vacua.
For the sake of clarity, let us stick to an explicit instanton example
\begin{equation}\label{bela}
 A_\mu(x) = -\im \frac{r^2}{r^2+R^2} U^{-1}(x) \partial_i U(x) \ ,
\end{equation}
where $R$ is an arbitrary length scale%
\footnote{Further comments on the parameter $R$ as quantifying the ``size'' of the instanton are given in Section \ref{moduli}.}.
It is manifest that for large $r$ the field satisfies the requirement \eqref{pure}.
Moreover, if we want to bring \eqref{bela} into the temporal gauge we have to perform a gauge transformation that for asymptotic time 
(asymptotic time implies asymptotic $r$) will return a configuration of the form \eqref{gaueq}.
The instanton \eqref{bela} then connects two vacua of the theory.

Although instantons are classical solutions that exist only in the Euclidean formulation of the theory%
\footnote{Where they represent zero-energy solutions; indeed self-duality implies the vanishing of the Euclidean Hamiltonian
density \eqref{eu}.}, they can be interpreted in Minkowski space-time as transition amplitudes
interpolating between distinct vacua related by a topologically non-trivial gauge transformation; 
the Euclidean derivation of instanton amplitudes can be
regarded in fact as the imaginary time continuation of the theory in its Minkowski formulation.
Imaginary-time methods for the computation of semi-classical real-time tunneling amplitudes are a standard technique.

\subsection{The \texorpdfstring{$\vartheta$}{} angle}

The gauge fixing procedure in the presence of non-trivial topological sectors may generate some doubts.
Take a specific gauge configuration defined on the Euclidean $\mathbb{R}^4$
\begin{equation}\label{gaggia}
 A_\mu = U^{-1} \partial_\mu U\ , 
\end{equation}
where $U(x)$ has non-trivial winding on the asymptotic space-time three-sphere.
A gauge fixing procedure is in general intended to remove the gauge redundancy and we could be tempted to discard \eqref{gaggia}
as a gauge equivalent representative of the trivial vacuum $A_\mu=0$.
Following the arguments in \cite{PhysRevLett.37.172}, we must specify that the gauge fixing procedure
removes from the functional integral over the gauge field configurations those which are related
by a topologically trivial gauge transformation%
\footnote{I.e. a transformation obtainable deforming continuously the constant gauge transformation $U(x)=\one$.}.
In other words, fixing the gauge prevents redundancy within the various topological sectors.
In fact, configurations belonging to different sectors cannot at all describe the same physical circumstance
and cannot therefore be redundant.
As we describe in the following, the topology has indeed phenomenological effects.
Sometimes in the literature people use the terms ``small'' and ``large'' gauge transformations to denote respectively
the proper gauge transformations and the topology changing ones%
\footnote{The same terminology has some other times a different meaning: ``small'' and ``large'' are
referred to local as opposed to global (called also ``rigid'') gauge transformations.}.

The quantum vacuum state is in general expected to be given by a functional of $A_\mu$ 
that is peaked on the classical vacuum. The spread of the vacuum functional is given by Heisenberg's indeterminacy of quantum fluctuations.
We can have a sketchy idea figuring a well whose bottom is the classical vacuum. 
However, the picture that emerged form the study of instantons is richer.
The various topological sectors of YM theory can be imagined as different wells arranged in a periodic lattice
whose period is measured by the elemental increment of the winding number.
Instantons themselves represent transitions from one well to another.
The vacuum state is sensitive to this periodic structure and therefore we have to represent the 
candidate fundamental state functional as follows
\begin{equation}\label{espansa}
 \Psi[A] = \sum_{n\in\mathbb{Z}} c_n \psi_n[A] \ ,
\end{equation}
where the component functionals $\psi_n$ are peaked around the vacuum with winding number $n$ and $c_n$ are coefficients.
As a physical state, the quantum vacuum has to be invariant with respect to small gauge transformations;
moreover, since it is stable by definition, it must be invariant with respect to the topology changing gauge transformations
as well.
The latter feature fixes the shape of the coefficients in \eqref{espansa} to be
\begin{equation}
 c_n = e^{\im n \vartheta}
\end{equation}
where $\theta$ is a parameter that spans a continuous one-dimensional family of vacua for the YM theory.
They are indeed usually called $\vartheta$\emph{-vacua}.
The $\vartheta$-vacua have a behavior that reminds us of Bloch waves in periodic potentials.
In this respect, $\vartheta$ parametrizes the ``conduction band'' of YM vacua and is analogous to the Bloch momentum%
\footnote{To have further details and comments we refer the reader to \cite{PhysRevLett.37.172,Jackiw:2005wp,weinberg1996quantum,Vandoren:2008xg}.}.

\section{Collective coordinates}
\label{moduli}

The global features of instanton solutions like, for instance, the instanton center position, are encoded in a set of parameters
usually referred to as ``collective coordinates'' or ``moduli''.
For a given value $k$ of the topological charge, the corresponding moduli space spanned by
the instanton collective coordinates is denoted
with ${\cal M}_k$ and contains all the instanton solutions associated to winding number $k$.

Let us have a direct look at the moduli of the simplest $k=1$ instanton example%
\footnote{This is the first instance of instanton studied in the original paper \cite{Belavin:1975fg}
by Belavin, Polyakov, Schwarz and Tyupkin. Indeed it is commonly referred to as BPST instanton.}.
Consider pure Euclidean Yang-Mills theory with gauge group SU$(2)$ in Landau's gauge, i.e. $\partial_\mu A_\mu^a=0$;
the index $a$ runs over the adjoint representation of the gauge group.
Take the gauge transformation
\begin{equation}\label{g1}
 U(x) = \frac{t\, \one + \im\, \bm{x}\cdot \bm{\sigma}}{r}\ ,
\end{equation}
where we notice that the gauge adjoint space is linked to the physical space; in other words, $\bm{\sigma}$ which is a vector in the
adjoint space of SU$(2)$ is multiplied by $\bm{x}$ which instead is a spatial vector. 
The topological non-trivial character of the instanton emerges form such relation between space and gauge representation.
Let us insert \eqref{g1} in the instanton solution \eqref{bela}.
Performing some not difficult passages, we obtain the following explicit form for the $k=1$ SU$(2)$ instanton
\begin{equation}\label{su2_inst}
 A_\mu^a(x,\bm{v}) = 2 G^a_{\ b}(\bm{v})\ \overline{\eta}^b_{\mu\nu} \frac{(x-x_0)_\nu}{(x-x_0)^2+R^2}\ ,
\end{equation}
where $\overline{\eta}^b_{\mu\nu}$ represents the anti-self-dual 't Hooft symbols defined in Appendix \ref{hooft}.
To go from \eqref{g1} to \eqref{su2_inst} we have used the properties of the 't Hooft symbols (see Appendix \ref{hooft})
and we have also manually inserted the matrix $G^a_{\ b}(\bm{v})$; this matrix represents a global SU$(2)$ gauge rotation%
\footnote{Being rigid gauge rotations a symmetry of the theory, the global gauge orientation is a relative concept and 
it has a well defined meaning only when we compare the gauge orientations of two objects such as two instantons,
or an instanton and an adjoint condensate breaking the global gauge invariance.}
and $\bm{v}$ is a vector on the basis of the Pauli matrices.
The field-strength corresponding to \eqref{su2_inst} is
\begin{equation}\label{fsinst}
 F^a_{\mu\nu} = - 4 \eta^a_{\mu\nu} \frac{R^2}{[(x-x_0)^2+R^2]^2}\ ,
\end{equation}
where we have used the 't Hooft symbols properties from which the self-duality of $F$ descends manifestly.
In \eqref{su2_inst} we can count $8$ parameters, namely $x_0^\mu$ , $\bm{v}=(v^1,v^2,v^3)$ and $R$;
they represent respectively the position of the instanton center in space-time, its overall gauge orientation and its size.
Indeed, the instanton field-strength \eqref{fsinst} becomes small whenever $|x-x_0|$ grows bigger than the size parameter $R$.

Note that the instanton classical action \eqref{BPS} is a function of the topological
charge alone; since the moduli do not influence the action, they parametrize flat directions of $S$.
Said otherwise, all the instanton solutions (i.e. the configurations satisfying the self-duality condition) 
corresponding to a certain value of $k$ participate
to the path integral with the same weight independently of the particular values of their collective coordinates.

So far, we have just looked at the simplest case of SU$(2)$, $k=1$ instanton;
more complicated solutions will present in general a higher number of moduli.
For instance, a multi-centered instanton possesses the parameters describing the positions of all the centers.
From the linearity of the Yang-Mills equations we have that the superposition of two classical solutions for the gauge field
is still a solution. We can therefore sum $1$-instanton solutions to obtain multi-instantons and
the winding number is an additive quantity.
To become aware of this possibility of adding instantons, let us consider the sum of two $k=1$ SU$(2)$ instantons like \eqref{su2_inst}
centered at a distance $D$ far larger than their size, namely
\begin{equation}
 D\gg R_i\ , \ \text{for} \ \ i=1,2\ .
\end{equation}
Defining the total field-strength $F=F_1+F_2$ obtained summing the two single instantons, we have
\begin{equation}
 {}^*F F \sim {}^*F_1 F_1 + {}^*F_2 F_2\ .
\end{equation}
The approximation is justified observing that the two field-strengths $F_1$ and $F_2$ are nowhere significantly 
different from zero at the same time. 
We can also repeat the Bogomol'nyi argument \eqref{BPS} for the composite $2$-instanton 
solution and again neglect the mixed ``$F_1 F_2$'' terms.
Eventually for the sum configuration we obtain 
\begin{equation}
 S_{\text{$2$-inst}} \sim 2 S_{\text{$1$-inst}} \ \ \ \text{and}\ \ \ k_{\text{$2$-inst}} = 2 \ .
\end{equation}
Extending the argument to the most general superposition of well detached instantons labeled by $I$, we have%
\footnote{It is possible to include into our analysis also the anti-instantons;
they correspond to negative values of the topological charge $k$.
Similary to what we have done in relation to multi-instantons, also the multi-anti-instantons
can be constructed starting from the $k=-1$ anti-instanton.
The explicit field configuration of the SU$(2)$, $k=-1$ anti-instanton is given by \eqref{su2_inst} where
the anti-self-dual 't Hooft symbol is substituted with its self-sual counterpart.}
\begin{equation}
 S_{\text{SUM}} = \sum_I S_I \ \ \ \text{and}\ \ \ k_{\text{SUM}} = \sum_I k_I \ .
\end{equation}

A possible generalization of the $k=1$ instanton \eqref{su2_inst} consists in considering
theories with higher rank gauge groups; in this respect, let us limit ourselves to 
special unitary gauge groups SU$(N)$.
The first natural guess to produce an instanton solution in SU$(N)$ gauge theory is to embed the SU$(2)$ instanton
\eqref{su2_inst} into an SU$(2)$ subgroup contained in the full SU$(N)$.
We can in fact explicitly consider the following embedding
\begin{equation}\label{guess}
 [A_\mu]_{N\times N} = \left(\begin{array}{cc}
                               0_{(N-2)\times(N-2)} & 0_{(N-2)\times 2}\\
                               0_{2\times (N-2)} & [A_\mu]_{2\times 2}
                              \end{array}\right)\ .
\end{equation}
This particular embedding is not the unique possibility.
Fortunately for us, there is a theorem firstly formulated by Bott which comes into play and helps us:

\noindent
{\bfseries Bott's Theorem:} Let $G$ be a simple Lie group containing SU$(2)$ as a subgroup.
Every map $S_3\rightarrow G$ defined on the three-sphere is homotopic to a map $S_3\rightarrow \text{SU}(2)$.

\noindent
In our case, a homotopy class of solutions is the set of all instantons characterized by the same value of the topological charge.
We can therefore read Bott's Theorem as follows: the kind of instanton solutions that we have built
starting from \eqref{guess} and \eqref{su2_inst} provides us with a representative
in any homotopy class of the SU$(N)$ instantons.

Having constructed an instanton representative for any value of the topological charge $k$ in SU$(N)$ gauge theory with general $N$,
let us count the number of its global parameters.
For $k=1$ we have again the center position $x_0^\mu$, the size $R$ and the three parameters $\bm{v}$ specifying
the orientation with respect to the SU$(2)$ of \eqref{guess}. They amount to $8$ parameters.
In addition, we have also to take into account the relative orientation of the SU$(2)$ 
subgroup within the total SU$(N)$ gauge group. 
This contributes a number of parameters coinciding with the
dimension of the coset space
\begin{equation}
 \frac{\text{SU}(N)}{\text{SU}(2)\times\text{U}(N-2) }
\end{equation}
that is
\begin{equation}
 \text{dim}[\text{SU}(N)] - \text{dim}[\text{SU}(2)\times\text{U}(N-2)] = 4N - 8\ .
\end{equation}
Putting things together, we have $4N$ collective coordinates for the $k=1$ case.
Since, as we have seen, particular multi-instanton configurations can be produced by summing $1$-instanton solutions
and the number of parameters is a topological feature (i.e. all the members of a topological sector
have the same number of parameters), we have that the generic multi-instanton with charge $k$
is specified by $4Nk$ collective coordinates.
This is the dimension of its moduli space,
\begin{equation}
 \text{dim}\left[{\cal M}_k^{\text{SU}(N)}\right]_{\text{YM}} = 4Nk\ ,
\end{equation}
where the pedex ``YM'' indicates that we are considering non-supersymmetric Yang-Mills theory.
Indeed, in the supersymmetric framework of Super-Yang-Mills (SYM), the bosonic moduli have a fermionic partner each
and the moduli space dimension is consequently doubled.

\subsection{ADHM construction}
\label{ADHMcon}

The moduli spaces of SYM instantons admit a particularly elegant and concise description called
ADHM construction after the names of the proposers, M. Atiyah, V. Drinfel'd, N. Hitchin, Y. Manin, \cite{Atiyah:1978ri}.
Technically, this is nothing other than a convenient way to parametrize the instanton moduli space.
In this section we give a brief description of the ADHM construction which, even though developed
in the context of field theory, can be (as we will see in the following) very naturally accommodated in the framework of D-brane models.

We describe the ADHM construction to build the most general self-dual solution
in an SU$(N)$ SYM gauge theory leaving some technical details to the Appendix \ref{ADHMprojection}.
A similar construction is available also for SO$(N)$ and Sp$(N)$ gauge theories but not
for exceptional gauge groups.
Let us introduce the complex matrix $\Delta_{\lambda i \dot{\alpha}}$ which constitutes the fundamental
object of the construction.
The index $\lambda$ is referred to as \emph{ADHM index} and it runs over the values $1,...,N+2k$, $N$
being the ``number of colors'' and $k$ the topological charge of the instanton solutions we are building.
The matrix $\Delta_{\lambda, i \dot{\alpha}}$ is therefore an $(N+2k)\times 2k$ matrix; $2k$ arises from the composition of
an instanton index $i=1,...,k$ and an anti-chiral index $\dot{\alpha}=1,2$.
At this level the instanton index is nothing other than a label corresponding to the fundamental representation
of an auxiliary U$(k)$ group; as we will see in the following, in the D-brane instanton construction it instead emerges as a 
gauge symmetry group of the field theory ``living'' on the instanton branes.
Furthermore, we consider $\Delta$ to be a linear function of the space-time coordinates:
\begin{equation}
 \Delta_{\lambda i \dot{\alpha}}(x) = \text{a}_{\lambda i \dot{\alpha}} + \text{b}^\alpha_{\lambda i} x_{\alpha \dot{\alpha}}\ ,
\end{equation}
where we have adopted Hamilton's quaternionic notation for $x$.
We define the conjugate of $\Delta$ as follows:
\begin{equation}
 \overline{\Delta}^{\dot{\alpha}\lambda}_i 
= \overline{\text{a}}^{\lambda \dot{\alpha}}_i + \overline{x}^{\dot{\alpha} \alpha } \overline{\text{b}}_{i \alpha}^\lambda 
\equiv (\Delta_{\lambda i \dot{\alpha}})^* \ .
\end{equation}
Notice that the quaternionic and the ADHM indexes are sensitive to the upper or lower position, whereas the instanton indexes are not.

The components of the matrices $a$ and $b$ represent a redundant set of coordinates for the moduli space ${\cal M}_k$
of $k$-instantons. To appreciate this we have to complete the description of the ADHM construction.
The matrix $\overline{\Delta} = \Delta^\dagger$ is $2k\times(N+2k)$;
we assume
it to define a surjective map
\begin{equation}
 \overline{\Delta} : \mathbb{C}^{N+2k} \rightarrow \mathbb{C}^{2k}\ ,
\end{equation}
so that its kernel is $N$-dimensional.
Let us consider an orthonormal basis for $\text{Ker}[\,\overline{\Delta}\,]$ of $N$ vectors $U_{\lambda u}$ with $u=1,...N$.
By definition we have
\begin{equation}
 \overline{\Delta}^{\dot{\alpha}\lambda}_i U_{\lambda u} = 0\ ,
\end{equation}
and also the conjugate relation,
\begin{equation}
 \overline{U}^\lambda_u \Delta_{\lambda i \dot{\alpha}} = 0\ .
\end{equation}
The orthonormality property of the basis translates into:
\begin{equation}
 \overline{U}^\lambda_u U_{\lambda v} = \delta_{uv} \ .
\end{equation}
The ADHM recipe constructs the gauge field instantonic configuration from the matrices $U$ in the following way:
\begin{equation}\label{ADHMfield}
 A^\mu_{uv} = \overline{U}^\lambda_u \partial^\mu U_{\lambda v}\ .
\end{equation}
Observe that this is perfectly natural for $k=0$ where the $A^\mu$ field configuration
is obtained with a $U$ (in this case $U$ is an $N\times N$ matrix) gauge transformation%
\footnote{Note that here $U(x)$ is assumed to be obtainable deforming continuously the constant and everywhere equal to the identity gauge transformation.}
of the trivial vacuum.

In the $k>0$ case a further ingredient is needed, namely the so called ADHM constraint:
\begin{equation}\label{constra}
 \overline{\Delta}^{\dot{\alpha}\lambda}_i \Delta_{\lambda j \dot{\beta}} = \delta^{\dot{\alpha}}_{\dot{\beta}} f^{-1}_{ij}\ .
\end{equation}
This condition restrains the redundancy of the
moduli space parametrization contained in the components of $\Delta$.
Notice that here we are assuming $ \overline{\Delta} \Delta$ to be invertible;
in other terms, in addition to the already stated surjectivity of $\overline{\Delta}$ we further assume
the map
\begin{equation}
 \Delta : \mathbb{C}^{2k} \rightarrow \mathbb{C}^{N+2k}
\end{equation}
to be injective.
As shown in Appendix \ref{ADHMprojection}, the just described set of conditions implies the following relation:
\begin{equation}
 P_{\lambda}^{\ \omega} \equiv U_{\lambda u} \overline{U}^\omega_u
                        = \delta^\omega_\lambda - \Delta_{\lambda i \dot{\alpha}} f_{ij} \overline{\Delta}^{\dot{\alpha}\omega}_j\ ;
\end{equation}
this expression defines the projector operator $P$ on the null space of $\overline{\Delta}$.
We are now able to prove that \eqref{ADHMfield} is indeed associated to a self-dual field-strength,
\begin{equation}\label{selfservice}
 \begin{split}
  F_{\mu\nu} &= \partial_\mu A_\nu - \partial_\nu A_\mu + [A_\mu,A_\nu] \\
             &= \partial_{[\mu} (\overline{U}\partial_{\nu]}U) + (\overline{U}\partial_{[\mu}U)(\overline{U}\partial_{\nu]}U) \\
             &= \partial_{[\mu} \overline{U} (1 - U \overline{U})\partial_{\nu]}U \\
             &= \overline{U} \partial_{[\mu} \Delta f \partial_{\nu]} \overline{\Delta} U \\
             &= \overline{U} b\, \sigma_{[\mu} f \overline{b}\, \overline{\sigma}_{\nu]} U \\
             &= \overline{U} b\, \sigma_{[\mu}\overline{\sigma}_{\nu]} f \overline{b}\, U \\
             &\propto \overline{U} b\, \sigma_{\mu\nu} f \overline{b}\, U \ .
 \end{split}
\end{equation}
The self-duality of $F_{\mu\nu}$ in \eqref{selfservice} is a direct consequence of the self-duality of $\sigma_{\mu\nu}$.

\section{Phenomenological relevance}
The main influence to gauge theories due to instantons concerns anomalous symmetry breakings.
In this regard, in the next subsection we describe in some detail the so called U$(1)$ problem
related to the anomalous axial current in QCD with matter. 
However, it should be mentioned at the outset that the chirality violation induced by instantons
in QCD can be read in analogy to the anomalous violation of the baryon/lepton number associated
to electro-weak instantons \cite{'tHooft:1976up,'tHooft:1976fv}

As a general feature, the inclusion of instanton effects
yields infrared divergent contributions corresponding to the large-size regime of instantons%
\footnote{Namely $R\gg 1$ for the explicit example \eqref{su2_inst}. }.
Nevertheless, there are occasions in which the infrared problem is cured.
For instance, in QCD deep-inelastic scatterings with high photon virtuality $q^2$
it has been argued that $1/\sqrt{q^2}$ can play the r\^{o}le of a dynamical infrared cut-off
for (gluon) instanton size \cite{Khoze:1991mx}.
The HERA data about hadronic final states of deep-inelastic scattering have a non-negligible sensitivity
to processes induced by QCD instantons \cite{Carli:1997kx,Carli:1997tw}; 
these processes can affect the hadrons structure functions \cite{Kochelev:1997ma}.

Besides, in a spontaneously broken theory, the expectation value $\phi_0$ for the
Higgs field yields as well an effective cut-off $1/\phi_0$ (related to the associated exponential massive fall-off)
curing the corresponding instanton infrared divergence%
\footnote{The details are explained briefly in \cite{Shifman:1999mv} and thoroughly in \cite{'tHooft:1976fv}.}.
This is the key feature that makes it possible to have quantitative results for instanton effects
within the electro-weak sector of the Standard Model.
A significant consequence regards the baryon/lepton number violation; in fact, this quantity does not receive any contribution from the perturbative 
part of the theory but could be affected by (electro-weak) instantons.

Experimentally, the baryon number conservation has proven so far to be well satisfied as
the instanton induced effects are too small to be appreciated%
\footnote{For details on the instanton phenomenological implications to baryon decay we refer to \cite{Vandoren:2008xg}.}.
Also, a neat signal of instantons from deep-inelastic scattering is still matter of work in progress.
In all, (as far as the author's awareness reaches) the direct observation of instantons is still lacking.


\subsection{The U\texorpdfstring{$(1)$}{} problem}

In QCD the quark part of the Lagrangian density is
\begin{equation}\label{act_qua}
 {\cal L}_{\text{quark}} = - \sum_i \overline{\psi}_i \slashed{D} \psi_i\ ,
\end{equation}
where the index $i$ runs over the flavors.
Let us consider just the lightest two among them, i.e. up and down;
the flavor group is then U$(2)$.
We can rewrite \eqref{act_qua} expliciting the right and left-handed parts of the spinors, namely:
\begin{equation}\label{rl_hand}
 {\cal L}_{\text{quark}} = - \sum_i \left(\overline{\psi}_i^R \slashed{D} \psi_i^R - \overline{\psi}_i^L \slashed{D} \psi_i^L \right)\ ,
\end{equation}
where we are assuming the masses to be null.
In this way the flavor rigid symmetry $\text{U}_R(2)\times \text{U}_L(2)$ becomes manifest.
It is possible to reorganize the flavor symmetry group considering its vector and axial parts
which correspond respectively to two associated U$(2)$ groups,
\begin{equation}
 \text{U}_V(2) \times \text{U}_A(2)\ .
\end{equation}
Having just reorganized the flavor group, we have not affected the theory itself;
nevertheless, this picture proves to be more convenient to confront the phenomenology
and the physical interpretation.
In fact, the traceless part of the vector subgroup, i.e. $\text{SU}_V(2)$,
is a symmetry which is realized in Nature and we classify
the hadrons with respect to it. Its trace $\text{U}_V(1)$ corresponds to the baryon number approximate symmetry%
\footnote{The baryon number conservation is exact at perturbative level whereas it is approximate in the full theory.}.
The axial traceless part, $\text{SU}_A(2)$, is spontaneously broken.
From Goldstone's theorem we know that for any spontaneous symmetry breaking there has to be an
associated boson; in this case the multiplet of pseudo-scalars composed by the pions and the $\eta$ meson can be interpreted as
Goldstone's bosons corresponding to the breaking of $\text{SU}_A(2)$. 
The $\text{U}_A(1)$ has a story on its own; this Abelian symmetry
is violated in Nature. In fact, if instead it were realized, it would lead to a doubling of mesons (with opposite parity)
that has never been observed.
However, there is no good candidate particle to represent the Goldstone boson associated to
a spontaneous breaking of $\text{U}_A(1)$.
Historically this question has been referred to as the U$(1)$ \emph{problem}.

The axial Abelian current associated to $\text{U}_A(1)$ possesses an Adler-Bell-Jackiw anomaly and
so it is not conserved.
This feature can be explained including instantons into our analysis.
It is interesting to observe (following 't Hooft \cite{'tHooft:1976fv}) that the U$(1)$ problem
is one occasion in which an explicit symmetry violation is a necessary consequence of
the first-principle study of relativistic quantum field theory instead of being required solely in accordance with phenomenological reasons.
Indeed, in an Euclidean background containing instantons, the integrated four-divergence of the anomalous
axial current is non-vanishing and it can be shown to be related to the instanton topological charge $k$,
\begin{equation}\label{ABJ}
 \int  d^4x\ \partial_\mu j_\mu^{(A)} = N_f k\ ,
\end{equation}
where $N_f$ is the number of flavors.
In this sense the introduction of instantons solves the U$(1)$ problem because it accounts for the non-conservation
of the anomalous chiral current $\text{U}_A(1)$.
In other terms, non-trivial gauge backgrounds ``source'' the axial anomalous current.
No spontaneous breaking does occur and no Goldstone's boson is therefore needed.
Equation \eqref{ABJ} has to be considered in Euclidean space-time because here is where instantons are defined.
However, it emerges from instantons which, as we have seen in section \ref{vacua}, can be regarded as Minkowski space-time
tunneling processes between different topological gauge vacua%
\footnote{We have described these amplitudes in \ref{vacua}.
Notice that in the literature there are other occasions where quantum tunneling processes can be
described by means of Wick rotated classical solutions of the equation of motion (for instance 
in the framework of instantonic methods to study decay problems), see \cite{coleman1988aspects,Mclaughlin:1972ws}}.

\section{Instantons in supersymmetric theories}
Instantons are main characters on the stage of supersymmetric gauge theories%
\footnote{For a wide and deep review of such panorama see \cite{Bianchi:2007ft} and \cite{Shifman:1999mv}.}.
Many supersymmetric theories possess a continuous degeneracy of inequivalent vacua;
these vacuum configurations correspond to \emph{flat directions} of the superpotential.
Usually perturbative contributions at any order do not affect the vacuum moduli space being the flat direction pattern
protected by supersymmetry. 
Instantons can be thus the leading contribution in lifting the superpotential flat directions.
Their presence reduces the amount of supersymmetry and can consequently modify qualitatively the perturbative vacuum structure%
\footnote{See for instance \cite{Shifman:1999mv}. An example of non-perturbative generated superpotential is described
in \cite{Argurio:2007vqa} and references therein.}.

\subsection{Extended \texorpdfstring{${\mathcal N}=2$}{} SUSY}

In relation to instanton calculus, in the present thesis the attention is especially focused on 
the ${\cal N}=2$ supersymmetric framework.
Extended supersymmetry (i.e. ${\mathcal N}>1$) opens dramatic technical possibilities in relation to instanton calculus
such as Nekrasov's localization method. 
Moreover, for ${\mathcal N}=2$ SU$(2)$ theory the instanton calculations offer a non-trivial check of the celebrated
Seiberg-Witten duality.
Such a check corroborates both the SW duality and the localization techniques themselves.
Indeed, it must be remembered that we still lack a full first-principle derivation of Nekrasov's method
which, so far, can be regarded as a prescription.
It nevertheless revolutionized the field allowing researchers to perform explicit multi-instanton calculations 
that would be otherwise out of the computational reach.

The most general low-energy two-derivative ${\cal N}=2$ effective model is completely determined by an analytic function $\cal F$,
called \emph{prepotential}.
The prepotential $\cal F$ depends only on the vector multiplets and because of ${\cal N}=2$
non-renormalization behavior it receives corrections only at the one-loop perturbative level and from the non-perturbative sector.

\subsection{Seiberg-Witten duality}

Seiberg and Witten studied systematically the relation between the perturbative and non-perturbative
regimes of the effective (macroscopic) theory describing pure ${\cal N}=2$ Super-Yang-Mills theory 
with gauge group SU$(2)$ in the broken U$(1)$ phase, \cite{Seiberg:1994rs}.
They determined the complete expression of the prepotential for this effective model from electro-magnetic 
duality arguments. In this context the electro-magnetic duality 
is usually referred to as \emph{Seiberg-Witten duality}. For a pedagogical review see \cite{Bilal:1995hc}.

%
%

\chapter{D-brane Instantons}
\label{dinst}
The first studies of the non-perturbative sector of gauge theory performed employing string methods date back to the second half of the 1990s
when the seminal papers \cite{Witten:1995im,Douglas:1995bn,Douglas:1996uz} have been published.
There it was established a connection between non-perturbative configurations in string models, i.e. D-brane setups, 
with non-perturbative objects in the corresponding low-energy effective field theory description, namely gauge instantons.

In Subsection \ref{tadpole} we observed that the presence of D-branes can generally affect the tadpole expectation values,
hence D-brane setups are associated to non-trivial vacuum configurations in which
the fields (corresponding to the vertex operators whose tadpoles are non-null) assume non-trivial profiles.
We can reasonably expect that the D-brane vacua are in some relation with non-trivial vacua
of the corresponding low-energy effective field theory. 
In particular, we are interested in finding D-brane models whose low-energy regime
reproduces the instantonic non-perturbative sector of the underlying gauge theory.

A paradigmatic example which will be the pivot of our analysis is represented by Type IIB D$3$/D$(-1)$ brane models.
D$(-1)$ branes (usually called \emph{D-instantons}) are totally localized objects whose world-volume is a point%
\footnote{As usual, a D$p$-brane world-volume has $p$ spatial directions plus one which is temporal.
Accordingly, a D$(-1)$ brane has a zero-dimensional (i.e. point-like) world-volume.}.
At low energy the SU$(N)$ supersymmetric theory%
\footnote{A stack of $N$ D$3$-branes supports a U$(N)$ gauge theory. 
For $N>1$ we have the decomposition U$(N) =$ SU$(N)\times$U$(1)$ where the Abelian factor is associated to the trace.
The running of the Abelian coupling constant associated to the U$(1)$ trace part and the running of the non-Abelian coupling associated to 
SU$(N)$ are different. Actually, the Abelian theory is IR-free, while the non-Abelian theory is UV-free.
As a consequence, since we deal with low-energy effective theories, we are interested in a regime in which the Abelian
coupling constant is likely to be negligible with respect the non-Abelian one. 
This sort of decoupling is what we understand whenever we neglect the ``center of mass'' part of the gauge group and
consider the stack as simply supporting a theory with gauge group SU$(N)$ instead of the full U$(N)$.} describing a system of $N$ coinciding D$3$-branes and $k$ D$(-1)$-branes
accounts for\footnote{This framework has been introduced in 
\cite{Witten:1995gx,Douglas:1995bn,Douglas:1996uz,Douglas:1996du,Sen:1996sk,Green:1997tv,Green:1997tn,Green:1998yf,Green:2000ke,Billo:2002hm};
for a review on the topic we refer the reader to \cite{Blumenhagen:2009qh}.} 
the fluctuations around an instanton background of topological charge $k$. 

In general, a D-brane instanton setup has (at least) two kinds of branes:
the \emph{gauge} and the \emph{instanton} branes.
The world-volume  of the gauge branes contains the four-dimensional physical space-time and it hosts the gauge theory of 
which we intend to study the non-perturbative sector.
The instanton branes, instead, host an auxiliary field theory which accounts for the instanton moduli ``dynamics''
(the instanton moduli has been described in Section \ref{moduli});
these branes, as they are associated to instantons, must have a world-volume which is completely localized from the 
physical space-time perspective.
For instance, in the D$3$/D$(-1)$ case, the D$(-1)$ (being here the instanton branes) 
are localized already in ten-dimensional space-time and, 
\emph{a fortiori}, also from the four-dimensional viewpoint. 
Conversely the D$3$ (here gauge branes) world-volume coincides with the physical four-dimensional space-time.
Another instance of D-brane instanton setup is furnished by D$3$/D$7$ models%
\footnote{An example of D$3/$D$7$ models is studied for example in \cite{Billo':2010bd}.}
where the D$3$'s play the r\^{o}le of the instanton branes and, in order
to be localized from the $4$-space-time perspective, their
world-volume has to extend along the internal (i.e. orthogonal to the physical space-time) directions.
In the D$3/$D$7$ case, the world-volume of the D$7$'s contains the physical space-time as a proper subset; therefore, to obtain
a phenomenological model, one needs to compactify the extra dimensions.
Let us anticipate that both in the cases in which there is the necessity%
\footnote{With ``necessity'' here is meant the need of compactifying extra dimensions for the sake of obtaining a phenomenological theory.} 
of compactification and in the cases in which
there is not, the internal space geometry plays at any rate a crucial r\^ole.
As we will see, the internal geometry arrangement and the internal space symmetry behavior of the branes
represent the crucial features distinguishing between ordinary and stringy instantons.
In the ordinary case, gauge and instanton branes share the same internal space characteristics while in the stringy case
they do not.

To become fully aware that appropriate D-brane systems reproduce the field theoretical instantons,
one has to study carefully the D-brane induced background profiles for the fields.
On the computational level, one considers the tadpole amplitudes and attach to them the propagators of the corresponding
fields; taking then the Fourier transform and considering the limit of great distance from the source (i.e. the branes),
one can actually recover the non-trivial profile of the background and show that it matches precisely 
with the leading term in the large-distance expansion of the field theory instanton solution in the singular gauge 
\footnote{The singularity in the field profile does not lead to singularity of any physical
(i.e. gauge invariant) quantity such as, for example, the action density. In instanton treatments it is usual to use singular expressions that can be
transformed, by means of singular gauge transformations, to perfectly well-behaved configurations, the latter, of course, 
continue to satisfy all the instanton defining properties (e.g. 
finite-action Euclidean solution, asymptotic pure gauge behavior, selfdual or anti-self-dual behavior,...), \cite{rajaraman1982solitons}.}
, \cite{Billo:2002hm}.
The D-brane description of gauge theory instantons is complete.
All the features of standard instanton calculus are accommodated into the string framework.
A particularly significant example is represented by the ADHM construction:
From a purely field theory viewpoint the ADHM construction is an elegant but rather technical and obscure 
way of constructing the instanton moduli space (we have introduced it in Subsection \ref{ADHMcon});
from the D-brane perspective, instead, the ADHM construction emerges naturally from the interactions
of the string modes attached to the instanton branes and the auxiliary $k$-instanton group U$(k)$ of the ADHM construction 
coincides with the gauge group of the field theory defined on the instanton branes.

Henceforth, we mainly stick to the D$3/$D$(-1)$ models; here, to obtain the gauge theory living on the D$(-1)$ branes,
we perform the zero-dimensional reduction of the Euclidean SUSY $\sigma$-model living on a generic brane:
The (Euclidean) path-integral involving all the modes associated to strings attached to the D-instantons defines
the instanton collective coordinate integral or, equivalently, the D-instanton partition function, \cite{Dorey:2002ik}.
A zero-dimensional gauge theory is usually called a \emph{matrix model}; 
only the interaction terms involving no derivatives survive the ultimate dimensional reduction, 
in fact, being the world-volume of D$(-1)$ branes point-like, it constitutes a degenerate manifold
that is spanned by no coordinates with respect to which one could take derivatives.

The effective gauge theory describing the low-energy regime of open strings and branes emerges as the first significant term
in the full D-brane Dirac-Born-Infeld expansion.
The subleading terms present a higher number of derivatives%
\footnote{There is an interesting open question in relation to self-dual configurations: these could represent
solutions of the would be total DBI action, i.e. solutions of the full string model instead of being solutions of only its effective low-energy approximation;
to have some comments on this see \cite{Billo:2009gc}.}.



\section{D-instanton models, a closer look}
\label{OrbiOrie}

As already mentioned, the D-brane context offers a particularly natural environment to treat instantons.
It provides us a framework that allows us to give an intuitive interpretation
of the ADHM construction;
all its ingredients, i.e. the ADHM moduli, their constraints and their
U$(k)$ symmetry, are described with the gauge theory living on instantonic branes
and with the dynamics of the string modes stretching between instanton and gauge branes,
\cite{Dorey:2002ik,Billo:2002hm}.
In this section we plunge into a more detailed excursion through the technical features and essential ingredients
of D-brane instanton constructions.
Although keeping sensitive to general aspects, we will often be concerned to a $\mathbb{C}^3/\mathbb{Z}_3$ 
orbifold model with the addition of an orientifold; such a model constitutes the specific setup on which our research has been performed
and it will be defined throughout the following sections.

We consider exact string backgrounds that admit a conformal field theory (CFT) treatment and,
in particular, \emph{orbifold backgrounds}; they can be seen as (almost) flat configurations obtained as specific limits of more complicated and curved backgrounds. 
Nevertheless (as we will see explicitly) the orbifolds, and more specifically their singular points, encode some important features of the 
curved backgrounds of which they represent the limiting case, especially with respect to the breaking/preserving of supersymmetries.

The orbifold/orientifold backgrounds are examples of non-compact space-times,
hence it is not necessary to address global tadpole cancellation problem.
This is instead unavoidable whenever one works in a compact space-time containing charged objects.
Indeed, what we refer as the tadpole cancellation problem can be intuitively thought of as the fact that in a compact space-time 
the flux-lines has to connect charges of opposite sign and, because of Gauss' theorem, the total charge has therefore to vanish. 
In a non-compact background, instead, the flux-lines being generated by a charged object can ``go to infinity'' without ending on 
another and oppositely charged object. 
This allows us to consider setups with a net total charge.
The results which we obtain in an orbifold/orientifold non-compact framework can have a ``local'' 
relevance in relation to models having a compact internal space as well. 
Indeed, as long as we work locally (i.e. we consider just a subspace of the whole internal compact manifold),
we can disregard global issues as the charge balance, \cite{Argurio:2007vqa}.

\subsection{\texorpdfstring{$\mathbb{C}^3/\mathbb{Z}_3$}{} orbifold background}

An orbifold background\footnote{Introductory
treatments of orbifolds are on many textbooks such as \cite{Becker:2007zj,polchinski2001string}.}
${\cal M}_{10}/\Gamma$ is the result of a quotient operation of the 
ten-dimensional space ${\cal M}_{10}$ with respect to a discrete group $\Gamma$ of isometry transformations acting 
only on the \emph{internal space} ${\cal M}_{I}$%
\footnote{The internal space is the manifold formed by the directions which are orthogonal to the four-dimensional physical 
space-time ${\cal M}_{\text{ph}}$, so
\begin{equation}
 {\cal M}_{10} = {\cal M}_{\text{ph}} + {\cal M}_{I}\ .
\end{equation}
For us, ${\cal M}_{\text{ph}}$ coincide with the D$3$ world-volume.}.
The orbifold ${\cal M}_{10}/\Gamma$ is usually indicated with just ${\cal M}_{I}/\Gamma$ since the physical space-time remains untouched by 
the action of the group $\Gamma$ of internal isometries.
The points of ${\cal M}_{I}/\Gamma$ represent the orbits%
\footnote{Remember that the orbit of a point $x\in {\cal M}_{I}$ is the set 
comprehending $x$ itself and its images under all the elements of the group $\Gamma$ of discrete isometries.} 
of the points in ${\cal M}_I$ under the action of the elements in $\Gamma$.
The name orbifold is actually a contraction of \emph{``orbit manifold''}.

Let us specialize the treatment to orbifold constructions in the presence of a stack of coinciding D$3$-branes and D$(-1)$ instantons
on top of them.
We parametrize the extended directions of the D$3$-branes with the first four 
coordinates of the ten-dimensional space-time as in Table \ref{tab:d3d-1}.
\begin{table}[ht]
\begin{center}
\begin{tabular}{c|cccc|cccccc}
\phantom{\vdots}
&0&1&2&3&4&5&6&7&8&9 
\\
\hline
\phantom{\vdots}D3&$-$&$-$&$-$&$-$&$\times$&$\times$&$\times$&$\times$&$\times$&$\times$\\
\phantom{\vdots}D(--1)&$\times$&$\times$&$\times$&$\times$
&$\times$&$\times$&$\times$&$\times$&$\times$&$\times$\\
\end{tabular}
\end{center}
\caption{Arrangement of the D-branes; the symbols $-$ and $\times$ denote respectively 
Neumann and Dirichlet boundary conditions for the open strings attached to the branes.}
\label{tab:d3d-1}
\end{table} 
The internal space is instead spanned by the coordinates labeled with $4,...,9$ and we organize them in three complex coordinates as follows%
\footnote{As it will shortly emerge, the complex notation is convenient to define the orbifold action.}:
\begin{equation}\label{zs}
 z^1 = X^4 + \text{i} X^5\ , \ \ \ \
 z^2 = X^6 + \text{i} X^7\ , \ \ \ \
 z^3 = X^8 + \text{i} X^9\ .
\end{equation}
We start with an internal manifold which is isomorphic to $\mathbb{R}^6 \sim \mathbb{C}^3$,
that is to say, just flat six-dimensional space.
String models can be however defined on more structured internal spaces, for instance on non-trivial\footnote{Actually
$\mathbb{R}^6$ is a particular example of Calabi-Yau manifold.} Calabi-Yau manifolds
(e.g. K3).
Oftentimes, the explicit form of the metric for Calabi-Yau manifolds is not available and it is impossible
to give a description of the string dynamics.
An important exception is furnished by orbifolds which represent particular singular limits of Calabi-Yau manifolds.
The orbifold projection arising from the quotient ${\cal M}_{I}/\Gamma$ yields a singular orbit space
whenever a point of the original manifold ${\cal M}_{I}$ is left invariant by the action of the 
non-trivial elements in $\Gamma$.
As opposed to the non-singular points in ${\cal M}_{I}/\Gamma$ where the local differential structure
is identical to the one ``around'' the corresponding points in ${\cal M}_{I}$,
at a singular point in ${\cal M}_{I}/\Gamma$ it is impossible to define a consistent tangent space and the metric, as well,
is singular.
Because of the lack of a well-defined metric, at such singularities the General Relativity description fails.
Conversely, after the introduction of new string states
called \emph{twisted states} (see Section \ref{twist}), string theory admits a consistent description
of the dynamics also in the presence of the orbifold singularities%
\footnote{From the supergravity point of view, there are situations in which there exist a characteristic length, usually referred to as the \emph{enhan\c{c}on}, 
``around a singularity'' where the brane probes become tensionless so that the supergravity description itself looses its validity, \cite{Johnson:1999qt}. 
The singularity remains outside the supergravity treatment. In an $ADS$/CFT-like perspective is interesting to mention that the enan\c{c}hon 
corresponds to the scale where the dual gauge coupling diverges, i.e. the non-perturbative dynamically generated scale $\Lambda$, \cite{Bertolini:2001gq}.}.

To have an intuitive idea, one can consider a two-dimensional cone as arising from the orbifold
projection of the complex plane under the action of the cyclic group $\mathbb{Z}_3$,
\begin{equation}
 \mathbb{Z}_3 = \left\{g,g^2,g^3=\one\right\}\ ,
\end{equation}
with $g$ the group generator.
Let us represent $\mathbb{Z}_3$ on the complex plane with discrete rotations around the origin of the polar coordinates by an angle of $\frac{2\pi}{3}$.
After the projection, the origin, which is actually left invariant by rotations, is mapped to the singular tip of the cone.
There are however two caveats to be mentioned to avoid confusion.
At first, there exist conical singularities which cannot be obtained by orbifolding a plane%
\footnote{This occurs when the \emph{deficit angle} is not expressible as $2\pi (n-1)/n$ with $n\in\mathbb{N}$.
In the explicit example just described in the main text, the deficit angle amounts to $4\pi/3$ (i.e. $\pi- 2\pi/3$)
corresponding to $n=3$.}
Secondly, there are also instances of manifolds with conical singularities where
the manifold itself is describable as a cone only locally, i.e. in the vicinity of the singularity;
this has to be opposed to the orbifolds of the type just described that instead yields global cones.

The stringy instanton model that we are going to describe in the following sections is built on a
$\mathbb{C}^3/\mathbb Z_3$ orbifold where $\mathbb{C}^3$ indicates the internal flat manifold in complex coordinates \eqref{zs}.
We assign the following action of $\mathbb{Z}_3$ on the internal space%
\footnote{Note that, since the coordinate $z^3$ is unaffected by the orientifold action, the orientifold itself could be thought of
as a $\mathbb{C}\times \frac{{\mathbb C}^2}{\mathbb Z_3}$; we however maintain the $\frac{{\mathbb C}^3}{\mathbb Z_3}$ notation.}:
\begin{equation}
 \label{gorb}
 g~:~~\begin{pmatrix}z^1 \cr  z^2 \cr z^3 \end{pmatrix}
 ~\to~
 \begin{pmatrix}\xi\,z^1 \cr  \xi^{-1}\,z^2 \cr z^3 \end{pmatrix}
\end{equation}
where $\xi=e^{\frac{2\pi\im}{3}}$.
Notice that, because of the complex notation \eqref{zs}, the transformation (\ref{gorb}) can be easily seen as a rotation of $\frac{2\pi}{3}$
on the $z^1$-plane together with a rotation of $-\frac{2\pi}{3}$ on the $z^2$-plane;
and the group generator $g$ can be then represented by
\begin{equation}
 \label{gorb1}
R(g) = e^{+\frac{2\pi\im}{3}J_1}\,e^{-\frac{2\pi\im}{3}J_2} \ ,
\end{equation}
where $J_i$ is the generator (in the vector representation) of the complex rotations on the plane spanned by the coordinate $z^i$. 
The expression \eqref{gorb1} is particularly convenient to define the orbifold action on any kind of field just
by choosing the corresponding representation for the rotation operators $J$. 
In Subsection \ref{3sing} we will follow this approach to study the orbifold transformation of fields carrying spinor indexes.
From the analysis of the spinor transformations emerges that half of the original ten-dimensional background supersymmetry
is broken by the $\mathbb{C}^2/\mathbb{Z}_3$ orbifold.
From the viewpoint of the four-dimensional theory on the world-volume of the D$3$-branes, the orbifold
preserves $2$ supersymmetries of the original $\mathcal N=4$ theory\footnote{We remind the reader that a stack of D$3$ branes on the flat 
ten-dimensional background is described at low energy by ${\cal N}=4$ SYM theory.}. 

Since the orbifold acts on the internal manifold ${\mathcal M}_I$, a brane placed at a specific point of ${\mathcal M}_I$ is
mapped to an image brane placed at the corresponding transformed point.
In a consistent treatment we have to include into the model all the images of the branes.
This proliferation of branes and their images is avoided if the branes themselves are placed at the orbifold singular points;
these points are in fact image of themselves under the action of all the elements of the orbifold group.
As we will see, the branes placed on the orbifold singularities can be associated to irreducible representations of the orbifold group;
in this case, they are called \emph{fractional branes}%
\footnote{The attribute ``fractional'' is opposed to ``regular''; both names will be commented in the next section.}.

\subsection{Orbifold transformation of Chan-Paton indexes, quiver diagram and fractional branes}
\label{twist}

So far we have considered only the action of the orbifold on the coordinates.
An important aspect which proves to be crucial for the developments we are to study,
consists in the careful analysis of non-trivial orbifold transformations for the Chan-Paton degrees of freedom.

Consider a D$3$-brane located at a generic point of the internal manifold and its two images under the orbifolds $\mathbb{Z}_3$;
the CP structure accounting for this set of branes is a $3\times 3$ matrix%
\footnote{A similar and more detailed analysis of the $\mathbb{Z}_2$ case is given in \cite{Bertolini:2001gq}.}.
As a brane is transformed, the corresponding CP label is transformed as well; in other terms, an open string attached to a brane will be attached
to its image after the orbifold transformation. The representation of $\mathbb{Z}_3$ on this CP structure is then
\begin{equation}
 G(\one) = \left( \begin{array}[]{ccc}
                   1 & 0 & 0\\
                   0 & 1 & 0\\
                   0 & 0 & 1    
                  \end{array}\right)\ , \ \ \ 
G(g)     = \left( \begin{array}[]{ccc}
                   0 & 1 & 0\\
                   0 & 0 & 1\\
                   1 & 0 & 0    
                  \end{array}\right)\ , \ \ \ 
G(g^2)   = \left( \begin{array}[]{ccc}
                   0 & 0 & 1\\
                   1 & 0 & 0\\
                   0 & 1 & 0    
                  \end{array}\right)\ .
\label{regu}
\end{equation}
The matrices \eqref{regu} realize the so-called \emph{regular} representation of $\mathbb{Z}_3$, namely
\begin{equation}
 [{\cal R}(a)]_{bc} = \delta_{(a\cdot b),\, c}\ ,
\end{equation}
where $a,b,c$ are generic elements of $\mathbb{Z}_3$.
As any multi-dimensional representation of an Abelian group, the regular representation is reducible and
the matrices \eqref{regu} can be correspondingly diagonalized,
\begin{equation}
H(\one) = \left( \begin{array}[]{ccc}
                   1 & 0 & 0\\
                   0 & 1 & 0\\
                   0 & 0 & 1    
                  \end{array}\right)\ , \ \ \ 
H(g)     = \left( \begin{array}[]{ccc}
                   1 & 0 & 0\\
                   0 & \xi & 0\\
                   0 & 0 & \xi^{-1}    
                  \end{array}\right)\ , \ \ \ 
H(g^2)   = \left( \begin{array}[]{ccc}
                   1 & 0 & 0\\
                   0 & \xi^{-1} & 0\\
                   0 & 0 & \xi    
                  \end{array}\right)\ ,
\label{regudia}
\end{equation}
We remind the reader that $\xi=e^{2\im\pi/3}$.
In the diagonal entries we can recognize the three irreducible representations of $\mathbb{Z}_3$,
\begin{equation}
 \begin{array}{ccc}
  R_1(\one)=1 & R_1(g)=1 & R_1(g^2)=1\\ 
  R_2(\one)=1 & R_2(g)=\xi\ & R_2(g^2)=\xi^{-1} \\
  R_3(\one)=1 & R_3(g)=\xi^{-1} & R_3(g^2)=\xi
 \end{array}
\end{equation}

In the NS open-string sector the massless states are expressed by
\begin{eqnarray}\label{NSh}
 \bm{A}^\mu = X_A \psi_{-1/2}^{\mu} |\bm{p} \rangle\\
 \bm{\Phi}^I = X_\Phi \psi_{-1/2}^{I} |\bm{p} \rangle 
\end{eqnarray}
where the index $\mu$ is associated to the extended directions of the D$3$-branes while
$I$ labels the three complexified internal directions; $\bm{p}$ represents the center of mass momentum along the D$3$-branes.
The orbifold acts trivially on the $\mu$ directions and according to \eqref{gorb} on the internal space indexes.
On the generic CP factor $X$ the orbifold generator $g$ acts according to
\begin{equation}
 \label{gCPH}
g~:~~X~\to~H(g)\,X\,H^{-1}(g) \ .
\end{equation}
The orbifold projection consists in retaining in our model only the states which are overall invariant with respect to the orbifold transformation;
the states surviving such projection must then satisfy the following condition:
\begin{equation}\label{condaqui}
 \bm{A}_\mu = H(g) \bm{A}_\mu H^{-1}(g) \ , \ \ \ \ \ \ \bm{\Phi}^I = (\xi)^I H(g) \bm{\Phi}_I H^{-1}(g) \ ,
\end{equation}
where $(\xi)^I$ denotes the phase $\xi$ to the $I$-th power.
The orbifold conditions \eqref{condaqui} constrain the modes \eqref{NSh} to have the following CP structures:
\begin{equation}
 \mathbf{A}_\mu = \left(\begin{array}{ccc}
               A_{\mu(11)} & 0 & 0\\
               0 & A_{\mu(22)} & 0\\
               0 & 0 & A_{\mu(33)}
              \end{array}\right) ~,\ \ \ \ \ \ \ \ \ 
\mathbf{\Phi}^3 = \left(\begin{array}{ccc}
               \Phi^3_{(11)} & 0 & 0\\
               0 & \Phi^3_{(22)} & 0\\
               0 & 0 & \Phi^3_{(33)}
              \end{array}\right) ~,
\label{Amuphi3}
\end{equation}
and
\begin{equation}
 \mathbf{\Phi}^1 = \left(\begin{array}{ccc}
               0& \Phi^1_{(12)}& 0\\
               0 & 0& \Phi^1_{(23)}\\
               \Phi^1_{(31)} & 0 & 0
              \end{array}\right) ~,\ \ \ \ \ \ \ \ \ 
\mathbf{\Phi}^2 = \left(\begin{array}{ccc}
               0& 0 & \Phi^2_{(13)}\\
               \Phi^2_{(21)} & 0 & 0\\
               0 & \Phi^2_{(32)} & 0
              \end{array}\right) ~.
\label{Phi1Phi2}
\end{equation}

As usual the vacuum expectation values of the internal scalars account for displacements in the corresponding direction of the branes themselves.
For example, if we had a non-vanishing VEV for the field $\Phi^1_{(12)}$, it would mean that the branes $1$ and $2$ are at a distance $<\Phi^1_{(12)}>$
along the internal direction labeled with $1$.
If we consider all vanishing VEV's for the scalars $\Phi$ then all the branes are placed at the origin on top of each other;
observe that the origin is also a singular point of the orbifold rotations.
It is possible to ``diagonalize'' the brane system (in the sense of \eqref{regudia}) and interpret the three diagonal branes as \emph{fractional branes}%
\footnote{Following \cite{Diaconescu:1997br}, the fractional branes are interpreted as object bound to the orbifold singular locus but free
to move along the four-dimensional physical space-time (and $z^3$ which, since it is left invariant by the orbifold action, spans the invariant locus).},
each of them associated to a diagonal factor of $H$.
In other words, we assign an irreducible representation to each fractional brane and,
pictorially, we describe this by means of a \emph{quiver diagram}%
\footnote{The word ``quiver'' in English indicates the sack carried on an archer's shoulder
to carry arrows. Apparently the presence of arrows (indicating oriented string modes)
suggested the fanciful name for the diagram.} possessing a node for any irreducible representation.
The arrows represent oriented string modes stretching between the branes placed at different nodes.
Alternatively, we could (but in the present analysis we do not) consider branes associated to the whole diagonalized regular representation;
these branes are commonly called \emph{regular branes}.

\begin{figure}
 \centering
 \includegraphics[width=65mm]{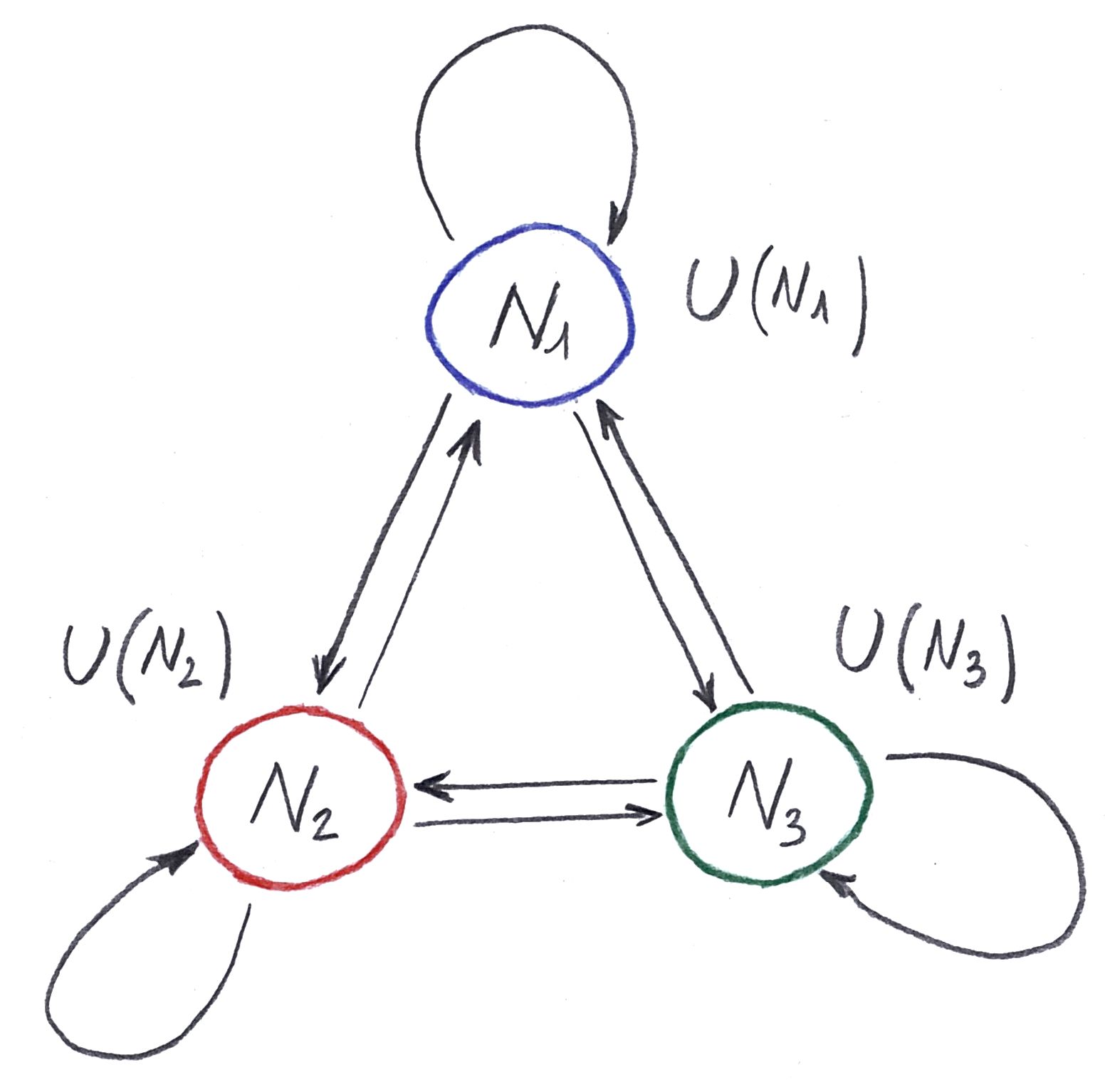}
 \caption{Quiver diagram of the $\mathbb{C}^3/\mathbb{Z}_3$ theory before orientifold projection;
it describes a configuration of $N_1$, $N_2$ and $N_3$ fractional D3 branes. The arrows starting and ending on the
same node represent $\cN=2$ vector multiplets in the adjoint representation of the $\mathrm{U}(N_i)$
groups. The arrows between different nodes represent bi-fundamental chiral multiplets
which pair up into $\cN=2$ hypermultiplets.}
 \label{Fig:1}
\end{figure}

We can have a general setup containing a generic number of fractional branes on any node.
The branes transforming into a particular irreducible representation will be pictorially ``placed'' on the corresponding node of the quiver diagram%
\footnote{Notice that also the D-branes corresponding to different nodes are nevertheless coinciding in space-time. 
Indeed, they are placed at the same singularity of the orbifold.}
and the number of fractional D$3$ branes occupying the $i$-th node (corresponding to the irreducible representation $R_i(g)$) is indicated with $N_i$.
In our $\mathbb{Z}_3$ case we have the Diagram \ref{Fig:1}.
Now the CP factors are $(N_1+N_2+N_3)\times(N_1+N_2+N_3)$ matrices and we generalize \eqref{gCPH} as follows:
\begin{equation}
 \label{gCP}
g~:~~X~\to~\gamma(g)\,X\,\gamma^{-1}(g) \ ,
\end{equation}
where $\gamma(g)$ is the matrix encoding the D$3$-brane assignment to the quiver nodes
\begin{equation}\label{d3cp}
 \gamma(g) = 
\begin{pmatrix}
\one_{N_1}&0&0\cr  
0&\xi\,\one_{N_2}&0\cr
0&0&\xi^{-1}\,\one_{N_3}
\end{pmatrix}
\end{equation}
with $\one_{N_i}$ denoting the $N_i\times N_i$ identity matrix. 
At low energy, this D-brane model is described with a gauge theory having gauge group 
$\mathrm{U}(N_1)\times \mathrm{U}(N_2) \times \mathrm{U}(N_3)$.

Repeating a similar reasoning for the D$(-1)$ branes we have that the
$\mathbb Z_3$ orbifold generator $g$ acts on the instantons with a matrix $\gamma^{(\text{inst})}(g)$ which has the same 
form as $\gamma(g)$ in (\ref{gCP}) but where the $N_i$'s are replaced with the $k_i$'s, namely
\begin{equation}\label{iscp}
 \gamma^{(\text{inst})}(g) = 
\begin{pmatrix}
\one_{k_1}&0&0\cr  
0&\xi\,\one_{k_2}&0\cr
0&0&\xi^{-1}\,\one_{k_3}
\end{pmatrix}
\end{equation}

In the presence of the $\mathbb{C}^3/\mathbb Z_3$ orbifold, the usual boundary conditions for the closed-string modes in the internal directions, i.e.
\begin{equation}
 X^I(\sigma+\pi) = X^I(\sigma)\ , \ \ \ \ \sigma \in [0,\pi]\ ,
\end{equation}
must be generalized to
\begin{equation}\label{twist_bou}
 X^I(\sigma+\pi) = r_I(h)\, X^I(\sigma)\ , \ \ \ \ h \in \Gamma = \mathbb{Z}_3 \ .
\end{equation}
where $r_I$ (the index $I$ is not summed) represents the orbifold action according to the assignment \eqref{gorb}.
Note that we have a set of boundary conditions for any element $h$ of the orbifold group,
therefore, for $h\neq \one$, we are defining new closed-string sectors; these sectors are usually called \emph{twisted sectors}.
Obviously, there are as many twisted sectors as non-trivial elements of the orbifold group.
It is possible to show that the fractional branes carrying an
irreducible representation of the orbifold group source the corresponding twisted closed-string modes \cite{Bertolini:2001gq}.

Again, also in relation to the closed-string sectors, the orbifold projection retains only the invariant modes.
Notice that, on the conformal theory computational level, the operator product expansion of the vertex operators ``surviving''
the orbifold projection closes. Indeed, the product of invariant operators is still an invariant operator.


\subsection{Orientifold}

The exotic instantons have in general some extra neutral fermionic zero-modes 
in addition to the zero-modes that are associated to the breaking of translations in superspace (i.e. the standard space-time translations
and their ``super'' partners). 
Let us anticipate what will be seen in detail in the following:
The presence of extra fermionic zero-modes (that we denote here with $\lambda$) is an exotic feature in contrast with the ordinary instanton case.
Indeed, on the computational level, the extra fermionic zero-modes emerge because the exotic configurations lack some bosonic moduli (related to the instanton size).
These bosonic moduli are needed to saturate the fermionic degrees of freedom $\lambda$; the exotic action does not contain such ordinary interaction terms involving $\lambda$ and in fact
it is completely independent of them.
Since a fermion is represented by a Grassmann variable, fermion zero-modes in the action renders the partition function integral null.
In such instances the exotic non-perturbative instanton do not have any effect on the low-energy field theory.

There are nevertheless some models in which also the ``dangerous'' fermionic zero-modes are projected away.
Specific backgrounds such as those involving orientifold projections, 
can eliminate the additional zero-modes \cite{Argurio:2007vqa} and then the exotic instanton partition function can yield finite 
contributions to the low-energy interactions and couplings.

In its basic definition, with \emph{orientifold projection} we mean the ``gauging'' of the world-sheet parity $\omega$.
In this case the orientifold operator $\Omega$ coincides with the world-sheet parity $\omega$,
\begin{equation}
 \Omega = \omega \ .
\end{equation}
Implementing an orientifold projection we discard all the variant states under the operator $\Omega$.
In ten-dimensional space-time, in the absence of any D-brane, the orientifold projection can be described as a ten-dimensional
extended object called O$9$-plane. 
Intuitively it ``acts as a mirror'' implementing new conditions for the string modes and essentially identifying the right and left-movers.

Take the generic bosonic string mode
\begin{equation}\label{modomodo}
X^M(z,\overline{z}) = X_L^M(z) + X_R^M(\overline{z})
\end{equation}
where we have split the left.moving (holomorphic) and right-moving (anti-holomorphic) parts.
The world-sheet parity $\omega$ action exchanges them
\begin{equation}
\omega [X^M(z,\overline{z})] = X^M(\overline{z},z) = X_L^M(\overline{z}) + X_R^M(z)
\end{equation}
In this case, the orientifold projection will select the states such that $X_L=X_R$.

The introduction of D-branes into the game makes the orientifold more complicated.
In the presence of toroidal compact directions, it is possible to see D-branes as arising from T-duality operations of ten-dimensional space;
more specifically, in the open-string sector, Neumann boundary conditions are transformed into Dirichlet boundary conditions by T-duality and we know that
D-branes correspond in fact to hyper-surfaces implementing the Dirichlet boundary conditions in their transverse space.
From the study of the interplay of world-sheet parity and T-duality we can understand how to orientifold a D-brane model.

For the sake of simplicity, let us consider a ten-dimensional model having one compact spatial direction (say $9$)
and we consider T-duality with respect to it.
T-duality amounts to a space-time parity operator acting only on the right-moving sector.
So the T-dual of \eqref{modomodo} is
\begin{eqnarray}\label{tidu}
 &X'^M(z,\overline{z}) = T_9 [X^M(z,\overline{z})]  &= X_L^M(z) + X_R^M(\overline{z})\ \ \   \text{for}\ M\neq 9\\
 &X'^9(z,\overline{z}) = T_9 [X^9(z,\overline{z})]  &= X_L^9(z) - X_R^9(\overline{z})\ \ \   
\end{eqnarray}
T-duality has introduced an asymmetric treatment with respect to the right and left sectors and 
the orientifold operator $\omega$ does not commute with $T_9$.
Asking the commutation between the to operators to hold, we must specify the definition of orientifold in the T-dualized perspective as follows: 
we regard the action of $\Omega$ on the generic mode $Y$ as the joint effect on the function $Y$ ``itself'' and on its arguments
(i.e. the interchange of holomorphic and anti-holomorphic dependence due to $\omega$), namely
\begin{equation}
\Omega[Y(z,\overline{z})] = Y_\Omega'(\overline{z},z) 
\end{equation}
In this fashion, we can assign the following transformation rules:
\begin{equation}
 X^M_\Omega = X^M 
\end{equation}
and
\begin{eqnarray}
& X'^M_\Omega &= X'^M \ \ \ \text{for}\ M\neq 9\\
& X'^9_\Omega &= -X'^9
\end{eqnarray}
Thus, in the T-dual framework, the orientifold operator has to be defined as the product of the world-sheet parity $\omega$ and the parity operator ${\cal I}_9$ in the $9$ direction,
\begin{equation}
 \Omega'= \omega {\cal I}_9
\end{equation}
In this way we have
\begin{equation}
 T_9 \Omega = \Omega' T_9 \ .
\end{equation}
The T-duality in the $9$ direction can be thought of as adding a D$8$ brane extended along the directions $M\neq9$.

Extending what we have described explicitly, it is not difficult to understand that in a setup containing D$3$-branes the orientifold operator has to be
\begin{equation}
 \Omega = \omega \,(-1)^{F_R}\,{\mathcal I}_{456789}
\label{OrientOperator}
\end{equation}
where ${\mathcal I}_{456789}$ is the parity operator of the internal space and $F_R$ represents the right-moving fermion number;
some comment is still necessary about the factor $(-1)^{F_R}$.
In \eqref{tidu} we have seen that the T-duality operation along the $9$ direction reverse the sign of the bosonic right-moving modes;
relying on superconformal invariance we have an analogous behavior also for the right-moving fermionic partners, so
\begin{equation}
 \tilde{\psi}^9(\overline{z}) = - \tilde{\psi}^9(\overline{z})\ .
\end{equation}
As noted in \cite{polchinski2005string} (to which we refer for further details), this implies that the chirality of the right-moving
Ramond sector is reversed by the $T_9$ duality operation; indeed, the associated raising and lowering operators 
\begin{equation}\label{chiro}
 \tilde{\psi}^8 + \im \tilde{\psi}^9\ \ \ \leftrightarrow \ \ \ \tilde{\psi}^8 - \im \tilde{\psi}^9\ ,
\end{equation}
are interchanged under $T_9$. This chirality change explains that from $T$-duality with respect to one (or any odd number of directions)
relates Type IIB string models with Type IIA.
If we now consider doing another $T$-duality operation also with respect to the $8$ direction (supposing it to be compact)
we have that under $T_8 T_9$ the right-moving fermions transforms as
\begin{equation}\label{chiro2}
 \tilde{\psi}^8 + \im \tilde{\psi}^9\ \ \ \leftrightarrow \ \ \ -\tilde{\psi}^8 - \im \tilde{\psi}^9\ .
\end{equation}
Collecting what just said, we have that two $T$-duality connect Type IIB again with Type IIB, but the right moving fermions in the $T$-dualized
directions acquire a minus sign. It is not difficult to imagine that if we $T$-dualize with respect to all the six internal coordinates,
we have that all the right-moving fermions along the internal directions acquire a minus.
Therefore, to have an orientifold operator commuting with $T$-duality we must insert a factor compensating for this additional signs;
in the case of $T$-duality in the whole internal space we have just to count the number of right-moving fermions and add a corresponding number of minus sign;
this is precisely the effect of the factor $(-1)^{F_R}$ in the definition of the orientifold operator.

Pictorially, in the framework of our $\mathbb{C}^3/\mathbb{Z}_3$ orbifold model we add also an orientifold 
O$3$ plane whose world-volume coincides with the four space-time directions of the D$3$-brane world-volume.
In other terms, the O$3$ is on top of the D$3$ stack.

The orientifold acts on the Chan-Paton structures as well and its effect on the generic CP factor $C$ is given by
\begin{equation}
 \label{omegaCP}
\Omega~:~~C~\to~\gamma(\Omega)\,C^{T}\,\gamma(\Omega)^{-1} \ ,
\end{equation}
where $\gamma(\Omega)$ is an invertible matrix representing $\Omega$ on the CP space.
Notice that the CP factor is transposed because of the world-sheet parity which interchanges the open-string endpoints.
The concurrent presence of an orbifold and an orientifold projection requires that their representations
on the CP indexes satisfy the following consistency condition \cite{Gimon:1996rq,Douglas:1996sw} (it is discussed
in Subsection \ref{commutcond}):
\begin{equation}
 \gamma(h)\, \gamma(\Omega) \,\gamma(h)^T = \gamma(\Omega)
\label{consistency}
\end{equation}
where $h$ indicates the generic element of the orbifold group.
The condition \eqref{consistency} must hold for both the D$3$ and the D$(-1)$ CP structures.
Observe that \eqref{consistency} amounts to requiring the commutation of the orientifold
and orbifold operations in any CP sector.
As we will see explicitly in Appendix \ref{shape}, according to \eqref{consistency}, the matrix $\gamma(\Omega)$ can be chosen to be either symmetric or
antisymmetric. 
Our choice will be anti-symmetric for the D$3$ branes, so the matrix $\gamma_-(\Omega)$ representing the orientifold on the D$3$ CP factor is
\begin{equation}
 \label{gamma-}
\gamma_-(\Omega) = 
\begin{pmatrix}
\epsilon&0&0\cr  
0&0&\one_{N_2}\cr
0&-\one_{N_2}&0
\end{pmatrix}
\end{equation}
where $\epsilon$ represents the $N_1\times N_1$ totally anti-symmetric matrix obeying $\epsilon^2=-1$. 
Observe that, as a consequence of the skew shape of $\gamma_-(\Omega)$, we must choose $N_1$ to be even and $N_2=N_3$.

As described in \cite{Gimon:1996rq,Argurio:2007vqa}, for consistency reasons,
the orientifold representation on the instantonic CP structure (i.e. the $(k_1+k_2+k_3)\times(k_1+k_2+k_3)$ matrices)
has to be chosen with the opposite symmetry property with respect to $\gamma_-(\Omega)$%
\footnote{This consistency requirement arises from the study of the string modes with Neumann-Dirichlet mixed boundary conditions
stretching between the two kinds of branes (D$3$ and D$(-1)$ in our case); in particular, from the analysis of the half-integer modes expansions of such strings
and their vertex operators, it can be observed that the orientifold eigenvalue on the members of the mixed sector acquires
an extra minus sign that has to be compensated choosing opposite symmetry properties for the orientifold matrix on the Chan-Paton structures associated 
to the two different kinds of branes. We refer the reader to \cite{Gimon:1996rq} for further details.}.
Since we have chosen $\gamma_-(\Omega)$ to be anti-symmetric, we must choose the orientifold matrix on the instanton to be
symmetric; we indicate it with $\gamma_+(\Omega)$. 
Taking a generic $(k_1+k_2+k_3)\times(k_1+k_2+k_3)$ instanton CP factor $C$, we have
\begin{equation}
 \label{omegaCPinst}
\Omega~:~~C~\to~\gamma_+(\Omega)\,C^{T}\,\gamma_+(\Omega)^{-1}
\end{equation}
where
\begin{equation}
 \label{gamma+}
\gamma_+(\Omega) = 
\begin{pmatrix}
\one_{k_1}&0&0\cr  
0&0&\one_{k_2}\cr
0&\one_{k_2}&0
\end{pmatrix}~.
\end{equation}
Observe that, differently from $N_1$ (which is associated to a totally anti-symmetric matrix), 
$k_1$ does not need to be even since the identity $\one_{k_1}$ can have any dimensionality.
However, since the $(23)$ block has skew structure, $k_2$ and $k_3$ have to be equal has already happened for $N_2$ and $N_3$.

Recalling the explicit form of the orbifold matrix (\ref{d3cp}) and \eqref{iscp},
it is not complicated to check that our choices $\gamma_{\pm}(\Omega)$ satisfy the orbifold/orientifold
commutation condition (\ref{consistency}) both for D$3$ and D$(-1)$ CP indexes.

\subsection{Orbifold-orientifold commutation condition}
\label{commutcond}

Let us consider an Abelian orbifold group $G$; the representation of $G$ on the Chan-Paton factors involves commuting matrices $\gamma(g)$, then in particular we have
\begin{equation}
[\gamma(g),\gamma^{-1}(g)] = 0
\end{equation}
The orientifold action on the Chan-Paton space is represented with a matrix $\gamma(\Omega)$.
The simultaneous presence of an orbifold and an orientifold projection posits a consistency question:
we need to require that the two projections commute. 
Let us show a quick argument to support this consistency requirement.
Suppose that $P_1$ and $P_2$ represent two projection operators 
that, by definition, are idempotent (i.e. $P_i^2=P_i$). 
Furthermore, assuming that they do not commute, we have:
\begin{equation}\label{PP}
  P_1 P_2 \neq P_2 P_1\ .
\end{equation}
Multiplying both members of \eqref{PP} by $P_1$ both from the left and from the right and using the idem-potency property we obtain
\begin{equation}
  P_1 P_2 P_1 \neq P_1 P_2 P_1 \ ,
\end{equation}
which is clearly inconsistent.

The application of the orientifold and orbifold projections in the two possible orders on a prototype
modulus $\lambda$ gives explicitly:
\begin{eqnarray}
&\Omega g :& \lambda \rightarrow \gamma(g) \, \lambda \, \gamma^{-1}(g) \rightarrow 
\gamma(\Omega) \, (\gamma^{-1})^T(g) \, \lambda^T \, \gamma^T(g) \, \gamma^{-1}(\Omega)\\
&g \Omega  :& \lambda \rightarrow \gamma(\Omega) \, \lambda^T \, \gamma^{-1}(\Omega) \rightarrow 
\gamma(g) \, \gamma(\Omega) \, \lambda^T \, \gamma^{-1}(\Omega) \, \gamma^{-1}(g)
\end{eqnarray}
Therefore we have the following commutation condition:
\begin{equation}
[\Omega,g] = 0 \Rightarrow \left\{
\begin{array}{l}
	\gamma(\Omega)(\gamma^{-1})^T(g)=\gamma(g)\gamma(\Omega)\\
	\gamma^T(g)\gamma^{-1}(\Omega)=\gamma^{-1}(\Omega)\gamma^{-1}(g)
\end{array}\right.
\end{equation}
The two equations on the right are equivalent; 
to see this it is sufficient to invert both members of the second equation taking in account that\footnote{Notice that this is always
true
\begin{equation}
 A A^{-1} = 1 \ \ \ \Rightarrow \ \ \
 (A^{-1})^T A^T = 1 \ \ \ \Rightarrow \ \ \
 (A^{-1})^T = (A^{T})^{-1}
\end{equation}} 
\begin{equation}
(\gamma^{-1})^T = (\gamma^T)^{-1}
\end{equation}
Moreover, the two equations are equivalent to the consistency condition
\begin{equation}\label{consistencia}
\gamma(g)\gamma(\Omega)\gamma(g)^{T} = \gamma(\Omega)
\end{equation}
as given in \cite{Gimon:1996rq}.

\section{BRST structure and localization}
The ADHM construction parametrizes the instanton moduli space with a redundant set of variables which are 
restrained to satisfy the ADHM constraints \eqref{constra}.
In the string framework, the content of the construction emerges from the open strings attached to the D-instantons;
the ADHM constraint results from the equations of motion of the instanton moduli.
Notice that, as the moduli are indeed non-dynamical degrees of freedom, the equations of motion are in fact
algebraic relations.
The quantum treatment of a system described by a redundant set of variable appropriately constrained is of course a
central problem in theoretical physics in general.
The gauge fixing question presents in these terms.

The mathematical tools that we employ in our instanton computations have in fact a stringent analogy with the 
modern BRST gauge fixing approach.
More specifically, an appropriate combination of the supersymmetry charges defines an anti-commuting operator under which the theory shows
a well defined BRST structure. 


\subsection{Localization formula}

Consider a manifold $\cal M$ having complex dimension $l$ and assume there is a group $G$ acting on $\cal M$ whose
action is encoded in the field $\xi$ as follows%
\footnote{Here we follow closely what described in \cite{Morales}.},
\begin{eqnarray}
 &\xi &= \xi^m (x) \frac{\partial}{\partial^m x}\\
 &\delta_\xi x^m &= \xi^m(x)
\end{eqnarray}
where $x$ spans the manifold $\cal M$. One can define the \emph{equivariant external derivative}
\begin{equation}
 Q_\xi \doteq d + i_\xi 
\end{equation}
satisfying
\begin{equation}
 Q^2_\xi = d i_\xi + i_\xi d = \delta_\xi
\end{equation}
with $d$ denoting the exterior derivative; $i_\xi dx^i = \delta_\xi x^i$ represents the contraction with the vector $\xi$
and $\delta_\xi$ is the Lie variation along the field $\xi$.

Consider a form $\alpha(x)$ defined on the manifold $\cal M$ that is \emph{equivariantly closed}, i.e. it has null $Q_\xi$-variation,
\begin{equation}
 Q_\xi \alpha = 0\ .
\end{equation}
The \emph{localization theorem} states that the integral of $\alpha$ on $\cal M$ is computable by considering the fixed points $x_0^s$
of the action of $G$ on $\cal M$,
\begin{equation}
 \xi^i(x_0^s) = 0 \ .
\end{equation}
More specifically, we have the \emph{localization formula}
\begin{equation}\label{vitaloca}
 \int_{\cal M} \alpha = (-2\pi)^l \sum_s \frac{\alpha(x_0^s)}{\det^{1/2}Q^2_\xi(x_0^s)}\ ,
\end{equation}
where $Q^2_\xi (x_0^s)$ is the matrix corresponding to the map from and to the tangent space of $\cal M$ induced by the vector field $\xi$,
\begin{equation}
 {Q^2}^i_j = \partial_i\xi^j : \bm{T}[{\cal M}] \rightarrow \bm{T}[{\cal M}]
\end{equation}
Intuitively, the localization formula \eqref{vitaloca} can be thought of in analogy to the integral of a total derivative;
this integral receives contribution only at the boundary or in the presence of singular source/pit points.
It is possible to generalize the localization formula to the case where $\cal M$ is a super-manifold;
the generalization leads to a result analogous to \eqref{vitaloca} where the determinant is promoted to a super-determinant.
To have further details we refer the reader to \cite{Bruzzo:2002xf}.

\section{Graviphoton background}
In the framework of the multi-instanton equivariant calculus,
an essential ingredient is the so called $\Omega$-deformation (here $\Omega$ is not to be confused with the orientifold operator!).
The moduli action is deformed by a U$(1)\times$U$(1)$ transformation which acts on some of the moduli
and preserves the ADHM constraints.
Such deformation represents a necessary step to perform the actual computation of
the instanton partition function.
The U$(1)\times$U$(1)$ deformation is parametrized
by a single phase $\epsilon$; the two U$(1)$ transformations are complex conjugate (so, explicitly, $e^{\im\epsilon}$ and $e^{-\im\epsilon}$)
and actually not independent. They act respectively on the chiral and anti-chiral spinor indexes of the ``internal Lorentz group''%
\footnote{Henceforth, with \emph{internal Lorentz group} we indicate the SO$(4)\sim\text{SU}(2)\times\text{SU}(2)$ associated to the rotations
with respect to the (real) internal coordinates labeled with $4,5,6,7$.}.

The $\epsilon$-deformation plays the r\^{o}le of a background regulator and has a clear interpretation in terms of string modes.
Indeed, in \cite{Billo:2006jm} it has been shown explicitly that the deformation is equivalent to considering a constant but non-null 
Ramond-Ramond closed-string background associated to the self-dual part of the graviphoton field $\cal F$. 
In this picture, the $\epsilon$ parameter represents the VEV of the graviphoton itself
and the effect of the graviphoton regulation is interpretable as the introduction
of a constant curvature in the ambient ten-dimensional space.
On the computational level, the effects of the graviphoton on the instanton moduli action are obtained studying
mixed open/closed disk amplitudes (see \cite{Billo:2006jm} for details).

When regarded simply as a regulator, the graviphoton background is turned on to actually perform the instanton calculations;
the results are eventually considered in the zero-graviphoton limit%
\footnote{Note that, before the zero-graviphoton limit, the terms in the graviphoton represent gravitational corrections to 
the flat-background case.}.
However, since the graviphoton is the field accounting for gravitational interaction in the ambient space,
the terms in the partition function that depend on $\cal F$ describe the gravitational effects on the instanton dynamics.

An alternative interpretation reads the $\epsilon$-deformation 
as arising from a non-trivial metric, called $\Omega$-background, on the instanton moduli space. 
The $\Omega$-background framework agrees with the RR graviphoton interpretation only at linear order in $\epsilon$ and then
it is not able to accommodate the higher gravitational corrections to the instanton action. However, in the literature the 
$\epsilon$-deformation is still often referred to as $\Omega$-background.

\section{Topological twist}
\label{topo}

As firstly noted in \cite{Witten:1988ze}, ${\cal N}=2$ Super-Yang-Mills theory can be reformulated in such a way
that proves suitable for the study of topological and co-homological properties. Indeed, the mathematical structure
emerging from the topological twist constitutes an essential step fro the localization techniques we will employ in the
following section when performing actual instanton computations. 
Let us here describe the precise meaning of topologically twisting our model.

The Lorentz symmetry of the D$3$-branes world-volume, since we adopt Euclidean metric,
is represented by proper four-dimensional rotations, i.e. to the group SO$(4)$.
At the level of the algebra, we have that SO$(4)\sim\text{SU}(2)\times\text{SU}(2)$
corresponding to the quaternionic expression of the SO$(4)$ vector indexes. 
Customarily, the two SU$(2)$ factors are denoted with SU$(2)_L$ and $\text{SU}(2)_R$ (L means left and R right)
and to them we associate the indexes $\alpha$ and $\dot{\alpha}$ respectively.

The internal space, orthogonal to the D$3$-branes, is constituted by $6$ real directions which we have parametrized with
three complex coordinates \eqref{zs}.
The orbifold transformation on the internal space given in \eqref{gorb}
leaves the $z^3$ direction invariant while $z^1$ and $z^2$ transform non-trivially.
Let us concentrate on the latter couple of coordinates; as they correspond to $4$ real directions,
the original rotation symmetry of the theory contains an SO$(4)$ factor associated to them;
we will refer to this SO$(4)$ as the \emph{internal Lorentz group}. Again, the internal Lorentz group can be ``split'' into two SU$(2)$ factors
that correspond to two indexes which will be denoted with $a$ and $\dot{a}$.

The topological twist that we consider in our analysis,
consists in the substitution of the original Lorentz group $\mathrm{SU}(2)_L\times\mathrm{SU}(2)_R$
with the twisted version $\mathrm{SU}(2)\times\mathrm{SU}(2)'$ where $\mathrm{SU}(2) = \mathrm{SU}(2)_L$
and $\mathrm{SU}(2)'=\diag\big(\mathrm{SU}(2)_R,\mathrm{SU}(2)_I\big)$.
$\mathrm{SU}(2)_I$ represents the SU$(2)$ factor of the internal Lorentz group associated to the index $a$%
\footnote{The identification of $\mathrm{SU}(2)_I$ with the SU$(2)$ labeled by $a$ is a matter of choice;
we could as well have taken the other SU$(2)$ factor associated to $\dot{a}$.}.
As we will see explicitly, such identification allows us to define a peculiar linear combination of the supersymmetry charges
that, in turn, reveals a significant BRST structure of the model.

One could wonder whether the topological twist represents actually a constraint of the theory affecting the dynamics;
we refer to \cite{Bianchi:2007ft} for the details, however, let us comment the effect of the topological twist on our model.
We are identifying two SU$(2)$ symmetries of the original theory, reducing in fact the 
total symmetry group. As will be shown explicitly%
\footnote{See Equation \eqref{eta} and the following comments.}, from the point of view of the string modes in our model, 
the topological twist amounts to a simple reorganization of the fields. 
Indeed, twisting the field content of our model does not reduce the number of degrees of freedom and,
as far as our analysis is concerned, we can think to the topological twist as a sort of change of basis for the fields.

\chapter{Stringy Instantons}
\label{mio1}
After the discovery of the D-branes , the string formalism has proved to be a particularly natural environment
to study the non-perturbative sector of supersymmetric gauge theories.
In the previous sections, we have already stressed the technical possibility of describing throughly and effectively all
the features of the field theoretical instanton calculus.

One crucial step forward consists in concentrating on the string
models used for instantonic calculations and generalizing them.
In other terms, one can assume a string perspective and study D-brane setups generalizing the models that reproduce ordinary
instanton calculus.
We will rely on the details of the possible generalizations through the present chapter, but let us anticipate
that the string formalism context allows us to produce non-perturbative effects of new type.
They are commonly referred to as \emph{exotic} or \emph{stringy}.
Although they do affect the low-energy effective gauge theory corresponding to the D-brane model under study,
the exotic effects are in general not interpretable from a purely field theoretical viewpoint%
\footnote{\label{piede} The D$7/$D$(-1)$
exotic instantons in eight dimensions have a natural interpretation as the zero-size limit of ordinary instantons
of the eight-dimensional gauge theory living on the D$7$-branes. 
Some comments are in order. 
In eight dimensions the self-duality (or anti-self-duality) condition is not imposed
on the field-strength $F$ but instead on the tensor $F\wedge F$.
In the eight-dimensional case the solutions of the self-duality condition for $F$ are not solutions of the equations
of motion (as instead occurs in four dimensions).   
More precisely, if we consider an instantonic configuration (i.e.
solving the duality condition) we have that the corresponding equations of
motion instead of being zero are proportional to $(d - 4)R=4R$ where $R$ represents
actually the modulus associated to the ``size'' of the instanton configuration. 
It is then natural to see that in the $R \rightarrow 0$ limit the self-dual configurations solve the equations of
motion as well. 
In four dimensions (i.e. $d=4$) the equations of motion are again proportional to $(d-4)R$ which now is of course identically zero;
the exotic configurations have no clear field theoretical interpretation in the four-dimensional case, see \cite{Billo:2009gc}}.
Indeed, they possess a stringy nature; for instance, they introduce in the low-energy theory
non-perturbative effects which can depend explicitly on the string scale $\alpha'$.

The string framework offers both a natural and effective tool 
to treat ordinary instanton computations and possible generalizations.
The D-brane approach allows us to explore the instantonic
non-perturbative sector in an unprecedentedly wide and deep fashion.
It is even tempting to regard the 
D-brane description as the definitely environment for instantons in general.

\section{Motivations}
\subsection{Theoretical Significance}

The theoretical interest on stringy instantons is especially related to the study of the vacuum structure of both supersymmetric
gauge theories and of the associated string models themselves; the main attention is tributed to their 
important effects regarding supersymmetry breaking and moduli stabilization \cite{Camara:2007dy,Blumenhagen:2007sm}.
From a purely string theory point of view, the ordinary and exotic configurations are on the same footing
and the difference between the corresponding D-brane models are simply technical.
All the investigations aimed at the study of the string vacua encompass necessarily the entire panorama of non-perturbative features, so
the stringy instantons as well.

\subsection{Phenomenological Interest}

At the outset we have to stress again that in this first part of the thesis the focus is on the context of ${\cal N}=2$ supersymmetric theories.
The distance to current real experiments is still significant; supersymmetry itself has still to be observed.
Given this premise, it is nevertheless physically crucial to meditate on the actual phenomenological value of the stringy instanton calculus.
Maintaining a cautious attitude\footnote{In the scientific jargon, this cautious phenomenological attitude is usually referred to as
\emph{semi-realistic}.}, it is important to study in detail specific models in which new kind of exotic effects could arise.
Even though the particular models themselves will turn out not to be realized in Nature\footnote{Or, more likely,
if any experimental evidence belongs to a more or less remote future.}, they furnish the inspiring theoretical proof that 
dynamics in extra dimensions can generate and accommodate essential phenomenological features like Yukawa couplings in GUT models 
\cite{Blumenhagen:2007zk}, right-handed neutrino masses \cite{Blumenhagen:2006xt,Ibanez:2006da,Ibanez:2007rs} 
and see-saw parameters.
Indeed, the main phenomenologically appealing characteristic of stringy instantons relies in the introduction of a new scale 
(related to the string scale $\alpha'$) in the low-energy theory; this novel scale could offer the framework for solving naturalness 
problems or hierarchy questions\footnote{In the exotic cases obtained by compactifying configurations living in higher
dimensionalities, the compactification scale can as well enter into the game.}.



\section{Stringy instanton salient features}
The ordinary vs. stringy classification of instantons has a precise meaning in relation to the features of the
associated D-brane models. 
In an ordinary instanton configuration, the gauge and instanton branes%
\footnote{Gauge and instanton branes has been defined in \ref{dinst}.}
share all the geometric and symmetry characteristics in the internal space.
Instead, the exotic configurations have instanton branes that because of a different geometrical arrangement or because of different symmetry 
properties, have a different internal-space behavior with respect to the gauge branes.

Consider a generic D-brane model involving D$(3+p)$ and D$(p-1)$ branes being respectively the gauge and the instanton branes.
If $p>1$ we need to compactify the extra dimensions in order to obtain an effective four-dimensional low-energy field theory.
The compactification would require the presence, in the internal manifold, of a compact $p$-cycle $\cal C$ around which we wrap the
gauge D$(3+p)$ branes. 
In this case, an ordinary instanton is associated to Euclidean D$(p-1)$ branes completely wrapping the same internal cycle.

We have two main ways to modify the internal behavior of instanton branes.
We could consider the possibility of wrapping the D$(p-1)$ branes on another different $p$-cycle ${\cal C}'\neq {\cal C}$ 
or we can assign different symmetry properties between the gauge and instanton branes.
This second possibility will be the one considered in the following, indeed we will associate gauge and instanton branes to
distinct representations of the background orbifold action (see \ref{OrbiOrie}).

Another characteristic feature of stringy instantons, as opposed to ordinary ones, is that they lack
the bosonic moduli describing the instanton size%
\footnote{As as a comment, remember that in the eight-dimensional case mentioned
in the footnote \ref{piede} of the present chapter, the stringy instantons are interpreted as instanton configuration in the zero-size limit (sometimes referred to as \emph{small instantons}).}.
This is a direct consequence of the different internal behavior between instanton and gauge branes.
Some fermionic zero modes are then difficult to saturate. 
In the moduli integral giving rise to the instanton partition function an unsaturated Grassmann variable leads inevitably to an overall vanishing result.
Stringy instantons can produce effects only within systems in which the extra fermionic zero modes are either projected away by orientifold projections 
\cite{Argurio:2007vqa,Bianchi:2007wy} (as in our case) or lifted by means of background fluxes 
\cite{Blumenhagen:2007bn,Billo:2008sp,Billo:2008pg} or also with other mechanisms such as those described in \cite{Petersson:2007sc,Bianchi:2009bg}.

Eventually, another stringy instanton peculiarity resides in the instanton group structure.
In Section \ref{sec:stringy} we will directly see that the SU$(2)$ stringy instantons of the model at hand enjoy an SO$(k)$ symmetry structure.
Being this related to the structure of the quiver diagram after the orientifold projection (i.e. with the effects of the orientifold on the 
group structures on the various stacks of branes), the occurrence of orthogonal instanton group structure for unitary gauge theories arises also in 
the generalization of our model to SU$(N>2)$%
\footnote{Some more comments about other gauge groups and the corresponding instanton groups can be found in Section \ref{comme}.}.
This exotic result is in contrast with its ordinary instanton counterpart; in SU$(N)$ gauge theories the ordinary instanton group is in fact unitary.

\section{Stringy instantons in \texorpdfstring{${\mathcal N}=2$}{} theories}
\subsection{Localization techniques for exotic instantons}

The localization techniques are pivotal in the framework of D-brane instanton computations.
In the case of ordinary instanton configurations, when technically available, one can check
the results obtained with string tools against ordinary, field-theoretical computations \cite{Bruzzo:2002xf}.
By definition, the same kind of checks cannot be performed for exotic instantons because, in general, they have no
purely field-theoretical counterpart.

There are instances in which the difficulty in checking the applicability of localization techniques to exotic instantons
can however be surmounted by means of dualities. 
A significant example is the employment of the heterotic/Type $\text{I}'$\footnote{Type $\text{I}'$ 
string theory (sometimes referred to as Type IA as well) is the
T-dual of Type I theory on a ten-dimensional space-time in which there is (at least) one compactified
dimension, \cite{Becker:2007zj}
One can think as follows: Type IIA and Type IIB are T-dual of each other; Type I is an orientifold
projection of Type IIB; Type $\text{I}'$ is the T-dual of Type I and can be thought of as an orientifold
projection of Type IIA.}. duality, \cite{Billo:2009di,Fucito:2009rs,Billo':2010bd}.
This duality has been proposed in \cite{Polchinski:1995df} following a correspondence in the spectrum of the two
theories involved\footnote{To find more details on the checks of the duality itself, look at the references contained in \cite{Billo:2009di}.}.
On the heterotic side of the duality there are some quartic interactions
that do not receive any corrections beyond the $1$-loop order. This happens because
they are protected by supersymmetry.
The same couplings can be obtained from the dual Type $\text{I}'$ theory but in this context they receive
both perturbative and non-perturbative contributions.
The Type $\text{I}'$ non-perturbative computations are performed employing localization techniques and are then
confronted with the corresponding heterotic perturbative results.
This furnish a non-trivial test for the extension of localization methods to the stringy instanton calculus.

\section{Description of the \texorpdfstring{$\text{D}3/\text{D}(-1)$}{} stringy instanton model}
At the outset, it should be underlined that the model under consideration is the first setup containing stringy instantons effects 
directly (i.e. without the need of compactifications) in a four-dimensional field theory. 
Indeed the field theory is defined on the four-dimensional world-volume of the D$3$-branes.
In the preceding literature about stringy instantons, the setups always involved gauge theories defined on the eight-dimensional world-volume 
of D$7$-branes.

We analyze the low-energy gauge theory describing a stack of fractional D$3$-branes in the already described $\mathbb{C}^3/\mathbb{Z}_3$ orientifold background
of Type IIB superstring theory preserving ${\cal N}=2$ supersymmetry in four dimensions.
The strategy we follow is analogous to the one presented in \cite{Argurio:2007vqa} for the one-instanton case.
Namely, we study a system of fractional of D$3$-branes realizing an ${\cal N}=2$ SU$(N)$ gauge theory containing a hypermupltiplet
(i.e. the ``matter'' content of the model) in the symmetric representation of the gauge group.
The instantons are encoded in a second stack of D$(-1)$-branes populating a different node of the quiver diagram with respect to the D$3$-branes.
Sitting on different nodes, the gauge and instanton branes belong to different irreducible representations of the orbifold group.
In our model, this difference in the orbifold representation is the key feature which makes the instanton stringy.

The fermionic zero modes arising from the open strings with mixed Neumann-Dirichlet boundary conditions are projected out by the orientifold.
The removal of such zero-modes leads to non-vanishing results for the moduli integral and therefore to non-null stringy effects on the low-energy prepotential.

\subsection{D\texorpdfstring{$3$}{}-branes at the \texorpdfstring{$\mathbb{C}^3/\mathbb{Z}_3$}{} orbifold singularity (field content)}
\label{3sing}

The presence of the D$3$-branes brakes the ten-dimensional SO$(10)$ Lorentz\footnote{Remember that we have Euclidean signature, so
``Lorentz'' transformations are indeed proper rotations.} group splitting it to SO$(4)\times$SO$(6)$.
The fields carrying representations of the Lorentz group split accordingly. 
Moreover, notice that since the translation invariance along the direction which are orthogonal to the D$3$ branes are broken as well,
the original ten-dimensional Poincar\'e invariance is reduced to the four-dimensional Poincar\'e invariance of
the D$3$ world-volume.

The ten-dimensional vector $A_M$ splits into an SO$(4)$ vector $A_\mu$ and a SO$(6)$ vector, however the latter is
seen as a collection of six scalar from the four-dimensional theory point of view.
An anti-chiral ten-dimensional spinor $\Lambda$ decomposes as follows:
\begin{equation}
 \Big(\Lambda^{\alpha A}\,,\,\Lambda_{\dot{\alpha} A}\Big) \ ,
\label{decospin}
\end{equation}
where $\alpha$ and $\dot\alpha$ are respectively chiral and anti-chiral spinor indexes of SO$(4)$;
the lower and upper indexes $A$ are respectively chiral and anti-chiral spinor indexes of SO$(6)$%
\footnote{The original spinor has $2^{10/2}=32$ components which are now organized as $2\times 4 + 2\times 4 = 32$.}.

The field content of the gauge theory describing at low energy the fractional D$3$-branes
emerges from the orbifold/orientifold projection.
To study the orbifold transformations of a generic field, we have to consider the orbifold generator \eqref{gorb1}
in the representation of the field under consideration.
Once we know the transformation of the fields we project away the non-invariant components;
on a computational level this consists in retaining only the invariant components that satisfy the so-called orbifold/orientifold conditions.

To study the spinor orbifold transformations we have to use the SO$(6)$ spinor weights and then, from (\ref{gorb1}),
we obtain
\begin{equation}
 \label{gorbspin}
g~:~~\begin{pmatrix}\Lambda^{\alpha ---} \cr  \Lambda^{\alpha ++-} \cr \Lambda^{\alpha +-+}
\cr \Lambda^{\alpha -++} \end{pmatrix}
~\to~
\begin{pmatrix}\Lambda^{\alpha ---}\cr  \Lambda^{\alpha ++-} \cr \xi\,\Lambda^{\alpha +-+}
\cr \xi^{-1}\,\Lambda^{\alpha -++} \end{pmatrix}~~~\mbox{and}~~~
\begin{pmatrix}\Lambda_{\dot\alpha +++} \cr  \Lambda_{\dot\alpha --+} \cr \Lambda_{\dot\alpha -+-}
\cr \Lambda_{\dot\alpha +--} \end{pmatrix}
~\to~
\begin{pmatrix}\Lambda_{\dot\alpha +++}\cr  \Lambda_{\dot\alpha --+} \cr \xi^{-1}
\,\Lambda_{\dot\alpha -+-}
\cr \xi\,\Lambda_{\dot\alpha +--} \end{pmatrix}~.
\end{equation}
Since only half of the spinor components are left invariant by the orbifold transformations, 
the corresponding orbifold projection yields $\mathcal N=2$ supersymmetry.

In the bosonic sector, the invariance requirement translates into the following system of conditions:
\begin{subequations}
\begin{align}
 \mathbf{A}_\mu &= \gamma(g)\, \mathbf{A}_\mu \,\gamma(g)^{-1}~,~~~\phantom{\!(\xi)^I}
\mathbf{A}_\mu =- \gamma_-(\Omega) \,\big(\mathbf{A}_\mu\big)^T \,\gamma_-(\Omega)^{-1}~,
\label{vector_cond} \\
\mathbf{\Phi}^I &= (\xi)^I\,\gamma(g)\, \mathbf{\Phi}^I \,\gamma(g)^{-1}~,~~~
\mathbf{\Phi}^I =- \gamma_-(\Omega) \,\big(\mathbf{\Phi}^I\big)^T \,\gamma_-(\Omega)^{-1}~,
\label{scal_cond}
\end{align}
\end{subequations}
where $\bm{A}_\mu$ with $\mu=0,...,3$ is the four-dimensional vector and $\bm{\Phi}^I$ with $I=1,2,3$ denotes the complexified internal scalars
(associated to the internal complex coordinates defined in (\ref{zs})).
The orbifold conditions, encoded in the left equations of \eqref{vector_cond} and \eqref{scal_cond}, impose that $\mathbf{A}_\mu$ 
and $\bm{\Phi}^3$ have only diagonal entries while $\mathbf{\Phi}^1$ and $\mathbf{\Phi}^2$ must have only off-diagonal components, more 
specifically we have the structure \eqref{Phi1Phi2}.

The field content emerging from the orbifold projection is further restricted by the orientifold conditions.
More precisely, we have that 
\begin{equation}
 A_{\mu(11)} = \epsilon \big(A_{\mu(11)}\big)^T\!\epsilon\ , \ \ \ \ A_{\mu(22)}=-\big(A_{\mu(33)}\big)^T\ ,
\end{equation}
and
\begin{equation}
 \Phi^3_{(11)} = \epsilon \big(\Phi^3_{(11)}\big)^T\!\epsilon\ , \ \ \ \ \Phi^3_{(22)}=-\big(\Phi^3_{(33)}\big)^T
\end{equation}
From the conditions on the gauge vector field entries, we have that the gauge group
of the low-energy theory is $\mathrm{USp}(N_1)\times \mathrm{U}(N_2)$.
As we have just observed, ${\mathcal N}=2$ supersymmetry is preserved and
$\mathbf{A}_\mu$ and $\mathbf{\Phi}^3$ can be interpreted as the bosonic part
of the ${\mathcal N}=2$ adjoint vector multiplet.

Let us underline that the representation of the orientifold action on the CP structure lead us
to an identification of the nodes $2$ and $3$ of the quiver diagram.
Whenever we use the notation $A_{\mu(22)}$ and $A_{\mu(33)}$ (as well as 
$\Phi^3_{(22)}$ and $\Phi^3_{(33)}$) as if they were distinct it must be understood
that they are in fact identified. 
Hence it is convenient to introduce the compact notation
$A_{\mu(22)}\equiv A_{\mu}$ and
$\Phi^3_{(22)}\equiv\Phi$.

The off-diagonal part of the CP dressed scalar field yields the bosonic components of the
matter ${\cal N}=2$ hypermultiplets. 
In particular, the orientifold conditions on the entries of the fields $\mathbf{\Phi}^1$ and $\mathbf{\Phi}^2$
return
\begin{equation}
\Phi^1_{(12)} = -\epsilon \,\big(\Phi^1_{(31)}\big)^T~,~~ 
\Phi^1_{(23)} = \big(\Phi^1_{(23)}\big)^T~,~~ 
\Phi^2_{(13)} = \epsilon \,\big(\Phi^2_{(21)}\big)^T,~~
\Phi^2_{(32)} = \big(\Phi^2_{(32)}\big)^T~.
\label{phi12}
\end{equation}

\begin{figure}
 \centering
 \includegraphics[width=65mm]{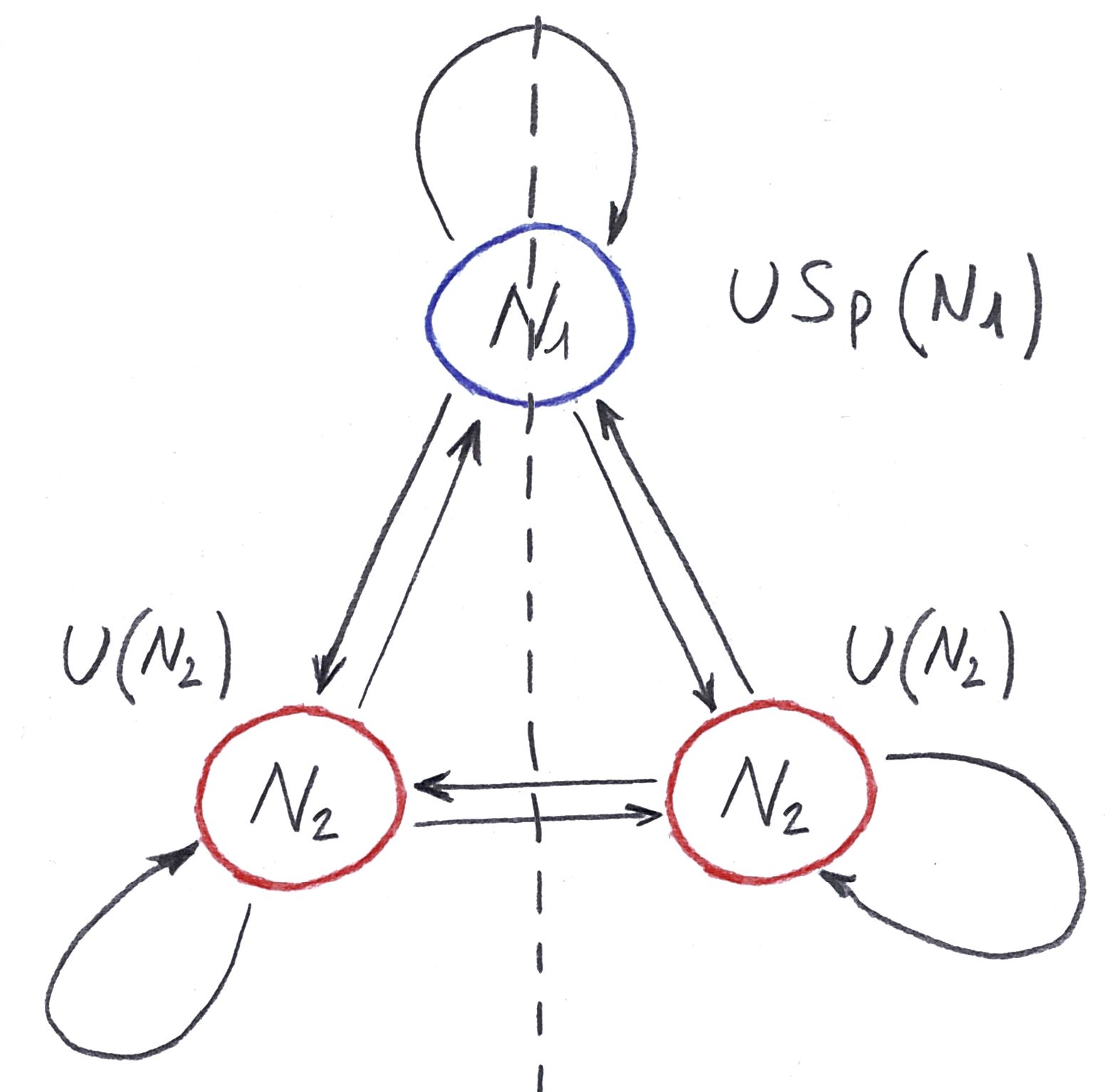}
 \caption{Quiver diagram of the orientifolded $\mathbb{C}^3/\mathbb{Z}_3$ theory. The dashed line represents symbolically the 
``mirror'' identification produced by the orientifold.}
 \label{quiori}
\end{figure}

In depicting the structure of the theory through the quiver diagram \ref{quiori} we 
can observe that any node corresponds to a stack of fractional branes leading to a factor in the composed gauge group
$\mathrm{USp}(N_1)\times \mathrm{U}(N_2)$ (remember that node $2$ and $3$ are identified).
Any oriented open string stretching from a node to another (or better from the stack of branes placed on a node to the stack of branes placed
at another node) possesses a pair of CP indexes transforming respectively in the fundamental and anti-fundamental representations associated to 
the starting and ending stacks. 
Note however that the anti-fundamental representation of USp$(N)$ coincides with the fundamental and that
the orientifold induced identification between nodes $2$ and $3$ of the diagram connects the fundamental representation
of U$(N_2)$ to the anti-fundamental representation U$(N_3)$ and vice versa.
From this considerations emerges the behavior of the fields listed in Table \ref{tab:matter}.
\begin{table}[ht]
\begin{center}
\begin{tabular}{|c||c|c|}
\hline
\phantom{\vdots}
field &$\mathrm{USp}(N_1)$&$\mathrm{U}(N_2)$ 
\\
\hline\hline
$\phantom{\vdots}\Phi^1_{(12)}$ & $\Yfund$ &$\overline{\Yfund}$
\\
$\phantom{\vdots}\Phi^1_{(31)}$ & $\Yfund$ &${\Yfund}$
\\
$\phantom{\vdots}\Phi^1_{(23)}$ & \begin{LARGE}${\cdot}$\end{LARGE} &${\Ysymm}$
\\
$\phantom{\vdots}\Phi^2_{(21)}$ & $\Yfund$ &${\Yfund}$
\\
$\phantom{\vdots}\Phi^2_{(13)}$ & $\Yfund$ &$\overline{\Yfund}$
\\
$\phantom{\vdots}\Phi^2_{(32)}$ & \begin{LARGE}${\cdot}$\end{LARGE} &$\overline{\Ysymm}$
\\
\hline
\end{tabular}
\end{center}
\label{tab:matter}
\caption{Matter content (more precisely, bosonic components of the ${\cal N}=2$ hypermultipltets) and their gauge representations.}
\end{table}

\subsection{D\texorpdfstring{$3$}{}-branes at the \texorpdfstring{$\mathbb{C}^3/\mathbb{Z}_3$}{} orbifold singularity (moduli content)}
\label{sec:dinst}

We turn the attention on the open strings attached to the D-instantons.
We proceed in analogy of the study of the D$3$ brane field content just performed in the preceding section.

The most general instanton configuration corresponds to $k_1$ D$(-1)$ branes lying on node $1$, $k_2$ D$(-1)$'s on node $2$ and $k_3$ D$(-1)$'s on node $3$. 
Again, because of the orientifold induced identifications we must restrict to $k_2=k_3$. 
The CP factor $Y$ of the generic open-string excitation stretching between two D-instantons
is a $(k_1+2k_2)\times(k_1+2k_2)$ matrix.

We organize the Neveu-Schwarz sector of the open strings connecting two instanton branes
in analogy to the sector of strings stretching between D$3$ branes, namely a four-dimensional
vector $a_\mu$ and three complex scalars $\chi^I$.
Notice that the ten real component fields, even if so arranged, are however all on the same footing;
in fact a D$(-1)$ instanton breaks completely the ten-dimensional Poincar\'e invariance and from the point of view
of its world-volume theory the fields $a_\mu$ and $\chi^I$ are non-dynamical scalars.
The $(a_\mu,\chi^I)$ representation is suitable to be interpreted
in terms of the parameters of the ADHM construction.

The orbifold/orientifold conditions for the D$(-1)/$D$(-1)$ moduli are
\begin{subequations}
\label{orien-1}
\begin{align}
 \mathbf{a}_\mu &= \gamma'(g)\, \mathbf{a}_\mu \,\gamma'(g)^{-1}~,~~~~\phantom{\!(\xi)^I}
\mathbf{a}_\mu =+ \gamma_+(\Omega) \,\big(\mathbf{a}_\mu\big)^T \,\gamma_+(\Omega)^{-1}~,
\label{amu} \\
\bm{\chi}^I &= (\xi)^I\,\gamma'(g)\, \bm{\chi}^I \,\gamma'(g)^{-1}~,~~~
\bm{\chi}^I =- \gamma_+(\Omega) \,\big(\bm{\chi}^I\big)^T \,\gamma_+(\Omega)^{-1}~.
\label{chiI}
\end{align}
\end{subequations}
As a confirmation of what was just observed, the orientifold condition for the ``vector'' $a_\mu$
differs from the orientifold condition for the proper vector $A_\mu$ in the sign and is instead formally
identical to the condition for the scalars $\Phi^I$'s.
This is precisely a consequence of the fact that while $A_\mu$ corresponds to Neumann-Neumann string modes,
both $a_\mu$ and $\Phi^I$ correspond to Dirichlet-Dirichlet modes.
Obviously this is true for the $\chi^I$'s as well.

Expliciting the orbifold conditions for $a_\mu$ we have:
\begin{equation}
 \mathbf{a}_\mu = \left(\begin{array}{ccc}
               a_{\mu(11)} & 0 & 0\\
               0 & a_{\mu(22)} & 0\\
               0 & 0 & a_{\mu(33)}
              \end{array}\right) ~,\ \ \ \ \ \ \ \ \ 
\bm{\chi}^3 = \left(\begin{array}{ccc}
               \chi^3_{(11)} & 0 & 0\\
               0 & \chi^3_{(22)} & 0\\
               0 & 0 & \chi^3_{(33)}
              \end{array}\right) \ .
\label{amuchi}
\end{equation}
The orientifold further constrains the degrees of freedom imposing
\begin{equation}
 a_{\mu(11)} = \big(a_{\mu(11)}\big)^T~,~~a_{\mu(22)}=\big(a_{\mu(33)}\big)^T~,~~
 \chi^3_{(11)} = -\big(\chi^3_{(11)}\big)^T~,~~\chi^3_{(22)}=-\big(\chi^3_{(33)}\big)^T~,
\label{amuchi1}
\end{equation}
Proceeding analogously for $\chi^I$ we have that the orbifold requires the following CP structures
\begin{equation}
 \bm{\chi}^1 = \left(\begin{array}{ccc}
               0& \chi^1_{(12)}& 0\\
               0 & 0& \chi^1_{(23)}\\
               \chi^1_{(31)} & 0 & 0
              \end{array}\right) ~,\ \ \ \ \ \ \ \ \ 
\bm{\chi}^2 = \left(\begin{array}{ccc}
               0& 0 & \chi^2_{(13)}\\
               \chi^2_{(21)} & 0 & 0\\
               0 & \chi^2_{(32)} & 0
              \end{array}\right) ~,
\label{chi1chi2}
\end{equation}
which are consequently constrained by the orientifold:
\begin{equation}
 \chi^1_{(12)} = -\big(\chi^1_{(31)}\big)^T~,~~\chi^1_{(23)}=-\big(\chi^1_{(23)}\big)^T~,~~
 \chi^2_{(13)} = -\big(\chi^2_{(21)}\big)^T~,~~\chi^2_{(32)}=-\big(\chi^2_{(32)}\big)^T~.
\label{chichi}
\end{equation}

From the conditions (\ref{amuchi1}) we deduce that the symmetry group (i.e. the gauge group of the zero-dimensional theory) on the D-instantons
is $\mathrm{SO}(k_1)\times\mathrm{U}(k_2)$, where the orthogonal factor refers to the first node
of the quiver and the unitary factor to the remaining two nodes that are identified with each other
under the orientifold projection.
This result is the D$(-1)/$D$(-1)$ counterpart of the D$3/$D$3$ result $\mathrm{USp}(N_1)\times\mathrm{U}(N_2)$.
Observe the important difference that, on node one, the anti-symmetric orientifold representation of the D$3$'s led to a symplectic factor
while the symmetric choice on the D-instantons led to a special orthogonal group.

Before turning the attention to the fermion modes, for which a similar general analysis could be performed,
we specialize the D-brane content of the model we want to study explicitly.
In this way we concentrate on the characteristic features of stringy instantons and simplify the treatment.

\subsection{Brane setups leading to stringy instantons}
\label{sec:stringy}

In the framework of the $\mathbb{Z}_3$ orientifold/orbifold model we have so far described,
we have ordinary or exotic configurations depending respectively on whether or not the D-instanton occupies a quiver node 
populated also by the stack of D$3$ gauge branes.
Nodes $2$ and $3$ are identified by the orbifold; in the $(N_1,N_2)$ notation we consider
a D$3$-brane configuration $(N_1,N_2)=(0,N)$ leading therefore to an SU$(N)$ gauge theory.
In this framework, a D-instanton configuration $(k_1,k_2)=(0,k)$ leads to an ordinary gauge instanton with instanton number $k$ and instanton group
$\mathrm{U}(k)$, see Table \ref{tab:inst}.
Conversely, a D-instanton configuration $(k_1,k_2)=(k,0)$ represents a stringy instanton with charge $k$ and instanton group $\mathrm{SO}(k)$.
Let us observe that the occurrence of an orthogonal instanton symmetry emerging within a theory with a unitary gauge group is an exotic feature.
\begin{table}[ht]
\begin{center}
\begin{tabular}{c|ccc|c|c}
\phantom{\vdots}
&D3's&$\oplus$&D(--1)'s&gauge group&instanton group
\\
\hline
\phantom{\vdots}gauge instantons~~~&$(0,N)$&$\oplus$&$(0,k)$&SU($N$)&U($k$)
\\
\phantom{\vdots}stringy instantons~~~&$(0,N)$&$\oplus$&$(k,0)$&SU($N$)&SO($k$)
\end{tabular}
\end{center}
\label{tab:inst}
\caption{D3 and D(--1) brane configurations and their associated symmetry groups corresponding to
gauge and exotic instantons.}
\end{table} 
In full generality, a D-instanton configuration in our $\mathrm{SU}(N)$ gauge theory can contain any superposition 
of $k_2$ ordinary and $k_1$ stringy instantons. 

Henceforth, the stringy instanton model at the center of our following study is obtained placing a stack of
$k$ D$(-1)$ branes on node $1$ of the quiver and two identified stack of $N$ D$3$-branes placed 
on nodes $2$ and $3$. 
The moduli content arising from such configuration lacks the bosonic moduli describing the instanton charges as 
a consequence of the different orbifold behavior between gauge and instanton branes. 
To study systematically the moduli content emerging from the open string modes having at least one
endpoint on a D$(-1)$ instanton we introduce the following classification:
\begin{itemize}
  \item {\bfseries Neutral Moduli:} associated to the open strings that start and end on a D$(-1)$. 
They encode the zero-dimensional ``gauge theory'' defined on the instanton world-volume.
  \item {\bfseries Charged Moduli:} corresponding to the open strings connecting D$(-1)$ and D$3$ branes.
\end{itemize}
The charged and neutral attributes are named in relation to the gauge group of the D$3$ gauge theory.
Indeed, the charged moduli appear inside interaction terms containing fields of the ${\cal N}=2$ vector multiplet
while the neutral moduli have a life ``on their own''.

\subsection{Neutral Moduli}

Remember that we have specialized the analysis to a D-brane system containing D-instantons only
on node $1$ of the quiver.
Since the neutral moduli are suspended between instanton branes, their CP structure can have
only the component $(11)$ different from zero.
On node $1$ we have arranged $k$ instanton branes, so the $(11)$ CP component is a $k\times k$ matrix.

Recalling the structures resulting from the orbifold projection \eqref{amuchi} and \eqref{chi1chi2},
we see that the complex scalar fields $\chi^1$ and $\chi^2$ do not have a $(11)$ component and
hence are forced to be null in our specific system.
Conversely, $a_\mu$ and $\chi^3$ present this diagonal component and, since the $(11)$
is the only CP entry to be non vanishing, it is convenient to simplify the notation as follows:
\begin{equation}
 a_{\mu(11)}\equiv a_\mu=\big(a_\mu\big)^T~,~~~~\chi^3_{(11)}\equiv \chi=-\big(\chi\big)^T~.
\end{equation}

The fermionic half of the neutral spectrum, after the orbifold projection, 
has only the components corresponding to the indexes $(\alpha---)$, $(\alpha++-)$, $(\dot\alpha+++)$
and $(\dot\alpha--+)$; they are in fact invariant under the action of $\mathbb{Z}_3$.
From the CP structure perspective, this is equivalent to being associated to ``diagonal'' entries.
As a consequence they can present a non-vanishing $(11)$ component and, adopting again a notation
well suited to conform with the ADHM organization of the fields, we denote the fermionic modes
with
\begin{equation}
 M^{\alpha a}\ \   \text{and}\ \ \lambda_{\dot{\alpha}a}\ .
\end{equation}
The upper index $a$ runs over the values $(---)$, $(++-)$, while its lower version assumes the 
values $(+++)$ and $(--+)$.
Again, being associated to open strings stretching from and to the stack of $k$ D-instantons on node $1$, 
also these fermionic modes are represented by $k\times k$ matrices.
The orientifold projection constrains them to be subjected to the following conditions:
\begin{equation}
M^{\alpha a} = +\big(M^{\alpha a}\big)^T 
~,~~~
\lambda_{\dot\alpha a} = -\big(\lambda_{\dot\alpha a} 
\big)^T ~.
\label{mlambda}
\end{equation}
These orientifold constraints follows directly from the choice of
the orientifold matrix $\gamma_+(\Omega)$ whose $(11)$ entry 
is the $k\times k$ identity matrix.
Furthermore, the orientifold operator (\ref{OrientOperator}), when acting on a ten-dimensional spinor, returns
the chirality of the spinor itself in the first four directions; 
this is the reason for the different signs in (\ref{mlambda}).

\subsection{Charged sector}

In the chosen D-brane configuration leading to exotic instantons, the
D-instantons and the D$3$ branes occupy different nodes;
the CP factors for the charged moduli have therefore non-vanishing entries
only among the off-diagonal components. 
In terms of orbifold transformation, being out of the diagonal means transforming non-trivially
under the action of the orbifold matrices $\gamma(g)$ and $\gamma'(g)$ given in \eqref{d3cp} and \eqref{iscp}.
More precisely the $3/(-1)$ strings have the following CP structure:
\begin{equation}
\begin{pmatrix}
\,0\,&\,0\,&\,0\,\cr
\star&\,0\,&\,0\,\cr
\star&\,0\,&\,0\,
\end{pmatrix}
\label{3-1}
\end{equation}
while the $(-1)/3$ have:
\begin{equation}
\begin{pmatrix}
\,0\,&\star&\star\cr
\,0\,&\,0\,&\,0\,\cr
\,0\,&\,0\,&\,0\,
\end{pmatrix}~.
\label{-13}
\end{equation}
The non-trivial behavior under the orbifold action due to the CP structure has to be compensated
with an opposite behavior of the vertex operators.
Now we come to a crucial point. 
In the Neveu-Schwarz sector, leading to bosonic modes, the GSO projected vertex operators
present an anti-chiral spinor index (notice that we are dealing with bosonic spinors)
with respect to the Lorentz group but are singlets with respect to the rotations in the internal space where the orbifold acts non-trivially.
This is a consequence of the mixed Neumann-Dirichlet boundary conditions.
In other terms, the vertex operator in the bosonic sector cannot behave in such a way to compensate
the CP structures \eqref{3-1} and \eqref{-13}.
The bosonic charged moduli are therefore projected out,
being their absence a hallmark of the exotic nature of the configuration.

Regarding the fermions emerging from the Ramond sector, instead,
the GSO projected physical states carry anti-chiral spinor indexes with respect to the internal space.
Indeed, recalling (\ref{gorbspin}), the $(+-+)$ and $(-++)$ components transform non-trivially
under the orbifold action.
It is then possible to build charged fermionic moduli whose vertex operators compensate for the CP
structure transformations leading to overall invariant states surviving the projection.
Adopting again an ADHM inspired notation, 
the physical charged moduli of the $3/(-1)$ sector are
\begin{equation}
\bm{\mu}^{+-+}=\begin{pmatrix}
\,0\,&\,0\,&\,0\,\cr
\,0\,&\,0\,&\,0\,\cr
\mu&\,0\,&\,0\,
\end{pmatrix} ~~~~\mbox{and}~~~~
\bm{\mu}^{-++}=\begin{pmatrix}
\,0\,&\,0\,&\,0\,\cr
\mu'&\,0\,&\,0\,\cr
\,0\,&\,0\,&\,0\,
\end{pmatrix}
 \label{mumu'}
\end{equation}
Since we are dealing with open strings connection D$3$ branes and D-instantons,
both $\mu$ and $\mu'$ are $N\times k$ matrices.
Eventually, the moduli belonging to the 
$(-1)/3$ sector, which correspond to open strings with opposite orientation, are related
to those of the $3/(-1)$ sector by the orientifold. In our case we have
\begin{equation}
 \begin{aligned}
  \bm{\bar\mu}^{+-+}&= \gamma_+(\Omega)\big(\bm{\mu}^{+-+}\big)^T\gamma_-(\Omega)^{-1}
= \begin{pmatrix}
\,0\,&+\mu^T&\,0\,\cr
\,0\,&\,0\,&\,0\,\cr
\,0\,&\,0\,&\,0\,
\end{pmatrix}~,\\
\bm{\bar\mu}^{-++}&= \gamma_+(\Omega)\big(\bm{\mu}^{-++}\big)^T\gamma_-(\Omega)^{-1}
= \begin{pmatrix}
\,0\,&\,0\,&-\mu^{\prime T}\cr
\,0\,&\,0\,&\,0\,\cr
\,0\,&\,0\,&\,0\,
\end{pmatrix}~.
 \end{aligned}
\label{barmumu'}
\end{equation}

\section{Moduli action}
The moduli action can be obtained in analogy with the low-energy gauge theory action of the D$3$ gauge theory.
Indeed it represents the action of the zero-dimensional gauge theory, i.e. matrix model, defined
on the point-like world-volume of the D$(-1)$'s.
The terms in the moduli action can be computed with a systematic analysis of the open-string scattering amplitudes.
In particular, we have to analyze the simplest topologies, namely the disk amplitudes, involving the instanton moduli\footnote{The original
papers in which the building of the effective action is actually performed from disk amplitudes are
\cite{Green:2000ke,Billo:2002hm}}.

For the sake of convenience, let us split the moduli action in three parts:
\begin{equation}
 S = S_{1} + S_{2} + S_{3}~,
\label{s}
\end{equation}
with
\begin{subequations}
 \begin{align}
S_{1} &=\frac{1}{g_0^2} \,\tr 
\Big\{\!\!-\frac{1}{4} \big[a^\mu,a^\nu\big]\,\big[a_\mu,a_\nu\big] 
- \big[a_\mu,\chi\big]\,\big[a^\mu,\overline{\chi}\big]
+ \frac{1}{2} \,\big[\overline{\chi},\,\chi\big]\, \big[\overline{\chi},\chi\big]\,  \Big\} ~,
\label{quartic}\\
S_{2} &=\frac{1}{g_0^2} \, \tr \Big\{ 2\, \lambda_{\dot{\alpha}a} 
\big[a^\mu,M_\beta^{\ a}\big]\, (\overline{\sigma}_\mu)^{\dot{\alpha}\beta} 
- \ii\, \lambda_{\dot{\alpha}a} \big[\chi,\lambda^{\dot{\alpha}a}\big] 
- 2 \ii\, M^{\alpha a} \big[\overline{\chi},M_{\alpha a}\big] \Big\} ~,
\label{cubic}\\
S_{3} &=\frac{1}{g_0^2} \, \tr \Big\{\!\!-\ii\, \mu^T\mu'\,\chi\Big\}
 \label{mixed}
 \end{align}
\label{ss}
\end{subequations}
Notice that $S_1$ and $S_2$ correspond respectively to quartic and cubic terms involving only neutral modes
while $S_3$ accounts for the terms involving the mixed or charged moduli.

A comment on the normalization of the action is in order.
We have collected a $1/g_0^2$ factor outside of the action where $g_0$ indicates the gauge coupling 
on the zero-dimensional Yang-Mills theory.
Being this field theory nothing but a low-energy effective description of string interactions,
the gauge coupling is related to the string coupling $g_s$. 
More specifically, the following relation holds (see \cite{Billo:2002hm}):
\begin{equation}
 g_0^2 = \frac{g_s}{4 \pi^3 \alpha'^2} ~.
\label{g0}
\end{equation}
Even though we have so far adopted a notation well adapted to conform with the standard ADHM analysis,
the dimension of the fields are not as in the ADHM construction.
In our formul\ae\ we are understanding in fact canonical dimension for the bosons, i.e. (length)$^{-1}$, and 
for the fermions as well, i.e. (length)$^{-3/2}$.
It is however possible to redefine the fields rescaling them with appropriate powers of the dimensionful
coupling $g_0$ in order to reproduce the standard ADHM dimension.
We do not perform such rescaling as it is not necessary nor convenient for our analysis.

We need to consider in depth the structure of the action  in order to be able to organize it in such a fashion
to apply the localization tools. 
The symmetry properties of the action are crucial in this respect. 
As a technical step which will prove manifestly convenient in the following, we express the $a_\mu$ quartic interaction terms
by means of auxiliary fields.
The introduction of auxiliary fields allows us to express the quartic interaction in terms of cubic interactions:
\begin{equation}
 S'_{1} 
=\frac{1}{g_0^2} \,\tr \Big\{\frac{1}{2} D_c D^c 
- \frac{1}{2} D_c \overline{\eta}^c_{\mu\nu}\, \big[a^\mu,a^\nu\big] 
- \big[a_\mu,\chi\big]\,\big[a^\mu,\overline{\chi}\big]   
+ \frac{1}{2} \big[\overline{\chi},\,\chi\big]\,\big[\overline{\chi},\,\chi\big] \Big\}
\label{quartic1}
\end{equation}
where the three auxiliary fields $D_c$ with $c=1,2,3$ are defined as
\begin{equation}
D^c = \frac{1}{2} \, \overline{\eta}^c_{\mu\nu} [a^\mu,a^\nu]~.
\label{eqD}
\end{equation}
The $\overline{\eta}^c_{\mu\nu}$ represent the anti-self-dual 't Hooft symbols given explicitly
in Appendix \ref{hooft}.
Observe that the definition of the auxiliary fields $D^c$ is indeed the algebraic equation of motion
one would find studying the variation of \eqref{quartic1}.
Substituting \eqref{eqD} into \eqref{quartic1} we can recover the original expression of the action without the $D^c$'s.

In view of the following developments we give another useful rewriting of the cubic action (\ref{cubic}). 
We indeed implement the already mentioned topological twist \ref{topo}
that concerns the identification of the internal spinor index $a$
with the space-time spinor index $\dot\beta$. 
In formul\ae\ we have:
\begin{equation}
\begin{aligned}
& \lambda_{\dot\alpha a}
~\to~\lambda_{\dot\alpha\dot\beta}\equiv\frac{1}{2}\, \epsilon_{\dot{\alpha}\dot{\beta}} \,\eta 
+ \frac{\ii}{2}\, (\tau^c)_{\dot{\alpha}\dot{\beta}}\, \lambda_c ~,\\
& M^{\alpha a}~\to~ M^{\alpha\dot\beta} \equiv\frac{1}{2}\, M_\mu \,(\sigma^\mu)^{\alpha\dot\beta}~. 
\end{aligned}
\label{eta}
\end{equation}
Note that, as anticipated in Section \ref{topo}, the topological twist implies a reorganization 
of the components of the moduli. Observe also that in  \eqref{eta} the number of degrees of freedom is not reduced 
by the topological twist.
Using the Equations (\ref{eta}), we rewrite the cubic action $S_{2}$ as follows
\begin{equation}
  S'_{2} =\frac{1}{g_0^2} \, \tr \Big\{\! \eta \, \big[a_\mu,M^\mu\big]  \!
+\lambda_c \,\big[a^\mu,M^\nu\big] \, \overline{\eta}^c_{\mu\nu} 
\!- \frac{\ii}{2} \,\eta\, \big[\chi,\eta \big] \!-  \frac{\ii}{2} 
\,\lambda_c \,\big[\chi, \lambda^c \big] \!- \ii M_\mu \big[\overline{\chi},M^\mu\big] \Big\} ~.
\label{cubic1}
\end{equation}
Eventually we replace the mixed action (\ref{mixed}) with
\begin{equation}
  S'_{3} =\frac{1}{g_0^2} \, \tr \Big\{\!\!-\ii\, \mu^T\mu'\,\chi+h^Th'\Big\}
\label{mixed1}
\end{equation}
where $h$ and $h'$ are charged auxiliary fields.
They do not interact with any other modulus and return the original mixed terms once put on-shell.
Even though this technical replacement looks trivial, it will prove to be convenient.

Having performed all the appropriate rewritings, the overall moduli action becomes:
\begin{equation}
 S' = S'_{1} + S'_{2} + S'_{3} \ .
\label{s'}
\end{equation}
We recall that it  is invariant under the D-instanton group $\mathrm{SO}(k)$ and the D$3$-brane
gauge group $\mathrm{SU}(N)$; $S'$ is invariant under the ``twisted'' Lorentz
group $\mathrm{SU}(2)\times\mathrm{SU}(2)'$ as well.
The moduli $a_\mu$ and $M_\mu$ transform in the
$(\mathbf{2},\mathbf{2})$, $\lambda_c$ and $D_c$ in the $(\mathbf{1},\mathbf{3})$ representations
of the ``twisted'' Lorentz group, while
all the remaining moduli $\chi$, $\overline{\chi}$, $\eta$, $\mu$, $\mu'$, $h$ and $h'$ are instead
singlets.

\section{BRST structure of the moduli action}
The topological twist opens the possibility of introducing a BRST-like structure with respect to which
the action (\ref{s'}) is invariant.
The identification of the indexes $a$ and $\dot{\beta}$ allows us to define the following ``singlet'' operator combining
the supercharges $Q_{\dot{\alpha}a}$ as follows
\begin{equation}
 Q = Q_{\dot{\alpha}\dot{\beta}} \epsilon^{\dot{\alpha}\dot{\beta}} \ .
\end{equation}
The charge $Q$ plays the role of the BRST transformation generator.
Note that being linear in the supersymmetry charges, it is fermionic as it is expected for any BRST-like operator.
Its action on the fields of the model is:
\begin{equation}
 \label{Q}
\begin{aligned}
& Q a^\mu  = M^\mu~,~~~ Q M^\mu = \ii\comm{\chi}{a^\mu}~, \\
& Q\lambda_c = D_c~,~~~ Q D_c  =  \ii\comm{\chi}{\lambda_c}~,\\
& Q\, \overline{\chi} = -\ii\eta~,~~~ Q\eta  = -\comm{\chi}{\overline{\chi}}~,~~~ Q \chi = 0~, \\
& Q \mu = h~,~~~ Q h = \ii\,\mu\,\chi~,\\
& Q \mu' = h'~,~~~ Q h' = \ii\,\mu'\,\chi~.
\end{aligned}
\end{equation}
From \eqref{Q} it is straightforward to prove the BRST invariance of the action; equivalently we can say that $S'$ is BRST-closed,
\begin{equation}
 Q\, S' = 0 \ .
\end{equation}

Note that the operator $Q$ is nihilpotent modulo an infinitesimal SO$(k)$ instanton rotation.
The instanton rotations are parametrized by the modulo $\chi$ and applying twice the operator $Q$ to any field we obtain
\begin{equation}
 Q^2 \, \bullet = T_{\mathrm{SO}(k)}(\chi) \, \bullet~,
\label{Q2}
\end{equation}
where $T_{\mathrm{SO}(k)}(\chi)$ generates the infinitesimal $\mathrm{SO}(k)$ transformation
in the appropriate representation (the one with respect to which the field $\bullet$ transforms)%
\footnote{Also the BRST-closure of the action can be thought of as a symmetry up to instanton rotations but, being $S'$ a scalar,
this assume a trivial connotation.}.
From (\ref{Q}) descends a clear BRST structure of the moduli content of our model;
we can actually arrange all the moduli except $\chi$ in BRST doublets
\begin{equation}
 (\Psi_0,\Psi_1) \ \ \ \text{with}\ \ \ Q\Psi_0 = \Psi_1 \ .
\end{equation}
See Table \ref{tab:rep} for the detailed account.
\begin{table}[ht]
\begin{center}
\begin{tabular}{|c||c|c|c|}
\hline
\phantom{\vdots}
$(\Psi_0,\Psi_1)$&$\mathrm{SO}(k)$ &$\mathrm{SU}(N)$&$\mathrm{SU}(2)\times\mathrm{SU}(2)'$
\\
\hline\hline
$\phantom{\vdots}(a_\mu,M_\mu)$ & $\Ysymm$ &$\mathbf{1}$&$(\mathbf{2},\mathbf{2})$
\\
$\phantom{\vdots}(\lambda_c,D_c)$ & $\Yasymm$
 &$\mathbf{1}$&$(\mathbf{1},\mathbf{3})$
\\
$\phantom{\vdots}(\overline{\chi},\eta)$ &$\Yasymm$ & $\mathbf{1}$
 &$(\mathbf{1},\mathbf{1})$
\\
$\phantom{\Big|}(\mu,h)$ & $\Yfund$ &$\mathbf{\overline{N}}$&$(\mathbf{1},\mathbf{1})$
\\
$\phantom{\Big|}(\mu',h')$ & $\Yfund$ &$\mathbf{\overline{N}}$&$(\mathbf{1},\mathbf{1})$
\\
\hline
\end{tabular}
\end{center}
\caption{Moduli in the stringy instanton configuration organized as BRST doublets; 
their transformation properties under the various symmetry groups of the model are accounted.}
\label{tab:rep}
\end{table} 

The total action is also BRST exact,
\begin{equation}
\label{S'1}
 S' = Q \,\Xi ~,
\end{equation}
where $\Xi$ is a fermionic quantity usually referred to as \emph{gauge fermion}.
Explicitly it is given by
\begin{equation}
 \Xi=\frac{1}{g_0^2} \, \tr \Big\{\ii M^\mu \big[\overline{\chi},a_\mu\big]
- \frac{1}{2}\, \overline{\eta}^c_{\mu\nu} \lambda_c \big[a^\mu,a^\nu\big]
+ \frac{1}{2} \,\lambda_c D^c 
- \frac{1}{2} \,\big[\chi,\overline{\chi}\big] \eta 
+ \mu^Th'\Big\}~.
\label{Xi}
\end{equation}
The exactness of the action can be proved directly using the transformation properties of the moduli and its
SO$(k)$ invariance.

The instanton partition function is given by the integral over the moduli space of the exponentiated action.
The moduli integration measure is a dimensionful quantity and to compute its dimension we can observe that,
as a fermion\footnote{We remind the reader that we use canonical dimensions for the fields throughout the analysis.} the BRST operator has dimension (length)$^{-1/2}$.
Therefore, within any doublet, the dimensions of the component fields are related as (length)$^\Delta$ and (length)$^{\Delta-1/2}$.
We then have that the moduli measure
\begin{equation}
 d\mathcal{M}_{k} = \,d\chi \prod_{(\Psi_0,\Psi_1)} d\Psi_0\,d\Psi_1
\label{measure}
\end{equation}
has the dimension
\begin{equation}
 \mbox{(length)}^{-\frac{1}{2}k(k-1)+\frac{1}{2}n_b-\frac{1}{2}n_f}~.
\label{dimmeas}
\end{equation}
The first term in the exponent of \eqref{dimmeas} corresponds to the unpaired modulus $\chi$ which belongs to the
anti-symmetric representation of $\mathrm{SO}(k)$, whereas $n_b$ and $n_f$ denote the number
of BRST doublets whose lowest component (i.e. $\Psi_0$) is respectively bosonic and fermionic.
The computation leading to \eqref{dimmeas} has been performed remembering that the differential of a fermionic (Grassmann) variable has opposite 
sign with respect to the variable itself\footnote{This is a direct consequence of Belavin's definition of the
integration over Grassmann variables,
\begin{equation}
 \int d\psi\, \psi = 1\ ,
\end{equation}
where $1$ is obviously dimensionless.}

Recalling the representation to which the various moduli belong (summarized in Table \ref{tab:rep}),
it is possible to verify that 
\begin{eqnarray}
 &n_b&=\frac{5}{2}\,k^2+\frac{3}{2}\,k\\
 &n_f&=\frac{3}{2}\,k^2-\frac{3}{2}\,k+2kN \ .
\end{eqnarray}
The measure $d\mathcal{M}_{k}$ then has dimension
\begin{equation}
 \left[d\mathcal{M}_{k}\right]=\mbox{(length)}^{k(2-N)}=\mbox{(length)}^{-kb_1}
\label{dimmeas1}
\end{equation}
where $b_1$ represents the coefficient of the $1$-loop $\beta$-function for our gauge theory (we are going to study it in the following, \ref{beta}).
It is important to remark that the minus sign in the exponent of (\ref{dimmeas1}) constitutes a hallmark of stringy instanton behavior.
Indeed, in the case of an ordinary instanton configuration the dimension of the moduli measure is%
\footnote{We refer to \cite{Dorey:2002ik} for a throughout treatment of the ordinary instanton case.} $\mbox{(length)}^{+kb_1}$.

\subsection{Moduli action in the presence of a graviphoton background}

Along the route taking us to the actual computation of the stringy instanton contributions to the 
low-energy effective theory, we have to introduce a constant graviphoton background.
This background deformation amounts to consider the model on a non-trivial curved background and, technically, the graviphoton 
can be regarded as a regulator.

We turn on a Ramond-Ramond three-form flux $F_{\mu\nu z^3}$ which has the first couple of indexes along the four-dimensional
space-time (i.e. the world-volume of the D$3$-branes) and the last (holomorphic) index along the (complexified) internal direction $z^3$
which is left invariant by the orbifold action \eqref{gorb}.
Such a background flux is not projected out by the orientifold because it is left invariant under the action of the orientifold operator \eqref{OrientOperator}.
Actually, $F_{\mu\nu z^3}$ is even under the world-sheet parity%
\footnote{The Ramond-Ramond $3$-form field-strength coincides with the external differential of the 
Ramond-Ramond two-form potential, $F_{\mu\nu\rho}=\partial_\mu C_{\nu\rho}$.
The latter, being a zero-mass mode of the closed-string sector, is associated to a term in the tensor decomposition 
of the ten-dimensional bi-spinor,
\begin{equation}\label{bispi}
 |\alpha> \otimes\, |\tilde{\beta}>
\end{equation}
The $\alpha,\beta$ indexes arise respectively from the right and left-moving Ramond vacua; 
in Type IIB \eqref{bispi} is decomposed on the tensors $C,C_{\mu\nu},C_{\mu\nu\rho\sigma}^{(+)}$ where $C$ and $C_{\mu\nu\rho\sigma}^{(+)}$
are associated to the symmetric part of the bi-spinor ($36$ components) and $C_{\mu\nu}$ is related to its anti-symmetric part ($28$ components)
with respect to the interchange of $\alpha$ and $\beta$. 
The world-sheet parity $\omega$ swaps the two anti-commuting spinors giving a minus sign. 
The anti-symmetry in $\alpha$ and $\beta$ compensates this and eventually the potential $C_{\mu\nu}$ and the
associated three-form strength are both even under $\omega$,} 
$\omega$, it is odd with respect to $(-1)^{F_L}$ as any mode in the Ramond-Ramond sector and it is 
odd also with respect to the internal space parity operator ${\cal I}_{456789}$ as any field having one internal index.
Let us simplify the notation indicating henceforth $F_{\mu\nu z^3}$ simply with ${\cal F}_{\mu\nu}$;
we organize its six independent components separating the self-dual and anti-self-dual parts,
\begin{equation}\label{caliF}
 {\cal F}_{\mu\nu} = -\frac{\im}{2} \overline{f}_c \eta^c_{\mu\nu} - \frac{\im}{2} f_c \overline{\eta}^c_{\mu\nu}.
\end{equation}
The coefficients $\overline{f}_c$ and $f_c$ (which are not complex conjugate of each other, the bar is just a notational feature)
belong respectively to the $(3,1)$ and $(1,3)$ representations of the (twisted) Lorentz group SU$(2)\times$SU$(2)'$.

In order to perform the computation (as it will become clear soon), we have to consider also the anti-holomorphic part of the graviphoton backgrounds,
namely $F_{\mu\nu \overline{z}^3} = \overline{\cal F}_{\mu\nu}$. As its holomorphic counterpart, it survives the orientifold projection.
By a holomorphicity argument, we will eventually show that the final results do not depend on $\overline{\cal F}_{\mu\nu}$ which can be
at last fixed to the most convenient value (see \ref{holo}).

The introduction of the three-form background fluxes follows the idea of exploiting all the symmetry features of the instanton moduli space
in order to be able to compute the instanton partition function. 
In other terms, we want to generalize the BRST structure asking that it closes (i.e. that $Q^2$ is nihilpotent) up to a generic infinitesimal symmetry
transformation of the moduli space and not only up to an instanton SO$(k)$ rotation.
The graviphoton will lead us to include twisted Lorentz transformations SU$(2)\times$SU$(2)'$, indeed ${\cal F}_{\mu\nu}$ parametrizes them.
To encompass the remaining SU$(N)$ gauge symmetry of the moduli space we have to include into the analysis also the D$3/$D$3$ fields and their interactions
with the instanton moduli. The SU$(N)$ gauge symmetry arises in fact from the D$3$-brane stack.
The field content of the theory living on the D$3$'s can be accounted for with an ${\cal N}=2$ chiral superfield $\Phi(x,\theta)$.
The actual computation of the interactions among $\Phi(x,\theta)$ and the instanton moduli is again performed studying carefully the disk diagrams involving
the associated vertex operators\footnote{We refer to \cite{Billo:2002hm,Billo:2006jm} for details.}.
Such interactions are described by a new term, namely
\begin{equation}
 \label{intPhi}
\frac{1}{g_0^2}\, \tr \Big\{\ii\,\mu^T{\Phi}(x,\theta)\,\mu'\Big\}\ ,
\end{equation}
that must be added to the moduli action (\ref{S'1}). 
As far as our following computations are concerned, it is sufficient to restrain the attention on the vacuum expectation value
of the superfield,
\begin{equation}
 \label{vev}
\phi = \langle {\Phi}(x,\theta) \rangle ~,
\end{equation}
and therefore the term which will be actually added to the action $S'_3$ is
\begin{equation}
 \frac{1}{g_0^2}\, \tr \Big\{\ii\,\mu^T\phi\,\mu'\Big\}~.
\end{equation}
The action itself acquires a dependence on $\phi$,
\begin{equation}
 S'_3(\phi) = S'_3 + \frac{1}{g_0^2}\, \tr \Big\{\ii\,\mu^T\phi\,\mu'\Big\}~.
\label{smixed2}
\end{equation}

Note that the introduction in the action of the D$3$ superfield renders the action itself dependent on the superspace coordinates.
This solves a possible doubt about the moduli integral corresponding to the center position of the instanton in superspace.
Without any explicit dependence on $x$ and $\theta$ they would be troublesome zero-modes.
On the computational level we discard the superspace dependence taking $\phi$ instead of the full-dynamical superfield but
we will show that the presence of the graviphoton regularizes the superspace integration.

The graviphoton forms $\cal F$ and $\overline{\cal F}$ are included into the analysis deforming the action and introducing their interactions with the moduli
that arise from the associated disk amplitude diagrams. 
Notice that in this case the diagrams involve both open and closed string vertex operators, the former are placed on the disk-boundary
whereas the latter are in its interior%
\footnote{We refer again to \cite{Billo:2002hm,Billo:2006jm} for details.}
The graviphoton interactions modify the quartic and cubic terms in the action, in particular we
deform (\ref{quartic1}) and (\ref{cubic1}) as follows:
\begin{equation}
 \begin{aligned}
S'_{1} & \rightarrow S'_{1}(\mathcal F,\overline{\mathcal F}) = 
S'_{1} +
 \frac{1}{g_0^2} \tr \Big\{\!\mathcal F^{\mu\nu}\!a_\nu\comm{\overline{\chi}}{a_\mu}+ 
\ii\,\overline{\mathcal F}^{\mu\nu}a_\mu\comm{\chi}{a_\nu}- 
\ii\,\overline{\mathcal F}^{\mu\nu}\! a_\mu \mathcal F_{\nu\rho}a^\rho
\Big\} ~,\\
S'_{2} & \rightarrow S'_{2}(\mathcal F,
\overline{\mathcal F}) = 
S'_{2} +
\frac{1}{g_0^2} \tr \Big\{\!\!-\frac{1}{2}\epsilon_{cde}\,\lambda^c\lambda^d f^e
-f_c\,\lambda^c\eta+\ii\,f_c\,D^c\overline{\chi}+\overline{\mathcal F}_{\mu\nu} M^\mu M^\nu \Big\}~.
 \end{aligned}
\label{squartic2}
\end{equation}

At last the deformed action encompassing the presence of Ramond-Ramond fluxes 
${\mathcal F}_{\mu\nu}$ and $\overline{\mathcal F}_{\mu\nu}$ as well as the vacuum expectation $\phi$ for the adjoint scalar of the gauge multiplet, is 
\begin{equation}
 S'(\mathcal F, \overline{\mathcal F}, \phi) = S'_{1}(\mathcal F,
\overline{\mathcal F})+S'_{2}(\mathcal F,\overline{\mathcal F})
+{S}'_{3}(\phi)~.
\label{smodfin}
\end{equation}
The extension of the action seems to have spoiled the BRST behavior of its undeformed version but
it is straightforward to prove that the new action is BRST invariant with respect to a generalized (or extended)
BRST transformation $Q'$.
Let us specify $Q'$ expliciting its action on the various moduli of the model:
\begin{equation}
 \label{Q1}
\begin{aligned}
& Q' a^\mu  = M^\mu~,~~~ Q' M^\mu = \ii\comm{\chi}{a^\mu}-\ii \,
{\mathcal F}^{\mu\nu} a_\nu~, \\
& Q'\lambda_c = D_c~,~~~~ Q' D_c  =  \ii\comm{\chi}{\lambda_c}+\epsilon_{cde}\,\lambda^df^e~,\\
& Q' \overline{\chi} = -\ii\,\eta~,~~~Q'\eta  = -\comm{\chi}{\overline{\chi}}~,~~~ Q' \chi = 0~, \\
& Q' \mu = h~,~~~~~~Q' h = \ii\,\mu\,\chi-\ii\,\phi\,\mu~,\\
& Q' \mu' = h'~,~~~~~ Q' h' = \ii\,\mu'\,\chi-\ii\,\phi\,\mu'~,
\end{aligned}
\end{equation}
More specifically, we have
\begin{equation}
 S'(\mathcal F, \overline{\mathcal F}, \phi) = Q'\,\Xi'
\label{s'q'}
\end{equation}
where the deformed gauge fermion $\Xi'$ is given by
\begin{equation}\label{xi'}
 \Xi' = \Xi + \frac{1}{g_0^2} \tr \Big\{ \ii\,f_c\,\lambda^c\overline{\chi}+
\overline{\mathcal F}_{\mu\nu}a^\mu M^\nu\Big\}
\end{equation}
with $\Xi$ defined in (\ref{Xi}).


The deformation of the action and the corresponding enlargement of the BRST structure let us exploit the complete symmetry
properties of the moduli space. Indeed, from the rules \eqref{Q1} we have that $Q'$ squares to zero up to an infinitesimal generic transformation
of the full symmetry group of the moduli space,
\begin{equation}
 Q'^2 \, \bullet = T_{\mathrm{SO}(k)}(\chi) \bullet\,
-\,T_{\mathrm{SU}(N)}(\phi)\bullet\, + T_{\mathrm{SU}(2)\times \mathrm{SU}(2)'}(\cF)\bullet
~.
\label{Q'2}
\end{equation}
The generators $T_{\mathrm{SO}(k)}(\chi)$, $T_{\mathrm{SU}(N)}(\phi)$ and $T_{\mathrm{SU}(2)\times \mathrm{SU}(2)'}(\cF)$
correspond to infinitesimal transformations of $\mathrm{SO}(k)$, $\mathrm{SU}(N)$ 
and $\mathrm{SU}(2)\times \mathrm{SU}(2)'$, 
parametrized respectively by $\chi$, $\phi$ and $\cF$, in the
appropriate representations. 

Let us give an explicit rewriting of the complete action which will be useful in the following:
\begin{eqnarray}
  S'(\mathcal F, \overline{\mathcal F}, \phi)=
\frac{1}{g_0^2}\!\!\!\!&&\!\!\!\!\tr \Big\{\eta\comm{a_\mu}{M^\mu}+
\lambda^c\comm{a^\mu}{M^\nu}\bar\eta^c_{\mu\nu}
-\frac{\ii}{2}\eta\comm{\chi}{\eta} 
-\ii\,M^\mu\comm{\overline{\chi}}{M_\mu}\nonumber\\
&-&\frac{1}{2}D_c\,\bar\eta^c_{\mu\nu}
\comm{a^\mu}{a^\nu}-\comm{a_\mu}{\overline{\chi}}\comm{a^\mu}{\chi}
+\frac{1}{2}\comm{\overline{\chi}}{\chi}\comm{\overline{\chi}}{\chi}
+\cF^{\mu\nu}a_\nu\comm{\overline{\chi}}{a_\mu}\nonumber\\
&-&\frac{1}{2}\lambda_c\,{Q'}^2\lambda^c
+\frac{1}{2}D_c\,D^c-\mu^T{Q'}^2\mu' +h^Th'-f_c\,\lambda^c\eta\nonumber\\
&+&\ii\,f_c\,D^c\,\overline{\chi}+\overline{\cF}^{\mu\nu}a_\mu\,{Q'}^2a_\nu +\overline{\cF}^{\mu\nu}M_\mu M_\nu\Big\}~.
\label{S'fin}
 \end{eqnarray}

\subsection{Classical part of the action}

So far we have not considered the classical part of the instanton action, namely
\begin{equation}
 S_{\mathrm{cl}}= -2\pi \,\im\, \tau\,k = \frac{2\pi}{g_s}\,k \ .
\label{sclass}
\end{equation}
It of course has to be taken into account as we will see in Section \ref{reno}.
The classical part of the action can be interpreted as the topological normalization of the bare D$(-1)$ disk amplitude with 
multiplicity $k$ without any vertex operator insertions \cite{Polchinski:1994fq,Billo:2002hm}. 
In \eqref{sclass} $\tau$ has been introduced to give a useful rewriting even though here it is simply related to the string coupling constant (i.e. the VEV of the dilaton).
Instead, whenever we have also a non-zero VEV for the Ramond-Ramond scalar $C_0$, $\tau$ is promoted to the full axion-dilaton combination $\tau=C_0+\frac{\ii}{g_s}$.

In the $\mathbb{C}/\mathbb{Z}_3$ stringy instanton model that we study explicitly the exotic character is given by a different symmetry 
behavior between gauge and instanton branes.
So, considering the same geometrical arrangement but simply changing the properties of the instanton branes with respect to the orbifold we can trade an exotic configuration for
an ordinary and vice versa. 
If the internal geometry arrangement of ordinary and exotic instanton branes is the same, they possess the same
classical action.
In our case the instanton branes are represented by D$(-1)$ brane having point-like world-volume.
In the general case (where instantonic brane can be extended along the internal directions) the classical contribution of the action is proportional
to the volume $V_{\cal C}$ of the internal cycle $\cal C$ being wrapped by the instanton branes. 
The classical weight accompanying the instanton contributions has an exponential shape where at the exponent we have the negative ratio of the over the
string coupling constant $g_s$. This feature is particularly interesting because we can observe that, calibrating the volume $V_{\cal C}$ of the internal cycle
wrapped by the instanton branes, we can have significant non-perturbative effects also when the coupling $g_s$ is small \cite{Blumenhagen:2009qh}.

\subsection{Holomorphicity of the partition function}
\label{holo}

In the context of ${\mathcal N}=2$ SYM gauge theory, the instanton moduli action depends only on $\phi$ and ${\cal F}$
and not on $\overline{\phi}$ and $\overline{\cal F}$%
\footnote{We remind the reader that $\phi$ and ${\mathcal F}$ represent respectively the VEV's of the scalar field belonging to the 
${\mathcal N}=2$ vector multiplet and the graviphoton field.}.
In other terms, when we consider only the instantonic configurations with $k>0$, the non-perturbative action is holomorphic.
As opposed to this, if we consider only anti-instantons ($k<0$) we have an anti-holomorphic non-perturbative action.
The holomorphicity properties of instantons and anti-instantons actions
are proven with co-homology arguments%
\footnote{The argument showing that the instanton action is independent of the anti-holomorphic quantities is formally analogous 
to the proof through BRST arguments that the 
YM action does not depend on the gauge fixing (or, more precisely, the action does not depend 
on the gauge fixing parameter); see \cite{Becchi:1996dt}.};
in order to perform the co-homology considerations, it is essential to rearrange everything
by means of the so-called topological twist. 
Such twist allows us to define $Q$ as in \eqref{Q}.
Remember that the action is $Q$-exact, namely
\begin{equation}
 S = Q\, \Xi\ .
\end{equation}
In the instanton action, the anti-holomorphic quantities $\overline{\phi}$ and $\overline{\cal F}$
are present only in the gauge fermion $\Xi$ 
but not in the $Q$-variations of the moduli%
\footnote{As a BRST operator $Q$ has fermionic character hence, when applied on a fermion like $\Xi$, it
returns a bosonic quantity, in the present case, the action $S$.};
in constrast, the holomorphic quantities $\phi$ and ${\cal F}$ appear explicitly in the action of $Q$, \eqref{Q1}.
As a consequence, any infinitesimal variation of the instanton partition function with respect to 
the anti-holomorphic quantities is given by a $Q$-exact term.
The terms $Z^{(k)}$ of the partition function
We have then
\begin{equation}
 Z^{(k)} = \int d{\cal M}_{(k)} e^{-S}
\end{equation}
\begin{equation}\label{invmes}
 \begin{split}
 \overline{\delta} Z^{(k)} &= \int d{\cal M}_{(k)} \overline{\delta}\left( e^{-S}\right)
                           = \int d{\cal M}_{(k)} e^{-S} \overline{\delta} S \\
                           &= \int d{\cal M}_{(k)} e^{-S} Q \left(\overline{\delta} \Xi\right) 
                           = \int d{\cal M}_{(k)} Q \left(e^{-S} \overline{\delta}\, \Xi\right) = 0
 \end{split}
\end{equation}
where $\overline{\delta}$ indicates the variation with respect a generic anti-holomorphic quantity which could be either $\overline{\phi}$
or $\overline{\mathcal{F}}$.
The last step in \eqref{invmes} is due to the fact that the moduli integration measure $d{\cal M}_{(k)}$ is BRST invariant;
as a consequence, the integral reduces to boundary terms at infinity where all the physical quantities are assumed to vanish.

\section[String scale and renormalization]{Explicit dependence on the string scale and renormalization behavior of exotic effects}
\label{reno}

In the D-brane models leading, at low-energy, to the ordinary $k$ instanton, the dimension of the moduli space integration measure
$d{\cal M}_k^{(\text{ord})}$ is
\begin{equation} \label{ord_dim}
 \left[d{\cal M}_k^{(\text{ord})}\right] = (\text{length})^{k b_1} \ .
\end{equation}
where $b_1$ is the $1$-loop coefficient of the $\beta$-function.
Equation \eqref{ord_dim} is the ordinary counterpart of the stringy result we have previously found in \eqref{dimmeas1}.

The contribution $Z_k$ to the total instanton partition function $Z$ contributed by the $k$ ordinary sector arises from the integration
over ${\cal M}_k^{(\text{ord})}$ of the exponentiated instanton action,
\begin{equation}\label{partk}
 Z_k \propto \int d{\cal M}_k^{(\text{ord})} e^{-S'({\cal F},\overline{\cal F},\phi)}.
\end{equation}
As a contribution to the partition function, being this dimensionless, we want to compensate the dimension carried by
${\cal M}_k^{(\text{ord})}$ with a dimensionful pre-factor having dimension $(\text{mass})^{kb_1}$.
More specifically, we introduce in front of the moduli integral a pre-factor $\mu^{kb_1}$ where $\mu$ is a reference mass scale;
in a stringy context $\mu$ is naturally related to the string scale, namely
\begin{equation}\label{ren_str_sc}
 \mu \sim \frac{1}{\sqrt{\alpha'}}\ .
\end{equation}
Since we are working with a non-Abelian gauge theory emerging as the low-energy limit of a string model, it is possible to think to $\mu$
as a renormalization reference scale related to the high-energy regime of the field theory (i.e. to its string UV completion).
In the pre-factor of \eqref{partk} we have to account for the classical part of the action \eqref{sclass}.
Eventually we have an overall pre-factor containing
\begin{equation}
 \label{exp_ren_eq}
 \mu^{k b_1} e^{-\frac{8\pi^2}{g_{YM}^2}k} = \Lambda^{kb_1}.
\end{equation}
where $\Lambda$ is another mass scale related to the high-energy renormalization scale $\mu$ through non-perturbative effects.
This can be regarded as a way to define the scale $\Lambda$ of a non-Abelian YM theory; $\Lambda$ is in fact generated by the (non-perturbative) dynamics of the model.
Indeed, \eqref{exp_ren_eq} is nothing other than the exponentiated renormalization group equation  
(see for instance \cite{Bianchi:2007ft}) where we are giving a precise interpretation of the $\mu$ mass scale in terms of the UV completion
of the field theory%
\footnote{The renormalization group equation can be obtained with a perturbative analysis (in the present case up to $1$-loop).}.

Let us see this point the other way around: We have that the stringy calculations brought into the game the string scale related to $\alpha'$.
In an ordinary instanton configuration, however, this scale is expressible completely in terms of the scale $\Lambda$.
In other words, we have that in the ordinary setups the factor of $\alpha'$ can be transmuted in to quantities in $\Lambda$ and,
once the results are completely expressed using the latter alone, we can forget the ``stringy'' origin of our calculations.

The same kind of analysis leads to a radically different outcome in the case of exotic configurations.
Let us repeat the expression for our specific exotic instanton whose moduli integration measure (derived in \eqref{dimmeas1}) is:
\begin{equation}\label{exo_beta}
 \left[d{\cal M}_k^{(\text{ex})}\right] = (\text{length})^{k(2-N)} = (\text{length})^{-kb_1}.
\end{equation}
Note that in \eqref{exo_beta} the exponent has opposite sign with respect to \eqref{ord_dim}. 
Therefore, the exotic counterpart of \eqref{exp_ren_eq} is
\begin{equation} \label{exo_dim}
 \mu^{-kb_1} e^{-\frac{8\pi^2}{g_{YM}^2}k}.
\end{equation}
In this case the renormalization group equation does not allow us to express the dimensionful pre-factor in terms of $\Lambda$ alone.
The exotic contribution cannot be expressed in terms of purely field theory quantities.
This is exactly the point of exotic configurations: they have an intrinsic stringy nature.
Exploiting \eqref{exp_ren_eq} in connection with \eqref{ren_str_sc},
we have that the exotic dimensionful pre-factor is proportional to
\begin{equation}
 (\alpha')^{kb_1} \Lambda^{kb_1}.
\end{equation}
the string scale $\alpha'$ will therefore appear manifestly in the exotic contributions to the low-energy prepotential
and couplings of the underlying gauge theory. 
The introduction of an explicit dependence on $\alpha'$ through the exotic sector is a general feature except for peculiar (conformal) cases; we discuss
in detail a conformal model in Section \ref{onformal}.

We noted that in our D$3/$D$(-1)$ model, the classical part of the action for ordinary and exotic instantons is the same.
This seemingly surprising feature is due to the fact that, the gauge and the instanton branes possess
the same internal geometry arrangement%
\footnote{We refer to \cite{Blumenhagen:2009qh} for details.}.

From now on we specialize the attention to models having no D$3$-brane on node $1$ (i.e. $N_1=0$) and $N_2=N_3=N$;
the low-energy regime of such D-brane models is accordingly described by a four-dimensional SU$(N)$ gauge theory
with matter in the symmetric representation.
The matter content arises from the modes stretching between nodes $2$ and $3$,
namely the complex fields $\Phi^1_{(23)}$ and $\overline{\Phi}^{\,2}_{(32)}$ in \eqref{Phi1Phi2} together with their
fermionic partners; they combine to form an $\mathcal N=2$ hypermultiplet in the symmetric representation of $\mathrm{SU}(N)$.

The $\beta$-function at $1$-loop depends on the field content of the field theory.
The fields contribute differently according to their transformation properties under the Lorentz and gauge group;
standard field theoretical computations (which we omit) lead to the following set of contributions
\begin{equation}\label{taaa}
	\begin{array}{cc|c}
	A_\mu& (\text{vector})& \frac{11}{3}C(R)\\
	\psi & (\text{spinor})& -\frac{2}{3}C(R)\\
	\phi & (\text{complex scalar})& -\frac{1}{3}C(R)
\end{array}
\end{equation}
where $C(R)$ is the Casimir constant of the representation $R$ of the gauge group under which the field transforms,
\begin{equation}
 \text{tr} [T^a_R,T^b_R] = C(R)\, \delta^{ab}\ .
\end{equation}
$T^a_R$ indicates the generators of the representation $R$. 
When the gauge group is SU$(N)$ it is possible to show that we have the following Casimir constants%
\begin{equation}
	\text{SU(N)} \rightarrow \left\{\begin{array}{rcl}
	C(\text{adj}) &=& N\\
	C(\text{fund}) &=& \frac{1}{2}\\
        C(\text{symm}) &=& \frac{1}{2}(N+2)\\
\end{array}\right.
\end{equation}
The field content of our ${\cal N}=2$ model consists in a vector multiplet in the adjoint representation of SU$(N)$ and an hypermultiplet
in the symmetric representation.
The former is formed by a vector field, two spinors and a complex scalar;
according to \eqref{taaa} its contribution to $b_1$ (i.e. the $1$-loop $\beta$-function) is
\begin{equation}
	b_1^{(\text{vec})} = \frac{11}{3} C(\text{adj}) + 2 \left(-\frac{2}{3}\right)C(\text{adj}) - \frac{1}{3}C(\text{adj}) = 2N\ .
\end{equation}
The hypermultiplet, instead, is composed by two spinors and two complex scalars in the symmetric representation of SU$(N)$;
its contribution to $b_1$ is then
\begin{equation}
	b_1^{(\text{hyp})}=2\left(-\frac{2}{3}\right) C(\text{symm}) + 2\left(-\frac{1}{3}\right) C(\text{symm}) = - (N+2)\ .
\end{equation}
The overall result amounts to 
\begin{equation}\label{beta}
	b_1 = N-2\ .
\end{equation}

\section{SU\texorpdfstring{$(2)$}{} conformal case}
\label{onformal}

So far we have considered a generic number $N$ of D$3$-branes placed on nodes $2$ and $3$ of the quiver.
It is however convenient to split the next developments into two and isolate the particular case $N=2$.
Indeed, the SU$(2)$ case, as opposed to any other value $N>2$, leads to a low-energy conformal field theory
Notice, in fact, that because of \eqref{beta} we have a vanishing $1$-loop beta function.
Since (because of ${\cal N}=2$ non-renormalization theorems), $b_1$ does not receive any other perturbative correction
beyond $1$-loop order, it represents the full, perturbative $\beta$-function.
In addition \eqref{dimmeas1} relating the instanton moduli dimension with the $\beta$-function,
yields that the exotic instanton measure for $N=2$ is dimensionless independently of $k$%
.
Conformality has significant effects, the main among these is the fact that also exotic instantons (even though stringy in nature)
do not introduce any dependence on the $\alpha'$ scale in the effective field theory.
Rephrasing the point, since no new scale is introduced, the theory remains conformal also taking into account the exotic non-perturbative sector.


\subsection{Localization limit}
\label{sec:stringyprepot}

The stringy instanton contributions to the prepotential of the low-energy effective ${\cal N}=2$ field theory are directly related to the 
instanton partition function; the computation of the latter constitute here our first aim.
We describe here a preliminary and essential step to its explicit evaluation, namely the \emph{localization limit}.

The contributions to the partition function $Z$ arising from the various topological sectors are additive and
we can expand $Z$ with respect to the value of the topological charge $k$,
\begin{equation}\label{partok}
 Z= \sum_{k=0}^\infty q^k Z_k \ ,
\end{equation}
where $q$ is generally a dimensionful parameter given by
\begin{equation}\label{q}
 q \doteq \mu^{b_1} e^{2\pi \im \tau} \ .
\end{equation}
Since the overall partition function $Z$ is a dimensionless quantity, its $k$-th term $Z_k$ has dimension $[\mu]^{-k b_1}$
In the special SU$(2)$ case (where $b_1=0$) we have
\begin{equation}
 q^{(\text{SU}(2))} =  e^{2\pi \im \tau} \ ,
\end{equation}
and all the $Z_k^{(\text{SU}(2))}$ are dimensionless as well.

Any term $Z_k$ in \eqref{partok} arises from the moduli integral of the instanton belonging to the $k$-sector, namely
\begin{equation}
 Z_k= \cN_k \int d\cM_k \,\ee^{-S'(\mathcal F, \overline{\mathcal F}, \phi)} \ 
\label{partfunc}
\end{equation}
where $\cN_k$ is a normalization constant.
Note that the classical factor is not present in \eqref{partfunc} because it has already been included in the definition of $q$ \eqref{q}.

We perform a set of rescalings on the variables in order to put the integrals \eqref{partfunc} in a form allowing the direct computation.
At first we rescale the open-string instanton moduli in the following way
\begin{equation}
\begin{aligned}
 &(a_\mu,M_\mu)\to\frac{1}{y}\,(a_\mu,M_\mu)~,~~(\overline{\chi},\eta)\to \frac{1}{y}\,
(\overline{\chi},\eta)~,\\
 &(\lambda_c,D_c)\to{y}^2\,(\lambda_c,D_c)~,~~(\mu,h)\to {y}^2\,(\mu,h)~,~~
(\mu',h')\to{y}^2\,(\mu',h')~,
\end{aligned}
\label{rescalings} 
\end{equation}
where $y$ is just a dimensionless scaling parameter.
Notice that the rescaling respect the BRST doublet structure.
Similarly, we scale also the anti-holomorphic part of the graviphoton field,
\begin{equation}
 \overline{\cF}_{\mu\nu}\to z\,\overline{\cF}_{\mu\nu}~.
\label{rescF}
\end{equation}
The possibility of performing such rescalings is intimately related to supersymmetry and the BRST structure of the theory.
In relation to \eqref{rescalings}, the equal treatment within any doublet guaranties that the moduli measure $d\cM_k$ is insensitive to the rescalings;
indeed, in $d\cM_k$ we have an equal number of bosonic and fermionic degrees of freedom which, upon rescaling, behave in opposite ways and compensate
each other%
\footnote{Consider the Berezin integral over a Grassmann variable $\psi$
\begin{equation}\label{beretin}
 \int d\psi \psi = 1 \ .
\end{equation}
Let us now consider the rescaling $\psi'=\alpha \psi$. Since the same expression \eqref{beretin}
has to hold also for the rescaled Grassmann variable $\psi'$, we have
\begin{equation}
 d\psi' = \frac{1}{\alpha} d\psi\ .
\end{equation}
};
the possibility of performing the rescalings \eqref{rescalings} without affecting the moduli integral is related to supersymmetry
(also before the BRST interpretation).
For \eqref{rescF} the BRST interpretation of $Q$ is instead crucial.
Our freedom in rescaling $\overline{\cal F}$ is due to the holomorphicity property of the action;
this property, as shown in Subsection \ref{holo},
is a direct consequence of the co-homological structure of the model.

We are allowed to choose the scaling parameters $y$ and $z$ in the most convenient way
in view of the computation of the partition function $Z_k$.
Specifically, we take the following limits
\begin{equation}
 y\to\infty~,~~z\to\infty~~~~\mbox{with}~~\frac{z}{y^2}\to\infty~.
\label{limit}
\end{equation}
The moduli action (\ref{S'fin}) becomes
\begin{equation}
 \begin{aligned}
  S'(\mathcal F, \overline{\mathcal F}, \phi)=\tr \Big\{&\!\!
-\frac{s}{2}\lambda_c\,{Q'}^2\lambda^c
+\frac{s}{2}D_c\,D^c-s\,\mu^T{Q'}^2\mu' +s\, h^Th'-t\,f_c\,\lambda^c\eta\\
&+\ii\,t\,f_c\,D^c\,\overline{\chi}+u\,\overline{\cF}^{\mu\nu}a_\mu\,{Q'}^2a_\nu +u\,\overline{\cF}^{\mu\nu}M_\mu M_\nu\Big\}~+\cdots~,
 \end{aligned}
\label{S'rescaled}
\end{equation}
where we have introduced the following coupling constants
\begin{equation}\label{cou}
 s=\frac{y^4}{g_0^2}~,~~~t=\frac{y}{g_0^2}~,~~~u=\frac{z}{y^2\,g_0^2}~.
\end{equation}
Note that the couplings \eqref{cou} tend to $\infty$ in the limit (\ref{limit}), actually in \eqref{S'rescaled} we have indicated with dots the subleading terms.

The technical manipulations performed so far have a very nice results, they led us to (\ref{S'rescaled}) which is quadratic in the moduli%
\footnote{Stating that the terms in \eqref{S'rescaled} are ``quadratic'' in the moduli,
we are not considering the modulus $\chi$ parameterizing the instanton rotations; actually $\chi$ is contained in the explicit
expression of the $Q'$-variations of the moduli. As we will see shortly, the integration with respect to $\chi$ will be the last one, and
it needs a different treatment and particular care.}.
The integrals over the moduli are therefore of Gaussian type and can be straightforwardly evaluated.
Since we are considering  a limiting case in which the action possesses only quadratic terms,
it can sound as we performed a semi-classical (i.e. approximate) evaluation, but it is not so.
Observe that, since we are allowed to take the limits \eqref{limit} (because of SUSY and co-homological arguments),
the subleading terms are infinitely small and the computations involving (\ref{S'rescaled}) lead to exact results.

\subsection{Details of the partition function integral computation}

In order to compute explicitly the $Z_k$ integral \eqref{partfunc}, we take the non-dynamical background $\cF_{\mu\nu}$
along the Cartan directions of $\mathrm{SU}(2)\times \mathrm{SU}(2)'$;
this amounts to consider 
\begin{equation}
 f_c= f\,\delta_{c3}~,~~~\bar f_c = \bar f\,\delta_{c3}~,
\label{fc3}
\end{equation}
in the self-dual/anti-self-dual parametrization of the graviphoton given in \eqref{caliF}.
The expanded matrix form of the graviphoton background is therefore
\begin{equation}
 \mathcal F = -\frac{\ii}{2}\,\bar f\,\eta^3 -\frac{\ii}{2}\, 
f\,\overline{\eta}^3 =-\frac{\ii}{2}\begin{pmatrix} 
      0& ~(\bar f+ f)&0&0  \cr
     -(\bar f+f)&0&0&0 \cr
      0&0&0& ~(\bar f- f)\cr
      0&0&-(\bar f- f)&0
     \end{pmatrix}~.
\label{Fcart}
\end{equation}

After having inserted (\ref{fc3}) into (\ref{S'rescaled}), the action contains the fermionic modulus $\eta$ only in the
term 
\begin{equation}
 -\text{tr} \left\{ t f \lambda ^3 \eta \right\} 
\end{equation}
We can then integrate simultaneously over $\eta$ and $\lambda_3$: This yields a factor $t\,f$ and,
since $\lambda_3$ is a Grassmann variable, all the other terms containing it do not give further contribution to the moduli integral.
The counterpart of this is given by the integration over $D_3$ and $\overline{\chi}$;
the modulus $\overline{\chi}$ appears solely in the term 
\begin{equation}
 \im \text{tr} \left\{ t f D^3 \overline{\chi} \right\}\ ,
\end{equation}
so that the Gaussian integration with respect to $D_3$ and $\overline{\chi}$ yields a factor $1/(t\,f)$.
The overall result of the integrations on $\lambda_3$, $D_3$, $\eta$ and $\overline{\chi}$ produces 
just a numerical factor that can be reabsorbed within the normalization constant $\cN_k$ outside the partition function. 

We have still to integrate upon the BRST doublets $(a_\mu, M_\mu)$,
$(\mu,h)$, $(\mu',h')$ and $(\lambda_{\hat c},D_{\hat c})$ where $\hat c=1,2$ before 
facing the last $\chi$ integral.
Performing the corresponding Gaussian integrations we obtain
\begin{equation}
\begin{aligned}
 &\int (d\lambda_{\hat c} dD_{\hat c})~\ee^{\tr \{\frac{s}{2}\lambda_{\hat c}\,{Q'}^2\lambda^{\hat c}
-\frac{s}{2}D_{\hat c}\,D^{\hat c}\}}\,\times\,\int (d\mu dh)\, (d\mu'dh')~
\ee^{\tr \{s\,\mu^T{Q'}^2\mu' -s\, h^Th'\}}\\
&\times\,\int (da_\mu dM_\mu)~\ee^{-\tr \{
u\,\overline{\cF}^{\mu\nu}a_\mu\,{Q'}^2a_\nu +u\,\overline{\cF}^{\mu\nu}M_\mu M_\nu\}}
~\sim~ \cP(\chi)\,\times\,\cR(\chi)\,\times\,\frac{1}{\cQ(\chi)}~
\end{aligned}
\label{int}
\end{equation}
where we have defined the following quantities:
\begin{subequations}
\begin{align}
 \cP(\chi)&\equiv\Pf_{\big(\Yasymm,\mathbf{1},\,(\mathbf{1},\mathbf{3})'\big)}\big(Q'^2\big)~,
\label{pchi0}\\
\cR(\chi)&\equiv
\mathrm{det}_{\big(\Yfund,\,\mathbf{\overline{N}},\,(\mathbf{1},\mathbf{1})\big)}\big(Q'^2\big)~,
\label{rchi0}\\
\cQ(\chi)&\equiv
\mathrm{det}^{1/2}_{\big(\Ysymm,\,\mathbf{1},\,(\mathbf{2},\mathbf{2})\big)}\big(Q'^2\big)~.
\label{qchi0}
\end{align}
\label{PRQ}
\end{subequations}
The sub-labels attached to the Pfaffian and determinant symbols indicate the representations of the symmetry group
(namely instanton, gauge and Lorentz group) with respect to which the associated moduli transform. 
Remember that the net result of a double $Q'$-variation is an infinitesimal symmetry transformation%
\footnote{In the first line of \eqref{PRQ}, $(\mathbf{1},\mathbf{3})'$ means that the component of
the BRST pair $(\lambda_c,D_c)$ along the null weight must not be considered, since it has
been already accounted for in dealing with the $\lambda_3$, $D_3$integration.}.
In other terms, $Q'^2$ is represented by a matrix whose explicit structure depends on which kind of modulus it acts on,
i.e. on the representations to which it belongs.

In \eqref{int} we have a proportionality symbol because we are again neglecting all the numerical factors which we absorb in the overall normalization constant ${\cal N}_k$.
Eventually, the integral corresponding to the $k$-th term $Z_k$ in the partition function topological expansion can be cast in the following form
\begin{equation}
 Z_k= \cN_k\,\int \Big\{\frac{d\chi}{2\pi\ii}\Big\}\,\frac{\cP(\chi)\,\cR(\chi)}{\cQ(\chi)}
~.
\label{zk1}
\end{equation}
Notice that the result does not depend on the coupling constants $s$, $t$ and $u$ defined in \eqref{cou}.
Equation \eqref{zk1} corresponds to the localization formula \eqref{vitaloca} where
\begin{eqnarray}
 &\alpha(X) &= e^{S[X]}\\
 & X_0 &= 0\\
 &\alpha(0) = 1
\end{eqnarray}
The variable $X$ here indicates a point in the moduli space ${\cal M}_k$ and the transformation group underlying our equivariant approach
is the full symmetry group of ${\cal M}_k$,
\begin{equation}
 \underbrace{\text{SU}(2)\times\text{SU}(2)'}_{\text{Lorentz}} \times \underbrace{\text{SU}(N)}_{\text{gauge}} \times \underbrace{\text{SO}(k)}_{\text{instanton}}
\end{equation}
whose unique fixed point on ${\cal M}_k$ is $X_0=0$ where all the moduli are null.

Before performing the last step concerning the integration over $\chi$ one has to regularize the integral (\ref{zk1}).
The singular behavior emerges for two reasons: firstly, we have a divergence whenever the denominator ${\cal Q}(\chi)$ in the integrand vanishes and,
secondly, for asymptotic values of $\chi$ the integrand itself remains finite giving another diverging contribute.
Both the troublesome features are solved following Nekrasov's prescription (to have some explicit examples see for instance \cite{Billo:2009di,Billo:2010mg})
that consists in a ``complexification'' of the naively ill-defined integral \eqref{zk1}.
$\chi$ is promoted to a complex variable.
and the singularities along the real axis are given a small imaginary part so that they are displaced away from the integration path.
The integral path itself is ``closed'' at infinity in the upper complex plane and interpreted as a contour integral.
This latter step consists in assuming that the contribution at infinity is unimportant.
It has to be underlined that Nekrasov's prescription is a recipe lacking a first principle derivation;
in spite of this, many non-trivial examples have been treated in the literature and whenever checks are possible Nekrasov's prescription
led to correct results.

\subsection{Explicit computations for the smallest instanton numbers}

\subsubsection{$k=1$}
The instanton partition function $Z_1$ for $k=1$ (i.e. the exotic $1$-instanton) is simple to compute.
Note indeed that when $k=1$ the $\lambda_c$ and $\chi$ moduli are absent and consequently the factor $\cP(\chi)$ is absent as well;
in particular there is no $\chi$ integration to be performed. 
The factors ${\cal P}(\chi)$ and ${\cal Q}(\chi)$ for $k=1$ are given by
\begin{equation}
 \begin{aligned}
  \cR(\chi) &\propto\,\det\phi~,\\
\cQ(\chi) &\propto\,\mathrm{det}^{1/2}{\cF}\,\propto\, E_1 E_2 \equiv \cE~,
 \end{aligned}
\label{RQ1}
\end{equation}
where we have defined $E_{1,2}$ as follows
\begin{equation}
\label{ef}
 E_1=\frac{f+\bar f}{2}~,~~~E_2=\frac{f-\bar f}{2}~.
\end{equation}
As anticipated, we are neglecting numerical factors because we absorb them into the overall normalization ${\cal N}_1$.
The $k=1$ result is
\begin{equation}
 Z_1 = {\cN}_1\, \frac{\det\phi}{\cE}~.
\label{z1}
\end{equation}
We remind ourselves that, as described in \cite{Billo:2009di}, the factor $1/\cE$ in \eqref{z1} is interpreted as the regularized volume of the four-dimensional $\cN=2$ superspace%
\footnote{For further details see \cite{Nekrasov:2002qd,Billo:2006jm}}.

\subsubsection{$k>1$}

As opposed to the preceding $k=1$ case, now the integration over $\chi$ has to be explicitly performed.
We exploit the $\mathrm{SO}(k)$ invariance of the integrand in (\ref{zk1}),
and we consider the variable $\chi$ along the instanton group Cartan's sub-algebra generated by $H^i_{\mathrm{SO}(k)}$.
As the $\chi$ represents the integration variable, its choice along the Cartan direction must be compensated with
the introduction  of the so called Vandermonde determinant $\Delta(\chi)$,
\begin{equation}
 \label{chicartan}
 \chi ~~\to~~ \vec\chi \cdot \vec H_{\mathrm{SO}(k)}~= \sum_{i=1}^{\mathrm{rank}\, \mathrm{SO}(k)}
\chi_{i} \,H^{i}_{\mathrm{SO}(k)} ~.
\end{equation}
The partition function integrals for $k>1$ become
\begin{equation}
\label{Zkred}
Z_k = \cN_k \int \prod_{i} \Big(\frac{d \chi_{i}}{2\pi\ii}\Big)~ \Delta(\vec\chi)\,
\frac{\cP(\vec \chi)\, \cR(\vec\chi)}{\cQ(\vec\chi)}~.
\end{equation}
Once more all numerical factors (in this case associated to the ``diagonalization'' of $\chi$) have been reabsorbed inside the overall normalization constant $\cN_k$.
exploiting the symmetry of the integrals with respect to the gauge rotations we can take, without losing the generality
of the treatment, the VEV of $\phi$ along the Cartan direction of $\mathrm{SU}(2)$, 
\begin{equation}
\phi =\frac{\varphi}{2}\,\tau^3~.
\label{phisu2}
\end{equation}

Now we examine the $k=2$ case, i.e. the exotic $2$-instanton; in this case we have
\begin{equation}
  \begin{aligned}
\cP(\vec \chi) & \propto\,-(E_1+E_2)~,~~~
\cR(\vec\chi) \propto\,
\big(\chi^2+\det\phi\big)^2~,\\
  \cQ(\vec\chi)& \propto\,\cE\,\prod_{A=1}^2(2\chi-E_A)(2\chi+E_A)~,~~~
\Delta(\vec\chi)=1~,
\end{aligned}
\label{ingrk2}
 \end{equation}
The precise derivation is in Appendix \ref{app:A}.
Putting \eqref{ingrk2} into the integral expression for $Z_2$ \eqref{partfunc}, we obtain
\begin{equation}
 Z_2=-\cN_2\,\frac{E_1+E_2}{\cE}\,\int\frac{d\chi}{2\pi\ii}~
\frac{\big(\chi^2+\det\phi\big)^2}{(4\chi^2-E_1^2)(4\chi^2-E_2^2)}~.
\label{z2}
\end{equation}
The $2$-instanton represents the smallest topological charge value leading to a non-trivial integration on $\chi$.
We follow Nekrasov's prescription and interpret the $\chi$ integral as a contour integral in the complex plane;
we ``close'' the contour in the upper complex plane and assume that no contributions come from the part of the contour at infinity.
The singularities corresponding to the zeros of the denominator are ``cured'' assigning a small imaginary part to the $E_A$'s.
Specifically, we choose%
\footnote{For details on the prescription and the assignments of the imaginary parts see \cite{Moore:1998et}.} 
\begin{equation}
 \label{imparts}
 \text{Im} E_1 > \text{Im} E_2 > \text{Im} \frac{E_1}{2} > \text{Im} \frac{E_2}{2} > 0~.
\end{equation}
The evaluation of the integral with the residue technique returns
\begin{equation}
 Z_{2} = \frac{\cN_2}{4\,\cE^2}\,\mathrm{det}^2\phi 
- \frac{\cN_2}{8\,\cE}\, \det\phi
- \frac{\cN_2}{64\,\cE} \big[(E_1^2+E_2^2)+\cE\big]~.
 \label{z2res}
\end{equation}

We proceed analogously to the next level, $k=3$. The algebra is more complicated and we just give the final result
\begin{equation}
 Z_{3} = \frac{\cN_3}{12\,\cE^3}\,\mathrm{det}^3\phi 
- \frac{\cN_3}{8\,\cE^2}\,\mathrm{det}^2\phi
- \frac{\cN_3}{192\,\cE^2}\big[3(E_1^2+E_2^2)-5\cE\big]\,\det\phi~.
 \label{z3}
\end{equation}

The computations for $Z_4$ and $Z_5$ are more difficult because involve two integration over $\chi$;
more precisely the Cartan of SO$(k=4,5)$ has two Cartan generators and therefore two eigenvalues over which we have to integrate.
Some details are given in the Appendix \ref{app:A}.
The computations lead to
\begin{subequations}
 \begin{align}
  Z_4 &=\frac{\cN_4}{48\,\cE^4}\,\mathrm{det}^4\phi + \cdots~,\label{z4a}\\
  Z_5 &= \frac{\cN_5}{240\,\cE^5}\,\mathrm{det}^5\phi + \cdots ~,
\label{z5a}
 \end{align}
\end{subequations}
where we have omitted the terms with higher powers in $\cal E$.

\subsection{Last step: the computation of the exotic non-perturbative prepotential}

Let us consider again Equation \eqref{partok} that can be interpreted as the grand-canonical partition function%
\footnote{The first term $Z_0$ is fixed at $1$ corresponding to the fact that at $k=0$ the exotic instanton contribute a factor
of $1$, i.e. do not give any contributions.} 
(where the r\^{o}le of the ``particle-number'' is played by the topological charge that intuitively counts the number of $1$-instantons).

In the low-energy D$3$-brane action, the contributions of the exotic non-perturbative sector are encoded by the corresponding corrections
to the effective prepotential $F$; the prepotential is related to the logarithm of the grand-partition function.
Finally, it is possible to promote the VEV $\phi$ in $\cZ$ to the corresponding full dynamical superfield $\Phi(x,\theta)$.
The very last step to obtain the exotic non perturbative corrections consists in considering the limit of zero graviphoton background,
\begin{equation}
 \label{SeffPhi}
 S^{\mathrm{(\text{exotic})}} = \int d^4x\, d^4\theta\, 
F^{\mathrm{(\text{exotic})}}\big(\Phi(x,\theta)\big)
\end{equation}
where the exotic prepotential $F^{\mathrm{(\text{exotic})}}(\Phi)$ is
\begin{equation}
 \label{prep1}
 F^{\mathrm{(\text{exotic})}}(\Phi) = {\mathcal{E}}\, \log Z\Big|_{\phi\to\Phi,E_A\to 0}~.
\end{equation}
We have to tribute particular care to the moduli $a_\mu$ and $M_\mu$ whose traces correspond to the center position of the instanton in superspace,
namely to the supercoordinates $x$ and $\theta$.
Since we want to extract the centered moduli contribution only, we have to multiply $\log Z$ by $\cE$%
\footnote{As described in \cite{Billo:2009di}, a factor of $1/{\cal E}$ correspond to the integral over the moduli associated to the instanton center position.}.

Expanding \eqref{prep1} in powers of $q$, we have
\begin{equation}
 \label{Fexp}
 F^{\mathrm{(n.p.)}}(\Phi) = \sum_{k=1}^\infty F_k\, q^k~\Big|_{\phi\to\Phi,E_A\to 0}
\end{equation}
Comparing the two ``topological expansions'' \eqref{partfunc} and \eqref{Fexp} we can express the $F_k$ in terms of the $Z_k$,
\begin{equation}
 \label{Fkexp}
 \begin{aligned}
  F_1 & = \cE Z_1~,\\
  F_2 & = \cE Z_2 - \frac {F_1^2}{2\cE}~,\\
  F_3 & = \cE Z_3 - \frac{F_2 F_1}{\cE} - \frac{F_1^3}{6\cE^2}~,\\
  F_4 & = \cE Z_4 - \frac{F_3 F_1}{\cE} - \frac{F_2^2}{2\cE} -
          \frac{F_2 F_1^2}{2\cE^2} - \frac{F_1^4}{24\cE^3}~,\\
  F_5 & = \cE Z_5 - \frac{F_4 F_1}{\cE} - \frac{F_3 F_2}{\cE} - 
           \frac{F_3 F_1^2}{2\cE^2} - \frac{F_2^2 F_1}{2\cE^2} -
           \frac{F_2 F_1^3}{6\cE^3} - \frac{F_1^5}{120\cE^4}~.
 \end{aligned}
\end{equation}

The ending step of our prepotential computation consists in the removal of the graviphoton regulator,
considering the limit $E_A\to 0$.
We require that in this limit the prepotential is well-behaved and defined but, at a first look, the $F_k$ terms present some naive divergences.
Since the various $F_k$ for different values of $k$ correspond to different topological sectors (and in \eqref{Fexp} are consequently
multiplied by different powers of $q$), we cannot have compensations effects between them and at any $k$-level the prepotential must be either finite or null.
There is still a freedom which we are not exploiting, namely the definition of the normalization constants ${\cal N}_k$.
In spite of the fact that the residual freedom (one factor parameter for any value of $k$) is less than the number of the divergence cancellation
conditions that we must satisfy, this proves to be enough.
Of course this is highly non-trivial and can be a sound argument to support the whole procedure and in particular Nekrasov's prescription in the case
of exotic instantons.
On the computational level we proceed imposing the cancellation of the most divergent terms of $F_k$ and fix in this way the overall normalization 
$\cN_k$. 

In the $k=1$ case, remembering \eq{z1}, we have the following result for the corresponding term in the prepotential%
\footnote{The same result has been previously found in \cite{Argurio:2007vqa}} 
\begin{equation}
 \label{F1is}
  F_1 = \cN_1\,\det\phi~.
\end{equation}
Regarding the case $k=2$, from Equations (\ref{Fkexp}) and (\ref{z2res}) we get
\begin{equation}
 \label{mdivF2}
 F_2 = \left(\frac{\cN_2}{4} - \frac{\cN_1^2}{2}\right)\frac{\mathrm{det}^2\phi}{\cE}
- \frac{\cN_2}{8}\, \det\phi
- \frac{\cN_2}{64} \big[(E_1^2+E_2^2)+\cE\big]~. 
\end{equation}
Choosing
\begin{equation}
 \label{n2sol}
 \cN_2 = 2 \cN_1^2~,
\end{equation}
we see that the most divergent term disappears, we are then left with
\begin{equation}
 \label{F2sol}
 F_2= 
- \frac{\cN_1^2}{4}\, \det\phi
- \frac{\cN_1^2}{32} \big[(E_1^2+E_2^2)+\cE\big]
\end{equation}
which is finite in the zero-background limit $E_A\to 0$.
We go on analogously for $k=3$. We use \eq{z3} and put in (\ref{Fkexp}) the expressions just obtained for $F_1$ and $F_2$,
\begin{equation}
\label{F3div}
F_3  = 
\left(\frac{\cN_3}{12} -\frac{\cN_1^3 }{6}\right)\frac{\mathrm{det}^3\phi}{\mathcal{E}^2}  
+  \ldots ~.
\end{equation}
To get rid of the divergent terms we have to choose
\begin{equation}
 \label{n3sol}
 \cN_3 = 2 \cN_1^3~.
\end{equation}
Having done this, the other divergences within $F_3$ cancel,
\begin{equation}
 F_3 = \frac{\cN_1^3}{12} \,\det\phi~.
\label{F3is}
\end{equation}
For $k=4$, we use the partition function $Z_4$ given in Appendix \ref{app:A} 
and once more we require the cancellation of the most divergent term in $F_4$. 
This constrains $\cN_4 = 2 \cN_1^4$. The explicit result for $F_4$ is then
\begin{equation}
 \label{f4res}
 F_4 = -\frac{\cN_1^4}{32} \,\det\phi - \frac{\cN_1^4}{256} \big[(E_1^2+E_2^2)+\cE\big]~,
\end{equation}
which remains finite in the limit $E_A\to 0$.
In the case $k=5$, we have $Z_5$ in Appendix \ref{app:A}
; here the cancellation of the mostly divergent term in $F_5$ leads to $\cN_5 = 2 \cN_1^5$, after which we get 
\begin{equation}
\label{f5res}
 F_5 = \frac{\cN_1^5}{80} \,\det\phi~.
\end{equation}

Promoting the VEV $\phi$ to the full fledged dynamical superfield $\Phi(x,\theta)$ 
and considering the limit $E_A\to 0$, we eventually get 
\begin{equation}
 \label{Fupto5}
  \begin{aligned}
   F^{\mathrm{(n.p.)}}(\Phi)  = -\Tr\Phi^2\,\Big(\frac{\cN_1}{2}\,q - \frac{\cN_1^2}{8}\,q^2 +
\frac{\cN_1^3}{24}\,q^3-\frac{\cN_1^4}{64}\,q^4+\frac{\cN_1^5}{160}\,q^5 \ldots\Big)~,
  \end{aligned}
\end{equation}
where we used 
\begin{equation}
 \det\phi=-\frac{1}{2}\,\Tr\phi^2 
\end{equation}
that is a consequence of \eq{phisu2}.

\subsection{Resumming exotic contributions}
\label{resum}

The results obtained explicitly up to $k=5$ suggest a very interesting structure in the tail of exotic instanton
contributions in the model under analysis. 
This is a special feature of the SU$(2)$ exotic and conformal case. 
Conformality of the model means that the exotic contributions do not bring any dimensional scale;
in other terms, the corrections due to the non-perturbative sector are dimensionless and, as such, 
are all (i.e. for any value of the instanton number $k$) on the same footing.
Indeed it appears reasonable to conjecture the possibility of resumming the entire series of exotic contributions%
\footnote{Remember that ${\cal N}_1$ is a normalization constant whose numerical value can be in principle recovered performing the instanton computations
without neglecting overall factors as we indeed have done throughout the present analysis.}
(note that we are assuming to guess properly the structure of the terms for all values of $k$ and so also the ones that have not been computed directly)
in a closed functional form,
\begin{equation}
 \label{Fresummed}
   F^{\mathrm{(n.p.)}}(\Phi)  = -\Tr\Phi^2\,\log\Big(1+\frac{\cN_1}{2}\,q\Big)~,
\end{equation}
The possibility of resumming at all orders in $k$ allows us to conjecture a compact expression for the
non-perturbative redefinition of the coupling of the quadratic term%
\footnote{A similar closed form for exotic contributions in a D$(-1)/\text{D}(7)$ model has been conjectured in \cite{Fucito:2009rs}. }.
This is very interesting, because it would allow one to redefine the coupling and treat the theory classically;
the effects of the stringy non-perturbative corrections could in fact be encoded simply in the new redefined coupling.
This has also a meaning in terms of ``renormalization'' arguments; indeed we could claim that for SU$(2)$ the exotic non-perturbative sector
does not change the ``nature'' of the model but it simply redefines the fundamental constants.
The last observation renders the point especially interesting with respect to duality arguments and could be a signal of deeper structure.
A theory presenting the same formal structure in different regimes could in fact, in some sense, be dual to itself and possibly
its non-perturbative regime can be related in some insightful way to the perturbative regime.

\subsection{Comment to the SU\texorpdfstring{$(2)$}{} conformal case}

In the conformal $\mathrm{SU}(2)$ case, the integration measure of the instanton moduli integral is dimensionless.
This lays at the basis of the resummation of exotic effects that we have just presented in Subsection \ref{resum}
because all the contributions corresponding to different values of $k$ have the same (null) dimension%
\footnote{We can think to the present case in analogy with ordinary instantons in $\cN=2$ SU$(2)$ gauge theory with $4$ fundamental flavors.}.

There is an interesting observation in relation to supersymmetry.
As the orbifold study showed. we have ${\cal N}=2$ SUSY from the four-dimensional theory point of view;
nevertheless, in the case of SU$(2)$ the symmetric representation (under which the hypermultiplet transforms)
coincides with the adjoint representation (which is instead associated to the vector multiplet).
In this peculiar case, all the fields belong therefore to the same representation of the gauge group and
we can have a \emph{supersymmetry enhancement} from ${\cal N}=2$ to ${\cal N}=4$.
The model under consideration can therefore be thought of as a non-trivial realization 
an $\cN=4$ SU$(2)$ super Yang-Mills theory in four dimensions.
In this case, ordinary gauge instantons do not contribute to the 
(quadratic) effective action and this has been checked successfully also in our model.
Conversely, the stringy instantons do contribute and the computations carried out here show the explicit results.
The contribution of the exotic non-perturbative sector is (completely) accounted by
modifications of the effective prepotential; such corrections spoil at non-perturbative level the supersymmetry enhancement.

\section{SU\texorpdfstring{$(N\neq2)$}{} not-conformal case}
The explicit computations reported so far regard the specific case of SU$(2)$ gauge theory;
the same kind of calculations can however be performed in SU$(N)$ gauge theory with generic $N$ as we
have done in \cite{Ghorbani:2011xh}; we refer to it for a detailed description of the computations
in the generic $N$ case.
There we have found explicitly the non-perturbative stringy corrections 
to the $\mathcal N=2$ prepotential in SU$(N)$ gauge theory employing
a setup analogous to the setup described in the previous Sections and in \cite{Ghorbani:2010ks};
of course the number of gauge branes $N$ is different from $2$ while the kind of background involving
the orbifold $\mathbb C^2/\mathbb Z_3$ and the O$3$-plane is as in Section \ref{OrbiOrie}. 

The main difference between the $N=2$ case and the $N>2$ case 
is that in the latter the one-loop coefficient $b_1$ of the $\beta$-function assumes non-zero values and then
the corresponding gauge theory is not conformal.
As described in \eqref{partok} and following lines, the partition function $Z_k$ resulting from the $k$-instanton sector is dimensionful
and its dimension is given by
\begin{equation}
 [\mu]^{-k b_1}\ ,
\end{equation}
where $\mu$ is a renormalization mass scale linked to the string scale (see Section \ref{reno}). 
Of course, the total partition function $Z$, obtained from the sum over all the instanton $k$-sectors,
has to be dimensionless; indeed, within the sum yielding $Z$, the dimension of any $Z_k$ is compensated by an appropriate power of $q$ defined in (\ref{q}).

Thanks to the same prefactors in $q$, we obtain a total prepotential with the correct dimension, namely (length)$^2$. 
It should be stressed that only for the special case $N=2$ all the addenda $F_k$ contributing to the prepotential have the same dimension;
as a consequence, for $N=2$, they can be re-summed to give the logarithmic closed form conjectured in \eqref{Fresummed}.
Conversely, when $N>2$, the terms $F_k$ in the expansion \eqref{Fexp} have different dimensions 
for different values of $k$; the tail of non-perturbative corrections cannot therefore be summed
because the addenda are on a different dimensional footing.
In other terms, this situation can be rephrased stating that for $N>2$ the various terms in the $k$ prepotential expansion \eqref{Fexp}
account for stringy perturbative corrections to different couplings of the ${\cal N}=2$ effective model that emerges
in the low-energy regime of the D-brane setup.

On a more technical level, to compute the non-perturbative corrections coming from the stringy instanton charges up to $k=3$,
we take advantage of the properties of the elementary symmetric polynomials and their relations with the power sums;
for the details we refer to the appendices of \cite{Ghorbani:2011xh}.
The explicit formul\ae\ we obtain for general $N$ are in agreement with the special case $N=2$.

From the dimensional analysis of the measure of the moduli integral we understand that, for our quiver model, 
the conformality of the gauge theory occurs exclusively for SU$(2)$ gauge group.  
Instead, for $N>2$ a dimensionful pre-factor has to be introduced in front of the moduli integral 
in order to produce a dimensionless action%
\footnote{Remember that we consider natural units where $\hbar=1$.}.
The exotic or stringy nature of the corrections is evident as the moduli integral pre-factor depends explicitly on $\alpha'$;
the stringy instantons then introduce into the low-energy field model an explicit dependence on the string scale.

\section{Final Comments and Future Developments}
\label{comme}
\label{sec:concl}

At the end of this part of the thesis, a general conclusion emerging from the analysis we went through 
is that stringy instantons can contribute to field theories describing the low-energy regime of D-brane models.
More precisely, we performed the study in the context of ${\cal N}=2$ SUSY setups where 
the computations are feasible thanks to the powerful technical tools related to the BRST structure and
the consequent localization framework.

The specific model we considered in depth furnishes an example of four-dimensional field theory defined on the D$3$-world-volume
living in a $\mathbb{C}^3/\mathbb{Z}_3$ orbifold/orientifold background.
We have computed the exotic contributions to the prepotential for the lowest values of the topological charge in the SU$(N)$ ${\cal N}=2$
theory with one vector in the adjoint representation and one hypermultiplet in the symmetric representation of the gauge group.

There are two main lines of future research consisting in generalizing a similar analysis to gauge theories with orthogonal and symplectic gauge groups
and a systematic study of the logarithmic resummation of the SU$(2)$ stringy correction proposed in Subsection \ref{resum}.

\begin{itemize}
 \item The extension of our approach to study models with symplectic and orthogonal gauge groups is already work in progress.
In order to obtain such different gauge groups we can start again from our orbifold/orientifold background but 
we need to dispose appropriately the D-branes on the nodes of the quiver. 
In the present treatment we have placed the gauge branes on the nodes $2$ and $3$ of the quiver (see Section \ref{OrbiOrie});
these two nodes are identified
by the orbifold projection. In other words, the orientifold projection reduces an initial SU$(N_2)\times\text{SU}(N_3)$ (the two factors arise
from the $N_2=N_3=N$ D-branes of the stacks placed at the two nodes) to a single SU$(N)$ emerging from their identification. 
No further projection is considered.
Placing instead the gauge branes on node $1$ we obtain a different outcome; the orientifold actually projects the initial SU$(N_1)$ down 
to its orthogonal or symplectic subgroup according to the kind of symmetry we choose for the orientifold representation on the CP indexes.

 \item The main motivation for studying all such cases is that, as our analysis of SU$(N)$ has shown, the instanton group associated to stringy instantons
are different from the ordinary case. It is then interesting to analyze models realizing such novel instanton structures. 
Indeed, for SU$(N)$ gauge theory, the ordinary instanton group is U$(k)$ while for exotic instantons we have found 
SO$(k)$. Already from a naive analysis of the quiver we can understand that 
both in the symplectic and orthogonal gauge group instances the exotic instanton group is U$(k)$.
Actually, the brane setups producing orthogonal or symplectic theories can be thought of in analogy with the SU$(N)$ model
where we interchange the position on the quiver of D$3$ and D$(-1)$ branes; as for SU$(N)$ gauge theory we had the full SU$(2)$
preserved by the orientifold projection, in the same fashion for orthogonal and symplectic models we have the whole U$(k)$ instanton group
surviving the orientifold projection.

\item Another very interesting future point of interest regards the closed expression
we conjectured in the conformal case (see Subsection \ref{resum}). Indeed, from the explicit computations of the
stringy instanton contributions for the lowest values of $k$, we were able to conjecture
a resummed formula for the stringy corrections to the quadratic coupling of the theory, see \eqref{Fresummed}.
As already noted, this possibility of resumming the complete tail of stringy instanton corrections
could be related to some yet unknown structure in the model. 
Note that the possibility of handling all-order expressions is very interesting especially
in relation to the extension of the results at strong-coupling where, in general, an order by order analysis 
could turn ill-defined%
\footnote{The extension to strong coupling of results computed at weak coupling generally implies the analytic continuation of all-order expressions;
actually, the decomposition according to topological charge can be even meaningless from the strong coupling perspective.}.

\end{itemize}

\part{Holographic Superconductors}

\chapter{Holographic Techniques}
In the context of quantum string and field theory, the term ``\emph{holography}'' is usually
adopted to indicate the study or the application of gauge/gravity dualities.
Holographic ideas suggesting a correspondence between specific gauge theories and string models have been
considered since when 't Hooft observed that the large number of colors (i.e. large $N$) limit
of non-Abelian Yang-Mills theory admits a topological expansion resembling the amplitude expansion of string diagrams.
However, in 1997, Maldacena's $AdS$/CFT conjecture constituted a revolutionary breakthrough.
Not only is $AdS$/CFT itself a powerful and insightful theoretical tool which opened many new computational
possibilities, but it inspired various kindred duality relations that
populate the rich holographic panorama.

This second part of the thesis is tributed to the application of holographic techniques to the study of condensed matter systems 
and, in particular, superconductors. 
The interdisciplinary nature of the subject makes it tantalizing and difficult to treat at the same time.
Even more so, because an ambitious aim consists in trying to bridge the holographic (i.e. string-inspired) studies with the 
(more standard) research in condensed matter physics (and the corresponding scientific communities).
This is essential for several good reasons such as: 
\begin{itemize}
 \item It is crucial to understand precisely how far the stringy-inspired tools can be pushed
in describing real-world systems by means of holographic correspondences. 
 \item ``Cross-fertilization'' can prove extremely valuable in both directions. For the string community, it would increase the awareness
of the state of the art and frontier problems in the condensed matter panorama offering the essential possibility of accompanying a bottom-up approach to 
the top-down attitude; for the condensed matter community, fresh new ideas emerging in an apparently unrelated context can hopefully
shed new light on many non-perturbative and strong-coupling questions.

\end{itemize}

The present treatment is not self-contained and we often refer to existing literature and reviews.

\section{Formulation of the correspondence}
We start by observing that in a CFT it is not possible to define asymptotic states nor the ${\cal S}$ matrix.
This is intimately related to the definition of asymptotic states, and the consequent ${\cal S}$ matrix elements
interconnecting them, in fact, such definition relies on a large distance (i.e. asymptotic) limit.
Since conformal invariance contains scale invariance, it states the equivalence between large and short distances.
In an interacting CFT, the asymptotic fields cannot be approximate with free propagating waves.

The essential objects to be considered in a conformal field theory are the operators.
Of course, only the gauge invariant operators are physical observables.
We will see that the $AdS$/CFT correspondence claims the existence of a gauge invariant operator
for any field living in the dual theory, \cite{Skenderis:2002wp}.
Specifically, for a systematic study of the operators of the CFT we need a method to handle quantitatively the
correlation functions of the operators themselves.
The $AdS$/CFT offers a recipe to obtain analytically the correlation functions of the ``boundary'' CFT from the analysis of its gravitational 
or string dual.
We give just a sketchy picture of how to compute correlation functions in the framework of holographic correspondences;
the topic is very important but also treated widely and in depth in the existing literature, in particular we refer to \cite{D'Hoker:2002aw}.

In the context of a generic quantum conformal field theory, let us consider an operator ${\cal O}$ 
and the corresponding source $\phi$. Let us insert the source term in the
action,
\begin{equation}\label{sou_ter}
 {\cal S}_{\text{CFT}} + \int d^4x\ {\cal O} \cdot \phi\ ,
\end{equation}
where the product $\cdot$ is a symbolic way to keep the treatment as general as possible;
the tensorial or spinorial structure of the operator is indeed unspecified and then generic.
The source $\phi$ is a non-dynamical (i.e. background) field. 
As it is standard in field theory, the computation of expectation
values of $n$-point correlation functions of the operator ${\cal O}$ is performed considering
multiple functional derivations with respect to the source which is eventually put to zero\footnote{
We do not introduce the field theoretic techniques mentioned in the text, 
you can nevertheless find details about them on any field theory textbook, for instance
\cite{Ramond:1981pw}.}.
Let us define the generating functional for connected correlation functions,
\begin{equation}
 e^{W[\phi]} \equiv \langle e^{\int d^4x\ {\cal O}\cdot \phi\ } \rangle_{\text{CFT}}\ .
\end{equation}
The correlation function of $n$ insertions of the operator ${\cal O}$ is obtained computing
\begin{equation}
 \langle\ \underbrace{{\cal O} ... {\cal O}}_n\ \rangle_{\text{CFT}} \sim \left. \frac{\delta^n W[\phi]}{\delta \phi^n} \right|_{\phi=0}\ .
\end{equation}

The core statement of the $AdS$/CFT correspondence (in its strongest version)
claims the complete equivalence and therefore identification of the partition functions of the two dual theories,
e.g. ${\cal N}=4$ U$(N)$ SYM theory in $4$-dimensional Minkowski space-time and full Type
IIB string theory 
on $AdS_5\times S^5$ background with non-trivial $5$-form flux,
\begin{equation}\label{corre}
 {\cal Z}_{\text{CFT}}[\phi] = e^{W[\phi]} = \langle e^{\int d^4x\ {\cal O}(x)\cdot \phi(x)}\ \rangle_{\text{CFT}} 
= {\cal Z}^{\text{Type IIB}}_{AdS_5\times S^5}[\phi]\ .
\end{equation}
With $\phi$ we denote an arbitrary, non-dynamical function defined on the boundary;
its r\^ole in the two sides of the duality is different:
In the conformal field theory $\phi$  represents the source associated to the operator ${\cal O}$ while, in the string theory partition function, $\phi$
represents the ``boundary value'' of the dynamical bulk field%
\footnote{From now on the term ``bulk'' will refer to objects living in the higher-dimensional space-time as opposed to ``boundary'' objects living on the
conformal boundary of $AdS$.}  $\hat{\phi}$.

Many comments are here in order, the most important points in view of the subsequent analysis are illustrated in the following subsections.

\subsection{Conformal structure of the \texorpdfstring{$AdS$}{} boundary}
\label{colf}

Studies of gravitational theories living on $AdS$ backgrounds usually treat the asymptotic region as a boundary. 
The large radii $r \gg 1$ region of $AdS$ space-times is instead technically a \emph{conformal boundary} 
(look at \cite{Skenderis:2002wp,Henningson:1998gx});
this means that the bulk metric yields a boundary metric up to conformal transformations.
In other terms, the metric $g$ in the bulk does not induce a unique metric $\tilde{g}$ on the boundary;
the near-boundary expansion of $g$ is
\begin{equation}
 g \sim \frac{1}{r^2}\ g^{(0)} + ... 
\end{equation}
where the dots indicate terms that are subleading for $r \rightarrow \infty$.
In order to define a boundary metric $\tilde{g}$ we have to choose a scalar function $f(r,t,x^1,x^2,x^3)$ defined on the bulk which vanishes linearly
at the boundary\footnote{The function $f$ is usually called the \emph{defining function} of the boundary metric $\tilde{g}$.
We have to require that $\lim_{r\rightarrow\infty} f^2/r^2$ is non-vanishing at any point $(x^i,t)$ of the boundary manifold.},
\begin{equation}
 \tilde{g} = \lim_{r\rightarrow \infty} f^2\, g \ .
\end{equation}
The bulk metric defines therefore an equivalence class of boundary metrics;
the transformation linking two boundary metrics which are representatives of the same class is the multiplication by a scalar function,
i.e. a conformal transformation.
For instance, we can choose the simplest defining function $f=r$ and induce a metric on the boundary from 
\begin{equation}
 r^2 ds^2 \ ,
\end{equation}
where
\begin{equation}\label{AdSmetric}
 ds^2 = \frac{r^2}{L^2} \left(-dt^2 + \sum_{i=1}^3 dx_i^2 \right) 
+ \frac{L^2}{r^2} dr^2 + L^2 d\Omega_5^2
\end{equation}
is the $AdS_5 \times S^5$ metric.

Notice that, as long as the large radius behavior of the background is described by 
the $AdS_5 \times S^5$ metric \eqref{AdSmetric}, the definition of the conformal boundary
remains unchanged.
In other terms, the holographic correspondence itself can be generalized to gravitational models admitting
asymptotically $AdS_5 \times S^5$ vacua. Such generalization to asymptotically
$AdS$ spaces will be of direct interest to our studies; indeed, in order to model thermal dual field theory, we will introduce
black hole configurations which, however, for large values of the radius will approach the $AdS$ geometry%
\footnote{The case which we will directly investigate refers to $AdS_4\times Y/\text{CFT}_3$ duality, i.e. a lower dimensional case of
holographic correspondence. $Y$ represents here the ``internal'' compact manifold.}

\subsection{Effective supergravity description}

The correspondence \eqref{corre} claims the equivalence of a conformal quantum field theory, namely ${\cal N}=4$ SYM, with a specific string model,
i.e. Type IIB on $AdS_5\times S^5$ background.
Since full string theory computations can be a pretty complicated subject, one can legitimately wonder if, even though the CFT happens to be strongly coupled,
it is in any sense useful to try to use stringy computations to obtain information on the dual conformal field theory.
We have to notice that, restraining to particular regimes, we can exploit an effective description of the string side of the duality;
the computations performed in the effective framework could be practically feasible.

We are dealing with Type IIB string theory on an $AdS_5\times S^5$ vacuum; the
string dynamics is characterized by a unique dimensionful parameter, namely $\alpha'$, but the vacuum solution provides another 
dimensionful parameter that is the $AdS$ radius of curvature denoted with $L$.
A physically significant quantity is given by the ratio between these two parameters,
\begin{equation}\label{sugra_ratio}
 \frac{L^4}{\alpha'^2} \sim \frac{L^4}{l_s^4}\ ,
\end{equation}
where we have used \eqref{car_length} relating $\alpha'$ to the characteristic string length $l_s$.
In Equation \eqref{sugra_ratio} we have the ratio between the background scale and the scale of the strings living on it;
from the relations found studying the supergravity solution describing a stack of $N$ D$3$ branes (see Appendix \ref{clues} and
mainly Equation \eqref{sugra_brane3}),
we can rewrite \eqref{sugra_ratio} and ask
\begin{equation}\label{sugra_necc}
 \frac{L^4}{l_s^4} \sim g_s N \gg 1 \ .
\end{equation}
This is a necessary condition for describing the string model effectively with the corresponding supergravity theory.
Notice, however, that in order to legitimate a perturbative weakly coupled string and then supergravity description\footnote{The
string coupling constant $g_s$ can be thought of as measuring the likelihood for a string to ``break'' or for
two strings to merge.
More precisely, in string Feynman diagrams, any string splitting or fusion introduces a factor $g_s$ in the associated amplitude.
This generalizes the introduction of the coupling to any interaction vertex in a field theory Feynman diagram.
Higher loop string diagrams present more string splittings and fusions; they contain therefore higher powers of $g_s$.}
we need $g_s\ll 1$ and therefore $N$ large enough to satisfy \eqref{sugra_necc}.
Furthermore, in the far $g_s\ll 1$ and $N\gg 1$ regime, the supergravity description becomes even classical 
and fully quantum CFT correlation functions can be obtained from a dual on-shell analysis on the gravity side of the correspondence.
For an important part, the computations involved in the analysis of our holographic superconductor
consists in the study of the system of equations of motion (and its solutions) associated to the gravitational dual model.

In order to appreciate the holographic meaning of the condition \eqref{sugra_necc}, let us repeat a comment reported in \cite{Hartnoll:2011fn}.
For the simplest gravitational system described by the Lagrangian density
\begin{equation}
 {\cal L} = \frac{1}{2\kappa^2} \left( R + \frac{6}{L^2} \right)\ ,
\end{equation}
it is possible to show that the free-energy of Schwarzschild-$AdS$ solution
is given by
\begin{equation}
 F = - T \log Z = T S_{\text{on-shell}} = \frac{(4\pi)^3}{2\cdot 3^3} \frac{L^2}{\kappa^2} V_2 T^3\ ,
\end{equation}
where $V_2$ represents the volume of the spatial part of the boundary.
The coefficient of the free-energy scaling with respect to the temperature is related to the number of degrees of freedom of the system.
For large $N$ we have $L/\kappa^2 \gg 1$ so a large $AdS$ curvature reveals indeed (in the thermodynamic picture of the gravitational model) 
a large number of degrees of freedom.

\subsection{IR/UV connection and holographic renormalization}

Let us consider again the $AdS_5$ part of the metric \eqref{AdSmetric} introducing the new coordinate $z\doteq \frac{1}{r}$,
\begin{equation}\label{conz}
 ds^2_{AdS_5} = \frac{1}{z^2} \left[ \frac{1}{L^2} \left(-dt^2 + \sum_{i=1}^3 dx_i^2 \right)
                       + L^2 dz^2 \right] \ ,
\end{equation}
which is manifestly invariant under the following rescaling transformation
\begin{equation}\label{scaling}
 t \rightarrow \lambda t \ , \ \ \ 
 x^i \rightarrow \lambda x^i \ , \ \ \
 z \rightarrow \lambda z \ .
\end{equation}
The coordinates $x^i$ and $t$ span the boundary which is the base manifold of the CFT
while $z$ is the $AdS$ radial coordinate.
Note that, for $\lambda>1$, the scaling \eqref{scaling} sends a mode oscillating with period $T$ 
into a mode with longer period $\lambda T$ and then lower frequency.
Since the frequency is associated to the energy, we can interpret the $AdS$ coordinate $z$ (which under rescalings behaves as the period $T$) as
an inverse energy scale.

Any energy regime of the CFT is associated to a corresponding value of the $AdS$ radial coordinate.
In other terms, the radial $AdS$ direction can be interpreted as a renormalization scale coordinate.
This correspondence can be supported by explicit calculations; for instance, it can be shown that Green's functions or Wilson's loops
associated to a particular energy scale $E$, when computed in the dual gravity model, receive contributions mainly from the bulk region
corresponding to $z\sim 1/E$, \cite{Zaffaroni:2000vh}.
Given that $z$ can be regarded as an inverse energy scale, the near boundary region, i.e. $z\ll 1$,
is associated to the high-energy regime of the dual conformal field theory and, conversely,
the central $AdS$ region (or near horizon when the gravity model possesses a black hole) corresponds to
the low-energy regime of the CFT.

To have a well defined quantum field theory we have to renormalize the UV divergences
suffered by the correlation functions. 
As the divergences to be cured are a high-energy phenomenon (i.e. we are considering UV-divergences), 
in the dual picture they must correspond to some kind of problem arising in the near-boundary region.
Indeed, the field theory UV-divergences are dual to the divergence of the $AdS$ volume in the asymptotic region. 
Specifically, an $AdS$ radial shell defined by $\overline{z} < z < \overline{z} + \delta$ contributes to the
$1/(\overline{z}+\delta) < E < 1/\overline{z}$ ``regime'' of the CFT and it presents a diverging volume in the limit
$\overline{z}\rightarrow 0$.

In quantum field theory, to make precise sense of the computations, one needs to cure the singular behavior of the correlators by subtracting the 
divergent part; in other words, it is necessary to regularize the divergences and renormalize the theory considering specific limits
of the regularized theory.
We do not enter into the details of this procedure which in the dual holographic picture is usually referred to as \emph{holographic renormalization}.
For a detailed analysis we refer to the review \cite{Skenderis:2002wp}.

Since from the $AdS$ space viewpoint the boundary region represents asymptotic distance from the branes, it is associated to the
long wave-length physics of the gravitational system; in other terms, the small $z$ (i.e. large $r$) region encodes the IR regime of the gravity model.
The correspondence between the UV physics of the CFT and the IR physics of the dual gravitational system
is referred to as \emph{UV/IR connection}.

\section{Holography and thermodynamics}
\label{HoloThermo}

As already said the holographic correspondence states the identification between the partition functions
of the two connected dual models.
Considering (as we will do in our explicit computations) the regime in which the gravitational theory is accountable with a semi-classical approach
we are allowed to evaluate the partition function with the exponential of minus the on-shell action,
\begin{equation}\label{calza}
 {\cal Z} \sim e^{-S_{(\text{on-shell})}} \ .
\end{equation}
This is the leading classical contribution which could be refined considering quadratic semi-classical fluctuations
around the classical solution or, even further, considering the full path integral.
Sticking to the classical level, form \eqref{calza} we can compute straightforwardly the free-energy thermodynamic potential
\begin{equation}
 F = - T \ln {\cal Z} \sim T S_{(\text{on-shell})} \ ,
\end{equation}
were $T$ represents the temperature about which we comment in Subsection \ref{Hawk}.

The identification of the partition function of the two dual fellow models
suggests that the thermodynamic of the two sides of the correspondence is the same. 
Indeed, the identification is true also for intensive thermodynamical quantity
like the temperature (see next subsection) and the entropy.

\subsection{CFT at finite temperature and Hawking temperature of the gravitational dual}
\label{Hawk}

The standard way to treat quantum field theory at finite temperature prescribes to consider analytic continuation with respect to 
imaginary times. The imaginary time is given a compact extension and it is therefore periodic; 
the inverse of the period is identified with the temperature of the system%
\footnote{A naive way of thinking the introduction of finite temperature through a compact imaginary time is to 
think to Heisenberg indeterminacy principle: specifically, a compact dimension localizes to some extent the physics producing
indetermination of the corresponding dual quantity (``dual'' in the sense of coordinate/conjugate momentum).
A compact time produces indeterminacy in energy, and this quantum effect for the imaginary part of the time coordinate
can be exploited to mimic thermal fluctuations.}.

In a holographic correspondence, the gravitational model is higher-dimensional but it contains the space-time directions
on which the dual CFT is defined. In particular, in the coordinate systems adopted in \eqref{AdSmetric} or \eqref{conz}, 
the time direction coincides on the two sides of the duality.
Considering a compact Euclidean time on one side, implies naturally the same feature on the other; hence,
the temperature of the two dual descriptions coincides.
Let us enter into the detail as it is useful in the following.
Consider a generic $AdS$ black hole whose $t,r$ part of the metric has the following shape
\begin{equation}
 ds^2 \sim - a(r) \, b(r)\ dt^2 + \frac{dr^2}{b(r)}\ ,
\end{equation}
where $b(r_H)=0$, i.e. it vanishes at the horizon%
\footnote{An event horizon is actually a locus characterized by the vanishing of the $tt$ component of the metric.}.
If we require Euclidean regularity at the horizon%
\footnote{In the limit (if it exist) in which the horizon shrinks to a point, the regularity requirement
amounts to avoiding a conical singularity in the Euclidean $t-r$ plane.}
we want the Euclidean (i.e. $t\rightarrow \im \tau$) part of the metric to behave as the
flat polar coordinates $r,\vartheta$,
\begin{equation}\label{piano}
 ds^2_{\text{pol}} = dr^2 + r^2\ d\vartheta^2\ .
\end{equation}
Let us perform some simple passage on the Euclidean-time metric
\begin{equation}\label{forte}
 \begin{split}
 ds^2_{\text{Eucl}} &\sim  a(r)b(r)\ d\tau^2 + \frac{dr^2}{b(r)} \ \propto \ b^2(r)a(r)\ d\tau^2 + dr^2 \\
 &\sim \left[\frac{d}{dr}(b \, a^{1/2})\right]^2_{r=r_H} \, (r-r_H)^2\, d\tau^2 + dr^2 + ...
 \end{split}
\end{equation}
Comparing \eqref{piano} and \eqref{forte}, we have
\begin{equation}
 \vartheta\ \leftrightarrow \left|\frac{d}{dr}(b \, a^{1/2})\right|_{r=r_H}\tau\ .
\end{equation}
As a consequence, we have that Euclidean time is periodic with the period given by
\begin{equation}
 \tau \sim \tau + \frac{4\pi}{\left|\frac{d}{dr}(b \, a^{1/2})\right|_{r=r_H}}\ .
\end{equation}
The temperature is identified with the inverse period of the Euclidean time, then
\begin{equation}\label{temp}
 T = \frac{1}{4\pi} \left|\frac{d}{dr}(b \, a^{1/2})\right|_{r=r_H}\ .
\end{equation}

\section{Motivations}
\subsection{Theoretical interest}

The theoretical interest of $AdS$/CFT and kindred holographic correspondences is vast and deep.
Already by itself the correspondence relates theories that have been thought of as separate and unrelated.
This can lead to paramount theoretical development as, already at the intuitive level,
there is the possibility of thinking of a theory in terms of its dual that possibly offers
an easier approach to specific questions.
Useless to say that this is precisely the case in that $AdS$/CFT relates the strongly coupled regime on one
side of the duality to the weakly coupled regime on the other side and vice versa%
\footnote{Observe that a strong-weak correspondence has also an interesting ``philosophical'' implication:
indeed, it not only consists in a connection between apparently detached branches of theoretical physics, but it could affect also our idea
of what is more ``fundamental''. A general attitude in theoretical physics is that progressively higher-energy regimes are related to more fundamental dynamics
and constituents; this is the case for example in relation of the UV-free QCD. However, in a would be gravitational dual of QCD
the asymptotic freedom regime would be mapped to a strong interacting string model.}. 
A great source of fascination indeed relies in the fact that it opens a novel path to the analytical study of
strongly coupled field theories.
On the same line but in the opposite direction, the $AdS$/CFT correspondence could be regarded as a conjectured definition of quantum gravity
on a particular geometrical background.
The correspondence has been useful to study black hole physics (e.g. in relation to the information problem)
with dual field theoretical means.

This thesis orients the light-spot on the striking possibility of describing models of strongly coupled media presenting coupled electric-spin
properties (with particular focus to superconductors) by means of their dual picture involving hairy black holes, whereas, as we will briefly report here, 
other various applications are viable.

As a generic motivation for holographic research (and other topics in theoretical physics)
let us repeat a nice example mentioned by Sachdev in \cite{2011arXiv1108} related to the history of strongly interacting systems.
The work of Bethe in the early '$30$ opened the way to studying a wide range of quantum many body and strongly interacting systems
defined in two dimensions (the time and one spatial). 
Customarily these models are referred to as ``integrable systems'' as they have an infinite number of conserved quantities.
Since the integrability properties requires fine-tuning of the theory, the generic expectation 
is that they do not describe directly any real-world system.
The study of integrable models has however led to a deeper insight and comprehension of quantum many body dynamics in one spatial
dimension; this, in turn, proved essential (to make just an example) to the the development of the Tomonaga-Luttinger liquid modeling electrons in 
$1$-dimensional conductors such as carbon nanotubes.

In some respect, the holographic panorama is similar. 
In general, the systems of which we know precisely the gravitational dual are 
not directly phenomenologically relevant microscopic models. For instance,
we still not have a dual for the Standard Model or some of its sub-sectors.
In spite of this, $AdS$/CFT and kindred correspondences make us aware of some features of the strongly coupled regime of quantum field theories
that can be relevant beyond the particular model.
Of course much effort is tributed to the quest of holographic setups able to reproduce at the microscopic level as closely as possible some real-world system,
but this is not the only possibility of exploiting the holographic methods.
Indeed, it is possible to use them at the ``macroscopic'' level;
in this case, the quantum field theory of which we investigate the strong-coupling regime is an effective theory.
The study of the unbalanced holographic superconductor performed in this thesis represents an instance of this latter attitude.
Even though at this stage we work with an effective theory, this does not prevent as a future development to be able to 
embed the gravitational system in a consistent truncation of a full fledged string model.
This development would allow a precise identification of the dual degrees of freedom in terms of which the strongly
coupled regime of the field theory admits a weakly coupled treatment.

\subsection{Phenomenological applications}


The $AdS$/CFT and similar string-inspired strong/weak dualities provide efficient analytical tools for studying quantitatively
some quantum field theory models at strong coupling.
Optimistically, when we encounter a strongly coupled gauge theory, we can hope
to find some dual argument or gravitational framework to work with perturbatively in order to extract
some information about the original theory or at least gain some qualitative insight.
There is actually a wide variety of phenomenological topics which are 
described by a strongly coupled quantum field theory whose dual model is not completely 
unknown or mysterious.
In other cases, although the dual theory is obscurer, by means of careful analogies to model
possessing a known gravity dual, one can
gain some insight about the strongly coupled dynamics.
In this section we will briefly spend some words and indicate some bibliography about the main 
applications of holographic techniques
in order to tribute due attention to the far-ranging phenomenological relevance and motivations
of string/gauge duality studies.

\begin{itemize}
 \item {\bfseries Strongly coupled field theory at finite temperature:} 

The strong-coupling regime generally posits difficult practical issues as in relation to field theoretical methods
that often rely on perturbative techniques. 
A viable alternative is represented by numerical calculations performed on the lattice, however,
also the lattice approach can result poor in dealing with systems out of the thermodynamic equilibrium.
The finite temperature out-of-equilibrium dynamics of strongly coupled media is difficult to treat even numerically 
and the main source of trouble consists in the fact that it is problematic to define real-time quantities, such as correlation functions,
at finite temperature; a ``complex weighting'' $e^{-\im S}$ ($S$ is the action) in the partition function, necessary for real-time computations, 
renders the usual importance sampling exploited in lattice simulations troublesome%
\footnote{A similar problem affects QCD Monte Carlo simulations at finite density.
In the QCD partition function the quarks appears quadratically, it ts then possible to perform the ``Gaussian'' integral over them
obtaining a determinant which, at finite baryon density, is complex.
Again we face the problem implied by a complex weighting within an importance sampling computation; this is commonly referred to as the \emph{sign problem},
see for instance \cite{Philipsen:2008zz}.}.
As alternative methods generally suffer because of complicated issues, the $AdS$/CFT-like techniques are an interesting and significant 
possibility for the study of strongly coupled media and especially their dynamics beyond thermal equilibrium.
Note however that the lattice and the holographic methods, even if presented usually as alternative to each others, are not known to 
cross-fertilization; to have an instance see \cite{Evans:2007we}.

 \item {\bfseries QCD and quark-gluon plasma:} 
The quark-gluon plasma is the strong coupling deconfined phase of quantum chromo-dynamics;
so far, QCD belongs to the set of
quantum field theories whose dual is not unknown.
As a consequence, in relation to QCD, the results obtained with holographic means remain to some extent qualitative
and based on analogies with other theories whose dual is specified.
Nevertheless, especially because its tremendous
phenomenological significance, there has been much focused interest and theoretical work on QCD
also by means of $AdS$/CFT-like tools.

There are many important achievements about the phenomenology
of the QGP obtained with string-inspired techniques such as the modeling of 
mesons dynamics within the deconfined plasma
 \cite{CasalderreySolana:2008ne} \cite{Erdmenger:2010zm},
the computation of shear viscosity, 
bulk viscosity (related to compressibility and the propagation
of shock waves and sound in the medium) 
and some insight on the topic of QGP thermalization%
\footnote{We reccommend the review \cite{CasalderreySolana:2011us} and references therein.}
The instance of shear viscosity is particularly interesting,
indeed direct experiments at the relativistic heavy ion collider (RHIC) and at LHC
indicate that the QGP shear viscosity is very small.
Perturbative QCD computations yield large values for the shear viscosity and,
as already mentioned, real time studies are troublesome on the lattice;
$AdS$/CFT techniques applied on ${\cal N}=4$ SYM indicate a value \cite{Policastro:2001yc}
\begin{equation}\label{shear}
 \frac{\eta}{s} = \frac{\hbar}{4\pi k_B}\ ,
\end{equation}
for the ratio of the shear viscosity over the entropy density.
This low value is closer to the QGP measurements than the field theoretic results%
\footnote{The experimental windows for the QGP shear viscosity over entropy density ration is $0.08 <\eta/s< 0.3$
at a temperature of about $170 \text{MeV}$. Perturbative QCD computations lead to $\eta/s\sim 1$ while
holographic methods for ${\cal N}=4$ SYM yield \eqref{shear} which, in natural units, becomes $\eta/s=\/4\pi \sim 0.08$, see \cite{Gavin:2006xd}.}. 
Furthermore, the expression \eqref{shear} is universal for theories possessing
two-derivative gravity dual models, \cite{Kovtun:2004de}.
In general, even in relation to the non-strictly universal results, ${\cal N}=4$ SYM theory (whose gravity dual is Type IIB string theory on $AdS_5\times S^5$)
in the strongly interacting double scaling limit\footnote{This is another
name for the 't Hooft limit \eqref{dou_sca}.} regime is expected to reproduce 
strongly coupled QCD dynamics to a good approximation.
For the sake of brevity, we do not enter into detail but refer to the review paper \cite{CasalderreySolana:2011us}.

Another important topic for which holographic techniques are relevant is the study of the QCD phase diagram \cite{Evans:2011eu}.
The question is rather delicate because the QCD phase diagram shows a pronounced sensitivity to the parameters of the theory; even more so,
since we lack a precise dual for QCD, the extrapolation of results obtained for other model to QCD can be troublesome.
Given the importance of the subject, there has been much effort also from the numerical (i.e. lattice) front which, however, suffers at finite density.

We have just presented a list of $AdS$/CFT applications to QCD which does not exhaust the complete panorama;
a concise review on the employment of stringy techniques to QCD and in particular to the QGP
is \cite{Gubser:2009fc}.

\item{ {\bfseries Condensed Matter:}} Many condensed matter systems admit a description with quantum field theory
in the strong-coupling regime. 
In the next points, we give a list of some peculiar instances in which holographic techniques can offer valuable information and investigation methods.

As dualities connecting two theories, $AdS$/CFT and its analogs are usually significant in a two-fold way,
namely using computations on one side to obtain information about the dual theory.
This can be done in both directions.
A very significant and tantalizing point to underline relies on the research of the possibility of engineering condensed matter systems
described by quantum field theories whose dual is precisely known. 
Actually, such a finding would open the doors to an experimental test
on aspects of the dual quantum gravity. 
A hopeful observation is that, as opposed to high energy physics where the quantum field theory is essentially unique, 
the condensed matter realm offers a wide range of different models described by different theories.
In this regard technological developments such as meta-materials enlarge the panorama of possibilities.
Indeed, meta-materials are essentially formed by arrays of nanostructured elements which play the role of artificial atoms;
the possibility of engineering systems with exotic properties is then significantly enhanced.

\item{ {\bfseries Quantum phase transitions:} A quantum phase transition is a phase transition at zero absolute temperature
which is driven by quantum fluctuations instead of thermal fluctuations. 
Even though, strictly speaking, a quantum phase transition occurs only at $T=0$, there exists a nearby region in the phase space
for $T>0$ where the dynamics of the system is strongly affected by the presence of the quantum critical point.
This region goes under the name of quantum critical region.

A quantum phase transition occurs in a system at $T=0$ and at a precise point in the parameter space where
the parameters themselves attain their critical value. Such external control parameters could be, for instance, the magnetic field
or the pressure. If we move away from the quantum critical point at $T=0$ increasing the temperature we discover that the quantum critical region
widens. In other words, for low enough temperature, the extension (in parameter space) of the region of the phase diagram 
which is affected by the critical quantum dynamics increases with temperature.
At a first thought this phenomenon could sound counterintuitive, in that the quantum criticality extends its relevance in the parameter space in moving away from
the quantum critical point. To have an intution of this we can reason as follows: the quantum critical point separates two different phases or
two different ordered states of the system; the two phases have different, long-range excitations which at the critical point require a vanishing energy 
to be excited. If we consider a non-vanishing temperature we add a thermal noise with a characteristic energy $\epsilon$. The thermal backgorund
makes us uncapable of distinguishing between fluctuations whose energy cost is smaller than $\epsilon$. As a consequence, a fluctuation with finite energy
smaller than $\epsilon$ can be ``confused'' with a zero energy fluctuation. In this sense, moving away from the critical temperature, the critical region widens.
Let us notice, however, that this kind of reasoning is valid as long as the quantum fluctuations dominate over the thermal ones or, in other terms, as long as 
the quantum order is not spoiled completely by thermal noise.

The qualitative and quantitative description of a continuous quantum phase transition exploits the same theoretical framework
employed for normal continuous phase transitions. 
More precisely, a system in the proximity of a quantum critical point is characterized by a diverging coherence length and the behavior of the observables
is described by the corresponding quantum critical exponents. 
Strictly at quantum criticality, the coherence length is infinite and the system becomes scale invariant; this is the reason why 
the critical system can be described with a conformal quantum field theory.
This is also where holographic tools enter into the game, especially when the critical system happens to be strongly coupled.

For a detailed review containing also examples of $AdS$/CFT applied to quantum phase transitions consult for example \cite{Herzog:2009xv}}.

\item{ {\bfseries Non-conventional superconductors:}} 

Conventional superconductors are described with Bardeen, Cooper and Schrieffer theory (BCS), where superconductivity is explained as arising
from the condensation of a fermionic bilinear operator describing interacting pairs of electrons, the so called Cooper's pairs.
This interaction is mediated by the crystalline lattice and in particular by phonons describing its vibrational modes.
The BCS picture, however, does not exhaust neither explain all the superconductivity phenomena observed in Nature.
Indeed, there exist for instance superconductors in which the occurrence of interacting Cooper-like pairs is mediated by spin-spin interactions.

The BCS approach relies on the possibility of describing the system with weakly interacting degrees of freedom. 
An important class of non-BCS superconductors is constituted by all the instances in which the weak interacting picture is not suitable.
The onset of superconductivity in a strongly interacting medium is usually connected with a quantum phase transition;
we do not enter in this complicated subject, for a succinct review look at \cite{Hartnoll:2009sz} and references therein.
Let us just mention that this topic involves the physics of the so called \emph{heavy fermion} metals, where the effective mass of the
conducting electrons (because of strong interactions between conducting and valence electrons%
\footnote{This enhancement of the effective electron mass is referred to as \emph{Kondo effect} and arises
from hybridization of conducting electrons and ``fixed'' strongly correlated electrons generally accounted for as a lattice of magnetic
moments (usually called \emph{Kondo lattice})}.) is orders of magnitude above the bare electron mass,
and the cuprate or layered iron pnictides superconductors presenting a high critical temperature for the occurrence of superconductivity as their peculiar feature%
\footnote{The order of magnitude of the critical temperature is within $10 \text{K}$ and $100 \text{K}$.} (for a review on the layered iron pnictides 
superconductors we refer to \cite{2010AdPhy..59..803J}).

\item {\bfseries Non-Fermi liquid:} Non-Fermi liquids are systems possessing some features similar to Fermi liquids (like 
the presence of a Fermi-like surface) but at the same time they do not admit a quasi-particle description for the corresponding microscopic dynamics.
In other terms, the would-be quasi-particles are ill-defined because of some strong interaction.
On a formal level, the presence of a Fermi-like surface is associated to a pole in the fermion Green's function;
the value of the momentum at which the Green function diverges plays the role of the ``Fermi'' momentum.
For further details see for instance \cite{2011arXiv1108} and references therein.
\end{itemize}

Other phenomena that admit a holographic macroscopic description are forced ferromagnetic or spontaneous
anti-ferromagnetic (or spin waves) systems, \cite{Iqbal:2010eh}.
Models with an external DC current and the description of a holographic Josephson junction are given in \cite{Arean:2010xd,Siani:2011uj}.
Various reviews on this subject exist in the literature, we particularly reccommend \cite{Hartnoll:2011fn}, \cite{Denef:2009tp} and \cite{Hartnoll:2009sz}.

\subsection{Beyond conformality} 

Holography relates the conformal group of the gauge theory and the isometry group of the dual gravitational $AdS$ space%
\footnote{See Appendix \ref{cor_group} for details.}.
There are different possibilities of breaking conformality and, essentially, they all involve the introduction of a characteristic scale into the theory.
The dual gravitational picture will therefore present some additional features (associated to the RG-flow radial coordinate) affecting and modifying the
$AdS$ geometry.
Scale invariance relates different energy regimes of the theory and is mapped into shifts of the $AdS$ radial coordinate of the dual model.
Breaking the scale invariance corresponds to breaking radial translations invariance.

Let us consider the possibility of describing holographically a finite temperature system.
The characteristic thermal energy represents obviously an energy scale which breaks conformaility.
Gravity setups at finite temperature are characterized by the presence of a horizon emitting quantum Hawking radiation with a thermal spectrum.
In other terms, we expect to have black holes solutions that tend asymptotically for large radius to $AdS$ space.
Indeed, the boundary region is related to the high-energy regime where the temperature scale becomes neglectable
and conformality is effectively ``recovered''.
Observe that the presence of the black hole horizon introduces a sort of ``cut-off'' in the radial direction.

If we examine an $AdS$-Schwarzschild black hole, it has a horizon radius which is related to the temperature in such a way that
in the zero-temperature limit the horizon shrinks to a point and vanishes. 
This feature is not generic of all black hole solutions. 
If we consider the $AdS$-Reissner-Nordstr\"{o}m black hole (which generalizes the Schwarzschild one adding a total charge)
the presence of a net charge affects the horizon and in particular the horizon radius does not shrinks to zero
in the $T\rightarrow 0$ limit. The presence of a total charge for the black hole describes holographically a charge density of the dual field theory
and the fact that, also at zero temperature, the horizon has finite radius can be interpreted as the breaking of conformality
due to the energy associated to the presence of the charge density.


The field of conformality breaking in a holographic context has received much attention
and non-conformal generalization of $AdS$/CFT correspondence have gathered keen interest.
Viable approaches rely on mass deformation of conformal field theories (and of then of the RG flows)
or model in which (on the gravity side) there are D-branes wrapping non-trivial cycles (whose volume introduces a characteristic
scale into the theory) of the internal manifold (see \cite{Bertolini:2002xu} for further details).

\chapter{Minimal Holographic Description of a Superconductor}

\section{Superconductors, introductory remarks}
This section will be unavoidably brief if compared with the significance and wideness of the subject;
the purpose here is just to mention some crucial ideas that will be useful in the following sections.
For a thorough treatment we refer to the abundant literature on the topic of superconductivity.

\subsection{Historical account}

\begin{figure}[ht]
  \centering
  \includegraphics[scale=.35]{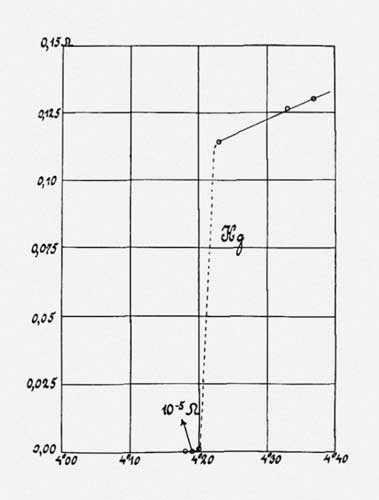}
  \label{figura}
  \caption{Conductivity of mercury at the superconductive transition (original plot taken from Onnes' Nobel Lecture \cite{superco}).}
\end{figure}
\emph{``...the experiment left no doubt that, as far as accuracy of
measurement went, the resistance disappeared. At the same time, however,
something unexpected occurred. The disappearance did not take place gradually 
but (compare Fig. \ref{figura}) abruptly. From $1/500$ ($\Omega$) the resistance at $4.2$ K drops
to a millionth part. At the lowest temperature, $1.5$ K, it could be established                                                     
that the resistance had become less than a thousand-millionth part of that at
normal temperature.
Thus the mercury at $4.2$ K has entered a new state, which, owing to its
particular electrical properties, can be called the state of {\bfseries superconductivity} \cite{superco}.''}.

Superconductivity was discovered by Kamerlingh Onnes in 1911 \cite{disco} in his Leiden laboratories.
It obviously appeared at once as an impressive breakthrough from the experimental (and later also technological)
viewpoint and also a challenging theoretical question; Kamerlingh Onnes itself started soon to study and measure the possibility of having superconductive
coils to produce unprecedentedly high magnetic fields. At the same time, without venturing to give
a theoretical explanation or interpretation, the observation of a threshold current beyond which superconductivity
was spoiled gave a first hint of the richness of the phase space structure of superconductors.

The measure of the goodness of the superconductors ``perfect DC conductivity'' has been soon and henceforth 
accurately tested. The best measurements exploit nuclear resonance techniques to detect the
variations of the field generated by persistently circulating currents;
in appropriate experimental conditions, it is not observable any decay of the persistence of superconductive currents for periods of time that,
quoting Ketterson and Song \cite{ketterson1999superconductivity}, ``are limited (only) by the patience of the observer''.
To have an idea about the orders of magnitude, precise measurements have returned a lower bound for the characteristic
decay time of persistent currents in a superconductor of about $10^5$ years,  \cite{tinkham2004introduction}.

Some years after the discovery of superconductivity, in 1933, Meissner and Ochsenfeld observed the perfect diamagnetism of superconductors;
this phenomenon has usually assumed the (probably) unfair name of Meissner effect.
Physically it consists in the expulsion of magnetic fields from the superconducting bulk.

A couple of years later, the London brothers gave an account of the perfect conductivity and diamagnetism
of superconductors by means of a phenomenological set of equations (the London equations)
but it is only in 1959 that Landau and Ginzburg firstly recognized the crucial role played 
by symmetry breaking in the superconductor physics. 
Abrikosov showed that from Landau and Ginzburg model is possible to predict theoretically the existence of two
categories of superconductors, namely Type I and II%
\footnote{The two type of superconductors differ in the sign of the energy cost of domain walls 
separating superconducting from non-supercoduction portions of the material; 
such difference yields qualitatively distinct behaviors at the phase transition.
For more details we refer to \cite{tinkham2004introduction}.}

In 1957, in the western side of the iron curtain, Bardeen, Cooper and Schierffer (BCS) gave
the first microscopic description of the superconducting mechanism emerging from phonon-mediated 
attractions between electrons. Meanwhile, in the eastern block, Bogoliubov was studying microscopic models
for describe superconductivity; one of his main results led to the description of the BCS vacuum by means
of canonical transformations%
\footnote{For a simple pedagogical model (in italian) describing BCS theory and Bogoliubov transformations we refer to \cite{teobech}.}.


As superconductivity is a phenomenon related to symmetry breakdown,
below a critical temperature, a system can develop an ordered phase in which a charged field condenses leading
to superconductivity.
The essential features of superconductivity can be studied in general without entering into the details of the microscopic dynamics of
any particular model, \cite{Weinberg:1986cq}.
A wide class of superconducting media are described with quantum U$(1)$ gauge field theories.
Indeed, the gauge structure of the quantum field theory itself, leads to some general properties
as infinite DC conductivity, the Meissner-Ochsenfeld effect, the flux quantization, the Josephson effect
and, more generally, many aspects of the magnetic behavior of superconductors.
In Subsection \ref{DCinf} and Appendix \ref{MEI} we respectively concentrate on the first two features in the above list.

Since the discovery of superconductivity, the phenomenon remained limited to the extremely low temperature region (around and below the liquid helium region
$\sim 4.2$ K).
The experimental investigation has maintained the focus on the metallic elements or alloys till the mid $1980$'s when a crucial breakthrough has been
accomplished by Bednorz and M\"{u}ller \cite{Bednorz_Muller_1986}: They discovered a superconducting transition around $T_c \sim 30$ K
on a specific oxide, namely $\text{La}_{1.85}\text{Ba}_{0.15}\text{Cu}\text{O}_4$, containing copper and lanthanum and doped with barium%
\footnote{They were awarded the Nobel prize in Physics in 1987; here is the link to the corresponding press release \cite{hightc}.}.
A surprising fact is that such kind of oxides are often, at higher temperature, almost insulators.
The discovery received stark attention both because the experimental and technological possibilities offered by high temperature superconductivity
are paramount and also because temperature values close to $30$ K where believed to be the theoretical bound for superconducting phenomena (according
to a weak-coupling analysis). The idea of some unconventional (i.e. non-BCS) superconducting phenomenon started therefore to be caressed in the scientific community%
\footnote{At present, the primate for the highest $T_c$ (at ambient pressure) goes to mercury barium calcium copper oxide ($\text{HgBa}_2\text{Ca}_2\text{Cu}_3\text{O}_{8}$), at around $135$ K 
\cite{primate}.
Applying a higher pressure, the transition temperature can been further increased to values slightly above $150$ K (see again \cite{primate}).}.

Later, in 2008, another class of high-$T_c$ superconductors was discovered \cite{Kamihara_2008}, namely layered Fe-based compounds;
the actual discovery occurred with $\text{LaFeAsO}_{1-x}\text{F}_{x}$ with doping parameter $x\sim 0.11$; this material presents a transition 
temperature $T_c \sim 26$ K.
Soon afterwards, other Fe-based materials have been investigated and higher values of $T_c$ have been found.

Even before any deeper observation, these high $T_c$ values call for an explanation beyond the standard BCS framework.

\subsection{London equation}

In this section we sketch the original phenomenological argument on which the introduction of London equation is based.
Let us consider a metallic superconductor in which a ``cloud'' of electrons moves within a crystalline array.
Below a critical temperature, superconductivity arises and we indicate with $n_s$ the density of electrons
participating to the superconductivity phenomenon.
The supercurrent is defined as
\begin{equation}\label{Londres}
 \bm{j}_s = n_s e\, \bm{v} \ ,
\end{equation}
where $e$ is the electron charge and $\bm{v}$ is the mean velocity of the superconducting electrons.
In a quantum context, the observables are of course substituted with the mean values of
the corresponding operators.
The velocity is obtained dividing the canonical momentum by the mass of the carriers,
\begin{equation}
 \bm{v} = \frac{1}{m} \left(\bm{p} - \frac{e}{c} \bm{A} \right) \ ,
\end{equation}
so that in general we have
\begin{equation}\label{quasiLondon}
 \bm{j}_s =   \frac{n_s e}{m} \left(\bm{p} - \frac{e}{c} \bm{A} \right) \ .
\end{equation}

As the London brothers did first, let us observe that in a periodic crystal Bloch's
theorems does hold. 
Therefore the quantum solutions are expressible as the product of plane waves times functions
sharing the same periodicity as the crystalline array encoded in the potential (These are known as Bloch's
solutions or Bloch's waves).
Postulating that the superconducting state is the ground state of the system,
Bloch's theorem implies that the momentum of the plane wave has to be zero.
Indeed for any solution with $p\neq0$ we can find a corresponding lower energy solution with $p=0$.
In a superconductor, equation \eqref{quasiLondon} reduces to
\begin{equation}\label{londra}
 \bm{j}_s = -  \frac{n_s e^2}{m c}  \bm{A}  \ ,
\end{equation}
which is actually the renown London equation.
It is customary in the literature (and we will adopt this convention as well) to denote as the first and second London equations
the relations obtained from \eqref{londra} taking respectively a time derivative and the curl on both sides.

\subsection{Infinite DC conductivity}
\label{DCinf}

In this Subsection we focus on the relation between the symmetry features of the gauge field theory description of
a superconductor and its phenomenological properties, in particular, on the DC superconductivity. 
Whenever we describe the superconducting medium with a quantum U$(1)$ gauge field theory,
the action is invariant under the gauge transformations
\begin{eqnarray}\label{gauge}
 & A_\mu(x) & \rightarrow A_\mu(x) + \partial_\mu \alpha(x) \, \\
 & \psi(x)  & \rightarrow e^{\text{i} q \alpha(x)} \psi(x) \ ,
\label{rigauge}
\end{eqnarray}
where we have assumed the presence of a single fermion species $\psi$ with electric charge $q$. 
The gauge parameter function $\alpha(x)$ is arbitrary and specifies the particular gauge we consider.
At a fixed point $\overline{x}$ in space-time, the transformations \eqref{gauge} and \eqref{rigauge} correspond to a compact U$(1)$ phase symmetry;
indeed, the values $\alpha(\overline{x})$ and $\alpha(\overline{x}) + 2\pi/q$ are identified.

In general, in a superconductor, the gauge symmetry \eqref{gauge} is supposed to be broken by the condensation of some operator.
Suppose that, in a phase characterized by the spontaneous symmetry breakdown, the original local U$(1)$ symmetry is reduced to a discrete subgroup
$\text{Z}_n\in \text{U}(1)$. 
From Goldstone's theorem we have that the symmetry breaking leads to the appearance of a massless mode $G$ parameterizing the
coset group U$(1)/\text{Z}_n$. The field $G$ behaves as a phase and then, under a gauge transformation, it transforms as follows:
\begin{equation}\label{Gol_gau}
 G(x) \rightarrow G(x) + \alpha(x) \ .
\end{equation}
In addition, as it spans the coset group U$(1)/\text{Z}_n$, we identify
\begin{equation}
 G(x) = G(x) + \frac{2\pi}{n q}\ .
\end{equation}

Relying on symmetry arguments, the Lagrangian for the gauge and Goldstone's fields has the following general shape:
\begin{equation}
 {\cal L} = -\frac{1}{4} \int d^dx \ \left\{ F \cdot F + L_g [A - d G] \right\} \ ,
\end{equation}
where the form of the Goldstone part $L_G$ of the Lagrangian density depends on the specific model while
its functional dependence on $A - dG$ is a general feature descending from gauge symmetry. Note indeed that $A-dG$ is a gauge
invariant quantity.
The spatial electric current and charge density are given by\footnote{We are assuming Euclidean space-time here.}
\begin{eqnarray}
 & J^i & = \frac{\delta L_G}{\delta A_i(x)} \ , \\
 & J^0 & = \rho = \frac{\delta L_G}{\delta A^0(x)} = - \frac{\delta L_G}{\delta (\partial_t G)} \ .
\end{eqnarray}
The second equation states that $-\rho$ represents the canonical conjugate variable to $G$;
then, within a Hamiltonian description, the energy density $\cal H$ is a functional of $G$ and $\rho$.
The Hamilton equation for $\partial_t G$ is
\begin{equation}\label{hami}
 \partial_t G(x) = - \frac{\delta {\cal H}}{\delta \rho(x)}\ .
\end{equation}
Let us interpret physically this Hamilton equation:
The Hamiltonian ${\cal H}$ gives the energy density and, since $\rho(x)$ represents the charge density,
we have that the right hand side of \eqref{hami} gives the change in energy density implied by a change in the charge density.
This is the electric potential.
The time derivative of $G$ is then related to the potential
\begin{equation}\label{pote}
 \partial_t G(x) = -V(x) \ .
\end{equation}
Let us consider a stationary state in which there is a steady current flowing through the superconducting medium;
stationarity means that nothing depends on time and, in particular, $\partial_t G(x) = 0$.
The potential $V(x)$ is then forced by \eqref{pote} to be zero too; since we have stationary currents without
any difference of potential sustaining them, we are facing a zero resistance or infinite conductivity phenomenon.
As we are concentrating on the stationary properties, we are studying the DC conductivity, i.e. the 
limit of the conductivity $\sigma(\omega)$ for vanishing frequency $\omega$.

We have just showed the occurence of infinite DC conductivity basing our 
argument (originally suggested by Weinberg in \cite{Weinberg:1986cq}) on the simple assumptions of having an Abelian gauge symmetry
which is spontaneously broken to a discrete group; no details of the actual mechanism leading to the spontaneous breaking
have been actually specified. The moral consists in recognizing the value of the symmetry breaking itself in leading to the
description of the superconductor phenomenology independently of its microscopic origin.
In this sense we can understand better why phenomenological models \emph{\'a la} Ginzburg-Landau are able to describe accurately
the phenomenology of superconductivity even though they rely on crude approximations like, for instance, the description of
the Cooper pairs with a single bosonic field.

\section{Hairy BH and dual condensates}
There are various gravity effective models which are dual to a boundary theory describing a superconductor.
A general feature of such models is the presence of a charged black hole that becomes ``hairy'' at
low temperature. 
As usual, speaking of ``hair'' in relation to a black hole solution indicates the presence of
some field which develops a non-trivial profile.
The occurrence of a non-trivial hair in the bulk is generally dual to some kind of operator
condensation in the boundary theory; in other terms, the boundary operator develops
a non-vanishing vacuum expectaion values.

We will concentrate on a bulk model developing scalar hair.
This corresponds to a superconductor with a scalar condensate or, using the superconductor
jargon, an s-wave superconductor\footnote{The gravity models presenting non-trivial profile
for a vector field are dual to the so called p-wave superconductors. 
If the non-trivial profile is associated to a spin $2$ field we have instead a d-wave superconductor.}.
Notice that the scalar operator which condenses is associated to the Cooper's pairs; 
indeed, whenever the electrons pair in a singlet spin state with zero orbital 
angular momentum, the Cooper condensate is actually describable at the effective level
with a scalar field.

As we will describe in detail, the ingredients needed to build the simplest bulk
models dual to superconductors involve an Abelian gauge field minimally coupled 
to gravity and a scalar field with a generic potential $V(\psi)$.
Moreover, since we want to model superconductors in flat space-time,
we consider black hole configurations presenting planar horizons.

\subsection{Note on the holographic description of a superconductor}

What does it mean, on a practical level, to deal with a holographic model of a superconductor?

Duality itself is a concept of which it is easy to have an intuitive idea: namely, it is
the map of the degrees of freedom and dynamics of a certain model to the degrees of freedom
and dynamics of another model. 
The two descriptions are proved or conjectured to be equivalent.

Let us underline that usually, in a holographic context, we exploit duality to study 
some strongly coupled model of which the microscopic description is not available.
It is then reasonable to ask on what basis we claim to describe a superconductor.
The macroscopic holographic description allows one to handle expectation values and correlations
of the operators of the boundary theory. Even lacking the Lagrangian of the ``boundary'' model we 
can study in detail many dynamical feature and the general thermodynamic behavior.
It is from this study that some features of superconductor phenomenology arise.
In particular, as we will see in the following, the holographic superconductor shows 
a normal-to-superconducting transition in the electric response function;
indeed, in trespassing the critical temperature, we observe a novel diverging contribution to the DC conductivity%
\footnote{As the system under analysis enjoys transaltional invariance,
it possesses a diverging DC conductivity also in the normal (i.e. non superconductivity) phase.
Such effect is merely due to the lack of momentum relaxation and it is
not to be confused with authentic superconductivity.} and
such contribution is naturally interpreted as a superconducting phenomenon at strong coupling (see Subsection \ref{sition}).

At the outset, a general caveat regarding the terminology of the
holographic literature should be mentioned.
When, in relation to a holographic model, some physically suggestive 
(e.g. inspired by condensed mayyer) terminology is adopted, 
one has always to keep in mind that the actual description of a 
real-world system can be still far apart.
It is advisable to start with the moderate attitude that the holographic 
context offers treatable examples and toy models able to reproduce some 
phenomenologically interesting features (especially at strong coupling) 
but the whole of the holographic model under study
can detach in some other respects from the actual phenomenology. 
The model we develop in the following sections is no exception.
When we speak of a superconductor, we do not mean that we expect to be able
to interpret any single aspect of the model in terms of some real-world example. 
Of course, this would be however amply desirable and, indeed, we will try to do that, 
but a problematic attitude toward any particular feature is probably the best 
way of judging the real value of the model. To rephrase, we suggest a 
``bottom-up attitude'' in the confidence we tribute to the realism of any holographic model.

\subsection{Effective electromagnetic background and non-dynamical photons}
\label{nondyn}

In a quantum field theory picture, the possibility of neglecting the photon dynamics corresponds to the small relevance
of processes involving virtual photons.
It should be stressed that the non-dynamical photon approximation leads to an effective description in which
the underlying U$(1)$ gauge symmetry is treated as a U$(1)$ global symmetry%
\footnote{Note that the treatment of the electromagnetic symmetry as a global U$(1)$ matches with the prescription of the 
holographic dictionary connecting a boundary global symmetry with a gauge bulk symmetry.}.
The photon dynamics is negligible whenever the electromagnetic coupling can be regarded as small
and the Feynman diagrams containing internal photons are correspondingly suppressed.
In real systems, for example, the screening effects that occur in charged media
are a ubiquitous feature in condensed matter systems and, at the level of non-microscopic description,
they can lead to an effectively small electromagnetic coupling.

To avoid confusion we must underline a significant caveat:
The holographic description
is particularly suitable to treat strongly coupled media (as the dual gravitational model becomes weakly coupled). 
When we consider the non-dynamical photon approximation, we treat the system
as a strongly interacting medium weakly coupled to external photons%
\footnote{Comments on the non-dynamical photon approximation can be found in \cite{Amariti:2010hw}.}.
Notice that, also in the non-dynamical photon approximation,
the strong interactions within the holographic 
medium can still involve electromagnetic phenomena; they are however 
encoded in the macroscopic effective description and no photon-like 
degrees of freedom are manifest.
More precisely, the meaning of our non-dynamical photon approximation consists in working under the assumption that the microscopic degrees of freedom 
of the medium can be treated collectively as a plasma which interacts weakly with the external electromagnetic background field%
\footnote{Similar observations can be found in \cite{Herzog:2009xv}.}.

Within the non-dynamical external photon analysis of a system, the response to the variation of an external electromagnetic field
is described in terms of induced currents in the medium.
In other terms, the total electromagnetic field coincides with the background value sourcing the charged currents within the system.
There is a natural compatibility between the non-dynamical photon approximation and the linear response theory 
because the charged currents are weakly coupled to the external source and then effectively describable
at linear order.

As underlined in \cite{Herzog:2009xv}, superfluidity corresponds in general to a spontaneous symmetry breaking
of a global symmetry whereas superconductivity is associated to the Higgs mechanism of a local gauge symmetry.
Since, in a holographic framework, boundary global symmetries correspond to local bulk symmetries,
superfluidity in the boundary theory should correspond to ``superconductivity in the bulk''.
For the purpose of the computation of the conductivities, however, there is no crucial difference between a superfluid and a superconducting phase 
because we retain only the linear effects. 
Indeed, from a purely field theoretical viewpoint it is possible to show that the linear response of a system to external perturbations
is insensitive to the fact that we work with dynamical or non-dynamical photons (the dynamics of the photons is encoded in the subleading orders).
As a consequence, the non-dynamical photon approximation which approximates 
a superconductor with a superfluid allows us to describe linear response of the superconductor 
to external electromagnetic perturbations.
This is another argument supporting the validity, in our context, of the non-dynamical photon approximation.

In a holographic framework, in order to go beyond the non-dynamical photon approximation, one needs to develop the so called ``gauged 
$AdS$/CFT''; this term refers to the problem of defining a dual configuration to a boundary gauge symmetry;
this theoretical possibility constitutes still an open problem%
\footnote{In \cite{Domenech:2010nf} there is described an attempt to have a gauged holographic correspondence.}


There is still another significant observation which can made about the non-dynamical character of the photons. 
In the present thesis we concentrate especially on a superconductor in $2+1$ dimensions, namely
a superconducting layer. Given the ``infinitesimal'' thickness of the superconducting region,
there is no Meissner-Ochsenfeld effect.
In other terms, we are supposing that the thickness of the superconducting layer is much smaller than the characteristic penetration depth
of the magnetic field inside the superconductor%
\footnote{See Appendix \ref{MEI} for some detail on the Meissner-Ochsenfeld effect.}. 
In higer dimensional systems, the Meissner-Ochsenfeld effect is related to the photon dynamics, 
but independently of the dynamical or non-dynamical character of the photons, such effect
does not occur in $2+1$ dimensions.
In this sense, our holographic model in the non-dynamical photon approximation is able to reproduce the phenomenology
of a superconducting layer (analogous observations can be found in \cite{Arean:2010xd}).

\chapter[Holographic Spintronics]{Holographic Superconductors with two Fermion Species and Spintronics}
\label{mio2}

\section{Mixed spin-electric conductivities and spintronics}
The term \emph{spintronics} is a short version of ``spin transport electronics'' known under the name of ``magneto-electronics'' as well.
The subject of spintronics concerns the role of electron spin in condensed matter physics, especially in relation to transport properties.
Indeed, the purpose of spintronics aims at the study of systems with particular spin transport or spin-dependent transport properties with the
objective of understanding and designing devices exploiting the individual electron spin instead or in connection with their charge.
Spintronics is in contrast with usual electronics where only the electron charge or collective magnetization are exploited.

The first phenomenon relating current flows and electron spin is Anisotropic Magneto Resistance \cite{Tom} (ANM);
it was observed by Thompson in 1857 and (much) later (1975) it has been described in a model involving spin-orbit coupling \cite{citeulike:3503602}.
The phenomenon itself consists in a dependence of the resistivity of a ferromagnetic metal on the relative angle between the
magnetization and the current flow. The order of magnitude of the resistivity variation are (at room temperature) of a few percent points 
($\sim 5 \%$).

The first steps of what has been later called spintronics were moved by Mott who in 1936 proposed the model
know as ``two-current model'' \cite{mott1,mott2} to describe some spin-dependent features of the conduction properties of ferromagnetic metals
below the Curie temperature. 
Mott proposed that well below the Curie temperature the conduction electrons propagating in the ferromgnetic metal
undergo scattering processes without changing their spin orientation.
As a consequence the two-current model depicts the spin-up and spin-down currents as two independent currents and the overall
properties of the material arise from the parallel of the spin-up and spin-down circuits.
In its simplest version the two currents are totally independent, however Mott's model can be improved considering
a weak coupling between the spin-up and spin-down currents (for instance because of spin mixing phenomena).
A source of spin-mixing is, for instance, the electron-magnon scattering which could lead to spin flip.
Let us remind the reader that the magnon is a collective mode of an ordered magnetized medium and arises from the quantization of spin-waves;
it constitutes the analogous of phonons for elastic lattice vibrations.

In 1966 Fert studied in depth the spin-dependent conduction properties of doped alloys where the sensitivity
to the spin is due to impurities with strong spin-dependent cross-sections%
\footnote{For a introductory account to spintronics see \cite{fert}.}.
It is on the basis of these preliminary studies that in 1988 one of the main achievements of spintronics was discovered:
the Giant Magneto-Resistance (GMR) \cite{Baibich:1988zz,Binasch:1989zz}.
It consists in the large difference of resistivity through a device constituted by different layers depending on the anti-ferromagnetic or
ferromagnetic polarization of adjacent layers. One of the interesting points is given by the possibility of controlling 
easily the relative polarization of the layers by means of external fields.

In the spintronic context, the possibility of affecting magnetization patterns by acting on electric currents has received particular attention
in the last decade \cite{BER-96,JC1996L1,citeulike:5319403,PhysRevB.84.184408}.
Indeed, recent results showed that an electric current flowing in a ferromagnetic conductor drives magnetic textures such as domain walls and vortices.
This mixed electro-magnetic effect has been studied theoretically \cite{PhysRevB.33.1572} and proved experimentally  \cite{shinjo2009nanomagnetism}.
The generation of such spin motive force is described in analogy with the DC Josephson effect%
\footnote{This phenomenon is sometimes referred to as ``ferro-Josephson effect''.
The DC Josephson effects consists in the occurrence of an electric current flowing between two linked superconductors separated by a thin insulating layer even though no
external voltage is applied (see Appendix\ref{joseph}). The analogy between the DC Josephson effect and the spin motive force is
particularly suitable for the case in which the flow of an electric current exerts torque on a magnetization domain wall \cite{PhysRevB.33.1572}.
Note that the conductor magnetic system under consideration is in general not superconducting.}.
The effect can be microscopically described by means of a torque exchange interaction among the unpolarized spins of the conduction electrons
flowing through the localized spins of the magnetic pattern.
The opposite effect can also occur, namely moving magnetization patterns can
drive electric currents. 

At the core of spintronics there is the mixed electromagnetic effects interlacing spin and charge transport.
In this context our holographic approach investigates the strong-coupling extension of weakly coupled spintronics.
Indeed, as we will explain in detail later, our holographic unbalanced system presents a conductivity
matrix mixing electric and magnetic effects. 
Since the conductivity is defined as a linear response phenomenon, the mixed entries in the conductivity matrix
correspond to the fact that at linear order an external electric perturbation leads to a net spin current
and conversely an external magnetic perturbation can drive an electric current.
This being a general feature of spin-up
spin-down unbalanced systems.

In Section \ref{SuCo} we introduce the model that will be described in detail henceforth, namely the holographic unbalanced superconductor.
It possesses two fermion species associated to two independent chemical potentials;
the system is said to be unbalanced whenever the two chemical potentials (or Fermi energies) differ.
The holographic unbalanced superconductor is relevant to studying strong-coupling unbalanced superconductivity but also (especially in its normal phase)
as a strong-coupling generalization of Mott's two-current model.
An interesting observation concerning the superconducting phase of our system arises from interpreting it as a
model of a forced (as opposed to spontaneous) ferromagnet at strong coupling in analogy with that studied in \cite{Iqbal:2010eh}.

\subsection{Spintronics and information technology}

Electron transport and magnetization have been the two pillars of information technology till fifteen years ago.
With magnetization is here meant the magnetic property of a big numbers of microscopic elements as opposed to
the magnetic properties of the single electrons.
The magnetization has mainly been employed in high-density storages and the need to read and write on such memories 
requires an integration of magnetic devices in to electronic circuits. 
In other words, information has to be translated from electric current or voltages into magnetic properties and vice versa.

Initially Faraday's law has been the first method to write and read magnetically storaged information but especially
the reading process proves rather inefficient. 
It naively consists in moving a coil in the proximity of the magnetized bit.
The route to increase in efficiency and the possibility of significant decrease in device size
moves naturally the attention to ``spintronics''. 
Indeed, the reading process has been based on the current flowing through the magnetized bits, and
the property of the current flux depend on magnetization.
In the last thirty years big progress 
has been attained by exploiting in succession anisotropic magneto-resistance, giant magneto-resistance and
tunneling magneto-resistance. 
They are mentioned in historical order which is also the order of the efficiency/miniaturization potentiality%
\footnote{The present account has been made based on \cite{citeulike:5319403} we further information can be found.}.

\section{Superconductor with two fermion species}
\label{SuCo}
The superconductor models with two fermion species are relevant both for QCD contexts
and in the panorama of condensed matter physics%
\footnote{An ample review encompassing (also) two-species superconductors is \cite{Casalbuoni:2003wh}.}.
The two fermion species might have different chemical potentials and generally the resulting system is said to be \emph{unbalanced}.
In high density QCD and nuclear matter systems, the chemical potential mismatch can be due to mass or charge differences between
the quark species; in condensed matter systems, where usually the two fermionic species describe spin-up and spin-down electrons,
the imbalance can be induced, for example, by magnetic impurities.
The pairing mechanism leads to Cooper's pairs formed by two fermions of different species and
the pair is a singlet zero-spin state\footnote{Note that we will be concerned with an s-wave superconductor;
at weak-coupling (where Cooper's pair are well defined), in an s-wave superconductor, the electrons bind to form a Cooper pair without orbital angular momentum
and with their spins in opposition. In a p-wave superconductor, instead, there is $L=1$ angular momentum leading to a minus sign contribution
to the parity of the pair; in order two have overall antisymmetry, the electrons have to be in the triplet state. 
So far the experimental evidence of a p-wave superconductor is still matter of debate, while p-wave superfluidity is a well established 
result discovered in superfluid $^3\text{He}$.
There exist also holographic models for p-wave superfluidity, see for instance \cite{Gubser:2008wv,Ammon:2009xh,Ammon:2010pg};
for a model of a holographic imbalanced p-wave superfluid see \cite{Erdmenger:2011hp}.}.
The BCS analysis shows that at weak-coupling the properties of the two fermion superconductor
are strongly sensitive to the chemical potential imbalance between the two species (look at Subsection\eqref{Imb_BCS} and \cite{Casalbuoni:2003wh}).
One can naturally ask what happens at strong-coupling; a viable way of addressing the question is the holographic approach.


In a holographic context, as already mentioned, the chemical potential is associated to the boundary value of a bulk gauge field.
It is then natural to implement the second chemical potential with the introduction of another Abelian
gauge field in the bulk.
More precisely, we will associate an Abelian bulk gauge field $A$ to the mean chemical potential
$\mu$ of the two species considered together, and a second Abelian gauge field $B$ to their chemical potential
mismatch $\delta\mu$, namely
\begin{equation}
 \begin{split}
  \mu &= \frac{1}{2} \left(\mu_1 + \mu_2\right) \\
  \delta\mu &= \frac{1}{2} \left(\mu_1 - \mu_2\right) \ \ .
 \end{split}
\end{equation}

The condensation, i.e. the transition to the superconducting phase, is associated to the breaking of the U$(1)_A$ 
symmetry.
with a VEV of a scalar field $\psi$.
Such scalar field represents the condensate operator and has charge $q\neq 0$ under the field $A$
while is instead neutral with respect to $B$.
For the sake of concreteness, think again to the two fermionic species as spin-up and spin-down electrons\footnote{The spin picture
is used many times throughout the text but as stated at the beginning of the chapter the analysis is more general and not
specific to this ``condensed-matter'' scenario.}. 
All electrons have the same electric charge $q/2$ so that the Cooper pair
has charge $q$; from the spin point of view the pair is instead neutral (i.e. the two bound electron are in a singlet state)
and $B$ represents the ``magnetic'' driving field (see Subsection \ref{magnotta}).

In the gravitational dual perspective, according to the standard holographic dictionary, the asymptotic, near-boundary behavior of the bulk gauge fields $A$ and $B$,
\begin{equation}\label{asym_A}
 A(r) \underset{r\rightarrow\infty}{\sim} \mu - \frac{\rho}{r} + ...\ \ , \ \ \ \ \ \ 
 B(r) \underset{r\rightarrow\infty}{\sim} \delta\mu - \frac{\delta\rho}{r} + ...\ \ , 
\end{equation}
account respectively for the collective mean chemical potential $\mu$ and total electric charge density $\rho$ arising considering both the fermion species
and the chemical potential difference $\delta\mu$ and charge density imbalance $\delta \rho$.
Even though the scalar field $\psi$ is uncharged with respect to $B$,
it is not completely insensitive to its dynamics.
In the dual holographic picture this feature is obvious:
the presence of the field $B$ backreacts on the gravitational background on which $\psi$ itself fluctuates.
This important point will be further developed in the following; let us here pinpoint the crucial role of the metric
noting that it is insufficient to work in the \emph{probe approximation}: We have to consider the backreaction of all the fields to the background%
\footnote{Details on the probe approximation are given in Appendix \ref{probe}.}.

\subsection{The ``magnetic gauge field'' U\texorpdfstring{$(1)_B$}{}}
\label{magnotta}

In the previous section we have seen that the introduction of a potential mismatch is naturally accommodated
in the holographic framework by the introduction of a second gauge field in the bulk.
As we have recognized in Subsection \ref{nondyn}, our holographic treatment of the boundary theory
approximates the electromagnetic gauge symmetry with its global version U$(1)_A$.
Inverting the line of thought, we can wonder what the gauge symmetry whose global part corresponds to U$(1)_B$ is.
It should be stressed that our gravity model provides an effective description of the symmetries and order parameters (e.g. the condensate
whose dual is given by $\psi$) of the `` boundary'' field theory. 
In this sense U$(1)_A$ and U$(1)_B$, that we interpret as holographic duals of ``charge'' and ``spin'' currents respectively, can 
represent any couple of Abelian global symmetries enjoyed by the field theory.
It is then to the effective stage that we have to stick, where U$(1)_B$ introduces the possibility of describing two mismatched fermion species.
In a would be gauged version of the model, where both U$(1)_A$ and U$(1)_B$ becomes local in the ``boundary'' theory,
the description of electromagnetic interactions by means of two dynamical Abelian symmetries would posit interpretation questions.
Note however that the point is rather speculative, actually we will not consider gauged $AdS$/CFT correspondence and, as already mentioned,
the $AdS$/CFT gauging possibility itself is still rather obscure.

\section[Inhomogeneous superconductivity]{Unbalanced superconductors at weak-coupling and inhomogeneous phases}
\label{Imb_BCS}

\begin{figure}
 \centering
 \includegraphics[width=.65\linewidth]{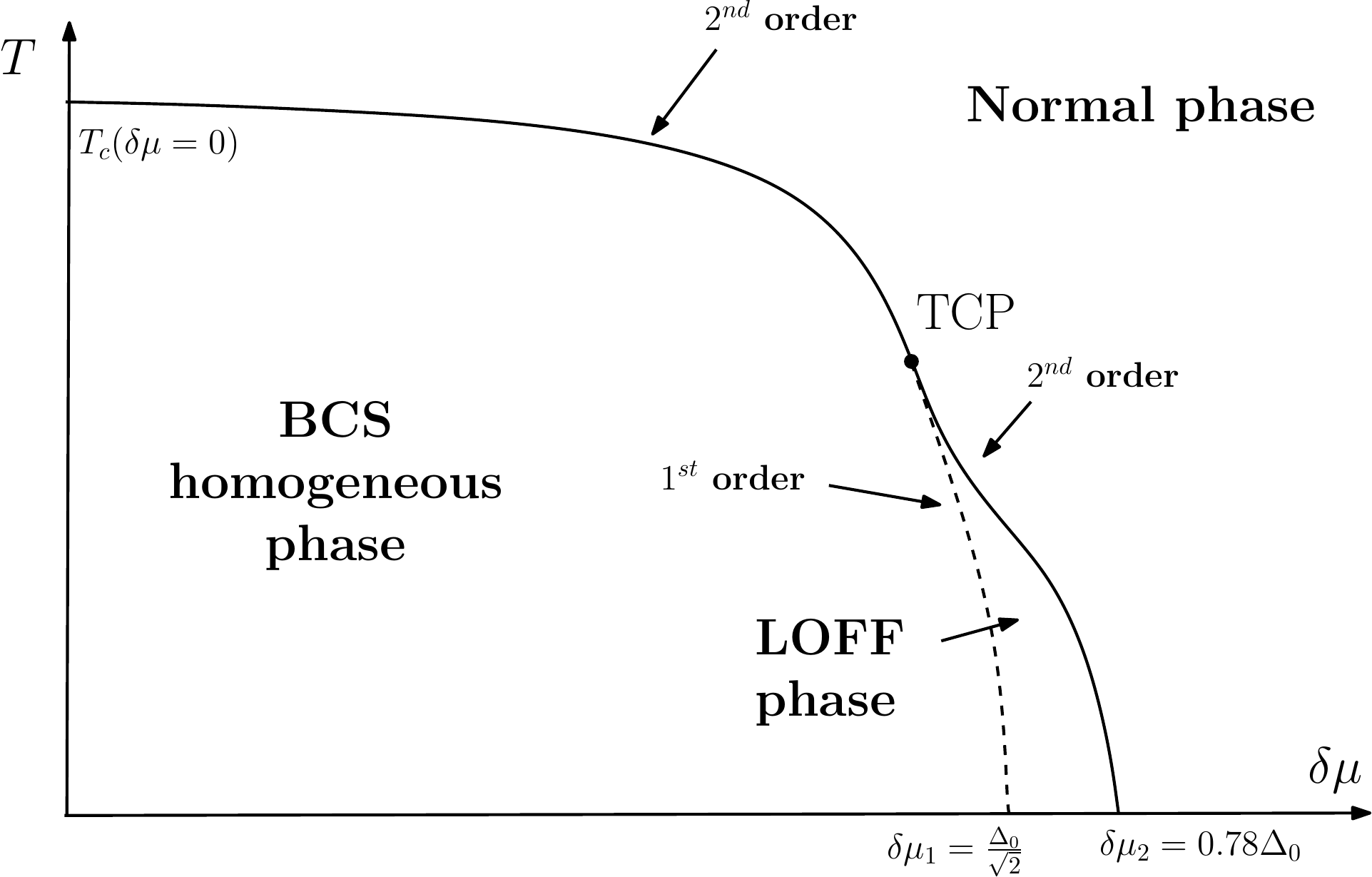}
 \caption{Phase diagram on the $(T,\delta\mu)$ plane for a generic weakly coupled superconductor.}
 \label{dia}
\end{figure}

Let us pinpoint some aspects of the unbalanced superconductor behavior at weak coupling
studied with the standard BCS approach; we will later compare these features with their strong-coupling counterparts investigated with holographic means.
There are two different possibilities for the superconducting phase: homogeneous phases and space-varying phases.
In the two cases the superconductor gap parameter is respectively a constant or a non-trivial function in space;
the same holds true for the condensate profile.

In the homogeneous case, the condensation occurs at a critical temperature $T_c$ which is a decreasing 
function of the potential mismatch $\delta\mu$.
The imbalance hinders the formation of the condensate, namely the more the system is unbalanced, the lower is its condensation temperature.
Furthermore, even at zero temperature, there is a maximal value for $\delta \mu$ above which the
homogeneous superconducting phase does not occur (Chandrasekhar-Clogston bound \cite{Chandra:1962,Clogston:1962zz}).

Considering the possibility of inhomogeneous phases, where the condensate is spatially modulated, 
one can observe the occurrence of Cooper's pairs with non-vanishing
total momentum\footnote{The condensate is related to the expectation value of a bosonic operator which, in a first-quantized picture,
can be regarded as the wave-function of Cooper's pairs. The inhomogeneous phase is related to a ``stationary wave'' configuration
and not to a ``superfluid-like'' net flow of Cooper's pairs.}. 
Actually, the typical wave-length of the spatial modulation corresponds in order of magnitude to the energy difference
between the two Fermi surfaces of the pairing fermions.
The weak-coupling analysis of the unbalanced superconductor shows that
an inhomogeneous superconducting phase at zero temperature can be energetically favored with respect to a homogeneous phase; 
this occurs in an interval of the potential mismatch $\delta\mu_1 < \delta\mu < \delta\mu_2$.
The inhomogeneous possibility is known as the Larkin-Ovchinnikov-Fulde-Ferrel (LOFF) phase \cite{LO,FF} and it presents a space-varying condensate
and gap parameter.
In the LOFF case, the spatial modulation is periodic and related to the modulus of the wave-vector $\vec{k}$ of Cooper's pairs.
The modulus $|\vec{k}|$ is determined by free energy minimization, its direction
corresponds instead to a spontaneous breaking of rotational symmetry.
Notice therefore that the LOFF phases break spontaneously both the translational and the rotational symmetries of the Hamiltonian.
It is possible to have also more complicated situations in which the condensate can be thought of as a superposition
of waves. The cases where the wave-vectors of the superimposed waves are linearly independent
are sometimes referred to as \emph{crystalline superconducting phases}.

When the system is unbalanced beyond the critical value $\delta\mu_2$,
the Fermi surfaces corresponding to the two fermionic species are too far 
apart and it is no longer energetically convenient to form Cooper's pairs;
both homogeneous and inhomogeneous superconducting phases are disfavored with respect to the normal phase.

The features that have been just described are summarized in the phase diagram \ref{dia}.
Concluding this brief section, let us remark that the experimental evidence for the occurrence of inhomogeneous phases is still uncertain.

\section{Holographic unbalanced superconductor: Dual gravity setup}
\label{strong}
In describing holographically the unbalanced strongly coupled superconductor we maintain an effective macroscopic attitude.
Indeed, we consider a bottom-up approach introducing the minimal 
set of ingredients able to reproduce the relevant phenomenological features 
of an $s$-wave unbalanced, unconventional (i.e. strongly coupled) superconductor in $2+1$ space-time dimensions%
\footnote{\label{ads4} Even though in the introductory remarks we have frequently referred to the original $AdS_5$/$\text{CFT}_4$ 
correspondence, here we employ its lower dimensional counterpart $AdS_4$/$\text{CFT}_3$.}.
The reason of choosing a $3$-dimensional space-time is related to the fact that, as a general feature, high $T_c$ superconductivity
occurs in layered materials.

In the dual, gravitational perspective, the bottom-up approach consists in working with effective low-energy 
approximations of a would-be full-fledged string model.
Although we study systems before knowing whether they could be consistently UV completed, we
nevertheless comment on the possibility of embedding our phenomenological
description into a string theory setup in Section \ref{UVcomp}.

The bottom-up approach constrains us to work in the large $N$ limit.
Indeed, going beyond such limit and considering lower values for $N$ requires to consider string theory corrections%
\footnote{As we are frequently referring to electrons and U$(1)$ electro-magnetic interactions, one could be confused by the large $N$ hypothesis.
Note however that the large $N$ SU$(N)$ group supported by the D-branes in the bulk mimics the strongly coupled dynamics of the dual medium. 
Our U$(1)_A$ and U$(1)_B$ currents arise instead from some other feature (such as flavor groups) of the would be string model, see Section \ref{UVcomp}.}.
Moreover, the lack of a precise stringy picture makes it difficult to account in detail for the microscopic
content of the boundary theory.
In other terms, we are able to give a description of the macroscopic observables of the CFT boundary theory 
without a detailed knowledge of the elementary degrees of freedom.
As we will see, the phenomenological models are nevertheless able to describe qualitatively and quantitatively 
some interesting dynamical features at strong coupling
which would be difficult (if not impossible) to study without the holographic tools.

The simplest holographic model describing an unbalanced superconductor at strong coupling corresponds to
the following Lagrangian density%
\footnote{One could consider more general kinetic terms for the field-strengths $F$ and $Y$, namely
\begin{equation}
 -\frac{1}{4} H(|\psi|) F^{ab} F_{ab}  -\frac{1}{4} K(|\psi|) Y^{ab} Y_{ab}\ ,
\end{equation}
where $H$ and $K$ are functions of the condensate field $\psi$.
Since we are dealing with an effective theory, we could have functions depending on any power of $\psi$.
The generalized versions would however correspond to non-minimal couplings between the scalar $\psi$
and the gauge fields. To take the simplest possibility, we are here concerned with the $H(|\psi|)=K(|\psi|)=1$ case only.}:
\begin{equation}\label{laga}
 {\cal L} = \frac{1}{2\kappa_4^2} \sqrt{-\text{det}\, g} \left[ R + \frac{6}{L^2} - \frac{1}{4}Y^{ab}Y_{ab}
-\frac{1}{4} F^{ab}F_{ab} - |\partial\psi - \im q A \psi|^2 - V(|\psi|) \right]\ ,
\end{equation}
where $F=dA$ and $Y=dB$ are the two field-strengths and $\kappa_4$ is the $AdS_4$ Newton gravitational constant.
All the fields appearing in the Lagrangian density are dimensionless, 
they have been rescaled in order to collect the factor in $\kappa_4$ outside and the charge $q$
is dimensionally an energy.
The complex scalar $\psi$ is manifestly charged under $A$ and neutral with respect to $B$.

The Lagrangian density \eqref{laga} represents a simple generalization of the one proposed in \cite{Hartnoll:2008vx,
Hartnoll:2008kx} for the balanced superconductor.
It admits the $AdS_4$ solution%
\footnote{See footnote \ref{ads4}.} of radius $L$ where all the fields except the metric are zero.
The finite temperature configurations correspond instead to black hole solutions which are still asymptotically $AdS_4$\footnote{This
is true for all charged/uncharged, hairy or not solutions. In the holographic language, the $AdS$ asymptotic geometry 
means that in the UV regime the boundary theory recovers the conformality.}.

The simplest, non-trivial choice for the scalar potential $V$ is
\begin{equation}\label{potcho}
 V(|\psi|) =  \frac{m^2}{L^2} |\psi|^2\ ,
\end{equation}
where $m$ represents the mass of the bulk scalar field $\psi$.
More specifically, the choice we adopt is
\begin{equation}\label{potamass}
 V(|\psi|) = - 2 |\psi|^2\ ,
\end{equation}
corresponding to $m^2=-2/L^2$.
Notice that even though the squared mass is negative it does not correspond to an instability;
the background we are considering is in fact $AdS_4$ and the mass value we have chosen 
is above the Breitenlohner-Freedman stability bound \cite{BrFr},
\begin{equation}
 m^2 L^2 \leq - \frac{9}{4}\ .
\end{equation}
As mentioned
in \cite{Hartnoll:2008vx}, $m^2=-2/L^2$ in our background corresponds to a ``conformally coupled'' scalar and
it represents a typical value arising from string theory embeddings of the effective setup%
\footnote{In particular, as an instance, it is possible to consider the truncation of ${\cal M}$ theory on $AdS_4\times S^7$
to ${\cal N}=8$ gauged supergravity.}.
As described in Appendix \ref{scala}, the $AdS$/CFT dictionary relates the mass of the bulk scalar field to the conformal 
dimension $\Delta$ of the corresponding dual operator:
\begin{equation}
 \Delta (\Delta - 3) = m^2 L^2.
\end{equation}

\subsection{Backreacted bulk dynamics}

Now we enter into the systematic study of the dual (classical) gravitational problem.
From the Lagrangian \eqref{laga} we obtain the following equations of motion%
\footnote{To have the equation in the case with generic dimensionality we refer to \cite{pinza}.}:
\begin{itemize}
 \item {\bfseries Einstein's equation}
\begin{equation}\label{eom1}
R_{ab}-\frac{g_{ab} R}{2} -\frac{3 g_{ab}}{L^2}=-\frac{1}{2}T_{ab}\,,
\end{equation}
where the energy-momentum tensor is given by
\begin{eqnarray}\label{stressenergy}
 T_{ab}&=& -
F_{ac}F_{\ \ b}^{c}-Y_{ac}Y_{\ \ b}^{c}+ \frac{1}{4}g_{ab}F_{cd}F^{cd}+ \frac{1}{4}g_{ab}Y_{cd}Y^{cd}\nonumber\\
 && + g_{ab}V(|\psi|)+ g_{ab} |\partial \psi-iqA\psi|^{2} \nonumber\\
&&-[(\partial_{a}\psi-iqA_{a}\psi)
(\partial_{b}\psi^{\dagger}+iqA_{b}\psi^{\dagger})+(a\leftrightarrow b)]\,,
\end{eqnarray}
\item {\bfseries Scalar equation}
\begin{equation}\label{eom3}
-\frac{1}{\sqrt{-g}}\partial_{a}[\sqrt{-g}(\partial_{b}\psi-iqA_{b}\psi)g^{ab}]
+iqg^{ab}A_{b}(\partial_{a}\psi-iq\, A_{a}\psi)+\frac{1}{2}\frac{\psi}{|\psi |}V^\prime (|\psi|)=0\,,
\end{equation}
\item {\bfseries Maxwell's equation for $A$}
\begin{equation}\label{eom2}
 \frac{1}{\sqrt{-g}}\partial_{a}(\sqrt{-g}g^{ab}g^{ce}F_{bc})=iqg^{ec}[\psi^{\dagger }(\partial_{c}\psi-iqA_{c}\psi)-
\psi(\partial_{c}\psi^{\dagger }+iqA_{c}\psi^{\dagger})]\,,
\end{equation}
\item {\bfseries Maxwell's equation for $B$}
\begin{equation}\label{eom5}
 \frac{1}{\sqrt{-g}}\partial_{a}(\sqrt{-g}g^{ab}g^{ce}Y_{bc})=0\,.
\end{equation}
\end{itemize}

We are interested in static and asymptotically $AdS$ black hole solutions to the system of equations of motion;
in accordance to this, the general ansatz we adopt for the metric is
\begin{equation}\label{metric}
 ds^2 = -g(r) e^{-\chi(r)} dt^2 + \frac{dr^2}{g(r)} + r^2 (d\vec{x}^2) \ .
\end{equation}
For the remaining fields we consider the following ``homogeneous'' (i.e. the functions depend only on the $AdS$ radial coordinate
and not on the spatial coordinates) ansatz:
\begin{equation}\label{homogeneous}
\psi=\psi(r), \quad A_a dx^a=\phi(r)dt, \quad B_a dx^a=v(r)dt\,.
\end{equation}

The ``black'' nature of the solution arises from the presence of an event horizon at $r=r_H$ in correspondence of the vanishing
of the $tt$ metric component, namely $g(r_H)=0$. 
Plugging the metric \eqref{metric} into the general formula for the black hole temperature we have derived in \eqref{temp},
we obtain
\begin{equation}\label{tempgen}
T=\frac{g^\prime(r_H)e^{-\chi(r_H)/2}}{4\pi}\,.
\end{equation}

Since $A_r$, $A_x$, and $A_y$ are null, their associated Maxwell equations imply that the phase of the complex scalar field is constant;
without any loss of generality, we can therefore take $\psi$ as a real quantity whose equation of motion is
\begin{equation}\label{equazione1}
 \psi^{\prime\prime}+\psi^{\prime}\biggl(\frac{g^\prime}{g}+\frac{2}{r}-\frac{\chi^\prime}{2}\biggr)
-\frac{V^\prime(\psi)}{2g}+\frac{e^{\chi}q^{2}\phi^{2}\psi}{g^{2}}=0\,,
\end{equation}
In light of the assumed ansatz, the Maxwell equation for the temporal component of the gauge field $A$ is
\begin{equation}\label{equazione2}
 \phi^{\prime\prime}+\phi^{\prime}\biggl(\frac{2}{r}+\frac{\chi^\prime}{2}\biggr)-\frac{2q^{2}\psi^{2}}{g}\phi
 =0\,,
\end{equation}
The remaining independent relations descending from Einstein's system are
\begin{equation}\label{equazione3}
 \frac{1}{2}\psi^{\prime 2}+\frac{e^{\chi}(\phi^{\prime 2}+v^{\prime2})}{4g}+\frac{g^\prime}{gr}+
\frac{1}{r^{2}}-\frac{3}{gL^{2}}+\frac{V(\psi)}{2g}+\frac{e^{\chi}q^{2}\psi^{2}\phi^{2}}{2g^{2}}=0\,,
\end{equation}
 \begin{equation}\label{equazione4}
 \chi^\prime+r\psi^{\prime2}+r\frac{e^{\chi}q^{2}\phi^{2}\psi^{2}}{g^{2}}=0\,,
\end{equation}
Eventually, the equation for the temporal component of $B$ becomes
\begin{equation}\label{equazione5}
 v^{\prime\prime} +v^{\prime}\biggl(\frac{2}{r}+\frac{\chi^\prime}{2}\biggr)=0\,.
\end{equation}

Notice that if we force $v(r)$ to vanish, we recover the standard holographic superconductor introduced in \cite{Hartnoll:2008kx}.
In the last equations we have again dealt with a generic potential, but in the following developments we will adhere to the particular
choice \eqref{potcho}.
To simplify the formul\ae\ we henceforth posit $L=1$ and also $2\kappa_4^2=1$ as we are allowed by the scaling symmetries of the equation of motions,
see  \cite{Hartnoll:2008kx} for details.

\subsection{Boundary conditions}

We want to proceed in solving the system of equations of motion and, to this end, we need to consider an appropriate set of boundary conditions.
Firstly, in order to obtain regular gauge field configurations, we have to impose that both the scalar potential $\phi$ and its $B$ analog $v$, 
vanish at the horizon, \cite{Hartnoll:2008kx}. Otherwise, we would have a non-trivial holonomy of the gauge fields around the imaginary
time circle; in case of horizon collapse, this would lead to a singular gauge connection.
At $r=r_H$, according to the event horizon definition, also the function $g$ vanishes so, as a whole, we have
\begin{equation}\label{IRcond}
 \phi(r_H)=v(r_H)=g(r_H)=0\,,\quad\mathrm{and} \quad  \psi(r_H),\chi(r_H) \enskip\mathrm{constants}.
\end{equation}
In agreement with \eqref{IRcond}, we have the following near-horizon expansions 
\begin{eqnarray}\label{IRseries}
&&\phi_{H}(r) = \phi_{H1} (r-r_H) + \phi_{H2}(r-r_H)^2 +\dots ,\label{crocco1} \\
&&\psi_{H}(r) = \psi_{H0} + \psi_{H1} (r-r_H) + \psi_{H2}(r-r_H)^2+ \dots ,\label{crocco2} \\
&&\chi_{H}(r) = \chi_{H0}+\chi_{H1} (r-r_H) + \chi_{H2}(r-r_H)^2+ \dots , \\
&&g_{H}(r) =  g_{H1} (r-r_H) + g_{H2}(r-r_H)^2 + \dots ,  \\
&&v_{H}(r) = v_{H1} (r-r_H) + v_{H2}(r-r_H)^2 +\dots.
\end{eqnarray}
The computational strategy consists in solving the system term by term until we gather enough boundary conditions to ``feed'' the numerical
computations.
Notice that, from the near-horizon analysis, we find the presence of $5$ degrees of freedom which parameterize the space of solutions:
\begin{equation}
 r_H\ , \ \ \
 \psi_{H0}\ , \ \ \
 E_{(A)}(r_H) \doteq \phi'(r_H)\ , \ \ \
 E_{(B)}(r_H) \doteq v'(r_H)\ , \ \ \
 \chi_{H0}\ ,
\end{equation}
where we have denoted with $E_{(x)}(r_H)$ the ``electric field'' at the horizon associated to the gauge field $x$.

Let us look at the same mathematical problem from the conformal boundary viewpoint.
The parameters at the horizon have a boundary counterpart.
We choose 
\begin{equation}\label{masschoice}
 m^2 L^2 = -2 
\end{equation}
which leads to the following asymptotic behavior for the scalar field $\psi$,
\begin{equation}\label{psiUV}
\psi(r)=\frac{C_1}{r}+\frac{C_2}{r^2}+\dots,\quad \mathrm{as}\quad r\rightarrow\infty\ ,
\end{equation}
where, as a consequence of the homogeneous character of our ansatz and the stationarity hypothesis, $C_1$ and $C_2$ are constants that do not depend
on the coordinates of the physical space-time. 
The choice for the mass \eqref{masschoice} led us to the asymptotic behavior \eqref{psiUV} where the two leading 
contributions are both normalizable%
\footnote{In the Lagrangian density \eqref{laga}, the terms involving the scalar are quadratic. Remembering the metric 
factor $\sqrt{-\text{det} g}$ which, according to \eqref{unosufullme} behaves as $r^2$ for large $r$, the normalizable terms are those behaving 
asymptotically as $r^{-a}$ with $a\geq 1$.};
we can therefore choose which between them plays the r\^{o}le of the source and which plays the r\^{o}le 
of the VEV of the corresponding operator (see \cite{Klebanov:1999tb}).
We will consider $C_1$ as the source and
\begin{equation}\label{opera}
 \langle {\cal O} \rangle = \sqrt{2} C_2\ ,
\end{equation}
as the operator representing the superconducting order parameter.
In \eqref{opera}, the $\sqrt{2}$ factor has been inserted to adhere to the usual convention, \cite{Hartnoll:2008kx}.
We furthermore put the source to zero, namely
\begin{equation}
 C_1=0\ ,
\end{equation}
so that the presence of a non-vanishing expectation for $\cal O$ (i.e. $C_2\neq 0$) will correspond to a spontaneous breakdown of the gauge symmetry%
\footnote{The ``spontaneity'' of a symmetry breaking consists in the presence of an unsourced VEV for the breaking operator.}

The vector fields at the boundary behave as
\begin{eqnarray}
\phi(r)&=&\mu-\frac{\rho}{r} +\dots \quad \mathrm{as}\quad r\rightarrow\infty\,, \label{phiUV}\\
v(r)&=&\delta\mu-\frac{\delta\rho}{r}+\dots\quad \mathrm{as}\quad r\rightarrow\infty\,,\label{vUV}
\end{eqnarray}
where $\mu$ and $\delta \mu$ represent respectively the chemical potential and the chemical potential imbalance and, similarly,
$\rho$ and $\delta \rho$ are respectively the charge density and its imbalance.
Note that the quantities $\mu$ and $\rho$ (or $\delta \mu$ and $\delta \rho$) are not independent and imposing from outside either
the values of the $\rho$'s or the values of the $\mu$'s corresponds to consider the canonical or grand-canonical description of the system.

Eventually the fields $g$ and $\chi$ have the following asymptotic behavior:
\begin{eqnarray}
 g(r)&=&r^2 -\frac{\epsilon}{2r} + \dots\quad \mathrm{as}\quad r\rightarrow\infty\label{gUV}\\
 \chi(r)&=&0 +\dots\quad \mathrm{as}\quad r\rightarrow\infty\,,
 \label{chiUV}
\end{eqnarray}
where we have imposed
\begin{equation}
 \chi \rightarrow 0\ \ \ \  \text{for}\ \ \ \   r\rightarrow \infty\ ;
\end{equation}
this follows from the requirement of having an asymptotic $AdS$ solution.


\subsection{Normal phase}

The normal phase is characterized by the absence of a non-vanishing expectation value for the condensate, so
\begin{equation}
 \langle {\cal O} \rangle = 0\ .
\end{equation}
In the bulk, the solution to the gravitational problem presents a vanishing scalar field $\psi$.
The metric corresponds to a Reissner-Nordstr\"{o}m-$AdS_4$ black hole charged under both the gauge fields $A$ and $B$; 
its metric is explicitly given by
\begin{eqnarray}\label{rnAdS}
ds^{2}&=&-f(r)dt^{2}+r^{2}(dx^{2}+dy^{2})+\frac{dr^{2}}{f(r)},\\
f(r)&=&r^{2}\bigg(1-\frac{r_{H}^{3}}{r^{3}}\bigg)+
\frac{\mu^{2}r_H^{2}}{4r^{2}}\bigg(1-\frac{r}{r_{H}}\bigg)
+\frac{\delta\mu^{2}r_H^{2}}{4r^{2}}\bigg(1-\frac{r}{r_{H}}\bigg)\,.\label{fr}
\end{eqnarray}
where the horizon radius $r_H$ refers to the black hole external horizon.
The profiles of the solution for the gauge fields are
\begin{eqnarray}
\phi(r)&=&\mu\bigg(1-\frac{r_{H}}{r}\bigg)=\mu-\frac{\rho }{r}\,,\label{phiN}\\
v(r)&=&\delta\mu\bigg(1-\frac{r_{H}}{r}\bigg)=\delta\mu-\frac{\delta\rho }{r}\label{vN}\,.
\end{eqnarray}
Repeating the analysis of Subsection \ref{Hawk}, we have that the temperature for our doubly charged RN black hole is
\begin{equation}\label{tempN}
T=\frac{r_H}{16\pi}\bigg(12-\frac{\mu^2+\delta\mu^2}{r_H^2}\bigg)\,,
\end{equation}
From \eqref{tempN} it is possible to express the horizon radius $r_H$ in terms of the thermodynamical variables of the system, namely
\begin{equation}\label{rhN}
 r_H=\frac{2}{3}\pi T+\frac{1}{6}\sqrt{16\pi^2T^2+3(\mu^2+\delta\mu^2)}\,.
\end{equation}

The AdS/CFT dictionary relates the free energy of the boundary theory with the on-shell value of the (regularized) dual action.
This emerges naturally from the formulation of the correspondence which identifies the two generating functionals, see Subsection \ref{HoloThermo}.
In our model we have the following explicit expression for the free energy
\begin{equation}\label{normalfree}
\omega_n=-r_H^3\bigg( 1+\frac{(\mu^2 + \delta\mu^2)}{4r_H^2}\bigg).
\end{equation}
Notice that employing Equation \eqref{rhN}, the free energy thermodynamic potential can be expressed in terms of $T,\mu$ and $\delta\mu$ alone.

When the temperature is lowered to $T=0$, the black hole solution becomes extremal and the degenerate horizon radius%
\footnote{Outer and inner horizons coincide at extremality.} is obtained considering $T=0$ into \eqref{tempN}, 
\begin{equation}
 (r_H^{(\text{ext})})^2=\frac{1}{12}(\delta\mu^2+\mu^2)\,.
\label{rhN0}
\end{equation}
The near-horizon geometry of the RN-$AdS_4$ black hole is $AdS_2\times R^2$ where the radii of the two solutions are related as follows%
\footnote{See Appendix \ref{nearhor} for details.}:
\begin{equation}\label{radiirel}
 L_{(2)}^2 = \frac{1}{6} L^2\ .
\end{equation}
This observation about the near horizon geometry is important in the study of the stability that we perform in Section \ref{stabi}.

At extremality, i.e. $T=0$, the charge density imbalance is given by
\begin{equation}
 \delta\rho = \sqrt{\frac{\mu^2+\delta\mu^2}{12}}
\delta\mu\,,
\end{equation}
where it is manifest that $\delta \rho$ vanishes as $\delta \mu$ does so. 
This behavior is in agreement with the weak-coupling unbalanced superconductor phenomenology%
\footnote{We rely further on this in Subsection \ref{ccb}.}.
The susceptibility corresponding to the charge imbalance is given by
\begin{equation}\label{deltacchio}
  \delta\chi
= \frac{\partial \delta\rho}{\partial\delta\mu}|_{\delta\mu=0} = \frac{\mu}{\sqrt{12}}\,.
\end{equation}
Note interpreting the field $B$ as associated to the ``spin of the electrons'' we have that $\delta \chi$
represents the magnetic susceptibility.

\subsection{A criterion for instability and hair formation}
\label{stabi}

Following and generalizing the approach proposed in \cite{Iqbal:2010eh} and \cite{Horowitz:2009ij},
we can find a criterion for the instability of the normal, non-supercondcting phase at $T=0$.
Such criterion can be expressed as a condition on the parameters $m$, $q$ and the ``external'' sources $\mu$, $\delta\mu$.
We let the complex scalar field $\psi$ fluctuate on the extremal U$(1)^2$-charged Reissner-Nordstr\"{o}m-$AdS$ background.
Recall the equation of motion for $\psi$ \eqref{equazione1} with background metric \eqref{rnAdS} and gauge fields given in \eqref{phiN} and \eqref{vN}
leading to the horizon radius \eqref{rhN0}.

In the near-horizon analysis, the scalar equation of motion reduces to an equation of motion for a scalar field of mass $m_{\text{eff}(2)}^2$ given 
by the following relation%
\footnote{To actually appreciate this, it is possible to repeat the same reasoning proposed in \cite{Horowitz:2009ij} to the context of our generalized model
containing an extra gauge field.}
\begin{equation}\label{massef}
 m_{\text{eff}\, (2)}^2 = m^2 - \frac{2q^2}{1+\frac{\delta\mu^2}{\mu^2}}\ ,
\end{equation}
on an $AdS_2$ background (hence the pedex $(2)$) having radius given by \eqref{radiirel}.
We recover an instability criterion asking that the effective mass \eqref{massef} is
below the near-horizon $AdS_2$ Beitenlhoner-Friedman bound, namely
\begin{equation}
 L_{(2)}^2 m^2_{\text{eff}\, (2)} = \frac{L^2}{6} m^2_{\text{eff}\, (2)} < - \frac{1}{4}\ ,
\end{equation}
leading to
\begin{equation}\label{semi}
\left(1+\frac{\delta\mu^2}{\mu^2}\right) \left(m^2+\frac{3}{2}\right)<2q^2\,.
\end{equation}
Notice that if $m^2 < - 3/2$, the RN solution at $T=0$ becomes unstable for any value of the
chemical potential ratio $\delta\mu/\mu$. The case we consider explicitly, $m^2=-2$, refers to this situation.
This observation implies that for $m^2 < - 3/2$ a superconducting phase developing non-trivial profile for $\psi$ is
always energetically favored at $T=0$. In other terms, for any value of the imbalance, we can have a superconducting phase
if we lower the temperature enough%
\footnote{A similar result emerged in the study of the instability of dyonic black hole charged both under an electric U$(1)$ and a magnetic U$(1)$, see \cite{Iqbal:2010eh}.}

Conversely, when $m^2 > - 3/2$, the normal phase becomes unstable at zero temperature only if the following condition is satisfied
\begin{equation}
 \frac{\delta\mu^2}{\mu^2} < 2 q^2 \frac{1}{m^2 + \frac{3}{2}} - 1 \ .
\end{equation}
Provided that $4q^2 > 2m^2 + 3$, we have in this case a bound on the imbalance above which the superconducting phase is unfavored at $T=0$.
This is analogous to the Chandrasekhar-Clogston bound occurring in unbalanced weakly-coupled superconductors%
\footnote{So, rephrasing the previous result, for $m^2 < - 3/2$ there is no Chandrasekhar-Clogston-like bound.}.

Let us comment on the $T=0$ phase transition which is expected in the presence of a CC-like bound.
Phase transitions at zero temperature are driven by quantum fluctuations instead of thermal fluctuations and
are accordingly named ``quantum phase transitions''.
In our system, crossing a CC-like bound by acting on the $\delta\mu/\mu$ ratio, would lead to a quantum phase transition;
the peculiarities of this transition, and in particular the kind of phase transition, are however not clear a priori.
An expected possibility is that the quantum transition associated to the crossing of the CC bound is of the
Berezinskii-Kosterlits-Thouless (BKT) type (see \cite{Kaplan:2009kr} and \cite{Jensen:2010ga}).
The BKT transitions are continuous transitions in which, as opposed to second order transitions, 
the order parameter vanish exponentially towards the critical point.
Indeed, in \cite{Jensen:2010ga} has been argued that a BKT phase transition can occur within a holographic context provided
that the model possesses two ``control parameters'' with the same dimension.
This is precisely what happens in our model; think to $\delta\mu$ and $\mu$ if working in the grand-canonical picture or to
$\rho$ and $\delta\rho$ if adopting the canonical picture.

A final remark is in order. According to the Mermin-Wagner theorem, no second-order phase transition can occur within systems 
with $2$ or less spatial dimensions. More precisely, in systems with $d\leq 2$ spatial dimension, with short ranged interactions and at finite temperature 
it is not possible to break spontaneously a continuous symmetry. In fact, the Goldstone modes associated to a hypothetic spontaneous breaking would
possess an infrared divergence and therefore the low-energy Goldstone quantum fluctuations would spoil the long-range order.
In our treatment, we deal with a $d=2$ system at finite temperature and, nevertheless, we refer to the scalar condensation
as a ``second order phase transition''. It should be precised that this means simply that the transition is continuous and it possesses
mean-field behavior (i.e. Landau critical exponents). In a holographic, large $N$ context, the scalar condesation is associated to the bulk violation of the near-horizon
Breitenlohner-Friedman bound; from the boundary theory perspective, the phase transition corresponds to the simultaneous condensation of $N$ operators.
This picture does not fit in the usual framework, the reasoning in terms of the Ginsburg-Landau approach and also the Mermin-Wagner theorem
are not strictly applicable to the holographic context.

\subsection{Chandrasekhar-Clogston bound at weak-coupling}
\label{ccb}

Let us consider the $T=0$ and $\delta\mu \ll \mu$ behavior of the BCS superconductor.
We expand the free energy Gibbs potential around $\delta\mu=0$%
\footnote{In other words, we are assuming \emph{analiticity} for the Gibbs potential in the 
grand-canonical ensemble.} and look at just the first terms,
\begin{equation}
 \Omega(\delta\mu)=\Omega(0)+\Omega(0)'\delta\mu +\frac12 \Omega''(0)\delta\mu^2 + {\cal O}(\delta\mu^3)\,.
\label{expa}
\end{equation}
Within the standard physical interpretation the number of particles is conjugate to the chemical potential;
an analogous relation holds for the quantities accounting for the imbalance, namely
\begin{equation}
\delta n\equiv n_u-n_d =
-\frac{\partial\Omega}{\partial\delta\mu}\,,\quad \delta\chi= \frac{\partial\delta
n}{\partial\delta\mu}=-\frac{\partial^2\Omega}{\partial\delta\mu^2}\,, 
\end{equation}
where $\delta n $ represents the ``population imbalance'' between spin-up and down electrons and $\delta \chi$ is the already mentioned 
``magnetic susceptibility'' \eqref{deltacchio}.
From the weak coupling BCS analysis, it is possible to show that the population imbalance corresponding to the normal phase at null temperature is given by
\begin{equation}
 \delta n^{(\text{norm})} \sim \rho\ \delta \mu\ .
\end{equation}
This leads to the following explicit Gibbs potential in the normal phase at $T=0$:
\begin{equation}
\Omega^{(\text{norm)}}(\delta\mu)\approx \Omega^{(\text{norm)}}(0)-\frac12\rho_F\delta\mu^2\,.
\end{equation}
Conversely, the superconducting BCS phase, again at $T=0$, presents a vanishing population imbalance.
The condensate, which we are assuming here homogeneous (i.e. non depending on the space coordinates and stationary), 
involves an equal number of spin-up and spin-down electrons. Indeed, any Cooper pair
is composed by an electron of each kind. Correspondingly, the free energy expansion in the superconducting phase is
\begin{equation}
 \Omega^{(\text{super})} (\delta\mu) \sim \Omega^{(\text{super})} (0) \ ,
\end{equation}
which is analogous to stating that the gap parameter of the superconducting phase is independent of the
imbalance $\delta\mu$.
It is easy to compare the Gibbs free energies corresponding to the normal and superconducting BCS phases
for the same $T=0$ and $\delta\mu\ll\mu$ thermodynamical situation
\begin{equation}
\Omega^{(\text{norm})}(\delta\mu)-\Omega^{(\text{super})}(\delta\mu)\sim \Omega^{(\text{norm})}(0)-\Omega^{(\text{super})}(0)-\frac12\rho_F\delta\mu^2.
\end{equation}
This comparison is intended to study which phase is energetically favored.
Relying on another BCS result, the difference between the free energies of the two phases at $\delta\mu=0$ is accounted for by
\begin{equation}
  \Omega^{(\text{norm})}(0)-\Omega^{(\text{super})}(0)=\rho_F\Delta_0^2/4\ ,
\end{equation}
where $\Delta_0$ represents the zero-temperature gap parameter. At non-zero $\delta\mu$ we have
\begin{equation}
\Omega^{(\text{norm})}(\delta\mu)-\Omega^{(\text{super})}(\delta\mu)\approx\frac{1}{4}\rho_F\Delta_0^2-\frac12\rho_F\delta\mu^2.
\end{equation}
We observe that the superconducting phase is favored whenever
\begin{equation}\label{CCb}
\delta\mu<\delta\mu_{1},\quad \delta\mu_{1}\equiv\frac{\Delta_0}{\sqrt{2}}\,.
\end{equation}
This relation is known as Chandrasekar-Clogston bound.

\subsection{Chandrasekhar-Clogston bound at strong-coupling}
\label{sucopha}

Repeating somehow the line of reasoning described in Section \ref{ccb} and extending it
to strong-coupling we can expect to find something analogous to the CC bound also in the strongly correlated regime.
Let us start from the study of the Gibbs potential in the normal phase (i.e. the doubly charged RN-black hole solution).
At $T=0$ and in the region of small $\delta\mu$ of the phase diagram (more precisely $\delta\mu \ll \mu$) we have the following
expansion for the Gibbs free energy
\begin{equation}
 \omega(\delta\mu)\approx \omega(0)-\frac12\frac{\mu}{\sqrt{12}}\delta\mu^2\,. 
\end{equation}
The normal phase result should be confronted with its counterpart in the superconducting phase.
However, the solutions of our system in the presence of non-trivial condensate are not known analytically and, 
in addition, the zero-temperature limit of the superconducting phase, namely the ground state of the unbalanced superconductor, is unclear%
\footnote{The determination of the holographic dual of the superconductor ground state is a delicate question. 
We indulge on this interesting topic in Subsection \ref{ground}.}.
We have therefore to resort to numerical computations and, in this fashion, study directly the emergence (or not) of a condensate.

\begin{figure}
 \centering
  \includegraphics[]{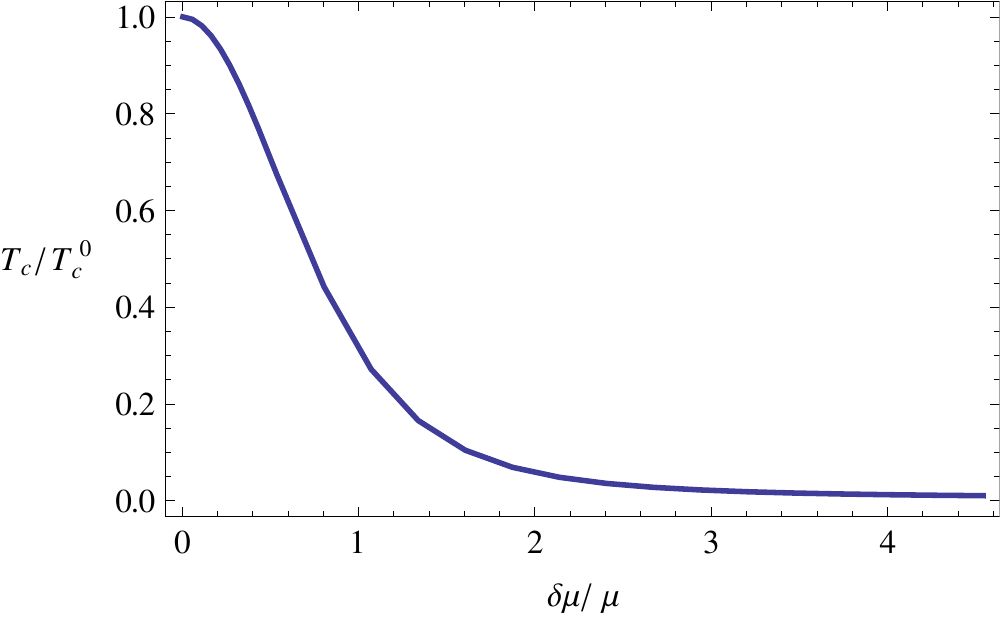}
  \caption{Critical temperature dependence on the chemical potential mismatch.}
  \label{phase}
\end{figure}

The numerical analysis shows that, for our unbalanced system, the superconductive condensation occurs for any value
of the chemical potential $\delta\mu$, provided the temperature is low enough. 
The critical temperature value depends on $\delta\mu$ and, specifically, for higher values of $\delta\mu$, the condensation occurs at lower values of the temperature,
look Figure \ref{phase}; the qualitative behavior of condensation, instead, does not change varying the value of the imbalance.
We have then not observed any Chandrasekar-Clogston bound. 
\begin{figure}
 \centering
 \includegraphics[width=.4\linewidth]{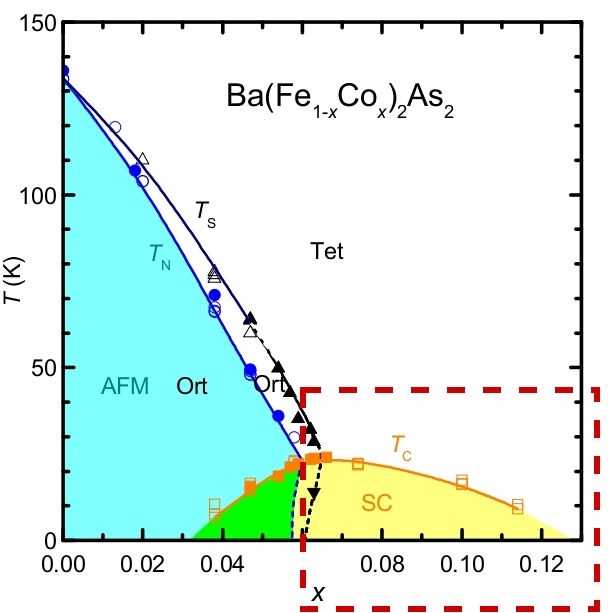}
 \caption{Figure taken from \cite{2010AdPhy..59..803J}. 
 Experimental phase diagram of a high-$T_c$ superconductor; SC indicates the superconductive phase while AFM denotes an anti-ferromagnetic phase.
 In the dashed box there is the line of $T_c$ for the superconductor-to-normal transition at different doping levels.}
 \label{cclike}
\end{figure}
Actually, the presence of a CC bound would translate in an intersection between the critical line and the $T_c/T_c^0=0$ axis,
i.e. $T_c(\delta\mu^*) = 0$ for a particular $\delta\mu^*$.
The behavior of the numerical results reported in Figure \ref{phase} suggests that the curve approaches the $T_c/T_c^0=0$ axis without intersecting it; 
if this extrapolation holds true for any value of $\delta \mu$ it implies that in our strongly coupled system there is no Chandrasekar-Clogston bound, namely
\begin{equation}
 T_c(\delta\mu) > 0\ ,\ \ \ \forall\ \ \delta\mu\ .
\end{equation}
In Figure \ref{cclike} is reported an experimental diagram taken from \cite{2010AdPhy..59..803J} where we put in evidence the line
of superconductor-to-normal transition at different doping levels $X$. 
To draw the comparison with our result reported in Figure \ref{phase} the doping level has to be related to the chemical potential imbalance;
this is natural if we consider doping with magnetic ``impurities''.
Note that the experimental plot does not show neither exclude the presence of a Chandrasekar-Clogston-like bound.

\subsection{The condensate}

The numerical analysis focused on the characterization of the condensate emerges from
a standard numerical study of the system of equations of motions of the dual gravitational model.
We underline that the results hold for strictly positive values of the temperature%
\footnote{We refer again to Subsection \ref{ground}.}.

Plugging the explicit metric \eqref{metric} into the formula for the temperature \eqref{temp} and using the near-horizon expansions
for the fields we find the following expression for the temperature,
\begin{equation}
 T = \frac{r_H}{16\pi L^2} \left[ (12 - 2 m^2 \Psi_0^2) e^{-\chi_0/2} 
     - L^2 \left( \frac{\phi_1}{r_b}\right)^2 e^{\chi_0/2}
     - L^2 \left( \frac{v_1}{r_b}\right)^2 e^{\chi_0/2}  \right]\ .
\end{equation}
We adopt the definition of the condensate (already given in \eqref{opera})
\begin{equation}
 \langle {\cal O} \rangle = \sqrt{2} \, C_2\ ,
\end{equation}
so we study the near-boundary behavior of the scalar field from which we extract the coefficient $C_2$ and then $\langle {\cal O} \rangle$.

\begin{figure}
 \centering
 \includegraphics[]{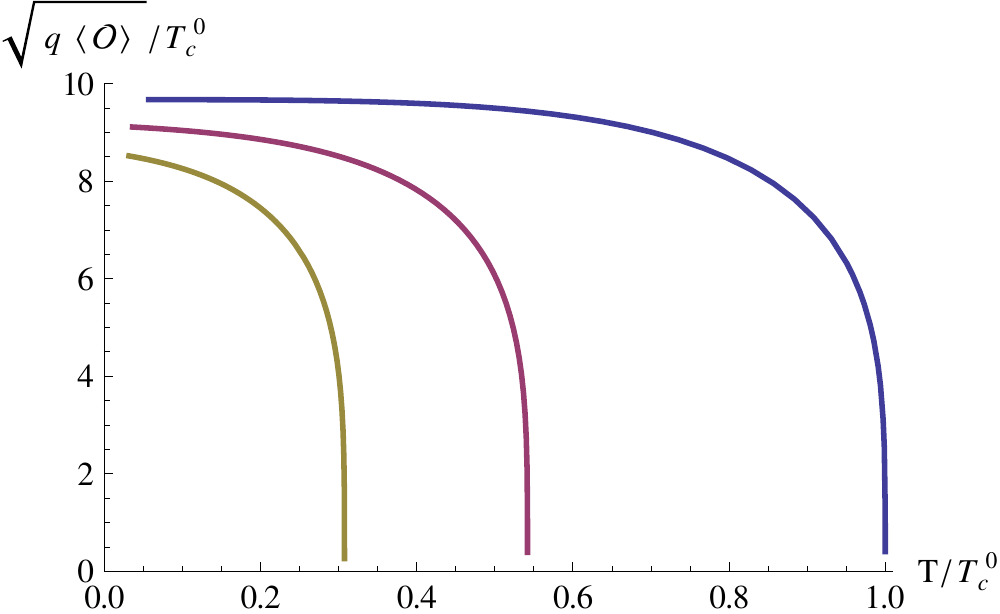}
 \label{ensati}
 \caption{Condensate as a function of temperature for $\mu=1$ and $q=2$; the plots refer to three different values of $\delta\mu$, namely (from right to left):
$0,1,1.5$; notice that as the mismatch increases, the critical temperature decreases.}
\end{figure}

For small values of the chemical potential mismatch, namely for $\delta\mu = 0.01$, and for different values of the external charge parameter,
we recover results which are in agreement with \cite{Hartnoll:2008kx}.
The qualitative shape of the condensates as the temperature is varied (see Figure \ref{ensati}) is again similar to the profiles
one recovers from a BCS approach to the standard superconductor.
We have also computed the condensate profiles for higher values of the chemical potential imbalance finding qualitative similar results.
As a general observation, in accordance with the intuitive expectation that the imbalance hinders the condensation,
we have that for higher chemical potential mismatch the critical temperature at which superconduction occurs is lower.
Though, the dependence of $T_c$ on $\delta\mu$ is not linear and has been already depicted in the ``phase diagram'' of Figure \ref{phase}.
From the viewpoint of the hologrpahic model at hand, let us notice that the bigger is the imbalance the bigger is the effective mass \eqref{massef};
then, for bigger values of $\delta\mu/\mu$ the condition for instability, though always satisfied, is met
with a smaller margin. This intuitively leads one to think that the $T=0$ doubly charged RN is less unstable
for bigger imbalance and the condensation requires a lower temperature.

\begin{figure}
    \centering
    \includegraphics[scale=.8]{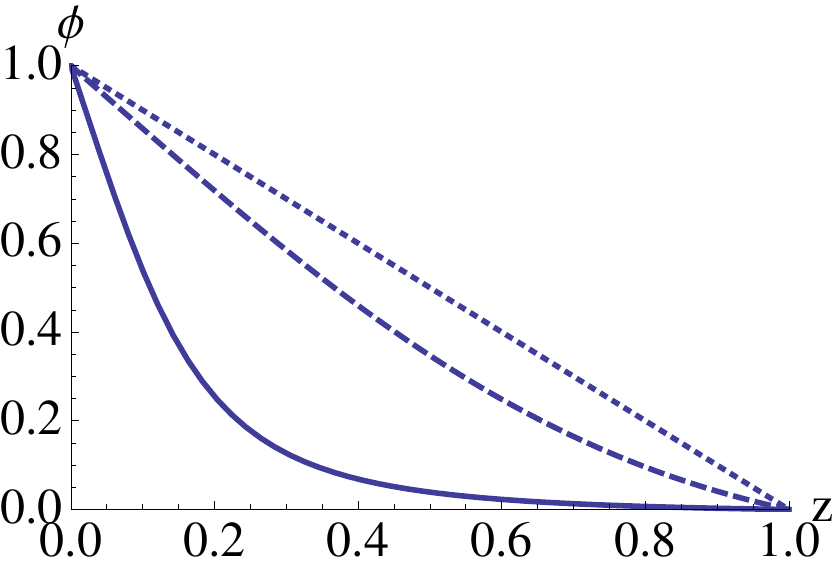}
    \includegraphics[scale=.8]{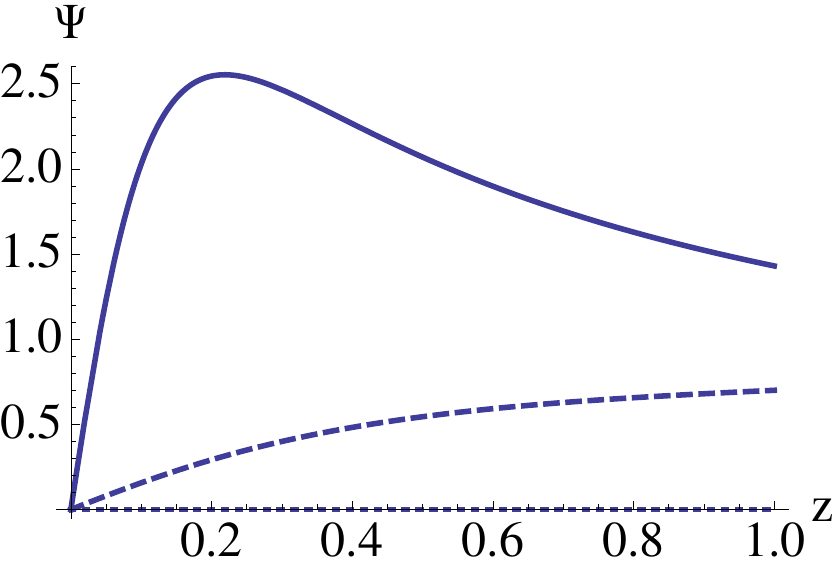}\\%
    \includegraphics[scale=.8]{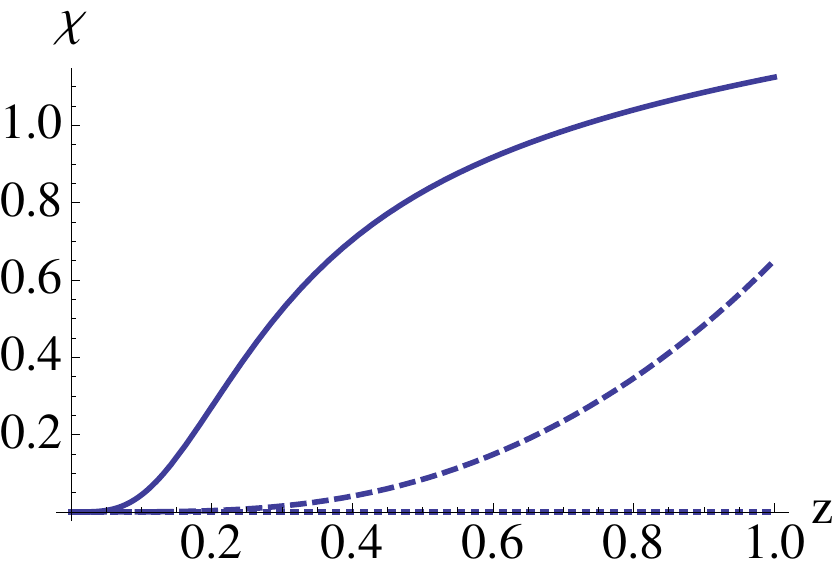}
    \includegraphics[scale=.8]{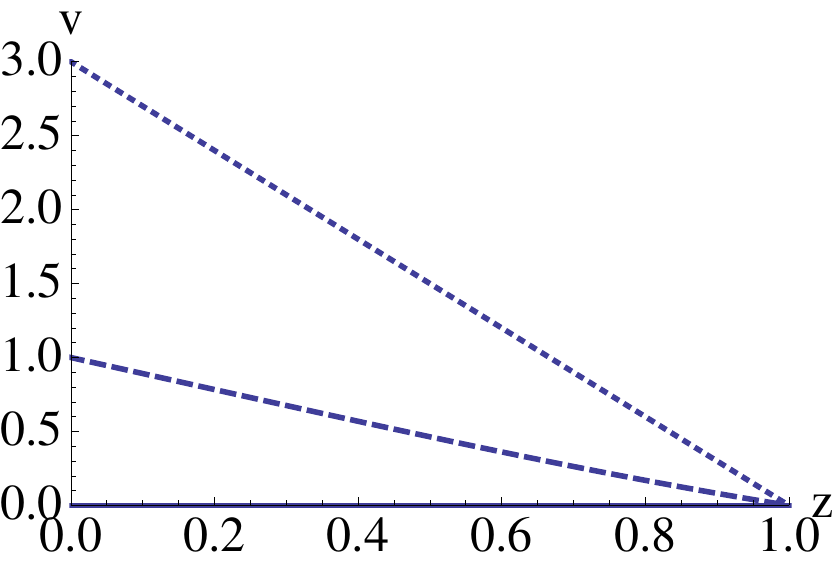}\\%
    \includegraphics[scale=.8]{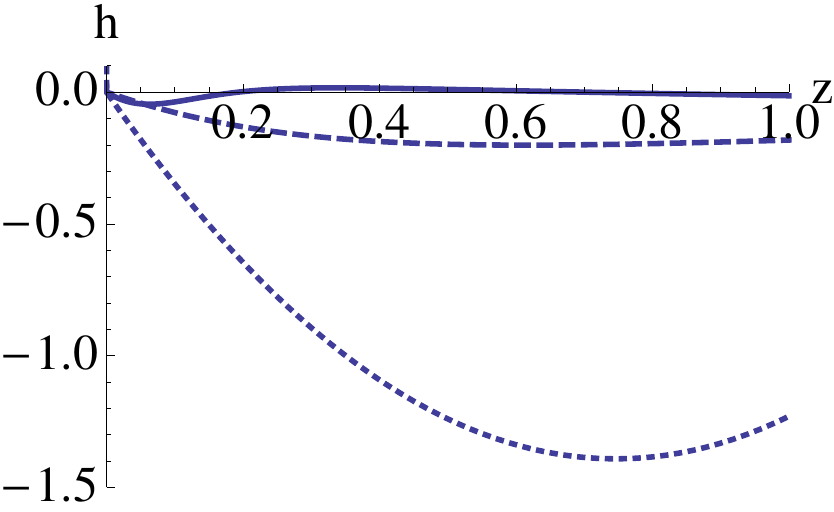}
    \caption{Background fields for $\mu=1$, $T\sim 0.027$ and $\delta\mu = 0$ (solid), 
   $\delta\mu = 1$ (dashed) and $\delta\mu = 3$ (dotted). At $z=0$ there is the conformal boundary while at $z=1$ there is the black hole horizon.}
   \label{bg}
\end{figure}

\subsection{A look to the ``unbalanced'' gravitational solutions}
\label{background}

In the Section \ref{fluctuations} we will perform a detailed study of the fluctuations around the solutions of the gravitational system.
Prior to this, it is useful to have a direct look at some features of the background and then we will let it fluctuate.
In Figure \ref{bg} we have the plot of the field profiles at fixed $\mu = 1$, fixed temperature $T\sim 0.027$
but varying $\delta\mu$. 
For the sake of computational convenience, we have defined the new dimensionless radial coordinate
\begin{equation}
 z \doteq \frac{r_H}{r}\ ,
\end{equation}
and adopted the following field redefinitions
\begin{eqnarray}\label{repsi}
 &\Psi(z) &\doteq \frac{1}{z} \psi(z)\\ 
 &h(z) &\doteq g(z) - \frac{r^2_H}{z^2} \label{reh}
\end{eqnarray}
In the plots of Figure \ref{bg}, the solid lines refer to $\delta\mu=0$, the dashed lines to $\delta\mu=1$ and the dotted lines to $\delta\mu=3$.
For $\delta\mu = 0$ and $\delta\mu=1$ the system is in the superconducting phase, while for $\delta\mu=3$ it is in the normal phase;
actually this can be guessed already from the plots noting that, for $\delta\mu=3$, the $\Psi$ profile is null.
More precisely, from \eqref{psiUV}, \eqref{opera} and \eqref{repsi} we have that the condensate is related
to the first derivative of the field $\Psi$ at the boundary, $z=0$; for the dotted line, the derivative of $\Psi$ at the boundary appears indeed vanishing.

At the boundary the value of $\phi$ represents the chemical potential $\mu$ which we are holding fixed to $1$;
$v(0)$ represents instead $\delta\mu$ that, in the plots of Figure \ref{bg}, assumes the values $0,1$ and $3$.
Note that being the scalar field $\Psi$ charged with respect to the gauge field $A$ (whose time component is denoted with $\phi$)
and uncharged with respect to $B$ (whose time component is $v$), we have that the shape of the $\phi$ profile is appreciably affected by the presence of $\Psi$
while $v$ (even though we change the boundary condition) preserves apparently the same qualitative linear shape.
For $\delta\mu = 0$ we recover the balanced holographic superconductor of \cite{Hartnoll:2008kx}%
\footnote{Even though the paper \cite{Hartnoll:2008kx} does not show the background explicitly, the code used by the authors is available on-line
on Herzog's personal web-page.}.

As an aside comment to the numerical computations, it should be underlined their delicacy.
The numerical solution of the system of equations of motion and the employment of the shooting method (see \ref{shoot})
can result in quite cumbersome and lengthy numerical evaluations. 
In many cases the process appears not to converge within a lapse of time compatible with work necessity (or Ph.D. student patience).
In order to obtain a solution having a specified set of values for the thermodynamical quantities
(a given temperature and chemical potentials), it is generally
sensible to move away varying step by step the parameters of a configuration on which the computation has already proved to be ``convergent''
instead of finding a new ``converging'' setup presenting the desired thermodynamic characteristics.
In other words, to explore the configuration space is usually convenient to move ``slowly'' in parameter space
in order to avoid waste of time and to maintain the situation as under control as possible.
In fact, the step-by-step approach allows us to compare the result we find with the result that has been just found at the previous step; 
we can thus monitor against possible troubles arising in the numerical computations.

\section{Fluctuations}
\label{fluctuations}
So far we have considered exclusively the thermodynamics of our system.
Let us now turn the attention on some of its out-of-equilibrium characteristics, in particular its transport properties.
The study of the transport features of the system emerges from the analysis of the linear response to variations of the external sources.
Of course, the validity of the linear response approach requires the source variations to be ``small'';
quantitatively, the terms beyond the linear one have to be neglectable with respect to it.
The coupling of the boundary theory with an external source is encoded in the action with a generic term of the following form, 
\begin{equation}
 \int d^Dx\ J^\mu A_\mu^{(0)} \ ,
\end{equation}
where $D$ is the dimensionality of the boundary, the space-time index $\mu$ runs over $\{0,1,..,D-1\}$, $J^\mu$ is the boundary operator representing the current
that is associated to the source $A_\mu^{(0)}$. 
The holographic prescription (usually referred as holographic dictionary) relates the source term $A_\mu^{(0)}$ to the boundary value of the
corresponding full-fledged bulk gauge field $A_\mu$.
Therefore, a small variation of the source corresponds, in the dual gravitational picture, to a small
variation of the associated dynamical gauge field boundary condition.
At the classical level, the study of the bulk system for small boundary variations consists in the analysis at linear order of the fluctuations
of the bulk fields around the background values.

We follow the lines described in \cite{Hartnoll:2008kx};
we consider a monochromatic solution ansatz, i.e. a time dependence for all the fluctuating fields of the type $e^{\im \omega t}$. 
Let us consider the linearized Einstein and Maxwell equations associated to the vector mode fluctuations along the $x$ direction%
\footnote{Because of rotational invariance, the choice of $x$ direction does not spoil the generality of the treatment.}:
\begin{equation}\label{maxa}
 A_x'' + \left(\frac{g'}{g}-\frac{\chi'}{2}\right) A'_x + \left(\frac{\omega^2}{g^2}e^\chi - \frac{2q^2\psi^2}{g}\right) A_x = 
 \frac{\phi'}{g} e^\chi \left(-g'_{tx} + \frac{2}{r} g_{tx}\right)
\end{equation}
\begin{equation}\label{maxb}
 B_x'' + \left(\frac{g'}{g}-\frac{\chi'}{2}\right) B'_x + \frac{\omega^2}{g^2}e^\chi B_x = 
 \frac{v'}{g} e^\chi \left(-g'_{tx} + \frac{2}{r} g_{tx}\right)
\end{equation}
\begin{equation}\label{grav}
 g'_{tx} - \frac{2}{r}g_{tx} + \phi' A_x + v' B_x= 0
\end{equation}
Here the prime represents the derivative with respect to the bulk radial coordinate $r$.
Notice that, since we consider the linearized version of the equations of motion, we are introducing an approximation in our computations.
The justification is that, as the boundary perturbations are small, the perturbed fields remain close to their background value;
higher-order terms in the fields within the equations of motions are then negligible.

Substituting \eqref{grav} into \eqref{maxa} and \eqref{maxb} we obtain:
\begin{equation}\label{fluctuA}
 \begin{split}
 A_x'' +& \left(\frac{g'}{g} - \frac{\chi'}{2}\right) A_x'
 + \left( \frac{\omega^2}{g^2}e^{\chi}  - \frac{2 q^2 \psi^2}{g} \right) A_x 
 - \frac{\phi'}{g} e^{\chi} \left[B_x v' + A_x \phi' \right] = 0
 \end{split}
\end{equation}
\begin{equation}\label{fluctuB}
 \begin{split}
 B_x'' +& \left[\frac{g'}{g} - \frac{\chi'}{2}\right] B_x'
 + \left[ \frac{\omega^2}{g^2}e^{\chi}  \right] B_x 
 - \frac{v'}{g} e^{\chi} \left[B_x v' + A_x \phi' \right] = 0
 \end{split}
\end{equation}
In this way we can deal with two (coupled) equations in which the metric fluctuations do not appear.
Observe, however, that the substitution has led to a system of equations in which the two gauge fields
$A$ and $B$ are mixed, in fact they appear in both the equations.
It is important to underline the r\^ole of the metric in such mixing, indeed, in the probe approximation
(where metric fluctuations are neglected) no mixing between the different gauge fields occurs%
\footnote{We describe the probe approximation for the holographic unbalanced superconductor in Section \ref{LOFF}.
In the full backreacted case, it is inconsistent to neglect the metric (vectorial) fluctuations as they are coupled
with the fluctuations of the gauge vector fields through the Einstein equation \ref{grav}.
In the probe approximation, the field $A$ is regarded as a perturbation of the background while $B$ belongs to the background itself;
so, in this approximated case, the problem of studying mixing between $A$ and $B$ fluctuations is ill-posed because the background 
is assumed by definition to be fixed.}.

In analogy to what we have done in dealing with the background, we perform a change of variable for the bulk radial coordinate
$r$ to $z=\frac{r_H}{r}$ where $r_H$ is the black hole horizon position.
Moreover, we adopt the field redefinitions \eqref{repsi} and \eqref{reh};
we then rewrite the fluctuation equations for the two gauge fields,
\begin{equation}
 \begin{split}
  A_x''(z) + 
 &A_x'(z) \left[\frac{2}{z} - \frac{\chi'(z)}{2} - \frac{2r_H^2 - z^3 h'(z)}{z(r_H^2+z^2h(z))}\right] \\
 &+ A_x(z) \frac{r_H^2}{r_H^2+z^2h(z)}\left(\frac{e^{\chi(z)} \omega^2}{r_H^2+z^2h(z)}  - 2 q^2 \Psi^2(z)\right) \\
 & \ \ \ \ \ \ \ \ \ \ \ \ \ \ \ \ \ \ \ \ - A_x(z) \frac{z^2 e^{\chi(z)} \phi'^2(z)}{r_H^2+z^2h(z)} 
 - B_x(z) \frac{z^2 e^{\chi(z)} \phi'(z) v'(z)}{r_H^2+z^2h(z)} = 0
 \end{split}
\end{equation}
\begin{equation}
 \begin{split}
  B_x''(z) + 
 &B_x'(z) \left[\frac{2}{z} - \frac{\chi'(z)}{2} - \frac{2r_H^2 - z^3 h'(z)}{z(r_H^2+z^2h(z))}\right] \\
 &+ B_x(z) \frac{r_H^2}{r_H^2+z^2h(z)}\left(\frac{e^{\chi(z)} \omega^2}{r_H^2+z^2h(z)}  \right) \\
 & \ \ \ \ \ \ \ \ \ \ \ \ \ \ \ \ \ \ \ \ - B_x(z) \frac{z^2 e^{\chi(z)} v'^2(z)}{r_H^2+z^2h(z)} 
 - A_x(z) \frac{z^2 e^{\chi(z)} v'(z) \phi'(z)}{r_H^2+z^2h(z)} = 0
 \end{split}
\end{equation}
The reason why we are using this rewriting for the fluctuation equations is that, for the background computation,
the introduction of the $z$ radial coordinate and the functions $h$ and $\Psi$ were particularly convenient.
As the study of the fluctuations relies on the background computation, it is better to stick to the same definitions.

In order to solve \eqref{fluctuA} and \eqref{fluctuB}, we assume the following near-horizon behavior ansatz
for the fluctuation functions:
\begin{eqnarray}
 & A_x(z) & = (1-z)^{\im \alpha \omega} \left[ a_0 + a_1 (1-z) + ... \right]\\
 & B_x(z) & = (1-z)^{\im \alpha \omega} \left[ b_0 + b_1 (1-z) + ... \right]\ .
\end{eqnarray}
We expand term-wise \eqref{fluctuA} and \eqref{fluctuB} in the proximity of the horizon, i.e. $z=1$;
the most divergent contributions behave as
\begin{equation}
 \frac{1}{(1-z)^2}\ .
\end{equation}
Imposing that at this level \eqref{fluctuA} and \eqref{fluctuB} are satisfied determines the value of $\alpha$.
We find two opposite possibilities that we call $\alpha_s^{(\pm)}$.
Since the differential equations we are solving are linear, we can consider the solutions
obtained combining the two possibilities we have found for $\alpha$, namely
\begin{eqnarray}
 & A_x(z) & = A_x^{(+)}(z) + A_x^{(-)}(z)\\
 & B_x(z) & = B_x^{(+)}(z) + B_x^{(-)}(z)\ ,
\end{eqnarray}
where
\begin{eqnarray}\label{vicinor}
 & A_x^{(\pm)}(z) & = (1-z)^{\im \alpha_s^{(\pm)} \omega} \left[ a_0^{(\pm)} + a_1^{(\pm)} (1-z) + ... \right]\\
 & B_x^{(\pm)}(z) & = (1-z)^{\im \alpha_s^{(\pm)} \omega} \left[ b_0^{(\pm)} + b_1^{(\pm)} (1-z) + ... \right]\ .
\end{eqnarray}
As the system shows two second-order differential equations, we need to impose $4$
initial conditions to determine one particular solution.
Let us fix the values of the coefficients $a_0^{(\pm)}$ and $b_0^{(\pm)}$,
in particular, since we want to consider just the ingoing wave, we set $a_0^{(-)}$ and $b_0^{(-)}$ to zero
(more comments on this choice are given in Subsection \ref{InOut}).
Hence, all the terms labeled with $(-)$ are consequently vanishing; we henceforth
simply ignore their existence.

In the literature of holographic superconductors, the constant term in the near-horizon ansatz for the fluctuations 
is usually chosen to be equal to $1$. 
In the case of only one gauge field $A$, the Maxwell equation in the bulk has, in fact, a scaling freedom for the gauge field; 
in the holographic framework, the physical quantities that, like correlation functions (and then conductivities), emerge essentially from the ratio of coefficients in
the near-boundary expansion of the bulk gauge field, are completely insensitive to the rescaling of the gauge field itself 
(indeed it affects the numerator and the denominator in the same way).
Our two-current model enjoys an analogous symmetry with respect to concurrent scalings of both fields $A$ and $B$ but
it is sensitive to their ratio; consequently we cannot scale the two gauge fields independently.
As we will see shortly we have to consider carefully this point in order to correctly compute the transport properties of our system.

\subsection{Ingoing/Outgoing solutions}
\label{InOut}

The fluctuation equations we obtained assumed that the fluctuation fields depend ``harmonically'' with respect to time, i.e. as $e^{\im \omega t}$.
Let us focus on the fluctuations of the field $A$, keeping in mind that the same argument could be repeated analogously for $B$.
Notice that in the linearized Maxwell equation \eqref{maxa} the time derivative appears only quadratically, 
therefore we are here insensitive to the sign of $\omega$. 
Moreover, the time oscillating factor $e^{\im \omega t}$ can be collected outside and disregarded.
In the linearized Einstein equation for the metric component $g_{tx}$, \eqref{grav},
there is no time derivative at all.
The fluctuation equation \eqref{fluctuA} obtained from the composition of the \eqref{maxa} and \eqref{grav} is therefore
insensitive to the sign of $\omega$.

Our solution ansatz for the fluctuation equation near to the horizon behaves like:
\begin{equation}\label{nearhorans}
 e^{\im \omega t}\,(1-z)^{\im \alpha \omega} (a_0 + a_1 (1-z) + ... )\ .
\end{equation}
In order to see in which direction along the radial coordinate the ``wave travels'' let us compare its value (close to the horizon) at two different times:
\begin{equation}
 e^{\im \omega t}\,(1-z)^{\im \alpha \omega} \sim e^{\im \omega t'}\,(1-z')^{\im \alpha \omega}\ .
\end{equation}
Taking the logarithm we get
\begin{equation}
 \ln (1-z) - \ln (1-z') = \frac{t'-t}{\alpha}\ .
\end{equation}
Using again the near-horizon assumption (i.e. expanding around $z=1$) we obtain:
\begin{equation}\label{nh_speed}
 \frac{z'-z}{t'-t} \sim \frac{1}{\alpha}\ .
\end{equation}
We notice therefore that the sign of $\alpha$ coincides with the sign of the wave propagation speed
along the radial direction.
Remember that the horizon is at $z=1$ and the boundary is at $z=0$;
going towards bigger values of $z$ means going towards the black hole. 
A positive $\alpha$ corresponds then to a wave traveling towards the center of the black hole,
i.e. an ingoing wave.

It is important to distinguish between ingoing and outgoing solution because of the prescription
we employ to compute holographically correlation functions for a Minkowskian boundary theory.
Indeed, to compute Minkowski retarded Green's function for the CFT boundary model we follow the recipe advanced in \cite{Son:2002sd} and,
according to the prescription, in order to study the CFT causal linear response one has to consider the ingoing fluctuation solutions.

From the general near-horizon ansatz \eqref{nearhorans}, we have that the fluctuation
solutions approach in the vicinity of the horizon a constant value for their modulus (related to the first coefficient in 
the expansion, i.e. $a_0$) whereas, at the same time, present a divergence in the phase.
Let us notice however that from \eqref{nh_speed} the wave propagation speed is related to
the exponent $\alpha$ and then, in the near-horizon limit, it tends to a constant finite value.

%
%
%

\subsection{Shooting Method}
\label{shoot}

In numerical analysis, the shooting method is a method for solving a boundary value problem
by reducing it to the solution of an initial value problem.
Let us try to clarify by means of an example.

For a boundary value problem of a second-order ordinary differential equation, the method is stated as follows.
Let
\begin{equation}
 y''(t) = f[t,y(t),y'(t)]\ ; \ \ \ \ y(t_0) = y_0 \ ; \ \ \ \ y(t_1) = y_1
\end{equation}
be the boundary value problem.
Let $y(t;a)$ denote the solution of the initial value problem
\begin{equation}
 y''(t) = f[t,y(t),y'(t)]\ ; \ \ \ \ y(t_0) = y_0 \ ; \ \ \ \ y'(t_0) = a
\end{equation}
Define the function $F(a)$ as the difference between $y(t_1;a)$ and the specified boundary value $y_1$
\begin{equation}
 F(a) = y(t_1;a) - y_1
\end{equation}
If the boundary value problem has a solution, then $F$ has a root, and that root is just the value of
$y'(t_0)$ which yields a solution $y(t)$ of the boundary problem.

\section{Conductivities}
The conductivities encode the linear response of the superconductor to perturbations of the external sources.
We consider only perturbations with zero momentum, i.e. trivial spatial dependence on the space coordinates%
\footnote{We will comment about the finite-momentum extension (which constitutes a future research direction) in Subsection \ref{opto}.}.
Exploiting rotational symmetry on the $x-y$ plane (i.e. the spatial sub-manifold of the boundary),
we can concentrate on the excitations and currents along the $x$ direction without spoiling the generality of the treatment.

The conductivity computations via holographic means are quite a standard procedure. The novelty of our analysis consists in
the concurrent presence of two gauge fields and their consequent mixing.
The two gauge fields correspond to two U$(1)$ gauge groups that through the $AdS$/CFT correspondence ``source''
two currents in the boundary theory. 
In this sense, our model could furnish the strong-coupling generalization of the two-current model proposed by Mott, 
and the mixing effects of the two currents are then read as spintronic features.
The two-current model refers to spin-up and spin-down electron currents flowing through a metallic ferromagnet,
we will oftentimes borrow the intuitive language of condensed matter systems.
It is appropriate to keep in mind however, that the model can have a larger relevance and some of its features
are completely general and not restricted to the condensed matter context.
The condensed matter interpretation is both interesting \emph{per se} as a way to investigate the physics of real
unconventional superconductors and as a source of intuitive insight of the holographic system at hand.
At the outset it should be mentioned that adopting the term ``holographic superconductor'' we do not claim that
all the holographic results have a clear and definite interpretation in terms of features of real superconductors;
nevertheless, the holographic framework offers an innovative environment in which crucial properties at strong-coupling can quantitatively studied.
More details on weak and strong points of the holographic description of superconducting systems will emerge in the analysis.

The boundary value of the bulk gauge field $A$ is interpreted as the electric field source or the electric external field $E_A$
in the boundary theory.
It provides the so-called electro-motive force, namely the force acting on electrically charged objects.
The external electric field induces a corresponding electric current $J_A$ through the medium and such response is
accounted for (at the linear level) by the electric conductivity $\sigma_{A}$. However, this is not the only effect we can obtain when exciting $E_A$;
indeed, in general, we can produce a spin current as well. This happens whenever the system reacts asymmetrically,
that is, spin-up and spin-down electrons behave somehow differently, and there is therefore a net transport of spin in response to an external electric 
perturbation.
We describe this electric-spin effect at linear order with the mixed conductivity $\gamma_{BA}$ encoding a spin current response to $E_A$.

The converse possibility is of course possible as well. When we excite the boundary value of the gauge field $B$ we 
source a \emph{spin-motive} force accounted for by the spin field $E_B$. This external field acts directly on the objects with spin producing
a spin current proportional to the spin-spin conductivity $\sigma_{B}$ but it could also, in general, induce an electric current.
We have again an ``off-diagonal'' component of the conductivity, namely $\gamma_{AB}$.

As we consider the possibility of mixed effects, it is natural to express the conductivities in a matrix form. Let us add to the picture also the thermal effects,
namely the thermo-electric and thermo-spin linear response of the system.
The relevant thermal quantities are the temperature gradient, which plays the r\^{o}le of the thermal external source, and the heat current flowing
through the system.
Again, the temperature gradient sources a heat flow proportional to the thermal conductivity $\kappa$ but, in general, it yields also
electric and spin transport.
Let us write explicitly the conductivity matrix,
\begin{eqnarray}\label{matrix}
\begin{pmatrix}  J^A \\ Q \\ J^B \end{pmatrix} = \begin{pmatrix}  \sigma_A & \alpha T & \gamma \\ \alpha T &
\kappa T & \beta T \\ \gamma & \beta T & \sigma_B \end{pmatrix} \cdot \begin{pmatrix}  E^A \\ -\frac{\nabla T}{T} \\ E^B \end{pmatrix} \, .
\end{eqnarray}
Following Onsager's argument (see appendix \ref{Onsa}), the symmetry of the matrix is a general 
feature of the response functions of the systems having time-reversal invariant equilibrium states.
Indeed, we defined $\gamma_{AB}=\gamma_{BA}\doteq \gamma$.

Let us observe that the two gauge fields $A$ and $B$ are not directly coupled in the bulk Lagrangian.
However, their fluctuations are coupled through the non-trivial fluctuations of the geometry;
so the $A$ and $B$ mixing is mediated by the metric.
When we excite $A$ and $B$, we look at vector fluctuations, i.e. fluctuations possessing a spatial index%
\footnote{Remember that we stick to the $x$ direction exploiting spatial rotational symmetry.}; such perturbations mix with the metric vector perturbations.
As both electric and spin currents carry momentum, they are naturally related to the $T_{tx}$ component of the
energy-momentum tensor describing the flow of momentum and energy through the system.
$T_{tx}$ is then both sourced directly by a temperature gradient (encoded holographically in the vector perturbation 
of the metric component $g_{tx}$, see Appendix \ref{QET}) and also whenever there is momentum transport sourced by electric or spin motive forces.

From the study of the fluctuations we have that the near boundary behaviors of the fluctuating bulk fields 
are given by the following asymptotic expansions:
\begin{eqnarray}\label{boufie}
 A_x(r) &= A_x^{(0)} + \frac{1}{r} A_x^{(1)} + ... \, ,\\
 B_x(r) &= B_x^{(0)} + \frac{1}{r} B_x^{(1)} + ... \, ,\\
 g_{tx}(r) &= r^2 g_{tx}^{(0)} - \frac{1}{r} g_{tx}^{(1)} + ...\, .
\end{eqnarray}
The solution of the Einstein equation for the fluctuations of the metric \eqref{grav} can be then expressed as
\begin{equation}
g_{tx} = r^2 \left( g_{tx}^{(0)} + \int_r^\infty \frac{\phi' A_x + v' B_x}{r^2} \right)\, ,
\end{equation}
so that
\begin{equation}\label{guno}
g_{tx}^{(1)} = \frac{\rho}{3}A_x^{(0)} + \frac{\delta\rho}{3}B_x^{(0)}\, .
\end{equation}

Substituting the Einstein equation into the Maxwell equations for the fluctuations of $A$ and $B$,
we find a system of two mixed equations \eqref{fluctuA} and \eqref{fluctuB}.
Here the metric does not appear explicitly any longer;
using the notation introduced in the near-boundary expansions \eqref{boufie}, we assume the following
linear ansatz
\begin{equation}
A_x^{(1)} =\im\, \omega\sigma_A A_x^{(0)} + \im\, \omega\gamma B_x^{(0)},\qquad B_x^{(1)} =\im\, \omega\gamma
A_x^{(0)} + \im\, \omega\sigma_B B_x^{(0)}\,, 
\label{onshAB} 
\end{equation}
Notice that, as already mentioned, the mixed conductivities are equal and denoted with a single symbol $\gamma$;
furthermore, as it will be clearer in the following, the coefficients in \eqref{onshAB} are proportional to the spin-electric conductivities.

As it is standard in field theory, the linear response to perturbations of the generic external source $\phi^b$
is encoded in the corresponding current $J^a$ and given by the associated retarded Green function (see Appendix \ref{Green}) 
\begin{equation}\label{grena}
 \delta \langle J^a \rangle =  \langle J^a J^b \rangle\ \delta\phi^b = G^R_{ab}\ \delta\phi^b\ \ ,
\end{equation}
where we are understanding that we have the following source/current term in the action
\begin{equation}
 \sum_a J^a \phi_a \ .
\end{equation}
The Green functions $G_R$ (or retarded correlators) are proportional to the corresponding conductivity
and can be computed in the holographic framework by studying the on-shell action of the gravitational dual system.
Since we aim at the computation of the linear response of the boundary system we have retained 
just the linear part of the equations of motion.
We then consider the on-shell action up to quadratic terms in the fluctuation fields.
The on-shell bulk action can be completely expressed in terms of contributions coming from the boundaries
of the bulk base manifold.
In order to do so, we must use the equations of motion for both the background fields and for
the fluctuation fields.
%
We have the following explicit expression for the on-shell bulk action
\begin{equation}\label{OS}
 S_{\text{O.S.}} = \int d^3x\, \left.e^{\chi/2} \left( -\frac{g}{2} e^{-\chi} A_x A'_x 
-\frac{g}{2} e^{-\chi} B_x B'_x - g_{tx}g'_{tx} 
+ \frac{1}{2}\left(\frac{g'}{g}-\chi'\right)g_{tx}^2\right)\right|_{r=r_\infty} \ .
\end{equation}
Note that we have just contributions coming from the upper radial limit $r_\infty$;
the contribution from the horizon corresponding to the lower radial boundary at $r=r_b$ vanishes
because both the background field $g$ and the vector fluctuations $g_{tx}$ are null at the
black hole horizon. Furthermore, all the fields are supposed to be vanishing at ``spatial''
and ``temporal'' infinity, i.e. when either $x$, $y$ or $t$ tend to plus or minus infinite.

The on-shell action \eqref{OS} is divergent for $r_\infty \rightarrow \infty$, its divergence being related to the
divergent volume of $AdS$ space.
To cure such divergence we apply the holographic renormalization procedure that consists
in regularizing the action with the introduction of appropriate counter-terms 
and than taking the limit of $r_\infty \rightarrow \infty$ of the regulated action.

The (standard) holographic renormalization procedure involves the introduction of the following three counter-terms 
(look at \cite{Hartnoll:2008kx,Liu:1998bu} and references therein):
\begin{equation}\label{ctGH}
 S_{\psi} = \int d^3x\, \left.\sqrt{-g_{\infty}}\ \frac{\psi^2}{L} \right|_{r=r_\infty}\ ,
\end{equation}
\begin{equation}\label{ctK}
 S_{\text{G.H.}} = -\int d^4x\, \sqrt{-\tilde{g}}\ 2 K  \ ,
\end{equation}
and
\begin{equation}\label{ctlambda}
 S_\lambda = \int d^3x\, \left. \sqrt{-g_{\infty}}\ \frac{4}{L} \right|_{r=r_{\infty}} \ .
\end{equation}
The term \eqref{ctK} is usually referred to as the Gibbons-Hawking boundary term (firstly introduced in \cite{GH}); 
$\tilde{g}$ is the metric induced on the $3$-surface consisting in a shell of constant radius and 
the scalar $K$ represents the extrinsic curvature.
The counter-term \eqref{ctlambda} represents a boundary cosmological constant\footnote{Notice that this term does not contribute to the
part of the action which is quadratic in the fluctuating fields; we mentioned it for completeness' sake but we will not
analyze it further.} and the metric $g_{\infty}$ is the ``boundary metric'', i.e. the metric induced by the bulk metric $g$ on the
asymptotic surface $r=r_\infty$ with $r_\infty \rightarrow \infty$, namely
\begin{equation}
 g_{\infty}=\lim_{r\rightarrow\infty}\tilde{g}(r).  
\end{equation}
Explicitly, the metric induced on a radial shell is given by
\begin{equation}\label{boumet}
 d\tilde{s}^2 = -g\, e^{-\chi} dt^2 + r^2 \left(dx^2+dy^2\right) + g_{tx} \left(dx\, dt + dt\, dx\right)\ ,
\end{equation}
The extrinsic curvature $K$ of a surface $r=const$ is defined as
\begin{equation}\label{extrins}
 K = g^{\mu\nu} \nabla_\mu n_\nu \ ,
\end{equation}
where $g^{\mu\nu}$ is the full metric (as opposed to $\tilde{g}^{\mu\nu}$)
and $n^\mu$ is the outward unitary normal vector to the surface.
Since it is defined to have unitary norm, the explicit expression if the normal vector $\bm{n}$
in the coordinate system ${t,r,x,y}$ is
\begin{equation}
 n^\mu = (0,1/\sqrt{g_{rr}},0,0)\ .
\end{equation}
As explained in Appendix \ref{appe}, being the extrinsic curvature \eqref{extrins} a covariant divergence,
it can be rewritten in the following way
\begin{equation}\label{exspa}
 K = g^{\mu\nu} \nabla_\mu n_\nu 
   = \frac{1}{\sqrt{-g}}\,\partial_\mu \left( \sqrt-{g}\ n^\mu\right) 
   = \frac{1}{\sqrt{-g}}\,\partial_r \left(\frac{\sqrt{-g}}{\sqrt{g_{rr}}} \right) \ .
\end{equation}

The regularized action is then:
\begin{equation}
 S_{reg} = S_{\text{O.S.}} + S_{\psi} + S_{G.H.} + S_{\lambda} \ .
\end{equation}
As anticipated, we want to analyze the term of $S_{reg}$ which, in the limit $r\rightarrow \infty$, is quadratic in the fluctuations,
\begin{equation}
 S_{quad} = \left.\lim_{r_\infty\rightarrow \infty} S_{reg} \right|_{{\cal O}(2)_{A_x,B_x,g_{tx}}}\ .
\end{equation}

From the study of the solutions of the background and fluctuation equations of motion we 
obtain the near boundary behavior of the fields%
\footnote{Remember that we are working in the case $\psi^{(1)}=0$; the same formul\ae\ can be found in \cite{Hartnoll:2008kx}.},
\begin{subequations}\label{boufiel}
 \begin{align}
 g &= \frac{r^2}{L^2} - \frac{\epsilon L^2}{2r} + ...\\
 \chi &= 0 \\
 \psi &=  \frac{1}{r^2} \psi^{(2)} + ...
 \end{align}
\end{subequations}
and of their derivatives
\begin{subequations}\label{bouder}
 \begin{align}
 g' &= \frac{2r}{L^2} + \frac{\epsilon L^2}{2r^2} ...\\
 \chi' &= 0 \\
  \psi &= -2 \frac{1}{r^3} \psi^{(2)} + ...
 \end{align}
\end{subequations}

Eventually the quadratic action can be expressed as follows
\begin{equation}\label{resuquad}
 S_{quad} = \int d^3x \left( \frac{1}{2}A^{(0)}_x A_x^{(1)}
+ \frac{1}{2}B^{(0)}_x B_x^{(1)}
- 3 g_{tx}^{(0)} g_{tx}^{(1)} -\frac{\epsilon}{2}g_{tx}^{(0)}g_{tx}^{(0)} \right) \ ,
\end{equation}
with $A_x^{(1)}, B_x^{(1)}, g_{tx}^{(1)}$ given in (\ref{guno}) and (\ref{onshAB}).
The details are given in Appendix \ref{appe}.

The entries of the conductivity matrix can be computed enforcing the holographic relations
\begin{eqnarray}\label{SA}
&& J^A =\frac{\delta S_{quad}}{\delta A_x^{(0)}}\,, \\
&& J^B =\frac{\delta S_{quad}}{\delta B_x^{(0)}}\,, \\\label{SB}
&& Q =\frac{\delta S_{quad}}{\delta g_{tx}^{(0)}} - \mu J^A  - \delta\mu J^B\,, \label{Qt}
\end{eqnarray}
where the following relations are to be employed%
\footnote{Details are given in Appendix \ref{QET}.} 
\begin{equation}\label{gra}
E_x^A = \im \omega(A_x^{(0)} + \mu g_{tx}^{(0)})\,,\quad E_x^B = \im \omega(B_x^{(0)} +
\delta\mu g_{tx}^{(0)})\,,\quad -\frac{\nabla_x T}{T}=\im \omega g_{tx}^{(0)}\,.
\end{equation}
We can thus get
\begin{eqnarray}
\sigma_A = \frac{J^{A}}{E^{A}}|_{g_{tx}^{(0)}=B_x^{(0)}=0} = - \frac{\im}{\omega} \frac{A_x^{(1)}}{A_x^{(0)}}|_{g_{tx}^{(0)}=B_x^{(0)}=0}\ , \nonumber \\
\gamma =  \frac{J^{B}}{E^{A}}|_{g_{tx}^{(0)}=B_x^{(0)}=0} = - \frac{\im}{\omega} \frac{B_x^{(1)}}{A_x^{(0)}}|_{g_{tx}^{(0)}=B_x^{(0)}=0}\ , \nonumber \\ 
\alpha T =  \frac{Q}{E^{A}}|_{g_{tx}^{(0)}=B^{(0)}=0} = \frac{\im \rho}{\omega} - \mu \sigma_A -\delta\mu \gamma \,,
\label{alphaT}
\end{eqnarray}
as well as
\begin{eqnarray}
\sigma_B = \frac{J^{B}}{E^{B}}|_{g_{tx}^{(0)}=A_x^{(0)}=0} = - \frac{\im}{\omega} \frac{B_x^{(1)}}{B_x^{(0)}}|_{g_{tx}^{(0)}=A_x^{(0)}=0}\ , \nonumber \\
\gamma =  \frac{J^{A}}{E^{B}}|_{g_{tx}^{(0)}=A_x^{(0)}=0} = - \frac{\im}{\omega} \frac{A_x^{(1)}}{B_x^{(0)}}|_{g_{tx}^{(0)}=A_x^{(0)}=0}\ , \nonumber \\
\beta T =  \frac{Q}{E^{B}}|_{g_{tx}^{(0)}=A_x^{(0)}=0} = \frac{\im \delta \rho}{\omega} - \delta \mu \sigma_B -\mu \gamma\, .
\label{betaT}
\end{eqnarray}

The off-diagonal conductivity $\gamma$ can be computed in two independent ways
\begin{equation}
\gamma=\sigma_A \frac{J^B}{J^A}|_{g_{tx}^{(0)}=B_x^{(0)}=0}=\sigma_B \frac{J^A}{J^B}|_{g_{tx}^{(0)}=A_x^{(0)}=0}\, .
\end{equation}
This possibility offers a non-trivial check for the numerical results. The test has been successfully passed by our numerics.

Eventually, we find that the thermal conductivity is given by
\begin{equation}\label{kappa}
\kappa T =\frac{\im}{\omega}\left(\epsilon + p -2\mu\rho -2\delta\mu\delta\rho\right) + \sigma_A \mu^2 + \sigma_B
\delta\mu^2 + 2 \gamma \mu \delta\mu\,,
\end{equation}
where the term in the pressure $p=\epsilon/2$ 
has been introduced to account for contact terms that have not been directly considered in the computations (see Herzog's review in \cite{Herzog:2009xv}). 

In order to compute a specific entry of the conductivity matrix
we have to ``excite'' the corresponding source fixing the other sources to zero.
The holographic dictionary relates a source to the boundary term of the corresponding field;
all such boundary fields must be put to zero except the boundary field whose response we are interested in.
To achieve this we have to choose the appropriate horizon conditions such that at the boundary we have
all the other sources to zero; this could be done by means of the shooting method (see Subsection \ref{shoot}).

\begin{figure}[t]
\begin{minipage}[b]{0.5\linewidth}
\centering
\includegraphics[width=78mm]{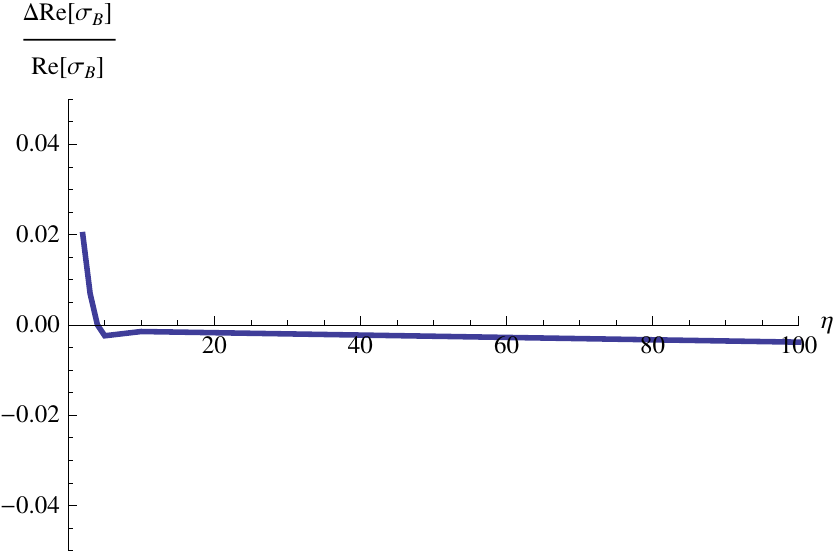} 
\end{minipage}
\begin{minipage}[b]{0.5\linewidth}
\centering
\includegraphics[width=78mm]{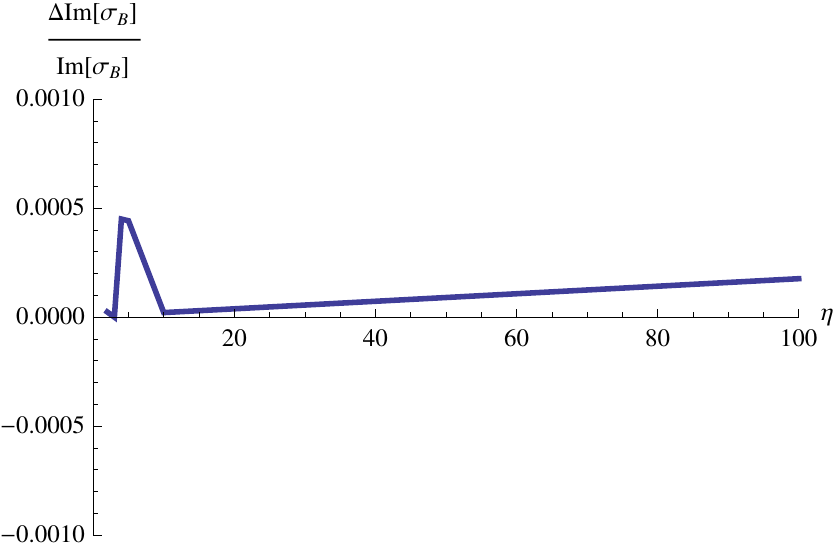}
\end{minipage}
\caption{Relative uncertainty on the computation of the real and imaginary parts of $\sigma_B$ as functions of the parameter $\eta$ defined in \eqref{eta}.}
\label{stab}
\end{figure}

It is possible to compute the conductivities (at $\nabla T =0$) with an alternative method, namely we consider the following equations
\begin{eqnarray}
J^A =\sigma_A E^A + \gamma E^B\,, \quad
J^B =\sigma_B E^B + \gamma E^A\,
\end{eqnarray}
for different arbitrary values of the horizon boundary conditions (i.e. $a_0$ and $b_0$ of \eqref{vicinor}) in order to obtain enough equations
to determine the three conductivities $\sigma_A$, $\sigma_B$ and $\gamma$. 
Since we are choosing arbitrarily the horizon condition (which would correspond to various physical sources configurations),
it is required to test the stability of the results upon different choices of the horizon terms.
This has been done systematically and the conductivity results proved stable; let us quantify this defining
the following parameter
\begin{equation}
 \eta = \frac{b_0^{(1)}/a_0^{(1)}}{b_0^{(2)}/a_0^{(2)}}\ ,
\end{equation}
where the labels $(1)$ and $(2)$ indicate two different arbitrary choices of the horizon terms%
\footnote{The test has been actually performed holding $a_0^{(1)}=a_0^{(2)}=b_0^{(1)}$ fixed to $1$ and letting $b_0^{(2)}$ vary}.
The conductivity (we present the results obtained for $\sigma_B$) has proved to be stable over a range of (at least)
two orders of magnitude, see Figure \eqref{stab}.

\subsection{Normal-phase conductivities}

The normal phase of the holographic model under study can be interpreted as describing a strongly coupled ``forced'' ferromagnet
(see \cite{Iqbal:2010eh}). The attribute \emph{forced} is opposed to \emph{spontaneous}, indeed in our model the ``spin'' density
represented by $\delta\rho$ is induced by the presence of an non-vanishing $\delta\mu$ accounting for an external magnetic field.

\begin{figure}[t]
\begin{minipage}[b]{0.5\linewidth}
\centering
\includegraphics[width=70mm]{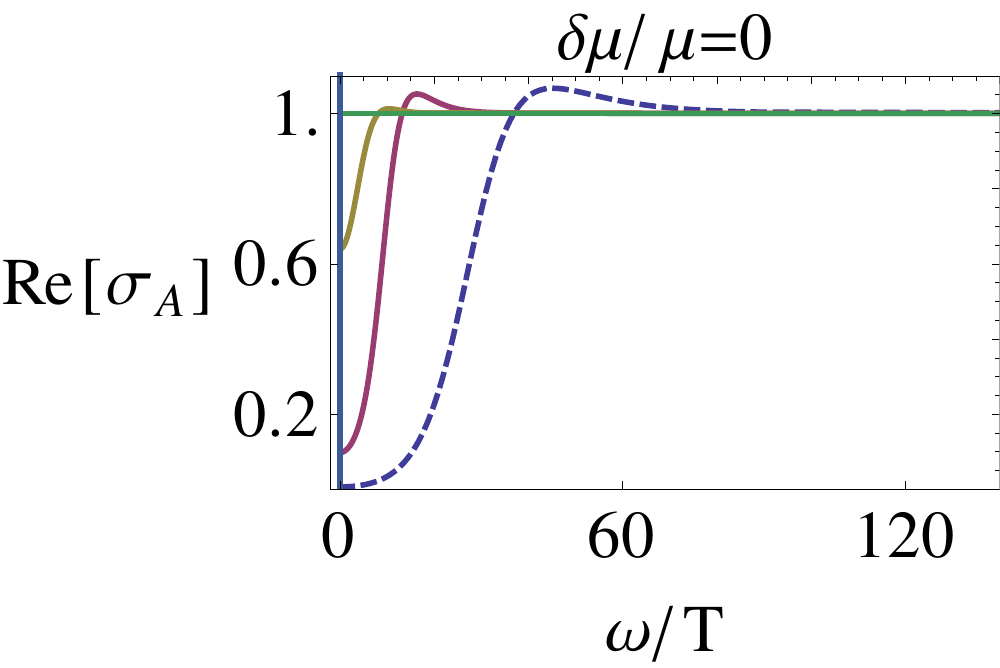}
\end{minipage}
\begin{minipage}[b]{0.5\linewidth}
\centering
\includegraphics[width=70mm]{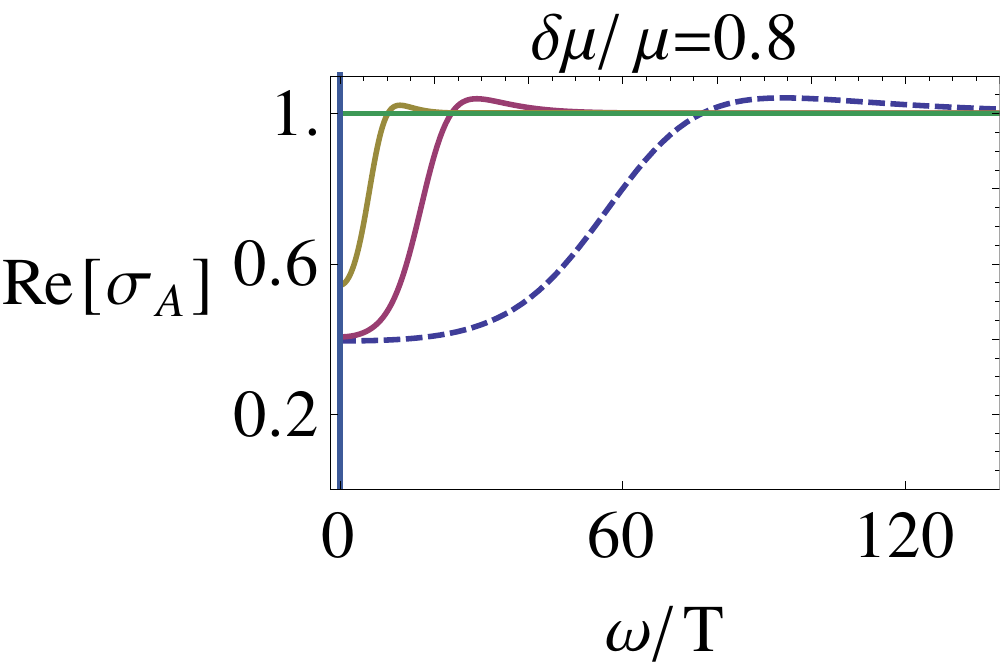}
\end{minipage}
\caption{The real part of the electric conductivity for $\delta\mu/\mu=0, 0.8$ (left plot, right plot) at $T_c$ (dashed curves) and $T>T_c$ (solid curves).}
\label{ReaNT0}
\end{figure}

We report in Figure \ref{ReaNT0} the plots of the real part of the optical, electric conductivity $\sigma_A(\omega)$ for different values of the imbalance and different 
temperatures. The frequency $\omega$ is dictated by the external field fluctuations.
The dashed lines represent the critical curves at $T=T_c$. 
The solid lines refers instead to progressively higher temperature, where the constant behavior is reached at high $T$.
This characteristic emerges in the holographic context in relation to the electro-magnetic duality
of the $4$-dimensional Einstein-Maxwell theory defined on $AdS_4$, see \cite{Herzog:2007ij}.
We further comment on this constant behavior (occurring for $\omega/T\gg 1$ and at any temperature value)
in Subsection \ref{moby}.

\begin{figure}[t]
\begin{minipage}[b]{0.5\linewidth}
\centering
\includegraphics[width=78mm]{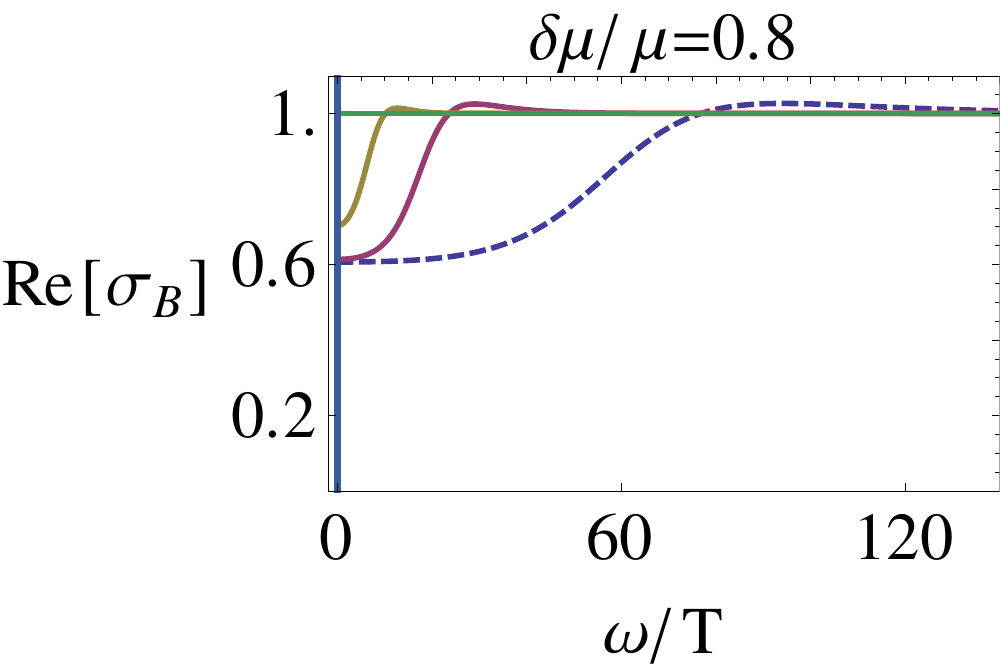}
\end{minipage}
\hspace{-0.2cm}
\begin{minipage}[b]{0.5\linewidth}
\centering
\includegraphics[width=78mm]{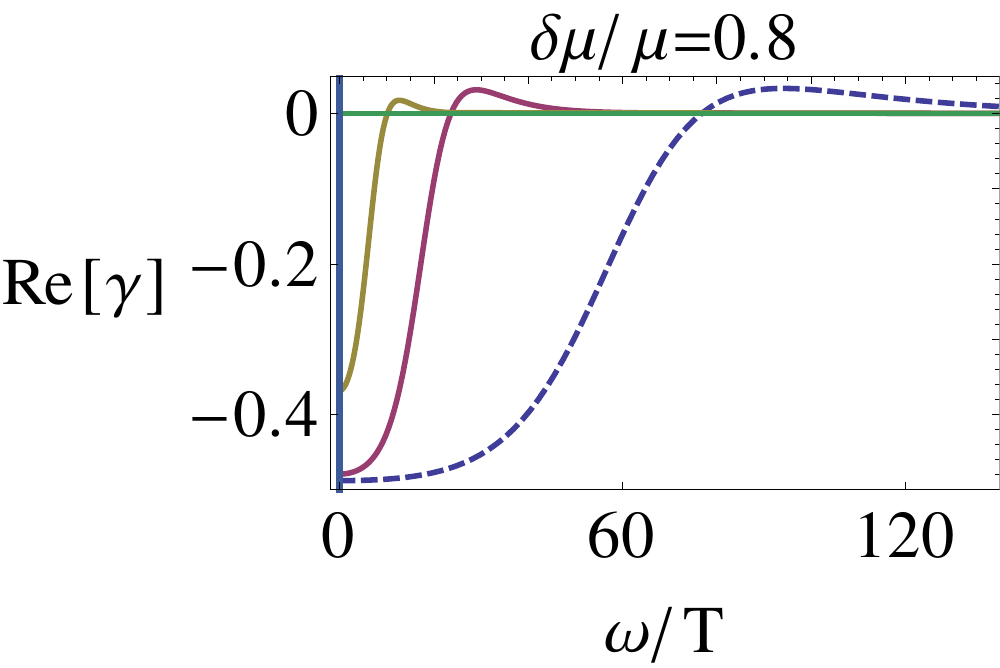}
\label{gama}
\end{minipage}
\caption{The real part of the ``spin-spin conductivity'' $\sigma_B$ (left) and of the ``spin conductivity'' $\gamma$ (right) for $\delta\mu/\mu=0.8$ at $T_c$ (dashed curves) and $T>T_c$ (solid curves).}
\label{RebNT}
\end{figure}

As the temperature is decreased, the conductivity is more and more depleted in the small frequency region.
The imaginary part of $\sigma_A$ (see the right plot of Figure \ref{uncha} to appreciate its qualitative behavior) has a pole at $\omega = 0$;
this corresponds (via a Kramers-Kroning relation, see Appendix \ref{causa}) to a delta function for Re$[\sigma_A]$ at the same point (the solid line at $\omega/T = 0$ in our plots).
The delta function at zero frequency (in the normal phase) is completely produced by the system translation invariance; 
in fact, in charged media, a DC external field causes an overall uniform acceleration (instead of a stable drift speed), 
and then an infinite DC conductivity%
\footnote{A comment about the depletion: because of a Ferrell-Glover-Tinkham sum rule,
the area under the curves representing the real part of the conductivity must be constant at different temperatures; the development of a delta function
at $\omega = 0$ is therefore compensated by a depletion of the conductivity at small frequency.}.
     

The left plot in Figure \ref{ReaNT0} refers to the balanced case and reproduces the results of \cite{Hartnoll:2008kx}.
The plot on the right instead refers to an unbalanced case; we se that, even though the qualitative shapes are the same as the
balanced setup, the low-frequency region of depletion manifests different characteristics.
In particular, the pseudo-gap is shallower and, correspondingly, the amplitude of the DC delta function is smaller.
Recall also that for the unbalanced case the critical temperature is smaller than its balanced counterpart.

Let us turn the attention on the ``spin-spin'' optical conductivity, that is $\sigma_B(\omega)$.
In the balanced $\delta\mu=0$ case its real part is a constant because the system does not contain a net overall spin and
its imaginary part is vanishing at any frequency.
The unbalanced case, instead, presents a $\sigma_B(\omega)$ which is qualitatively similar to $\sigma_A(\omega)$, see Figure \ref{RebNT}.
For $\delta\mu\neq 0$ also the ``spin-spin'' conductivity has a DC delta and a corresponding depletion region for small $\omega$.
These are not surprising features: essentially we obtain the conductivities from the near-boundary study of the gauge field fluctuation
equations of motion \eqref{fluctuA} and \eqref{fluctuB} which, in the normal phase (i.e. $\psi=0$), are symmetric with respect to the substitution
\begin{equation}
 \begin{split}
  \mu &\leftrightarrow \delta\mu\\
  A_x &\leftrightarrow B_x
 \end{split}
\end{equation}
This symmetry translates in the following relation between the conductivities
\begin{equation}\label{symsig}
 \sigma_A \left(\mu,\delta\mu, \frac{\omega}{T} \right) = \sigma_B \left(\delta\mu,\mu,\frac{\omega}{T} \right) \ ,
\end{equation}
that we have tested also numerically with an ${\cal O}(10^{-3})$ accuracy at least.
From \eqref{symsig} we have that the behaviors of $\sigma_A$ and $\sigma_B$ with respect to the ratio $\delta\mu/\mu$ are opposite.
Observe that Equation \eqref{symsig} in the particular case $\delta\mu=\pm \mu$, namely a perfectly polarized configuration 
where the spins are all oriented in the same direction (in our conventions we have respectively all ``spin-up'' for $+$ and all ``spin-down'' for $-$),
states that the electric ($\sigma_A$) and spin ($\sigma_B$) conductivities coincide. 
This happens because we have normalized the ``spin'' and electric charges of the electron both to one and therefore, in a perfectly polarized case, 
a charge flow corresponds always to an equally intense spin flow.

Relation \eqref{symsig} is one among other relations connecting the various components of the conductivity matrix;
these descend from symmetry characteristics of the system of fluctuation equations.
We will further rely on these relations in Subsection \ref{moby} where we parametrize the conductivity matrix in terms of a single 
$\omega$-dependent function $f$; this possibility suggests an interesting interpretation in terms of a \emph{mobility} function for individual carriers (see Subsection \ref{moby}).

The behavior of the mixed conductivity $\gamma$, plotted in Figure \ref{RebNT} (on the right), has again a qualitative shape similar to the $\sigma$'s;
note however that it presents negative values and, for large $\omega$, it saturates to zero%
\footnote{A comment of negative values for mixed conductivities is given in Appendix \ref{simple}.}.
Eventually, Figure \ref{RealphaTN} contains the thermo-electric and-spin-electric conductivities.
Obviously at $\delta\mu=0$ there is no spin transport and the spin-electric conductivity is vanishing on the entire range of $\omega$;
apart from this feature, $\alpha$ and $\beta$ behave similarly and, in particular, they are always negative.
\begin{figure}[t]
\begin{minipage}[b]{0.5\linewidth}
\centering
\includegraphics[width=78mm]{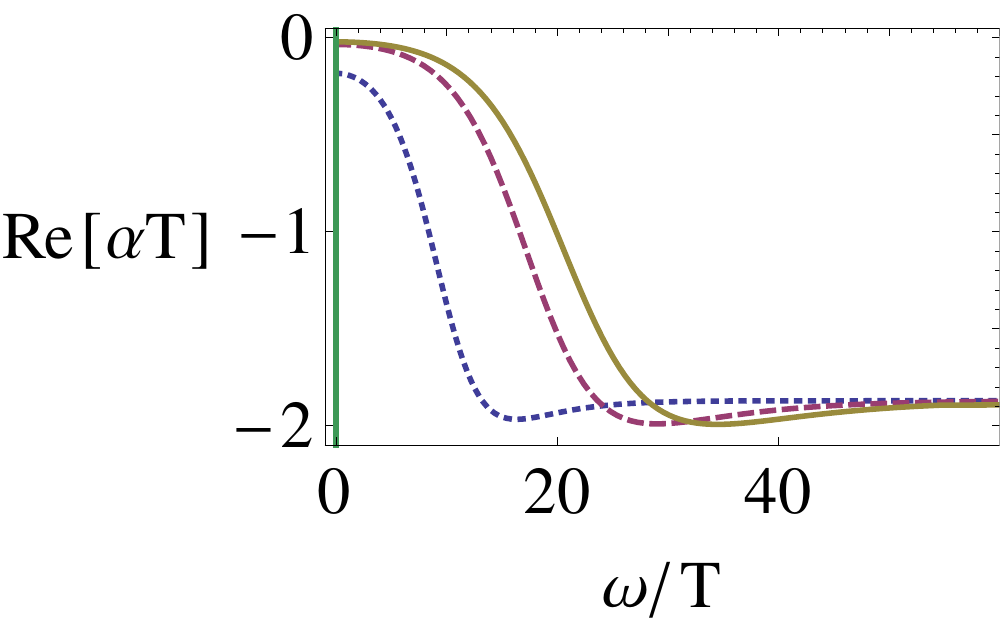}
\end{minipage}
\hspace{-0.2cm}
\begin{minipage}[b]{0.5\linewidth}
\centering
\includegraphics[width=78mm]{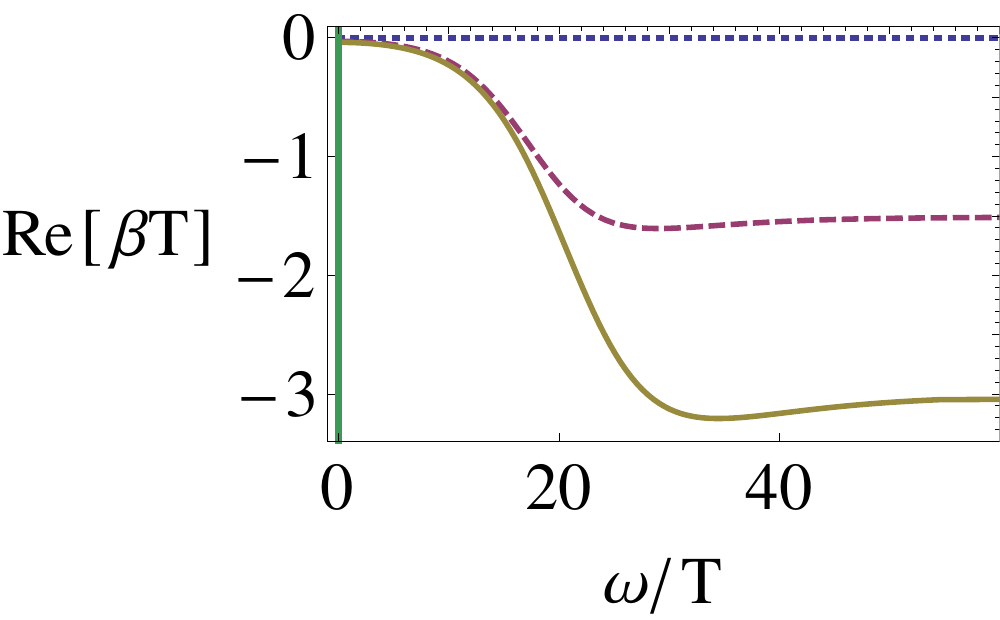}
\end{minipage}
\caption{The real part of the thermo-electric conductivity $\alpha T$ (left) and of the ``spin-electric'' conductivity $\beta T$ (right) for $\delta\mu/\mu=0, 0.8, 1.6$ (dotted, dashed and solid lines respectively) at fixed temperature.}
\label{RealphaTN}
\end{figure}

\subsection{Superconducting-phase conductivities}

\begin{figure}[t]
\begin{minipage}[b]{0.5\linewidth}
\centering
\includegraphics[width=78mm]{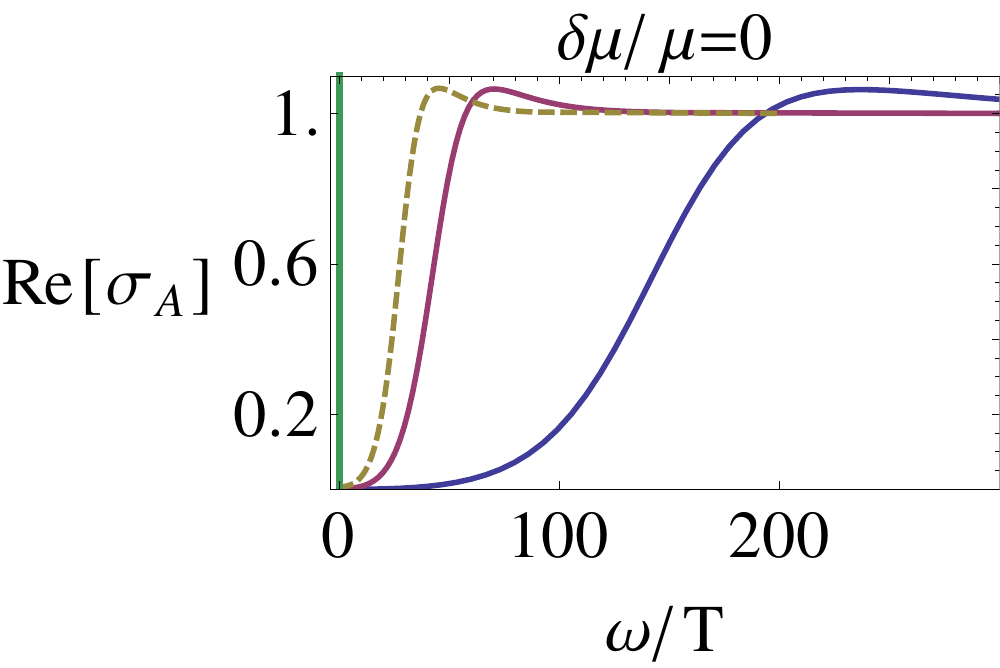}
\end{minipage}
\hspace{-0.2cm}
\begin{minipage}[b]{0.5\linewidth}
\centering
\includegraphics[width=78mm]{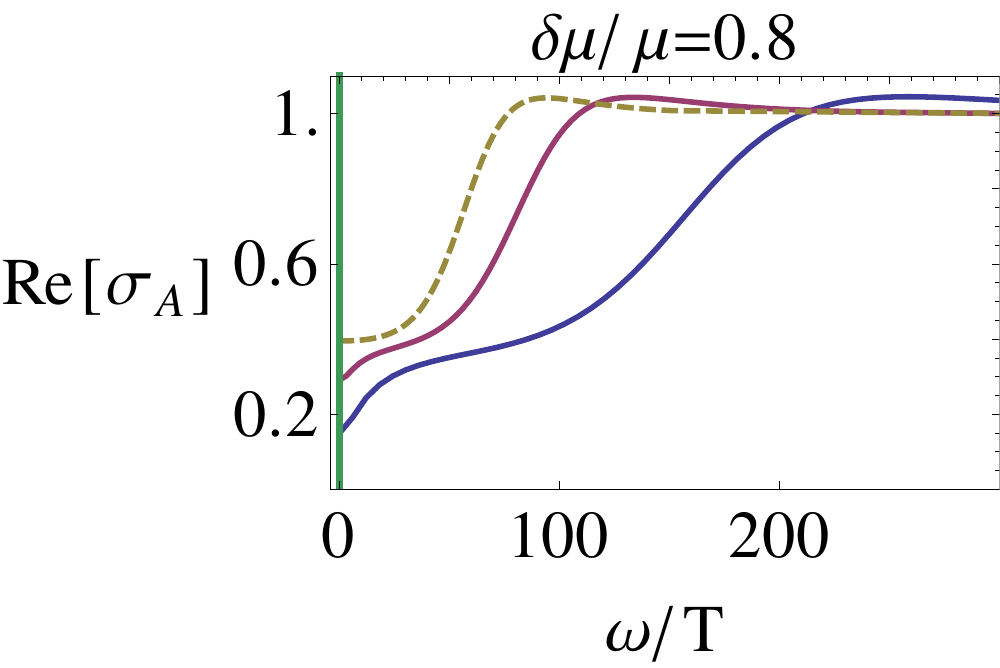}
\end{minipage}
\caption{The real part of the electric conductivity for $\delta\mu/\mu=0, 0.8$ (left plot, right plot) at $T_c$ (dashed curves) and $T<T_c$ (solid curves).}
\label{ReaT0}
\end{figure}

\begin{figure}[t]
\begin{minipage}[b]{0.5\linewidth}
\centering
\includegraphics[width=78mm]{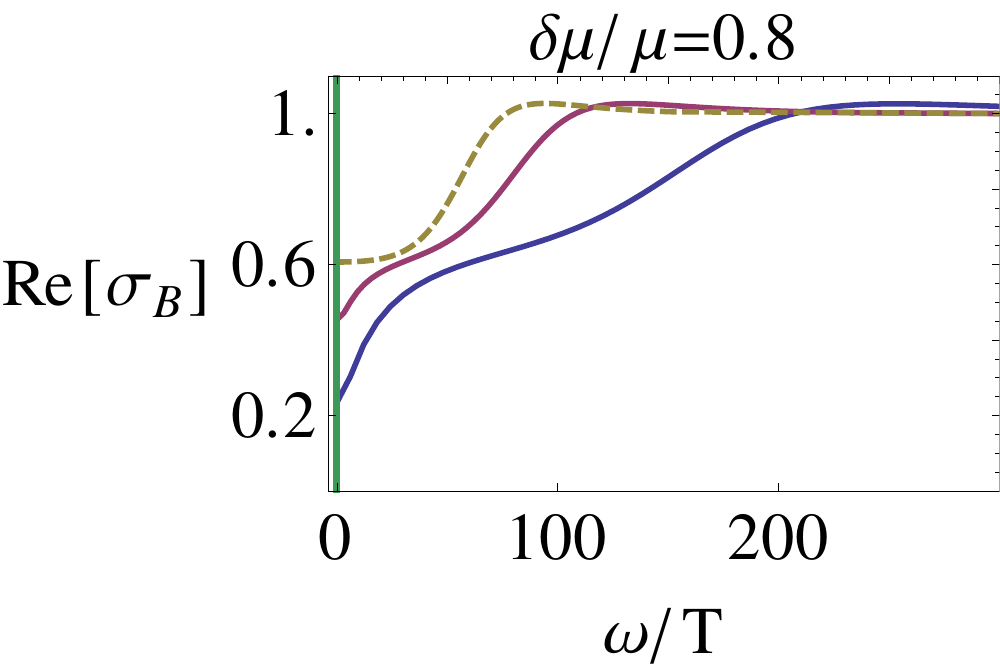}
\end{minipage}
\hspace{-0.2cm}
\begin{minipage}[b]{0.5\linewidth}
\centering
\includegraphics[width=78mm]{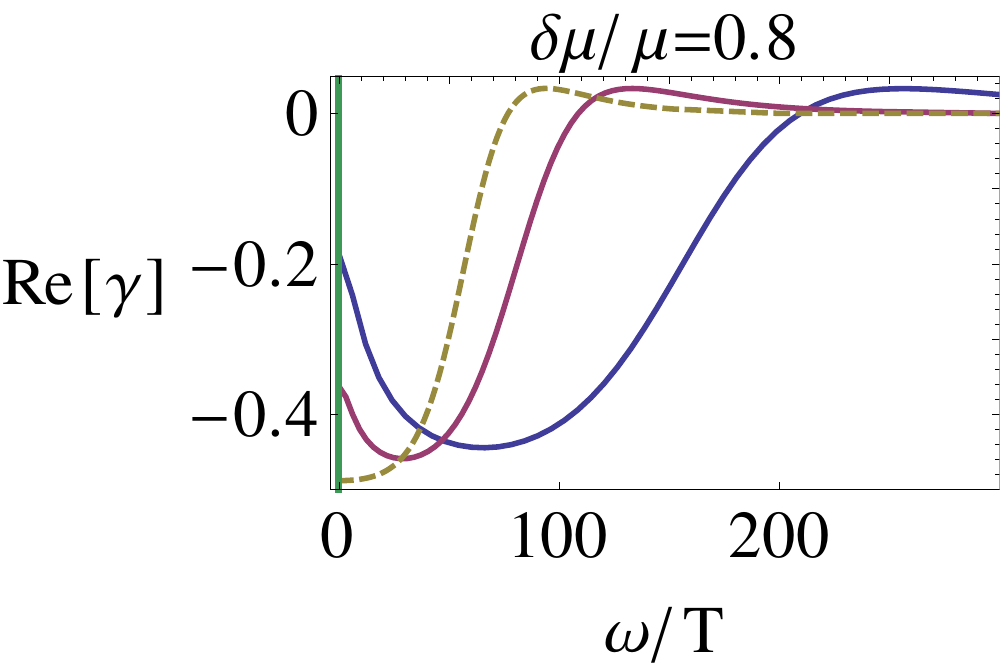}
\end{minipage}
\caption{The real part of the ``spin-spin conductivity'' $\sigma_B$ (left) and of the ``spin conductivity'' $\gamma$ (right) for $\delta\mu/\mu=0.8$ at $T_c$ (dashed curves) and $T<T_c$ (solid curves).}
\label{RegT}
\end{figure}

The overall qualitative behavior of the real part of the optical conductivities in the superconducting case (Figure \ref{ReaT0} and Figure \ref{RegT})
appears, at a first sight, very similar to the normal-phase
plots. The main and crucial difference relies in the amplitude of the $\omega=0$ delta function; we will analyze systematically this point in Subsection \ref{sition}.
Let us note, however, that the low-frequency behavior appears to be ``more structured''; the new features such the inflection points, have no clear interpretation
but, in general, can be thought of to be consequences of the presence of a new scale (and therefore different regimes) introduced by the chemical potential imbalance.
Indeed, for a completely uncharged black hole (see Subsection \ref{high}), we obtain constant, ``featureless'' conductivities; 
considering a chemical potential $\mu$, we observe the ``opening of a gap'' at low-frequency and the occurrence of a DC delta;
eventually, in the presence of both $\mu$ and $\delta\mu$, we observe the already mentioned, more complicated, behavior.

\begin{figure}[t]
\begin{minipage}[b]{0.5\linewidth}
\centering
\includegraphics[width=78mm]{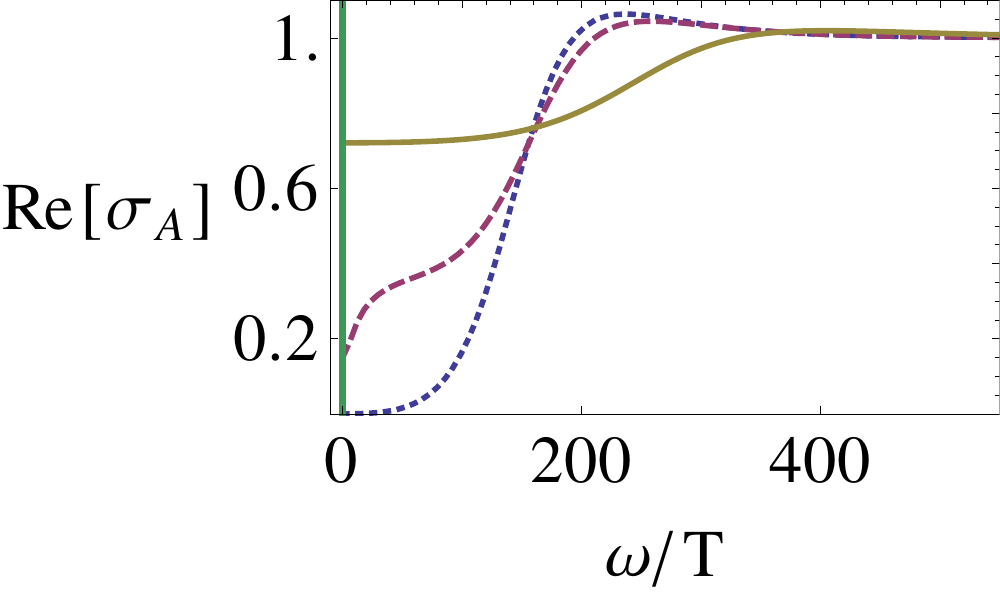}
\end{minipage}
\hspace{-0.2cm}
\begin{minipage}[b]{0.5\linewidth}
\centering
\includegraphics[width=78mm]{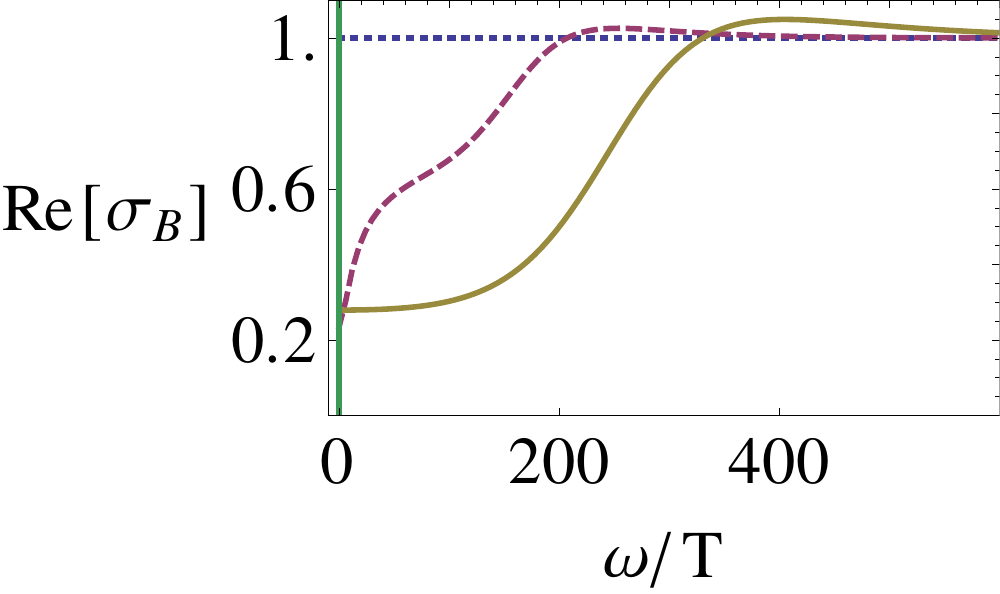}
\end{minipage}
\caption{The real part of the electric conductivity $\sigma_A$ (left) and of the ``spin-spin conductivity'' 
$\sigma_B$ (right) for $\delta\mu/\mu=0, 0.8, 1.6$ (dotted, dashed and solid lines respectively) at fixed temperature below $T_c$.}
\label{Rea}
\end{figure}

\begin{figure}[t]
\begin{minipage}[b]{0.5\linewidth}
\centering
\includegraphics[width=78mm]{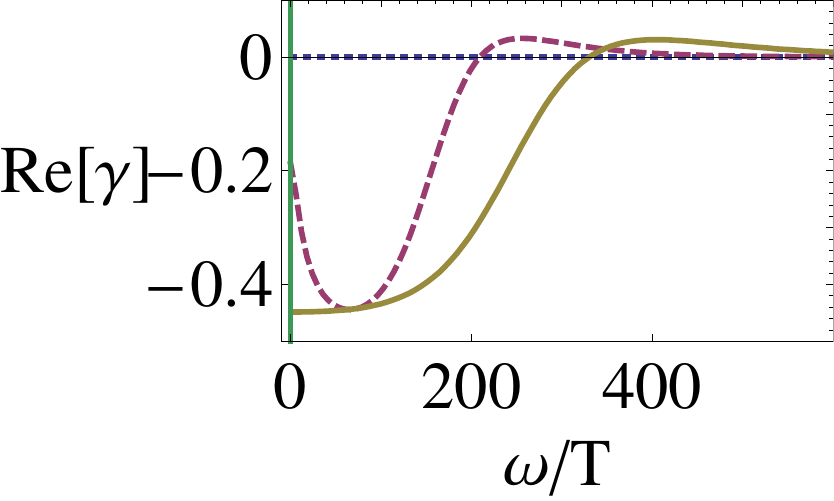}
\end{minipage}
\hspace{-0.2cm}
\begin{minipage}[b]{0.5\linewidth}
\centering
\includegraphics[width=78mm]{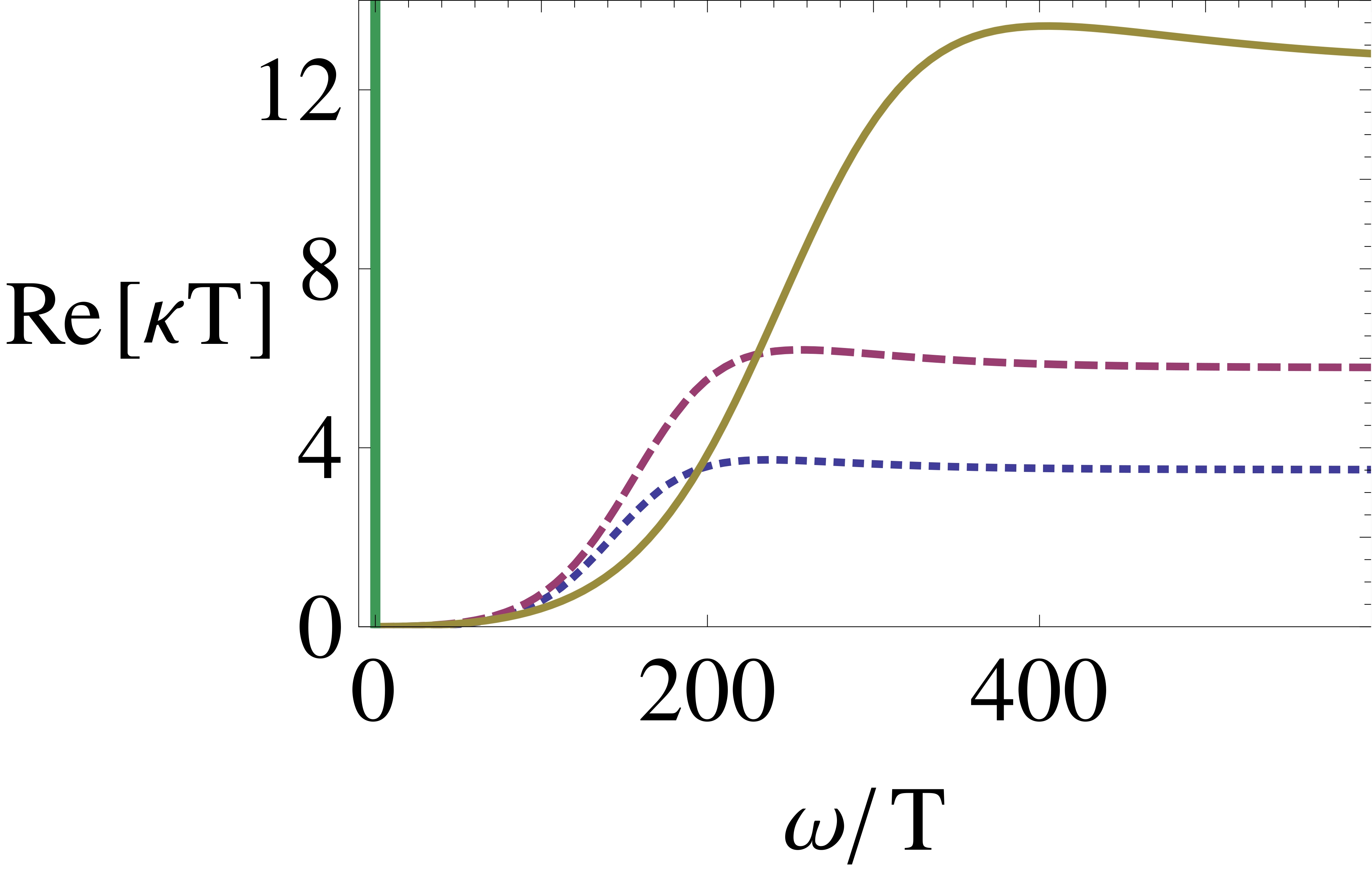}
\end{minipage}
\caption{The real part of the ``spin conductivity'' $\gamma$ (left) and of the thermal conductivity $\kappa T$ (right) for $\delta\mu/\mu=0, 0.8, 1.6$ 
(dotted, dashed and solid lines respectively) at fixed temperature below $T_c$.}
\label{Reg}
\end{figure}

We present in Figures \ref{Rea}, \ref{Reg} different plots of the $\sigma_A,\sigma_B,\gamma$ and $\kappa T$
conductivities of our system for different values of the imbalance $\delta \mu$ but at fixed temperature $T<T_c$.
The optical electric conductivity is characterized by the presence of a \emph{pseudo-gap} in the small frequency region just above the $\omega=0$ delta.
At large frequency it saturates at the unitary value as occurs in the normal phase%
\footnote{Further comments on the high-$\omega$ behavior of the conductivities is given in Subsection \ref{moby}.}.
We use the term ``pseudo-gap'' term to indicate the depletion at small frequency because the real part of the electric conductivity appears
(as far as numerical computations are concerned) exponentially small with respect to $T$ but not strictly null (even at zero $T$).
In this regard, we have to recall that the holographic model describes, strictly speaking, a superfluid where the U$(1)_A$ symmetry 
 broken by the condensate is global. The spectrum of the model is then without gap because it does not contain the Goldstone boson associated to the broken symmetry%
\footnote{Further comments can be found in \cite{Horowitz:2009ij} and in Subsection \ref{gap}.}.
In the numerical analysis, it is however possible to define a thermodynamic exponent $\Delta$ according to which
the conductivity of the ``bottom of the depletion region'' scales with temperature, in formula
\begin{equation}\label{expo}
 Re\left[\sigma_A^{(\text{pseudo-gap})}\right] \propto e^{- \frac{\Delta}{T}}\ .
\end{equation}

In Figure \ref{Rea} (right) we plot the behavior of the electric conductivity at fixed temperature $T<T_c$ but at different value of $\delta\mu/\mu$.
It is possible to notice that increasing the imbalance of the system the depletion region at small $\omega$ becomes less pronounced,
eventually disappearing for a high enough value of $\delta\mu$.
This is in agreement with the expectation that, increasing the imbalance, the system encounters a phase transition (a second order one)
to the normal phase; indeed, in the normal phase, the high $\delta\mu$ behavior consist in the disappearance of the pseudo-gap%
\footnote{We have not shown a corresponding plot at fixed temperature and increasing $\delta\mu$ in the normal phase;
the behavior consisting in a shrinkage of the depth of the pseudo-gap can be guessed comparing the left and right Figures \ref{ReaT0}.}.
Again, since the qualitative behavior is similar in both the normal and the superconducting phases, a careful care has to be tributed to discontinuous quantities through
the transition. As already mentioned, the clearest signal of superconductivity is given by a discontinuity in the thermal behavior of the amplitude of the DC delta (see Subsection \ref{sition}).


\subsection{Depletion at small \texorpdfstring{$\omega$}{} and the pseudo-gap}
\label{gap}

As it is manifest from the conductivity plots, the real part of the conductivities $\sigma_A$ and $\sigma_B$
show pronounced depletion regions at low values for $\omega$%
\footnote{We comment on a possible analogy between the low-frequency shape of our conductivities
with the quasi-particle expectation of a Drude-like model in Subsection \ref{moby}.}.
As we have already mentioned, an important question in the phenomenology of holographic superconductors concerned
the occurrence or not of a ``hard gap'' in the real part of the conductivity at low frequency.
The term ``hard gap'' indicates a region in which, at zero temperature, the conductivity vanishes exactly;
in \cite{Hartnoll:2008kx} it is numerically tested that for low temperature the low-frequency limit
(not zero because there we expect the DC delta) of Re$[\sigma(\omega)]$ is affected by thermal fluctuations
and behaves exponentially with respect to the temperature as in \eqref{expo}
where $\Delta/T\gg 0$ and $\Delta$ is a quantity proportional to the width of the gap $\omega_g$. The proportionality
factor is $1/2$ in the probe (i.e. large $q$ limit) and smaller for smaller charge (look at \cite{Hartnoll:2008kx}
and references therein).
The numerical approach cannot nevertheless solve the conceptual question of excluding or confirming a hard-gap for $T=0$
because there the numerics become unreliable and, even more importantly, the $T=0$ case can present qualitative new feature with respect to the low-temperature region 
(see Subsection \ref{ground}).

In \cite{Horowitz:2009ij} the numerical difficulty is circumvented employing an analytical approach
that shows that there is no ``hard-gap'' for any choice of the scalar potential $V(\psi)$.
Said otherwise, in none of the simplest (singly charged) models the conductivity occurs to be exactly zero
in the low energy regime, no matters which scalar potential is considered.
This conclusion is a consequence of the possibility of mapping the conductivity to a 
reflection coefficient ${\cal R}$ in a scattering problem obtained from the gauge field fluctuation equation
with an appropriate change of radial coordinate.
The exact vanishing of Re$[\sigma(\omega)]$ corresponds, fro the point of view of the associated scattering problem, to total reflection, $|{\cal R}|=1$.
As showed in \cite{Horowitz:2009ij}, the potential in the scattering problem vanishes
always at the horizon and it saturates to a finite height at zero temperature so that there is
always a non-null probability of transmission, and therefore no ``hard-gap''.

\begin{figure}[t]
\begin{minipage}[b]{0.5\linewidth}
\centering
\includegraphics[width=78mm]{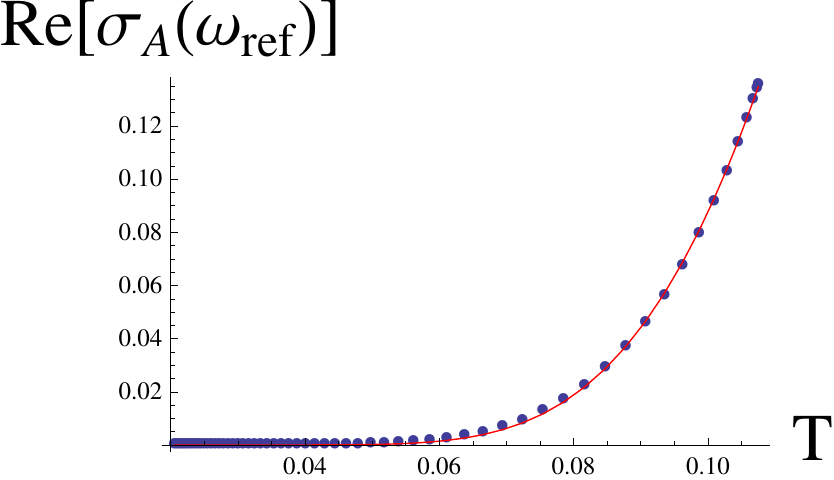}
\end{minipage}
\hspace{-0.2cm}
\begin{minipage}[b]{0.5\linewidth}
\centering
\includegraphics[width=78mm]{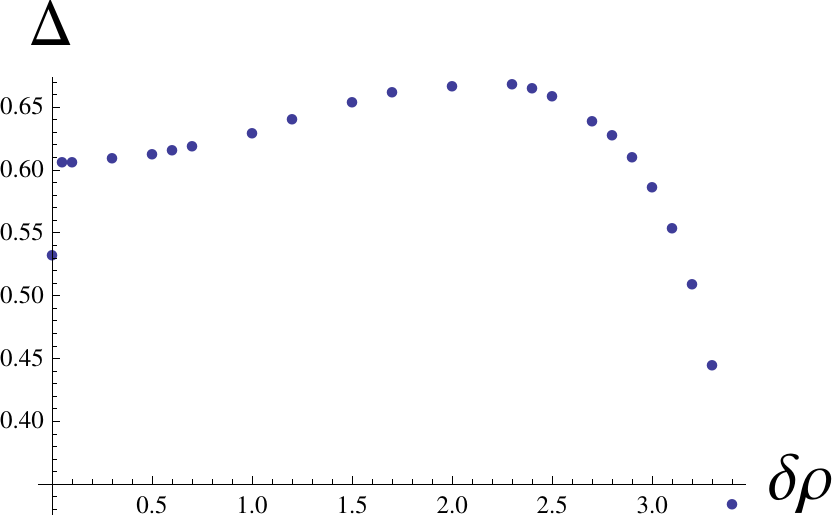}
\end{minipage}
\caption{On the left: exponential-like behavior of the real $A$ conductivity at small frequency as a function of $T$.
On the right: effective thermal exponent as a function of the imbalance.}
\label{dermo}
\end{figure}

In our unbalanced case we have performed a numerical study similar to \cite{Hartnoll:2008kx} where we analyzed the thermal behavior of
Re$[\sigma_A(\omega_{ref})]$, where $\omega_{ref}$ represents a reference frequency corresponding to the ``bottom of the depletion region''%
\footnote{In our computation we have chosen $\omega_{\text{ref}} = T$ which, in the considered range, is always well within the depletion region.}.
We have performed the analysis in the probe approximation because we wanted to study the behavior of $\Delta$ with respect to $\delta\rho$;
this requires a lot of calculations%
\footnote{The probe case is indeed much faster.} indeed, for any value of $\delta\rho$, we need to perform a scan in temperature to produce a series of points that, 
once fitted with a functional shape as \eqref{expo}, returns the value of $\Delta$. As an example, we plotted in Figure \ref{dermo} the points computed for $\delta\rho=0.7$
and the associated exponential fit.
There is an important caveat to be mentioned: we needed to adopt the probe approximation for practical reasons, but the results have to be regarded with attention 
because the low-temperature region can be in general troublesome for the approximation itself. In the future, the analysis has to be repeated in the full-backreacted 
case.

\subsection{Normal-to-superconductor transition}
\label{sition}

\begin{figure}[t]
\begin{minipage}[b]{0.5\linewidth}
\centering
\includegraphics[width=70mm]{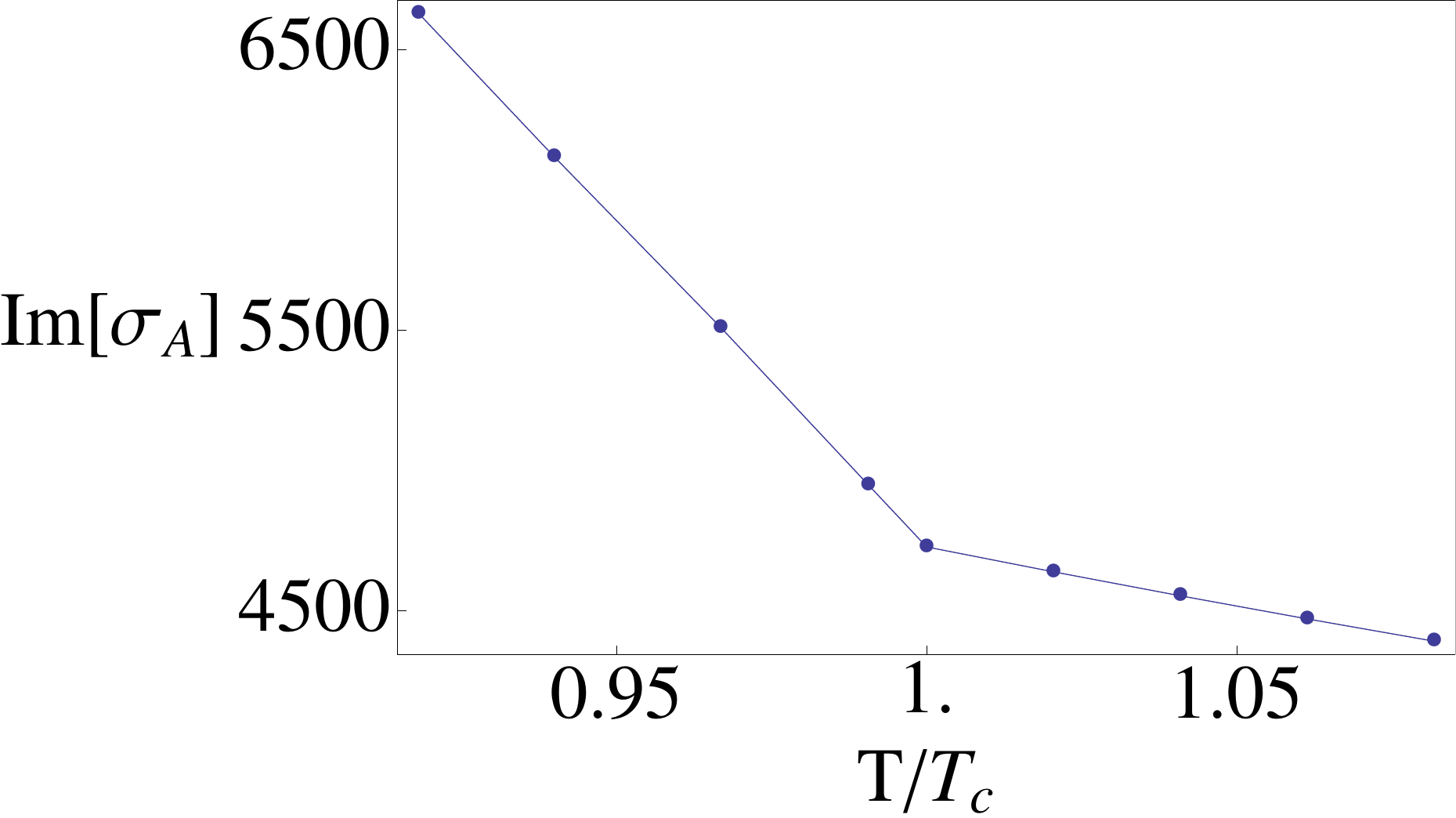} 
\end{minipage}
\hspace{-0.3cm}
\begin{minipage}[b]{0.5\linewidth}
\centering
\includegraphics[width=70mm]{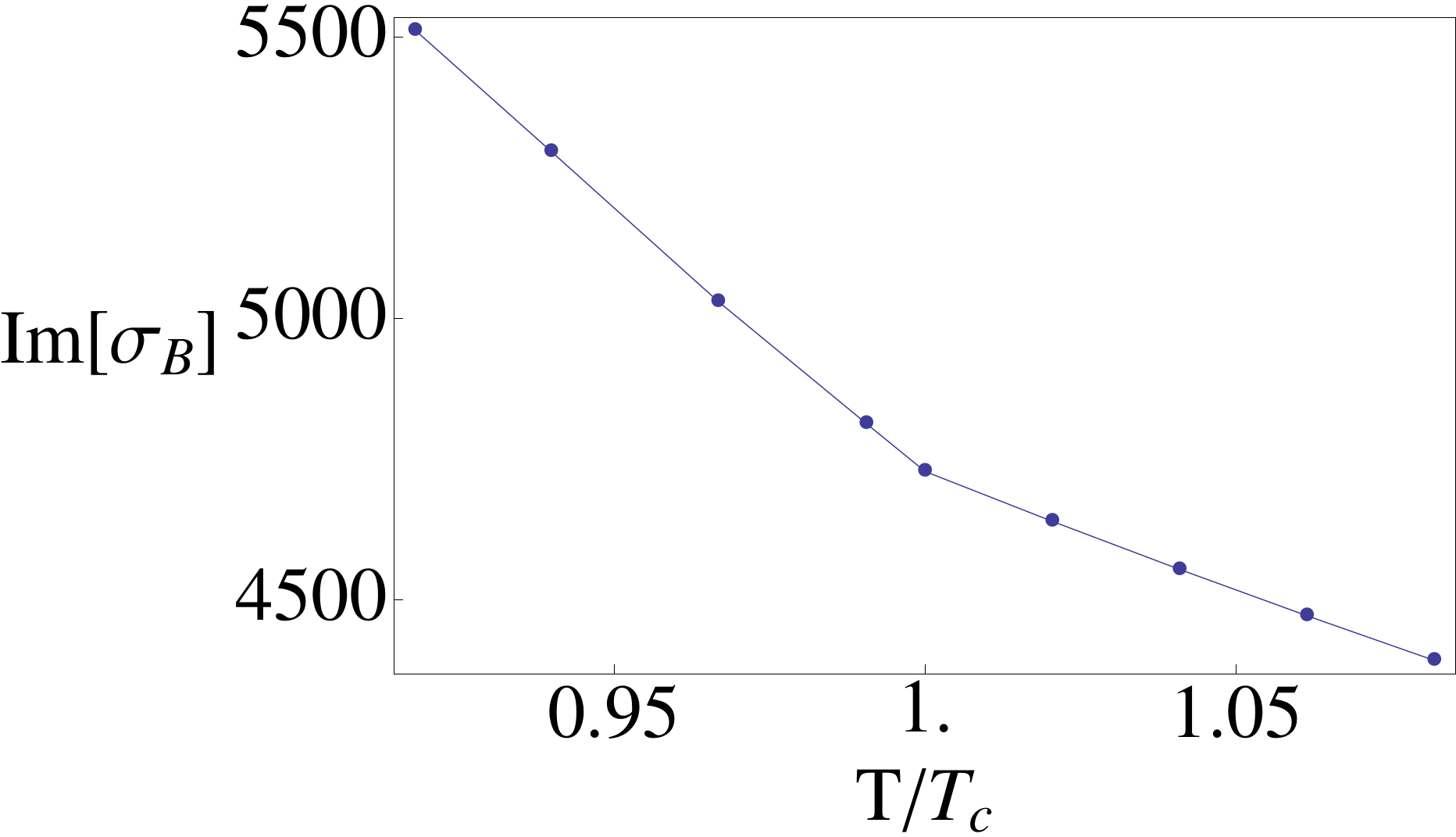}
\end{minipage}
\caption{The discontinuous behavior at $T_c$ in the imaginary part of the electric conductivity $\sigma_A$
(left) and of the ``spin-spin conductivity'' $\sigma_B$ (right), signaling a discontinuity
in the temperature dependence of the magnitude of the delta function at $\omega= 0$ in the corresponding DC
conductivities.}
\label{jump}
\end{figure}

We work on a model which enjoys translational invariance;
this feature alone leads to a divergent static conductivity (i.e. $\sigma(\omega=0)=\infty$) encoded in a delta function.
Such contribution represents the ideal version of the Drude peak that, in real materials, is
``smeared'' by impurities, defects and any other feature spoiling the translational invariance of the medium.
Notice that, in order to be sure of dealing with an actual superconductor, we must distinguish such divergent contribution to the DC conductivity
due to translational invariance from actual superconductivity.
To study this point it is essential to consider the static conductivity of our model and its behavior at the normal-to-superconductor
transition. 
Since it is impossible to deal numerically with a delta function, we have to study it indirectly.
One way to do this consists in exploiting the Kramers-Kronig relation which maps the delta at $\omega=0$ in the real part of $\sigma$
to a corresponding pole in the imaginary part at the same frequency. The residue of the pole corresponds
to the ``area'' (or amplitude) of the delta function.

The numerical analysis is focused on the study of $\text{Im}(\sigma)$ for a small value of the frequency
and its behavior in moving through the normal-superconductor transition. As shown in figure \ref{jump} there is a discontinuity
of the derivative of the DC conductivity at $T_c$.
This is interpreted as the key signal of superconductivity; a new contribution due to qualitative new feature
of the superconducting phase led to an abrupt change in the conductivity behavior.

Observe also that both below and above the transition the DC conductivity increase as the temperature is lowered.
This is a feature in agreement with the expectation and, moreover, below $T_c$ we have that its rate of increase
is greater. We read this as a new contribution to the conductivity due, to use the superconductive terminology,
to the condensation of ``carriers'' in a superconducting macroscopic state.

As anticipated, this feature at the transition is one of the crucial clues denoting superconductivity on our
holographic model.
It is interesting to observe that, in passing to the superconducting phase,
a novel DC delta contribution arises in the ``spin-spin'' $\sigma_B$ conductivity as well.
There is indeed no breaking of symmetry associated specifically to the gauge field $B$,
nevertheless, its supecronducting-like behavior is induced by its mixing with the electric conductivity.

\subsection{Pseudo-gap threshold characterization}

\begin{figure}[t]
\begin{minipage}[b]{0.5\linewidth}
\centering
\includegraphics[width=78mm]{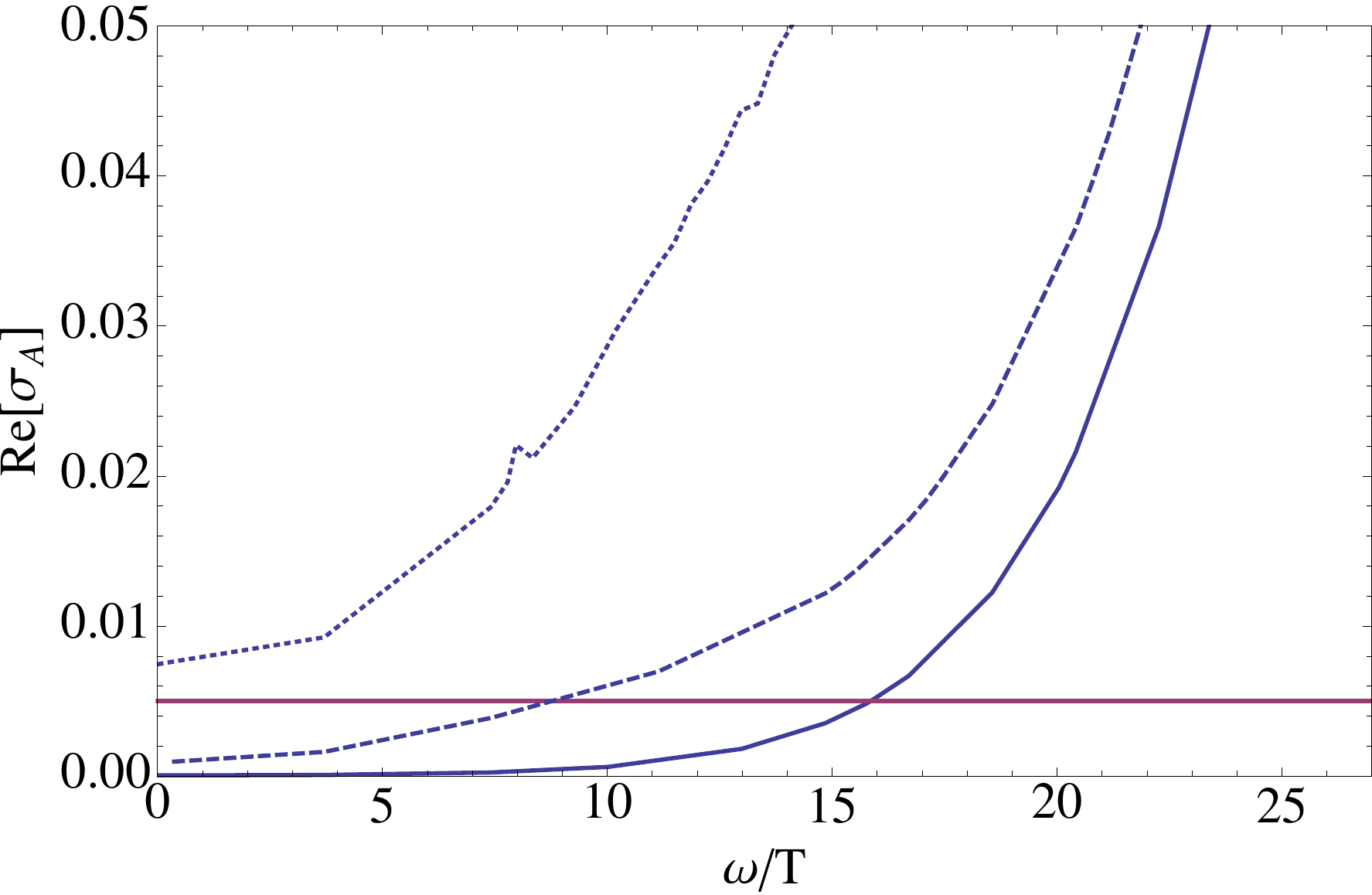}
\caption{Low frequency plots of Re$[\sigma_A]$ for $\delta\mu/\mu=0,0.3,0.7$ (solid, dashed, dotted lines respectively).
The horizontal line represents the threshold value (0.005) employed to define $\omega^{*}_{\text{pg}}/T$.
{\color{white} aaaaaaaaaaaaaaaaaaaaaaaaaaaaaaaaaaaaaaaaaaaaaaaaaaaaaaaaaaaaaaaaaaaaaaaaaaaaaaaaaaaaaaaaa
aaaaaaaaaaaaaaaaaaaaaaaaaaaaaaaaaaaaaaaaaaaaaaaaaa} }
\label{left}
\end{minipage}
\hspace{0.2cm}
\begin{minipage}[b]{0.5\linewidth}
\centering
 \includegraphics[width=78mm]{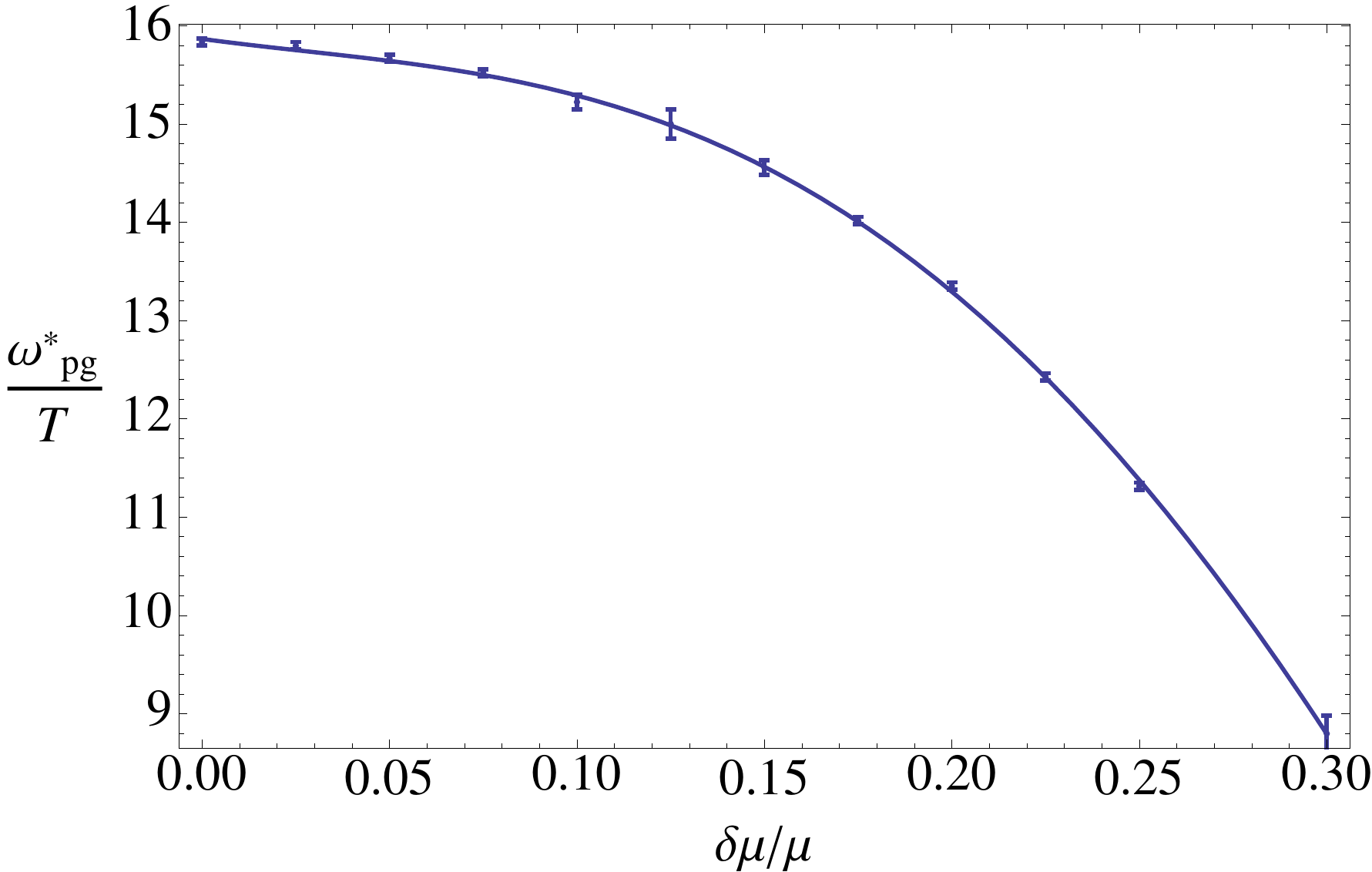}
  \caption{Plot of the threshold-pseudo-gap $\omega^{*}_{\text{pg}}$ with respect to the imbalance $\delta\mu$;
the error bars correspond to numerical uncertainty (which happened to be quite variable from point to point); the continuous line
emerges from a fit by means of a quartic polynomial functional shape.}
\label{gapp}
\end{minipage}
\end{figure}

In order to study quantitatively the depletion region of the $\sigma_A$ ``electric'' conductivity, 
we fix a small threshold value, namely
\begin{equation}\label{thres}
 \text{Re}[\sigma_A(\omega^*_{pg})] = 0.005\ ,
\end{equation}
and seek the value of $\omega^*_{pg}$ satisfying \eqref{thres} for different values of $\delta\mu/\mu$ and fixed (low) $T$.
In other terms, we evaluate the extent of the frequency interval where the real part of the conductivity is essentially vanishing;
we furthermore study the behavior of this pseudo-gap  
for different imbalance over chemical potential ratios (see Figure \ref{left}).
After having collected the values of the threshold frequency performing a scan in $\delta\mu/\mu$, 
we perform a fit to study precisely the behavior of $\omega^*_{pg}$ as a function of $\delta\mu/\mu$;
the results are reported in Figure \ref{gapp}.
The functional form employed in the fit procedure is a quartic polynomial;
the purpose is to show that the dependence of $\omega^*_{pg}$ on $\delta\mu/\mu$ is non-trivial.
This low-temperature, strong-coupling behavior has to be contrasted with its weak-coupling counterpart;
the weakly coupled (BCS) unbalanced superconductor presents, at $T=0$, a gap which is insensitive of $\delta\mu$,
then our fit has to be contrasted with a constant weak-coupling behavior.

In Figure \ref{gapp} the points are given an error bar accounting for numerical uncertainty.
The bars have different amplitudes for different values of $\delta\mu/\mu$; this is essentially
related to the ``numerical noise'' observed in the conductivity plots, see Figure \ref{left}.

\subsection{Mobility function for the carriers}
\label{moby}

\begin{figure}
 \centering
 \includegraphics[]{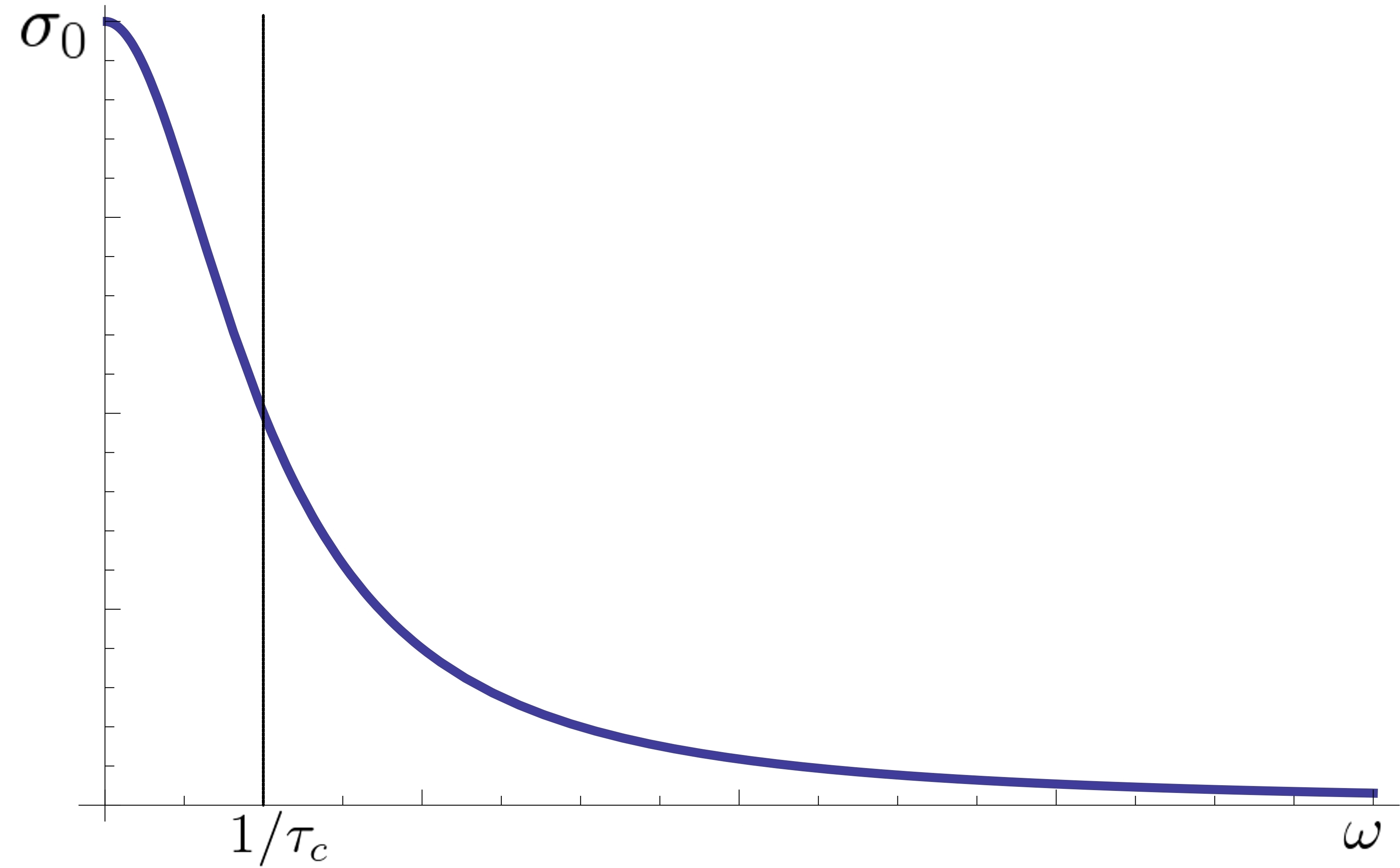}
 \caption{Conductivity arising from carriers with Drude behavior characterized by mean free-propagation time equal to $\tau_c$;
the symbol $\sigma_0$ denotes the DC value of the conductivity.}
 \label{dru}
\end{figure}

The linear response of the system to the perturbations of the external sources is described by the conductivity matrix;
its matrix character, with non-vanishing off-diagonal terms, accounts for mixed response.
For instance, an external electric field perturbation induces not only electric transport but, in general, also spin
and thermal transport as well. The spin-electric, thermo-electric and thermo-spin responses are indeed 
encoded in the off-diagonal entries of the conductivity matrix,
\begin{eqnarray}\label{superrelazione}
\hat\sigma = \begin{pmatrix}  \sigma_A & \alpha T & \gamma \\ \alpha T &  \kappa T & \beta T \\ \gamma & \beta T & \sigma_B \end{pmatrix}\ .
\end{eqnarray}
An interesting feature of our model is the possibility of encoding the normal-phase conductivity by means of a parameter function $f(\omega)$
in the following way
\begin{eqnarray}
\hat\sigma = \begin{pmatrix} f \rho^2+1 & \frac{i \rho}{\omega}-\mu(f \rho^2+1)-\delta\mu f \rho\,\delta\rho & f \rho\ \delta\rho \\ \frac{i \rho}{\omega}-\mu(f \rho^2+1)-\delta\mu f \rho\,\delta\rho &  \kappa T &  \frac{i \delta \rho}{\omega}-\delta\mu(f \delta\rho^2+1)-\mu f \rho\,\delta\rho  \\ f \rho\ \delta \rho & \frac{i \delta \rho}{\omega}-\delta\mu(f \delta\rho^2+1)-\mu f \rho\,\delta\rho  & f \delta\rho^2+1 \end{pmatrix} \, .\nonumber
\end{eqnarray}
where
\begin{equation}
\kappa T =\frac{\im}{\omega}\left(\epsilon + p -2\mu\rho -2\delta\mu\delta\rho\right) + (f\rho^2+1) \mu^2 + (f \delta\rho^2 + 1)
\delta\mu^2 + 2 f\delta\rho\,\rho\, \mu\, \delta\mu\,,
\end{equation}
These equations concern both the real and the imaginary parts of the conductivities.
The parametrization given by the introduction of $f$ offers a physically suggestive hint;
actually we can think of $f(\omega)$ as a mobility function for some carrier-like feature of our model%
\footnote{Note that, usually, the term ``mobility'' is employed to indicate $f\rho$; we adopt it to indicate just the $f$ part, i.e. the frequency dependent factor.}.
Let us focus on the components $\sigma_A$,$\sigma_B$ and $\gamma$: they (apart from the $+1$ contribution on which we comment in the following)
suggest the presence of a single kind of carrier population characterized by a density of charge $\rho$ with respect to $A$ and 
$\delta\rho$ with respect to $B$. 
Note that in $\sigma_A$ the quadratic dependence on $\rho$ is in accordance with the expectation that the response current
has to be quadratic in the carrier charge; indeed, the higher is the charge, the stronger is the coupling with the external field and,
in addition, if the carriers have a higher charge, their flow is associated to a bigger charge transport.
This is clear observing that the mixed terms $\gamma$ are proportional to $\rho\,\delta\rho$ where the two effect are distinct,
i.e. in $\gamma_{AB}$ $\delta \rho$ accounts for the coupling to the external field $B$ and $\rho$ accounts for the consequent transport of charge of type $A$.
The opposite can be said of $\gamma_{BA}$; this, however, leads to the same result $\gamma = f \rho\, \delta\rho$.

Let us underline that both the presence of a unique mobility function $f(\omega)$ and the observation of the mixed conductivity $\gamma$
indicate the presence of a single species of fundamental carriers.
This can sound surprising as we are describing a superconductor with two fermion species, but we have to remember that we are here speaking
of the would-be (Cooper-like) degrees of freedom of the strongly coupled regime. When we try a carrier-like description we suppose that
in the strongly coupled medium there is some kind of degree of freedom (which is distinct from the original weakly coupled fermions) that
admits the carrier-like interpretation%
\footnote{See for instance \cite{2011arXiv1108} where the transport properties of the superconductor phase of the Hubbard model is described
by a flow of vortexes.}
In a gravitational or bulk perspective, the uniqueness of the function $f$ can be related to the observation that the gauge fields
$A$ and $B$ are governed (in the normal phase) by the same kind of equation. Once a solution for $A$ is found, it is therefore natural that the same functional shape
works for $B$ as well.

The generalization of \eqref{superrelazione} to the superconducting phase seems possibly troublesome.
We expect that in the superconductor there is an additional component (the condensate) contributing to the conduction phenomenon.
Such component has to be neutral with respect to $B$ and its fundamental degrees of freedom would have an appropriate mobility function
in principle different from $f$. We can derive this ``mobility function'' but we lack a precise physical expectation
to confront it against.

We report here a qualitative speculation along the lines of the review \cite{2011arXiv1108};
In the instances in which the conductivity is due to quasi-particles (or particles) that, moving 
in a Brownian-like fashion, drift along the direction of the external field, an effective analysis can be encoded
in the following equation 
\begin{equation}\label{bolza}
 \frac{d\bm{v}}{dt} + \frac{\bm{v}}{\tau_c} = q \bm{E}\ ,
\end{equation}
where $\bm{E}$ is the external electric field, $q$ is the charge of the carriers, $\bm{v}$ the average velocity (i.e. the drift speed)
and $\tau_c$ represents the characteristic mean time between two interactions of a particle (with, for example, the lattice or an impurity).
Assuming harmonic time dependence, from \eqref{bolza} we have that the drift velocity (which is proportional to the current) is given by
\begin{equation}
 \bm{v} = \frac{q \bm{E}}{\im \omega + \frac{1}{\tau_c}}\ ,
\end{equation}
then, for the conductivity, we have
\begin{equation}
 \sigma = \frac{q }{\im \omega + \frac{1}{\tau_c}}\ .
\end{equation}
In a normal conductor $\tau_c$ is finite and at low frequency the conductivity is characterized by a peak (known as Drude's peak)
while for high frequency it tends to vanish; see Figure \ref{dru}.
In a medium in which the carriers flow freely the mean time between two interactions of a carrier with the surrounding environment
diverges, $\tau\rightarrow \infty$. The consequence is that the conductivity develops a pole $-\im q/\omega$ in its imaginary part.
The Kramers-Kronig relation connects the existence of a pole in the imaginary part of the conductivity to the presence of a delta in the real part 
in correspondence of the same value of the frequency. This argument supports what we have already studied specifically for our system in Subsection \ref{sition},
namely both a ``normal'' translational invariance and superconductivity give a DC delta contribution.


\begin{figure}
 \centering
 \includegraphics[width=.6\linewidth]{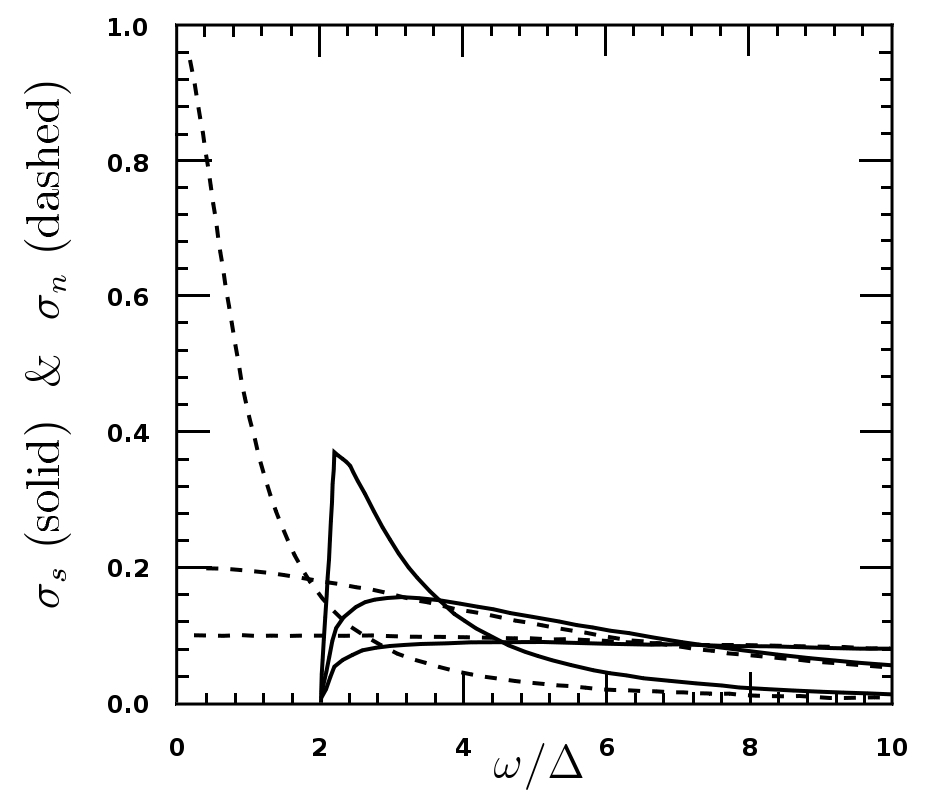}
 \caption{Figure taken from \cite{Chen_1993} showing the superconductive and normal optical conductivities.
The different plots correspond, from top to bottom, to increasing disorder in the system. }
\end{figure}

\subsection{High \texorpdfstring{$\omega$}{} behavior of the conductivities}
\label{high}

The high-frequency behavior of the real parts of the $\sigma_A$ and $\sigma_B$ conductivities that we have found needs some specific comments%
\footnote{The thermal conductivity $\kappa T$ behaves similarly but it does not tend to a unitary value for $\omega\gg 1$.}.
\begin{figure}[t]
\begin{minipage}[b]{0.5\linewidth}
\centering
\includegraphics[width=78mm]{{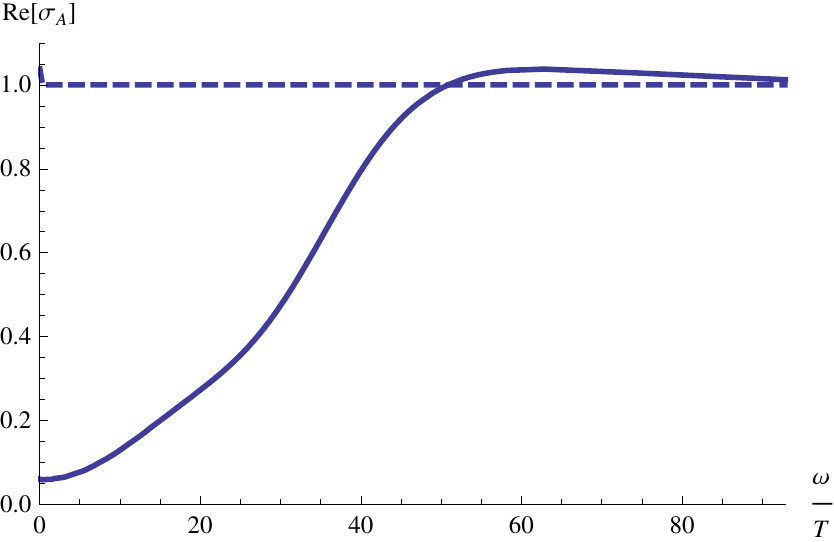}} 
\end{minipage}
\begin{minipage}[b]{0.5\linewidth}
\centering
\includegraphics[width=78mm]{{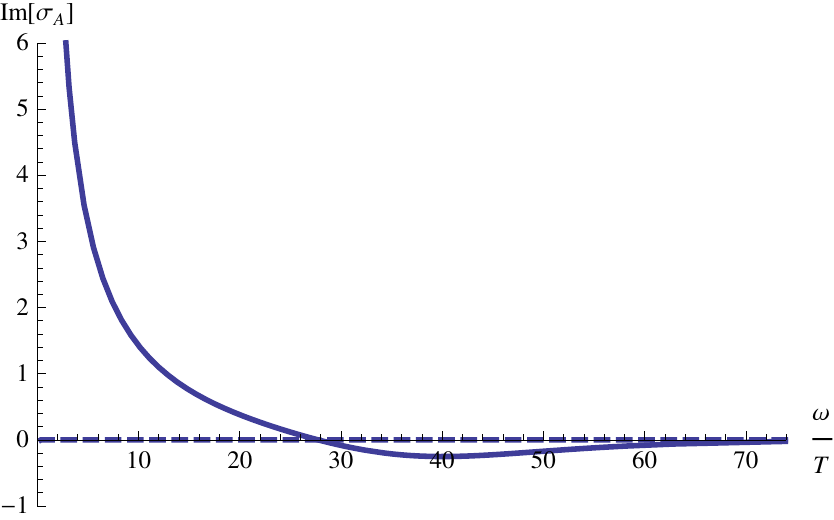}}
\end{minipage}
\caption{Real and imaginary parts of the conductivity of a singly charged ($\delta\mu=0$ and $\mu\neq0$) black hole represented with the solid lines
as opposed to a totally uncharged ($\delta\mu=\mu=0$) case represented by the dashed lines.}
\label{uncha}
\end{figure}
The $\omega \gg 1$ behavior of the conductivity is difficult to interpret within the quasi-particle, or individual carriers, picture
(look for example at Figure \ref{Rea}).
Let us have a closer look: raising the frequency, we find a pseudo-gap (``pseudo'' because the bottom of the gap is not strictly null) and then further increasing $\omega$
we observe a raise in the real part of $\sigma$ which reaches a unitary asymptotic value.
From the study of the uncharged limit of our system we find that in this case the conductivity (see Figure \ref{uncha})
has a real part which is approximately constant and stable on the value $1$ while the imaginary part remains approximately null.
In other terms, we are observing that the constant unitary contribution to $\text{Re}[\sigma_{A/B}]$ (see \eqref{superrelazione}) is due to a conduction phenomenon
that is already present in the uncharged black hole;
this is a general feature of holographic conductivity computations (see for instance \cite{Hartnoll:2008kx}) which,
as explained in \cite{Herzog:2007ij}, arises as a consequence of the electro-magnetic selfduality of the bulk theory on $AdS_4$%
\footnote{Roughly speaking, electro-magnetic duality swaps the roles of particles and vorteces; the vortex conductivity is
given by the inverse of the particle conductivity so, self-duality, implies that the conductivity is unitary.}.
Even though the theoretical origin of this constant conductivity is clear, it still lacks a precise microscopic interpretation.
In general, the possibility of having conductivity in a neutral system is not surprising as global neutrality can arise from the sum of contributions
of opposite charge or pair productions phenomena.
The flatness of $\sigma(\omega)$ is instead quite mysterious; indeed, an equal response to any frequency signals the lack of structure in the medium
or, equivalently, the lack of any characteristic value for $\omega$ or time scale%
\footnote{In the example of quasi-particle, the characteristic time $\tau_c$ produced a precise ``structure'' in the conductivity shape, namely the Drude peak.}

\begin{figure}
 \centering
 \includegraphics[width=.5\linewidth]{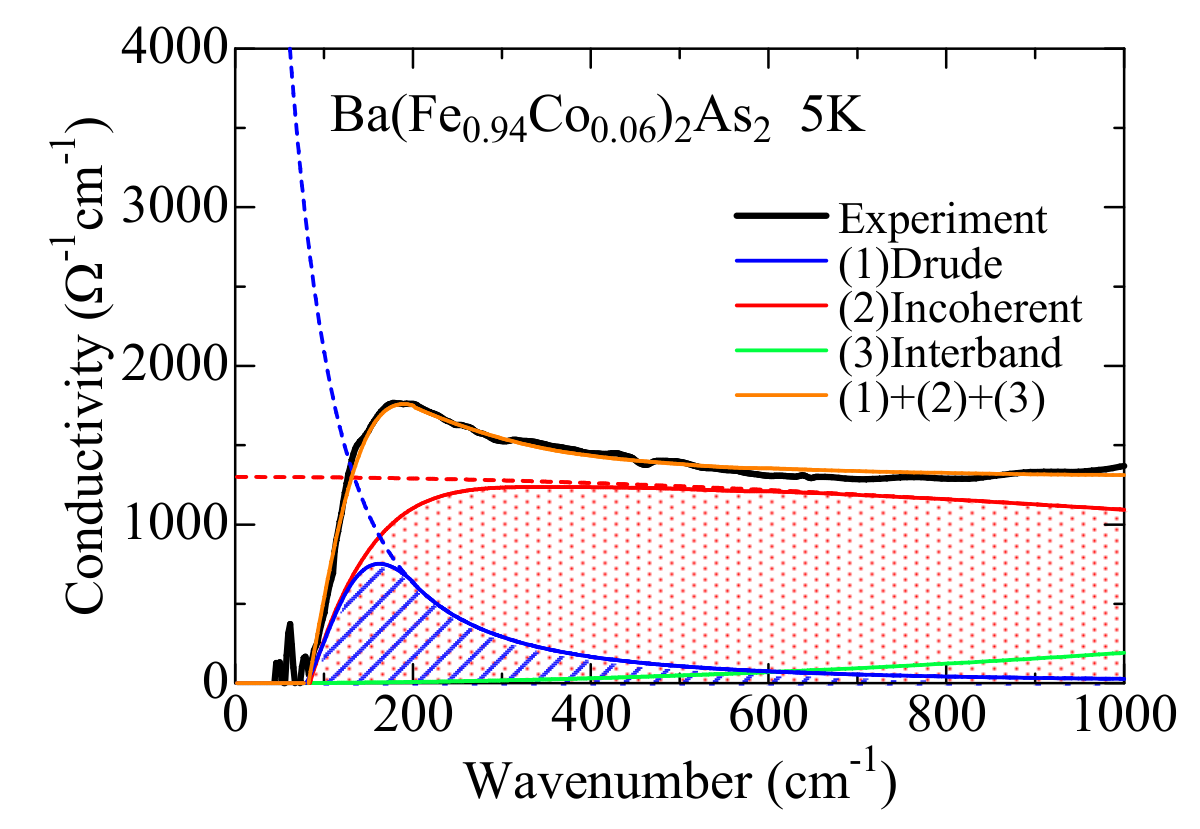}
 \caption{Figure taken from \cite{2010PhRvB..81j4528N}. The authors denote with ``incoherent'' an additional component contributing to the transport
and with ``interband'' the effects of the presence of multi-bands.}
 \label{band}
\end{figure}

It is interesting how a similar question is faced in some experimental papers, see for instance\cite{2010PhRvB..81j4528N}
from which the Figure \ref{band} has been taken: There, to recover the experimental plot (black line),
they have to add to their theoretical model a component (green line); this component is interpreted as describing the contributions
of successive bands becoming available (in terms of electron transport) as the energy is increases.
It would be nice to try to interpret our high-$\omega$ in a similar fashion and check if the holographic description
could possibly account for features like multi-band structure. 
At this stage, this is (in our model) just a speculation viable for future work;
much caution is advisable and, so far, no claim in this direction is maintained.

As already noted, in our system, the constant unitary contribution at high $\omega$ is represented by the $+1$ term in the $\sigma_A$ entry of \eqref{superrelazione};
a totally analogous feature occurs for $\sigma_B$ as well. 
The conductivity matrix \eqref{superrelazione} refers to the normal phase, but the constant ``uncharged-black hole''
contributions are present also in the superconducting phase.

\section{Non-homogeneous phases?}
\label{LOFF}
Let us study the possibility of inhomogeneous phases in our minimal, holographic model for an unbalanced superconductor.
In order to do so, we consider the so-called \emph{probe approximation} (introduced in \cite{Hartnoll:2008vx}) which consists
in rescaling the scalar $\psi$ and the $A$ gauge field by a factor of $1/q$ and consequently taking the limit of large charge $q\gg 1$.
The simplification yielded by the probe approximation resides in the fact that the backreaction of the Abelian Higgs model composed by
the scalar and the $A$ fields can be neglected 
(they are indeed regarded as probes); they are therefore treated as fluctuations on the fixed background of all the other fields of the model%
\footnote{Note that, being coupled to one another, the fields $\psi$ and $A$ must be assumed to be probes together. 
It would not be legitimate to consider, for instance, $\psi$ and $B$ as probes on a fixed $A$ background. }
Notice that the probe approximation is expected to be valid only for the range in the temperature $T$ where the condensate (which is related to the scalar field $\psi$)
is not too large. The probe approximation can be reliable in a region not too below the critical temperature $T_c$ where we have the onset of superconductivity;
the approximation is expected to fail in the low-temperature regime where the condensate (and correspondingly the bulk scalar field)
can assume large values.
Nevertheless, for the sake of studying a possible inhomogeneous phase we are interested in the near-$T_c$ region where a tri-critical point is expected 
(see \cite{Bigazzi:2011ak} and references therein).

In our framework, a inhomogeneous LOFF-like phase would be described by a gravity solution where the background is fixed whereas the scalar field acquires (spontaneously)
a dependence on the spatial coordinates. Let us restrict to the simplest case in which the spatial dependence is on only one coordinate and the functional shape is
either a complex plane wave or a real cosine%
\footnote{Similar inhomogeneous phases have already been considered for $p$-wave superconductors (i.e. where the condensate is vectorial)
in \cite{Donos:2011bh}, \cite{Donos:2011qt}, \cite{Donos:2011ff}, \cite{Nakamura:2009tf}.}
Our fixed background is a U$(1)_B$-charged asymptotically $AdS$ Reissner-Nordstr\"{o}m black hole (see Appendix \ref{probe} for some further detail), namely
\begin{eqnarray}
ds^{2}&=&-f(r)dt^{2}+r^{2}(dx^{2}+dy^{2})+\frac{dr^{2}}{f(r)}\,,\label{procione1}\\
f(r)&=&r^{2}\left(1-\frac{r_{H}^{3}}{r^{3}}\right)+\frac{\delta\mu^{2}r_H^{2}}{4r^{2}}\left(1-\frac{r}{r_{H}}\right)\,,\label{procione2}\\
v_{t}&=&\delta\mu\left(1-\frac{r_{H}}{r}\right)=\delta\mu-\frac{\delta\rho }{r}\,.\label{procione3}
\end{eqnarray}
We consider a plane wave ansatz of the following shape
\begin{equation}\label{planewave}
 \psi(r,x)=\Psi(r)e^{-ixk}\,,\quad A=A_t(r)dt + A_x(r)dx\,. 
\end{equation}
Plugging the assumed shape \eqref{planewave} into the equations of motion (\ref{eom2}) and
(\ref{eom3}) of the U$(1)_B$-charged RN-$AdS$ background
(\ref{procione1}), (\ref{procione2}), (\ref{procione3}), one obtains the Maxwell equations
\begin{eqnarray}
\partial_r^2A_t+\frac{2}{r}\partial_rA_t-\frac{2\Psi^2}{f}A_t=0\,,\\
 \partial_r^2A_x+\frac{f^\prime}{f}\partial_r A_x-\frac{2\Psi^2}{f}(k+A_x)=0\,,
\end{eqnarray}
and the scalar equation
\begin{equation}
\Psi^{\prime\prime}+\Psi^\prime\bigg( \frac{2}{r}+\frac{f^\prime}{f}\bigg)+
\Psi\bigg( \frac{A_t^2}{f^2}+\frac{2}{f}-\frac{(k+A_x)^2}{r^2f}\bigg)=0\,.
\end{equation}
These equations of motion admit a trivial solution where $A_x = -k$; its triviality is due to the fact
that $A_x = -k$ is obtainable by means of a bulk gauge transformation of the $A_x^{(0)}=0$ solution;
in fact
\begin{equation}
 A_x = U(x)^{-1} \left[A_x^{(0)} + \partial_x \right] U(x)\ ,
\end{equation}
with
\begin{equation}
 U(x) = e^{- \im k x }\ .
\end{equation}
Like any couple of gauge equivalent configurations, the two configurations $A_x = -k$ and $A_x^{(0)}=0$ correspond to the same physical situation, in particular
they correspond to a homogeneous phase with zero current (or zero net superfluid flow).

To study possible inhomogeneous phases we have therefore to seek for non-trivial solutions.
Moreover, as described in \cite{LO} and \cite{FF}, a necessary condition for having a LOFF ground state
is the existence of an absolute equilibrium state without any flow of superfluid current%
\footnote{Hence we are seeking a different configuration with respect to the configurations described in \cite{Basu:2008st}, \cite{Herzog:2008he},
\cite{Arean:2010zw} and \cite{Arean:2011gz} where instead a superfluid net velocity is present.}.
We have therefore to require a near-boundary behavior as
\begin{equation}\label{bouzeroco}
A_x(r\rightarrow\infty)\approx -k + \frac{J}{r} + ... \,, \quad{\rm with}\quad J=0\,. 
\end{equation}
while the UV asymptotic behavior for $\Psi$ and $A_t$ are the same as in the homogeneous case (see \eqref{psiUV} and \eqref{phiUV}).
Studying the equations of motion it is possible to show that non-trivial $A_x$ in accordance with the boundary requirement \eqref{bouzeroco}
are not admitted for any value of $m^2$.
Let us remind ourselves that the near-boundary behavior of the scalar is $\Psi \propto r^{-\lambda}$
where $\lambda$ represents the dual operator dimension given by
\begin{equation}
 \lambda = \frac{1}{2} \left(3 + \sqrt{9 + 4 m^2}\right)\ .
\end{equation}

Extending the ansatz \eqref{planewave} assuming that $A_t=A_t(x,r)$ and $A_x=A_x(x,r)$
and studying the associated system of equations of motion we reach a similar conclusion.
In the large $r$ asymptotic region Maxwell's equations imply separability $\partial_x\partial_r A_x = 0$
which, for consistency reasons, yields $A_x$ (and $A_t$ as well)
to lose its dependence on $x$; this brings us back to the previous, already considered situation.
Eventually, the possibility of a real cosinusoidal condensate (namely a two-plane waves solution)
\begin{equation}
 \psi = \Psi \cos(kr)\ ,
\end{equation}
can again be excluded at the level of equations of motion.
In conclusion, from the analysis of our system (in the probe approximation), appears impossible to have a inhomogeneous
LOFF-like phase.

\section{String embeddings and UV completion}
\label{UVcomp}
In order to seek for an UV completion of the phenomenological model considered so far,
one has to try to embed it into string theory or M-theory.
In other words, one has to study the possibility of truncating consistently such UV complete theories
and producing at low energy our effective model in all its features, namely both the field content and the interactions.
As it has already been mentioned, the investigation of a possible embedding concerns the microscopic structure
underlying the phenomenological model;
in particular, the origin of the two Abelian gauge groups and the representation to which the fermions belong.
For instance, we can have condensates composed by either fundamental or adjoint (or more general
$2$-index representations of some gauge group) fermions. 

Within the holographic framework, fundamental matter fields are introduced by considering
models having ``flavor'' D-branes. 
In other terms, following similar studies performed on holographic p-wave 
superconductors \cite{Erdmenger:2011hp,Ammon:2008fc,Ammon:2009fe},
we consider the possibility of embedding our model in probe flavor brane setups.
Being inspired by four-dimensional QCD-like models, it is for instance possible to
consider a non-critical five-dimensional string model possessing $N_c$ D$3$ and
$N_f$ space-time filling D$4$-anti-D$4$ branes, see \cite{Bigazzi:2005md,Casero:2007ae}.
In this setup, the low-energy modes of the D$3$-branes would correspond to the SU$(N_c)$ gluons, while the
``flavor'' D$4$-anti-D$4$ branes would provide the left and right handed fundamental flavor fields
(i.e. the quarks). 
Notice that this model contains also a complex scalar field (the would-be tachyon of the open
string stretching between branes and anti-branes) which transforms in the (anti)fundamental of
SU$(N_f)_L \times \text{SU} (N_f)_R$. The condensation of this complex scalar field
drives the breaking of the chiral symmetry down to SU$(N_f)$ and
therefore it is dual to the chiral condensate of fundamental fermions \cite{Sugimoto:2004mh,Bigazzi:2005md,Casero:2007ae}. 
The model, or at least the simplified version described in \cite{Bigazzi:2005md}, can provide $AdS_5$ flavored solutions at zero temperature
and densities with trivial tachyon and constant dilaton.  The corresponding finite temperature
versions have been studied in \cite{Casero:2005se,Bertoldi:2007sf}.

Within the just described context, our specific ``$AdS_4/CFT_3$'' model
in the case of fundamental fermions could have a natural string theory embedding 
involving a system of $N_c$ D2-branes and one (i.e. $N_f=1$) space-filling brane anti-brane pair;
this system has some features that match the 
characteristics of our holographic superconductor, namely:
\begin{itemize}
 \item the presence of two gauge fields associated respectively to two open-string modes, one
starting and ending on the brane, the other starting and ending on the anti-brane.
 \item the presence of a complex scalar field corresponding to an open-string mode connecting the brane to the anti-brane.
 \item the fact that the scalar field is charged under a combination of the two U$(1)$'s (hence playing the role of our U$(1)_A$)
and uncharged under the orthogonal combination (corresponding to U$(1)_B$).
 \item the fundamental fermions are associated to open-string modes connecting the space-filling brane and the space-filling anti-brane to the D2's.
 \item the scalar is naturally related to a fermion bilinear.
\end{itemize}
The scalar mode associated to the open strings stretching between the flavor brane and anti-brane can in general be tachyonic;
in such a circumstance it signals the instability of the brane/anti-brane pair. 
Notice, however, that in a curved space-time the effect of the curvature can lead to stable brane/anti-brane pairs.
The characteristics of the background geometry and the parameters of the model (like for instance the mass of the above mentioned scalar mode)
can be precisely given only in a full consistent string setup.



\section{General models and holographic fit}
As explained in Section \ref{strong}, we chose the simplest unbalanced holographic superconductor model, namely the model encoded in the Lagrangian density \eqref{laga}.
In \cite{Aprile:2010yb} the minimal balanced holographic superconductor has been generalized introducing  
functions of the scalar field $\psi$ as coefficients of the various terms in the Lagrangian density. 

A systematic study of the generalized model sheds light
on the various possibilities of the phenomenological holographic approach. 
The general framework consists in studying a system in the vicinity of a continuous phase transition
due to a U$(1)$ symmetry breaking. At the transition the correlation length diverges and the system
is expected to be describable by a strongly interacting conformal model. 
Universality of fixed points makes the physics around them to be ruled by few parameters which,
in a holographic general model, would coincide with the first terms in the $\psi$ expansions of the 
phenomenological coefficient functions appearing in the Lagrangian density.

The modification of higher terms in $\psi$ (whose VEV represents the order parameter of the system) can be read as a
perturbation of the critical region physics by irrelevant operators.
Indeed, in the vicinity of the critical point $\psi\ll 1$ and therefore the higher terms in $\psi$ are subleading.
The subleading terms do not change the kind of critical fixed point but are nevertheless able to influence the dynamics 
inside and especially outside the critical region.
For instance, they can be significant in relation to dynamical features such as transport properties.
A particularly interesting example concerns the introduction of higher terms that produce resonance peaks in the conductivity (see \cite{Aprile:2010yb});
As the temperature is raised, such peaks widen and their height diminishes; 
this behavior is suggestive of impurities, but at this stage caution prevents from giving any interpretation.
Recalling that Kubo's formula relates the linear response conductivity with the spectral density of states of the systems,
the presence of the peaks could offer interesting insight into the energy levels of the system itself.

In \cite{Aprile:2010yb} it is even considered
the possibility of having non-analytic coefficient functions; this opens the possibility of having critical exponents
differing from the standard Landau theory values. Non-analytic terms could only be due to quantum corrections suppressed
as $1/N$. In the large $N$ limit one expects standard mean-field theory behavior.

One important aspect of the general phenomenological holographic model relies on the possibility
of considering ``holographic fits'', i.e. 
the best choice of the functions parameterizing the Lagrangian density aimed to describe a particular real system.

A similar phenomenological extension of the couplings and kinetic terms could be repeated in the unbalanced case as well.

\chapter{Future Directions}

\section{Which ground state?}
\label{ground}
\begin{figure}[t]
 \centering
 \includegraphics[width=90mm]{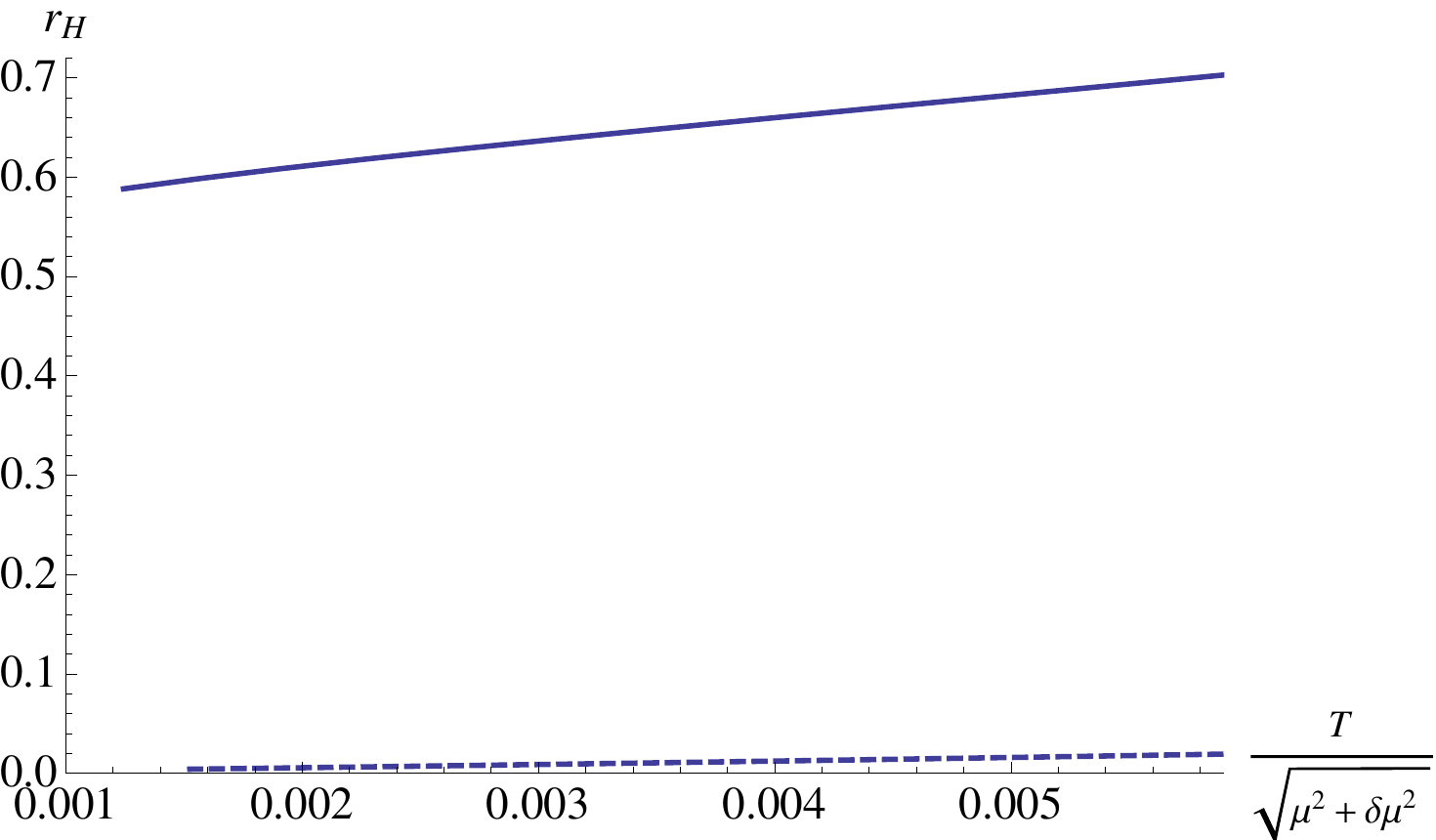}
 \caption{Low-temperature behavior of the black hole horizon. Both lines refer $\mu=1$ but different imbalance: $\delta\mu=2$ (solid line) and
$\delta\mu=0$ (dashed). }
 \label{low_T_horizon}
\end{figure}

The study of the ground state of a holographic model for a superconductor can be quite a tough matter.
It must be noted indeed that the problem to address is the one concerning the extremal limit of hairy black holes;
this is notoriously a delicate question. 
The main physical feature of interest consists in the presence or not of an extremal horizon at finite radius;
the former case actually implies a non-vanishing entropy at $T=0$ which is equivalent to a ground state degeneracy.
The Reissner-Nordstr\"{o}m black holes have such property while for Schwarzschild black holes the horizon vanishes in the
zero-temperature limit.

In the last couple of years some progress about the holographic superconductors ground state has been attained.
Specifically, in \cite{FernandezGracia:2009em} some no-hair theorems for black holes with 
flat or spherical horizon within gravity models containing a scalar field minimally coupled
to an Abelian gauge field, have been proved.
The most interesting result contained in \cite{FernandezGracia:2009em} is the theorem stating
that in the cases possessing negative cosmological constant and where the scalar and mass of the scalar field satisfy the 
bound
\begin{equation}\label{garciabound}
 m^2 - 2 q^2 > 4 |\Lambda|\ ,
\end{equation}
an ansatz as the one that is obtained disregarding the second gauge field in our equations \eqref{metric} and \eqref{homogeneous}, 
cannot admit extremal black hole solutions with non-zero area and non-trivial scalar hair.
Observe that the bound \eqref{garciabound} claims the impossibility of having hairy extremal black holes 
with finite horizons for $m^2<0$.
In the case in which $m^2<0$ and $q^2>|m^2|/6$ the paper \cite{Horowitz:2009ij} enriches the picture furnishing an explicit near-horizon ansatz
showing that the horizon shrinks and vanishes for $T\rightarrow 0$. 
Thinking about Reissner-Nordstr\"{o}m, one could wonder how a charged black hole could in fact have a vanishing horizon in the zero-temperature limit;
the answer relies on the presence of charged scalar hair which at $T=0$ contribute the entire total charge of the system leaving the
black hole ``depleted'' of its finite charge at $T\neq 0$.
In the absence of charged scalar hair, a charged black hole correspond to the Reissner-Nordstr\"{o}m picture with finite horizon at $T=0$.

So far we have discussed and argued in relation to the ground state of the standard holographic superconductor dual to a singly charged 
gravitational model. We would like to extend the analysis to our doubly charged system and gain information about its ground state.
As the scalar field of our model is charged under one gauge field and uncharged with respect to the other, we can expect to find
a finite horizon at $T=0$; indeed, from the second gauge field viewpoint, the charge of the black hole cannot be ``acquired'' by non-trivial hair
also if present. The expectation is supported by the fact that a finite charge for a black hole with vanishing horizon would
unavoidably lead to a divergence in the electric field.

At the same time, the the system of equations ruling our system is very similar to the standard singly charged system and it a priori not easy
to understand how a modification of the ansatz advanced in \cite{Horowitz:2009ij} can lead to finite $r_H$ as $T\rightarrow 0$.
We have performed a numerical analysis studying the behavior of the horizon radius $r_H(T)$ for low values of $T$;
this has to be regarded as a guide to the intuition but it cannot solve the problem of the strict $T=0$ configuration
which instead has to be treated analytically (even though in some approximation such as some near-horizon hypothesis).
Indeed, from a numerical perspective, the low temperature regime is usually delicate and numerics become here often unreliable; 
moreover, the finite, even though low, temperatures reached numerically can be qualitatively
different to the true $T=0$ case where a specific ansatz could prove to be necessary.

We show the results of the numerical approach in Figure \ref{low_T_horizon}.
The plots show that the case charged under the field $B$ (namely $\delta\mu\neq 0$) the trend of $r_H$ appears to intersect
at $T=0$ the axis at a finite value. Conversely, the uncharged (with respect to $B$) setup seems to present a vanishing horizon.


Let us conclude with the remark that no-hair theorems or a direct study of the features of the zero $T$ solutions
furnish information concerning the phase diagram of the dual model.

\section{Crystalline lattice and/or impurities?}
\label{impu}
In the holographic context, the fully back-reacted systems under study are in general translation-invariant.
This feature generates an infinite DC conductivity which has to be distinguished from the contribution of genuine superconductivity%
\footnote{This has been the subject of Subsection \ref{sition}.}.
Of course, a relevant phenomenologically-oriented development, would be the possibility of encompassing 
in our models features such as the presence of a crystalline lattice and ``heavy'' or static impurities.
They would break translational invariance allowing for momentum relaxation phenomena that prevent the occurrence of a non-superconductive DC delta
in the conductivity.

There are many approaches to this question in the literature; we here mention some of their characteristics and comment on possible improvements of our model
towards realistic DC resistivity.
To have momentum relaxation for the charge carriers, some additional degrees of freedom to which momentum transfer can occur must be present in the model.
Note that such degrees of freedom have to ``absorb'' momentum from the flowing carriers at a rate that is bigger than the momentum they return to the carriers.
There are two main ways to achieve this: one is to suppose that the additional degrees of freedom are heavy or fixed the other is to suppose that (even thou maybe light)
the additional degrees of freedom are quantitatively ``many more'' than the carriers such that as a whole they remain almost fixed even if receiving momentum from the carriers.
Note that both framework suggest the idea that the carriers constitute somehow a ``small'' perturbation of the total system and, in holographic terms,
this is related to the probe approximation.
This point can be made precise considering that in the probe approximation the metric is held fixed while its fluctuation would couple indeed to 
the energy momentum tensor.
Let us see in a slightly more detailed fashion the two momentum relaxation mechanism just introduced starting from the latter.

In \cite{Karch:2007pd} the authors perform a conductivity calculation over the frequency range and obtain a finite DC value.
The dual model consists in a system of $N_f$ ``flavor'' branes in the background of the $N_c$ ``color'' branes generating (in the low-energy picture) the black hole;
The charge we consider to mimic the electric charge is the diagonal part of U$(N_f)$ (i.e., in the flavor simile, the barion number).
The probe approximation in this case is realized by the assumption $N_f \ll N_c$; in other words we suppose that the flavor degrees of freedom, namely
the open-string modes involving the flavor branes evolve in the presence of a ``bath'' of far more numerous color degrees of freedom.
In the dual field theoretical picture, this translates in having a density of carriers flowing through a plasma whose density of degrees of freedom is far higher.
In this sense, momentum relaxation for flowing carriers is achieved; from the plasma viewpoint the momentum transfer is neglected.

The alternative way is to involve some ``fixed'' UV feature mimicking heavy impurities of a lattice. 
In \cite{Hartnoll:2012rj}, there is a description of how a UV behavior characterized by a momentum scale $k_L$ (playing the role of
the lattice momentum scale) influences the low-energy physics giving a temperature dependence to the DC conductivity resembling
the result of umklapp scattering in condensed matter systems%
\footnote{An umklapp scattering process results in the transformation of the momentum of the carrier from one Brillouin zone to another.
The neat result is a momentum transfer to the lattice; from the lattice viewpoint, this momentum transfer 
is usually simply neglected assuming large (or even infinite) lattice extension.}.
Following again \cite{Hartnoll:2012rj}, it is important to notice that the $\omega = 0$ resistivity is a quantity that, even though it is clearly IR,
it is nevertheless sensitive to UV features of the model; for instance, the presence of the lattice whose spacing introduces a ``high-energy'' scale%
\footnote{The low-energy regime is generically meant to regard the physics of collective excitation over many lattice spacings.}.
In a holographic context, where the renormalization flow or energy scale is encoded in the ``radial'' $AdS$ coordinate,
such a relation between high and low energy scales is translated in a dependence of the near-horizon behavior on the near-boundary physics.
The dual treatment has to be therefore complete and neither the large $r$ nor the small $r$ study can yield an exhaustive account.
In \cite{Hartnoll:2012rj} however no explicit example of such UV structure generating $k_L$ is described.

Observe that in our model there are some features which can be interpreted in connection to the presence of impurities.
The imbalance, indeed, is read as produced by impurities that have magnetically asymmetric effects.
Their influence is accounted for by the chemical potential;
even though it is natural to think to the imbalance as arising from spin-dependent scatterings of carriers on impurities,
no microscopic feature of such scatterings is implemented in the model. 
It would be very interesting to study the possibility of dynamical impurities or the lattice in our model,
however there seems to be a tough obstacle to face: on the one side the momentum-relaxation seems to require probe approximation
while a crucial point of our mixed spin-electric transport was the mediating role of the metric%
\footnote{Note that we have already commented on spatially inhomogeneous features within our system in relation to LOFF phases (see Section \ref{LOFF}).}.

On the subject of non-trivial transverse profiles for the chemical potential of inhomogeneous 
holographic superconductor, a basic example is given in \cite{Siani:2011uj}. 
The chemical potential transverse profile consists in a ``localized'' region in which $\mu$ is constant but has a different value from
the rest of the space; this is maybe a possibility of mimicking an electric impurity. 
In \cite{Siani:2011uj} it is noted that the macroscopic thermodynamics transport properties of the holographic 
medium are independent of the presence of the ``impurity''; this could maybe match with an extension at strong coupling of 
``dirty superconductors'' Anderson's theorem
\footnote{Anderson's theorem for ``dirty'' superconductors states that at weak-coupling the thermodynamical features of a superconducting 
system are insensitive to time-reversal-preserving and short-ranged perturbations.}.
An extension of the ideas of \cite{Siani:2011uj} to $\delta\mu$ in our model can possibly mimic a localized magnetic impurity.

\section{Finite momentum}
\label{opto}
A natural continuation to the research described in this second part is centered on the introduction
of finite momentum fluctuations into our model.
This conceptually simple extension, which could however prove to be delicate on the computational level,
allows us to study many significant and interesting aspects.
Let us just mention the salient among these.
On the lines followed by \cite{Amariti:2011dm} and \cite{Amariti:2011dj} it would be very interesting to study the optical properties
of our system%
\footnote{Another very interesting (but still at an early stage) future perspective would consists in continuing
along the line of \cite{Amariti:2010hw} and drawing a parallel between the excitons (strongly coupled electron-hole states)
and the ``mesons'' arising in flavored holographic setups. To have an idea see \cite{CasalderreySolana:2008ne}
and for a wider review of the topic of holographic mesons we refer to \cite{Erdmenger:2007cm}.}; 
this requires the knowledge of the electrical permittivity and the magnetic permeability which arise, in fact,
from the study of the response of our system to spatially modulated source variations.
Another possibly viable subject regards the characterization of the \emph{shot-noise} properties of our system;
as opposed to thermal noise, this kind of noise is related to quantum effects and it is very important in mesoscopic physics.
A review on the subject is \cite{2000PhR...336....1B}.

\appendix

\part{Appendices}


\chapter{Graviton}
\label{demogravi}
In the present appendix we describe an argument by Weinberg%
\footnote{We actually follow the lines of \cite{Nicodemi} which, in turn, is inspired by \cite{PhysRev.135.B1049}} 
which shows that a massless spin $2$ particle has to couple ``democratically'', i.e. with the same coupling constant, with all other particles.
To this end, let us consider scattering amplitudes associated to the emission or absorption of a 
spin $1$ or spin $2$ massless particle.
In the two cases, the amplitudes have to be of the following shapes
\begin{eqnarray}
 & {\cal A}_{\text{spin}\, 1} &= \epsilon_\mu M^\mu \\
 & {\cal A}_{\text{spin}\, 2} &= \epsilon_{\mu\nu} M^{\mu\nu}
\end{eqnarray}
where $\epsilon_\mu$ and $\epsilon_{\mu\nu}$ are respectively the polarization vectors of the spin $1$ and $2$ particles
while the tensors $M$ are given by combinations of the four momenta and spins of the other particles involved in the scattering process.
Gauge invariance implies that the tensors $M$ can always be taken in such a way that
\begin{eqnarray}\label{spinuno}
 & k_\mu M^\mu &= 0 \\
 & k_\mu M^{\mu\nu} &= 0 \label{spinduo}
\end{eqnarray}
where $k$ denotes the momentum of the massless particle.

Let us consider the amplitude associated to the scattering of $N$ massive particles;
the momentum conservation relation is
\begin{equation}\label{momcon}
 \sum_i \eta^{(i)} p_\mu^{(i)} = 0\ ,
\end{equation}
where the index $i$ labels the particles involved in the scattering process and $\eta^{(i)}$ is $+$ or $-$
for incoming and outgoing particles respectively.
We then consider the same amplitude with the additional emission of a massless spin $1$ or $2$ particle
whose momentum is $k$; in particular we consider the soft limit $k\rightarrow 0$.
In both the spin $1$ and $2$ cases the amplitude is dominated by the diagrams in which
the massless particle is attached to an external leg.
We have therefore $N$ (as many as the number of legs) diagrams to take into account whose associated amplitude is henceforth denoted with $M_{(i)}$ 
(with the appropriate tensorial structure).
For the spin $1$ case we can then write
\begin{equation}
 M^\mu = \sum_i M^\mu_{(i)}\ .
\end{equation}
Neglecting the spin labels, each $M^\mu_{(i)}$ will have the following form
\begin{equation}
 M^\mu_{(i)} = \eta^{(i)} \frac{g_{(i)} p_{(i)}^\mu}{2 p_{(i)}\cdot k} {\cal M}
\end{equation}
where $g_{(i)}$ represents the coupling associated to the interaction of the $i$-th particle involved in the scattering process
and the emitted spin $1$ massless particle.
The symbol $\cal M$ indicates the remaining part of the amplitude.
Requiring gauge invariance is equivalent to impose \eqref{spinuno}, so
\begin{equation}
 \sum_i \eta^{(i)} \frac{g_{(i)} p_{(i)}\cdot k}{2 p_{(i)}\cdot k} {\cal M} = 0
\end{equation}
which implies (for ${\cal M}\neq 0$)
\begin{equation}
 \sum_i \eta^{(i)} g_{(i)} = 0\ .
\end{equation}
This is just charge conservation.

An analogous argument can be repeated for the spin $2$ case.
We have similarly
\begin{equation}
 M_{(i)}^{\mu\nu} = \eta^{(i)} \frac{G_{(i)} p_{(i)}^\mu p_{(i)}^\nu}{2 p_{(i)}\cdot k} {\cal M}
\end{equation}
where $G_{(i)}$ denotes the coupling of the $i$-th particle with the spin $2$ particle.
Gauge invariance implies \eqref{spinduo} then, explicitly,
\begin{equation}
 \sum_i \eta^{(i)} \frac{G_{(i)} p_{(i)}\cdot k\, p_{(i)}^\nu}{2 p_{(i)}\cdot k} {\cal M} = 0\ ,
\end{equation}
that, for ${\cal M}\neq 0$, leads to
\begin{equation}\label{mono}
 \sum_{i} \eta^{(i)} G_{(i)} p_{(i)}^\nu = 0\ .
\end{equation}
Note that \eqref{mono} is not compatible with momentum conservation \eqref{momcon}
unless we have
\begin{equation}
 G_{(i)} = G \ \ \ \ \forall i\ .
\end{equation}
In conclusion, we have that the spin $2$ massless particle has to couple universally to all matter.

\chapter{'t Hooft Symbols}
\label{hooft}

The compact expression for 't Hooft symbols we adopt for the calculations described in the main text is
\begin{equation}
 \begin{split}
&  \eta^c_{\mu\nu} = -\delta^4_\mu \delta^c_\nu + \delta^c_\mu \delta^4_\nu + \epsilon^c_{\ \mu\nu}\\
&  \overline{\eta}^c_{\mu\nu} = \delta^4_\mu \delta^c_\nu - \delta^c_\mu \delta^4_\nu + \epsilon^c_{\ \mu\nu}
 \end{split}
\end{equation}
which leads to the following explicit matrices:
\begin{equation}
 \eta^1_{\mu\nu} = \left(\begin{array}{cccc}
                 0 & 0 & 0 & 1\\
                 0 & 0 & 1 & 0\\
                 0 & -1& 0 & 0\\
                -1 & 0 & 0 & 0
                \end{array}\right)\ \ \ \ \ 
 \overline{\eta}^1_{\mu\nu} = \left(\begin{array}{cccc}
                 0 & 0 & 0 & -1\\
                 0 & 0 & 1 & 0\\
                 0 & -1& 0 & 0\\
                 1 & 0 & 0 & 0
                \end{array}\right)
\end{equation}
\begin{equation}
 \eta^2_{\mu\nu} = \left(\begin{array}{cccc}
                 0 & 0 & -1 & 0\\
                 0 & 0 & 0 & 1\\
                 1 & 0 & 0 & 0\\
                 0 & -1 & 0 & 0
                \end{array}\right)\ \ \ \ \ 
 \overline{\eta}^2_{\mu\nu} = \left(\begin{array}{cccc}
                 0 & 0 & -1 & 0\\
                 0 & 0 & 0 & -1\\
                 1 & 0 & 0 & 0\\
                 0 & 1 & 0 & 0
                \end{array}\right)
\end{equation}
\begin{equation}
 \eta^3_{\mu\nu} = \left(\begin{array}{cccc}
                 0 & 1 & 0 & 0\\
                -1 & 0 & 0 & 0\\
                 0 & 0 & 0 & 1\\
                 0 & 0 & -1& 0
                \end{array}\right)\ \ \ \ \ 
 \overline{\eta}^3_{\mu\nu} = \left(\begin{array}{cccc}
                 0 & 1 & 0 & 0\\
                -1 & 0 & 0 & 0\\
                 0 & 0 & 0 & -1\\
                 0 & 0 & 1 & 0
                \end{array}\right)
\end{equation}

\chapter{ADHM Projector}
\label{ADHMprojection}

The ADHM matrix $\Delta_{\lambda i \dot{\alpha}}$ is an $(N + 2k) \times 2k$ matrix
where the ADHM index $\lambda$ runs over $1, ..., N + 2k$, the instanton index $i$ runs over $1,...k$
and the anti-chiral index $\dot{\alpha}$ runs over $1,2$%
\footnote{The value of $k$ represents the topological charge of the instanton.}.
Let us define the ADHM complex vector space ${\cal B} \sim \mathbb{C}^{N+2k}$ spanned by the ADHM index $\lambda$
and the complex vector space ${\cal A} \sim \mathbb{C}^{2k}$ spanned by $i \dot{\alpha}$.
We then consider the following maps:
\begin{eqnarray}
 \overline{\Delta} &:& {\cal B} \rightarrow {\cal A} \ ,\\
            \Delta &:& {\cal A} \rightarrow {\cal B} \ ,
\end{eqnarray}
where $\overline{\Delta}$ is surjective and $\Delta$ is injective.
As a consequence, we have
\begin{eqnarray}
 \text{dim}\left(\text{\bfseries Ker}[\,\overline{\Delta}\,]\right) &=& 
 \text{dim}\left({\cal B}\right) - \text{dim}\left({\cal A}\right) = N \\
 \text{dim}\left(\text{\bfseries Im}[\,\Delta\,]\right) &=&
 \text{dim}\left({\cal A}\right) = 2k
\end{eqnarray}
Note that the following relations hold
\begin{eqnarray}
 \text{\bfseries Ker}[\,\overline{\Delta}\,] \oplus \text{\bfseries Im}[\,\Delta\,] &=& {\cal B} \\
 \text{\bfseries Ker}[\,\overline{\Delta}\,] \wedge \text{\bfseries Im}[\,\Delta\,] &=& \{0_{\cal B}\} \\
\end{eqnarray}

The equations from which we start are:
\begin{eqnarray}
 \overline{U}^\lambda_u U_{\lambda v} &=& \delta_{u v} \label{UU}\\
 \overline{\Delta}^{\dot{\alpha} \lambda}_i U_{\lambda u} &=& \overline{U}^\lambda_u \Delta_{\lambda i \dot{\alpha}} = 0 \label{DU}\\
 \overline{\Delta}^{\dot{\alpha} \lambda}_i \Delta_{\lambda j \dot{\beta}} &=& \delta^{\dot{\alpha}}_{\ \dot{\beta}} f^{-1}_{ij}
 \label{DbD}
\end{eqnarray}
where the matrix $U_{\lambda u}$ represents a collection of $N$ linearly independent vectors in the ADHM space ${\cal B}$;
as a consequence of \eqref{DU}, they generate $\text{\bfseries Ker}[\,\overline{\Delta}\,]$.
Using \eqref{DU}, we define the projector operator
\begin{eqnarray}
 P_\lambda^{\ \mu} &\equiv& U_{\lambda u} \overline{U}^\mu_u\ , \\
 \overline{\Delta}^{\dot{\alpha} \lambda}_i  P_\lambda^{\ \mu} &=& 0\ ,
\end{eqnarray}
projecting the generic vector of the ADHM space ${\cal B}$ on $\text{\bfseries Ker}[\,\overline{\Delta}\,]$;
note that it is idempotent as a consequence of \eqref{UU} and, since $\overline{U}=U^\dagger$, $P$ is an
$(N+2k)\times(N+2k)$ matrix of rank $\text{dim}\left(\text{\bfseries Ker}[\,\overline{\Delta}\,]\right)=N$.

The projector operator on $\text{\bfseries Im}[\,\Delta\,]$ is then given by:
\begin{equation}\label{p_primo}
 P' = 1 - P\ ,
\end{equation}
where $P'$ is again an $(N+2k)\times(N+2k)$ but possessing rank equal to 
$\text{dim}\left(\text{\bfseries Im}[\,\Delta\,]\right)=2k$.
The generic $(N+2k)\times(N+2k)$ matrix of rank $2k$, can be expressed in the following form
\begin{equation}\label{gen_man}
 \Delta_{\lambda j \dot{\beta}} M^{\dot{\beta}}_{\dot{\alpha}} G_{ji} \overline{\Delta}^{\dot{\alpha} \lambda}_i\ ,
\end{equation}
where $M$ is a $2\times 2$ matrix and $G$ a $k\times k$ matrix and both have to be invertible,
\begin{equation}
 \text{det}\, M \neq 0 \ \ , \ \ \ \ \ \text{det}\, G \neq 0\ .
\end{equation}
Note that $M$ and $G$ account for the right number of parameters, i.e. $(2k)^2$.
We want \eqref{gen_man} to be the explicit matrix realizing the projector \eqref{p_primo},
therefore we have to impose the idempotency.
From this requirement and using \eqref{DbD} we obtain
\begin{eqnarray}
 M M &=& M \ ,\\
 G f^{-1} G &=& G\ ;
\end{eqnarray}
exploiting the existence of an inverse for both $M$ and $G$ we finally get
\begin{eqnarray}
 M^{\dot{\alpha}}_{\dot{\beta}} &=& \delta^{\dot{\alpha}}_{\dot{\beta}} \ ,\\
 G_{ij} &=& f_{ij}\ .
\end{eqnarray}
The eventual explicit result for the ADHM null projector $P$ is:
\begin{equation}
 P_{\lambda}^{\ \mu} = \delta_{\lambda}^{\ \mu} - 
\Delta_{\lambda i \dot{\alpha}} f_{ij} \overline{\Delta}^{\dot{\alpha} \lambda}_j\ .
\end{equation}

\chapter{Shape of the Chan-Paton Orientifold Matrix} 
\label{shape}

As a consequence of the orbifold/orientifold consistency condition \eqref{consistency}, 
the orientifold representation on the Chan-Paton indexes is constrained. 
Let us consider the D$3$ CP labels and remind the reader of the choice \eqref{d3cp} to represent the orbifold on them.
Furthermore, we also chose the orbifold to have anti-symmetric representation on the D$3$ CP labels, so
a priori we can have for $\gamma(\Omega)$ the following general form
\begin{equation}
\gamma_-(\Omega) = \left(\begin{array}{ccc}
	\epsilon_1&A&B\\
	-A^T&\epsilon_2&C\\
	-B^T&-C^T&\epsilon_3
\end{array} \right)
\end{equation}
where the $\epsilon$'s are anti-symmetric matrices. 
Putting the explicit expression for the orbifold/orientifold matrices into the consistency condition \eqref{consistency} we obtain
\begin{equation}
\left(\begin{array}{ccc}
	1&0&0\\
	0&\xi&0\\
	0&0&\xi^2
\end{array}\right)
\left(\begin{array}{ccc}
	\epsilon_1&A&B\\
	-A^T&\epsilon_2&C\\
	-B^T&-C^T&\epsilon_3
\end{array} \right)
\left(\begin{array}{ccc}
	1&0&0\\
	0&\xi&0\\
	0&0&\xi^2
\end{array}\right) = 
\left(\begin{array}{ccc}
	\epsilon_1&A&B\\
	-A^T&\epsilon_2&C\\
	-B^T&-C^T&\epsilon_3
\end{array} \right)\ .
\end{equation}
Performing a simple passage we have
\begin{equation}
\left(\begin{array}{ccc}
	\epsilon_1&\xi A&\xi^2 B\\
	-\xi A^T&\xi^2\epsilon_2&C\\
	-\xi^2B^T&-C^T&\xi \epsilon_3
\end{array} \right) =
\left(\begin{array}{ccc}
	\epsilon_1&A&B\\
	-A^T&\epsilon_2&C\\
	-B^T&-C^T&\epsilon_3
\end{array} \right)\ .
\end{equation}
The condition can be fulfilled only if
\begin{equation}
 A=0 \ , \ B=0 \ , \ \epsilon_2=0 \ , \ \epsilon_3=0  \ ,
\end{equation}
so the most general anti-symmetric and consistent CP orientifold representation is given by the following matrix
\begin{equation}
\gamma_-(\Omega)=\left(\begin{array}{ccc}
	\epsilon_1&0&0\\
	0&0&C\\
	0&-C^T&0
\end{array} \right)
\end{equation}

A completely analogous argument can be repeated in the symmetric case leading to
\begin{equation}
\gamma_+(\Omega)=\left(\begin{array}{ccc}
	s_1&0&0\\
	0&0&C\\
	0&C^T&0
\end{array} \right)
\end{equation}
where $s$ is a symmetric matrix.

\chapter{Details on the D-instanton Computations}
\label{app:A} 

Let us give an explicit expressions for $\cP(\chi)$, $\cR(\chi)$, $\cQ(\chi)$ introduced in 
(\ref{PRQ}) and appearing in the integrand of the instanton partition function (\ref{zk1}).
A very convenient way of computing the determinants and Pfaffians is to consider the weights of the
representations of the instantonic symmetry group $\mathrm{SO}(k)$, of the twisted Lorentz group $\mathrm{SU}(2)\times\mathrm{SU}(2)'$
and of the gauge group SU$(2)$ involved in the model.
We give here a list of the weights vectors for the groups relevant to our explicit computations in the main text.

\paragraph{Weight sets of SO$(2n+1)$:}
The group has rank $n$. We denote with $\ve{i}$ the versors in the $\mathbb{R}^n$ weight space, so
\begin{itemize}
 \item the set of the $2n+1$ weights $\vec\pi$ of the vector representation is 
 \begin{equation}
  \label{setvec1}
  \pm\ve i~,~~~~ \vec 0~~\mbox{with multiplicity $1$}~;
 \end{equation}
 \item the set of $n(2n+1)$ weights $\vec\rho$ of the adjoint representation (corresponding
to the two-index antisymmetric tensor) is 
\begin{equation}
 \label{setrho1}
 \pm \ve i \pm \ve j~(i < j)~,~~~~
 \pm \ve i~,~~~~
 \vec 0~~\mbox{with multiplicity $n$}~;
\end{equation}
\item the $(n+1)(2n+1)$ weights $\vec\sigma$ of the two-index symmetric tensor%
\footnote{Note that this is not an irreducible representation: it decomposes 
into the $(n+1)(2n+1)-1$ traceless symmetric tensor plus a singlet. One of the
$\vec 0$ weights corresponds indeed to the singlet.} are
\begin{equation}
 \label{setsymm1}
 \pm \ve i \pm \ve j~(i < j)~,~~~~
 \pm \ve i~,~~~~
 \pm 2 \ve i~,~~~~
 \vec 0~~\mbox{with multiplicity $n+1$}~.
\end{equation} 
\end{itemize}

\paragraph{Weight sets of SO$(2n)$:}
This group has rank $n$.
Denoting the versors in the 
$\mathbb{R}^n$ weight space with $\ve{i}$ we have
\begin{itemize}
 \item the set of the $2n$ weights $\vec\pi$ of the vector representation is 
 given by
 \begin{equation}
  \label{setvec}
  \pm\ve i~;
 \end{equation}
 \item the set of $n(2n-1)$ weights $\vec\rho$ of the two-index antisymmetric 
tensor is the following:
\begin{equation}
 \label{setrho}
 \pm \ve i \pm \ve j~(i < j)~,~~~~
 \vec 0~~\mbox{with multiplicity $n$}~;
\end{equation}
\item the $n(2n+1)$ weights $\vec\sigma$ of the two-index symmetric tensor%
\footnote{Again, this is not an irreducible representation, since it 
contains a singlet.} are
\begin{equation}
 \label{setsymm}
 \pm \ve i \pm \ve j~(i < j)~,~~~~
 \pm 2 \ve i~,~~~~
 \vec 0~~\mbox{with multiplicity $n$}~.
\end{equation}
\end{itemize}

\paragraph{Weight sets of SU$(2)\times$SU$(2)'$:} The representations of the twisted
Lorentz group relevant for our computations are the $(\mathbf{1,3})$ and the $(\mathbf{2,2})$
in which the BRST pairs $(\lambda_c,D_c)$ and $(a_\mu,M_\mu)$ transform respectively.
\begin{itemize}
 \item the weights $\vec\alpha$ of the $(\mathbf{1,3})$ representation are given by the
following two-component vectors
\begin{equation}
 (0,\pm1)~,~~~~(0,0)~.
\label{alpha1}
\end{equation}
In our conventions, the weight $(0,+1)$ is considered to be positive.
\item the weights $\vec\beta$ of the $(\mathbf{2,2})$ representation are given by the
following two-component vectors
\begin{equation}
\big(\!\pm 1/2,\pm 1/2\big)~.
\label{beta1}
\end{equation}
The weights $\big(\!\pm 1/2,+1/2\big)$ are considered positive in our conventions.
\end{itemize}

\paragraph{Weight sets of SU$(2)$:} In relation to our computations, the only relevant SU$(2)$ representation
is the fundamental; its two weights $\vec \gamma$ are simply given by $\pm1/2$. 

In order to evaluate the moduli integral and obtain the instanton partition function, 
a convenient choice consists in aligning the vacuum expectation value $\phi$ of the chiral multiplet along 
the Cartan direction of SU$(2)$, and the external Ramond-Ramond background $\cF$ along the Cartan directions of
$\mathrm{SU}(2)\times\mathrm{SU}(2)'$, in formul\ae\ we have
\begin{equation}
\phi = \vec{\phi} \cdot \vec{H}_{\mathrm{SU}(2)}~~~\mbox{and}~~~
\cF = \vec{f} \cdot \vec{H}_{\mathrm{SU}(2)\times\mathrm{SU}(2)'}
\label{phiF_cart}~.
\end{equation}
Comparing with Eq.s (\ref{Fcart}) and (\ref{phisu2}), we see that
\begin{equation}
 \vec\phi = \varphi~~~\mbox{and}~~~\vec f = (\bar f, f)~.
\label{fphi}
\end{equation}
Similarly, we exploit the SO$(k)$ invariance and arrange the $\chi$ moduli along the
Cartan directions, namely
\begin{equation} 
\label{chi_cart}
 \chi \rightarrow \vec{\chi} \cdot \vec{H}_{\mathrm{SO}(k)} = \sum_{i=1}^n \chi_i H^i_{\mathrm{SO}(k)}~.
\end{equation}
As the $\chi$ modulus compares in the instanton integration measure (as opposed to $\phi$ and $\cal F$ which represent ``external'' backgrounds
from the instanton viewpoint) we can rotate it along its Cartan direction at the price of introducing in the integral a Vandermonde determinant given by
\begin{equation}
 \label{vanderm}
\Delta(\vec\chi) =
\prod_{\vec\rho\not=\vec 0}\vec\chi\cdot \vec\rho 
=\left\{
 \begin{aligned}
&  \prod_{i<j} 
\big(\chi_i^2 -\chi_j^2\big)^2 & \mbox{for } k=2n~,\\
 &(-1)^n\prod_{i=1}^n \chi_i^2\prod_{j<\ell}\big(\chi_j^2-\chi_\ell^2\big) ^2& \mbox{for } k=2n+1~.
 \end{aligned}
 \right.
\end{equation}
We gathered all the necessary ingredients to write the explicit expressions for the functions
$\cP(\chi)$, $\cR(\chi)$, $\cQ(\chi)$. {From} \eq{pchi0},  
\begin{equation}
 \label{pchi}
\begin{aligned}
 \cP(\vec\chi) &= \prod_{\vec\rho}\prod_{\vec\alpha}^+\left(\vec\chi\cdot
\vec\rho -\vec f\cdot \vec\alpha\right)\\
&= \left\{
 \begin{aligned}
 & (-f)^n
\prod_{i<j}^n\left[(\chi_i+\chi_j)^2-f^2\right]\left[(\chi_i-\chi_j)^2-f^2\right]& \mbox{for } k=2n~,\\
 &f^n\prod_{i}^n\big(\chi_i^2-f^2\big)
\prod_{j<\ell}^n
\left[(\chi_j+\chi_\ell)^2-f^2\right]\left[(\chi_j-\chi_\ell)^2-f^2\right]
& \mbox{for } k=2n+1~.
 \end{aligned}
 \right.
\end{aligned}
\end{equation}
where the product over $\vec\alpha$ is limited to the positive weight $(0,+1)$;
this is the meaning of the superscript $+$. 
{From} \eq{rchi0}, we have 
\begin{equation}
 \label{rchi}
\begin{aligned}
 \cR(\vec\chi) = \prod_{\vec\pi} \prod_{\vec\gamma}\left(\vec\chi\cdot
\vec\pi -\vec\phi\cdot \vec\gamma\right) =\left\{
 \begin{aligned}
 & \prod_{i=1}^n \big(\chi_i^2+\det\phi\big)^2 & \mbox{for } k=2n~,\\
 & \det\phi\prod_{i=1}^n \big(\chi_i^2+\det\phi\big)^2 & \mbox{for } k=2n+1~.
 \end{aligned}
 \right.
\end{aligned}
\end{equation}
and finally from \eq{qchi0}, we have
\begin{equation}
 \label{qchi}
\begin{aligned}
 \cQ(\vec\chi) &= \prod_{\vec\sigma} \prod_{\vec\beta}^+\left(\vec\chi\cdot
\vec\sigma -\vec f\cdot \vec\beta\right)\\
&=\left\{
 \begin{aligned}
 &\cE^n\prod_{A=1}^2 \prod_{i=1}^n
\big(4\chi_i^2-E_A^2\big)
\prod_{j<\ell}\left[(\chi_j+\chi_\ell)^2-E_A^2\right]\left[(\chi_j-\chi_\ell)^2-E_A^2\right]& \mbox{for } k=2n~,\\
 &\cE^{n+1}\prod_{A=1}^2 \Bigg\{\prod_{i=1}^n
\big(\chi_i^2-E_A^2\big)\big(4\chi_i^2-E_A^2\big)~\times\\
&\hspace{40pt}\times
\prod_{j<\ell}\left[(\chi_j+\chi_\ell)^2-E_A^2\right]\left[(\chi_j-\chi_\ell)^2-E_A^2\right]\Bigg\}
& \hspace{-30pt}\mbox{for } k=2n+1~,
 \end{aligned}
 \right.
\end{aligned}
\end{equation}
where again the product over $\vec\beta$ is limited to the positive weights.

Using these explicit formul\ae\ and recalling from \eq{ef} that $f=(E_1+E_2)$, it is possible to find that at instanton number $k=2$ the partition function
(\ref{Zkred}) reads
\begin{equation}
 Z_2=-\cN_2\,\frac{E_1+E_2}{\cE}\,\int\frac{d\chi}{2\pi\ii}~
\frac{\big(\chi^2+\det\phi\big)^2}{(4\chi^2-E_1^2)(4\chi^2-E_2^2)}~.
\label{z2app}
\end{equation}
as reported in \eq{z2} of the main text. 
After having evaluated the $\chi$ integral as a contour integral in the
upper half complex plane with the pole prescription (\ref{imparts}), and summing the residues at
$\chi=E_A$ and $\chi=E_A/2$ for $A=1,2$, we finally obtain the result \eq{z2res}.
Proceeding in a similar way, at instanton number $k=3$ we find
\begin{equation}
 Z_3= -\cN_3\,\frac{\det\phi\,(E_1+E_2)}{\cE^2}\int\frac{d\chi}{2\pi\ii}~
\frac{\big(\chi^2-(E_1+E_2)^2\big)\big(\chi^2+\det\phi\big)^2}{
(\chi^2-E_1^2)(\chi^2-E_2^2)(4\chi^2-E_1^2)(4\chi^2-E_2^2)}~,
\label{z3app}
\end{equation}
from which the result given in \eq{z3} follows.

We conclude by giving the explicit expressions of the instanton partition functions at $k=4$
and $k=5$. They are
\begin{eqnarray}
 Z_{4} &= &\frac{\cN_4}{48\,\cE^4}\,\mathrm{det}^4\phi -\frac{\cN_4}{16\,\cE^3}\,\mathrm{det}^3\phi
 - \frac{\cN_4}{384\,\cE^3}\big[3(E_1^2+E_2^2)-19\cE\big]\mathrm{det}^2\phi
\label{z4app}\\
&&+\, \frac{\cN_4}{256\,\cE^2}\big[(E_1^2+E_2^2)-3\cE\big]\det\phi
+\frac{\cN_4}{4096\,\cE^2}\big[(E_1^2+E_2^2)-7\cE\big]\big[(E_1^2+E_2^2)+\cE\big]~,
 \nonumber
\end{eqnarray}
and
\begin{eqnarray}
 Z_{5} &= &\frac{\cN_5}{240\,\cE^5}\,\mathrm{det}^5\phi -\frac{\cN_4}{48\,\cE^4}\,\mathrm{det}^4\phi
 - \frac{\cN_5}{384\,\cE^4}\big[(E_1^2+E_2^2)-13\cE\big]\mathrm{det}^3\phi\nonumber\\
&&+\frac{\cN_5}{768\,\cE^3}\big[3(E_1^2+E_2^2)-17\cE\big]\mathrm{det}^3\phi
\label{z5app}\\
&&+\, \frac{\cN_5}{61440\,\cE^3}\big[15(E_1^4+E_2^4)-170 (E_1^2+E_2^2)+299\cE^2\big]\det\phi
~. \nonumber
\end{eqnarray}

\chapter{Bulk Massive Scalar Field}
\label{scala}
In the present appendix we follow the lines of \cite{Zaffaroni:2000vh}.
In order to perform the analysis of a generic bulk field $\hat{\phi}$ in the framework of an effective classical theory of gravity on $AdS_5\times S^5$, 
we have to consider the solutions of its equations of motion.
Depending on the nature of the field $\hat{\phi}$ and on its mass $m$, the behavior of the classical solution in the near boundary region is different.
Let us consider for instance a massive scalar. We introduce a new $AdS_5$ radial coordinate
$z= r_H/r$ and then identify the conformal boundary with $z=0$ and the horizon with $z=1$.
The metric \eqref{AdSmetric} for $AdS_5\times S^5$ becomes
\begin{equation}
 ds^2_{AdS_5\times S^5} = \frac{1}{R^2 z^2} (-dt^2 + \sum_{i=1}^3 dx_i^2) + \frac{R^2}{z^2} dz^2 + R^2 d\Omega^2_5\ .
\end{equation}
Let us consider just the $AdS_5$ part neglecting the compact directions,
\begin{equation}\label{AdS5}
 ds^2_{AdS_5} = g_{mn} dx^m dx^n = \frac{1}{R^2 z^2} (-dt^2 + \sum_{i=1}^3 dx_i^2) + \frac{R^2}{z^2} dz^2 \ ,
\end{equation}
where $m,n=1,...,5$ and we define $z\doteq x^5$.

Considering $R=1$, the action for the massive scalar $\hat{\phi}$ is
\begin{equation}\label{mass_sca_act}
 \begin{split}
 S_{\text{scal}} &\sim \int_{AdS_5} dx^5 \sqrt{-g} \left(g^{mn}\partial_m\hat{\phi}\, \partial_n\hat{\phi} + m^2 \hat{\phi}^2 \right)\\
&= \int_{AdS_5} dx^4 dz \frac{1}{z^{5}} \left(z^2 (\partial_z\hat{\phi})^2 + z^2 (\partial_\mu\hat{\phi})^2 + m^2 \hat{\phi}^2 \right) \ ,
 \end{split}
\end{equation}
and the corresponding equation of motion is given by
\begin{equation}
 \partial_z \left(\frac{1}{z^{3}} \partial_z \hat{\phi} \right) + \partial_\mu \left(\frac{1}{z^{3}} \partial^\mu \hat{\phi}\right)
 = \frac{1}{z^{5}} m^2 \hat{\phi} \ .
\end{equation}

Suppose at first that we seek solutions being independent on the $4$-dimensional space-time coordinates $x^\mu$ so
that the equation of motion simplifies as follows:
\begin{equation}
 \partial_z \left(\frac{1}{z^{3}} \partial_z \hat{\phi} \right) = \frac{1}{z^{5}} m^2 \hat{\phi} \ .
\end{equation}
We can find two independent solutions having a power behavior $z^\Delta$ with $\Delta$ satisfying%
\footnote{For a generic dimensionality $d$ of the boundary theory we have
\begin{equation}
 \Delta(\Delta-d) = m^2\ .
\end{equation}
See \cite{Klebanov:1999tb} and references therein.
}
\begin{equation}
 \Delta(\Delta - 4) = m^2\ .
\end{equation}
Therefore
\begin{equation}
 \Delta_\pm = 2 \pm \sqrt{4 + m^2} \ ;
\end{equation}
let us define 
\begin{equation}\label{Del_def}
 \Delta_+ \equiv \Delta \geq 2 \ , \ \ \ \ \Delta_- = 4 - \Delta \leq 2\ 
\end{equation}
where the equalities hold for $m^2=0$%
\footnote{Remember that on $AdS$ space we can have negative values of $m^2$ as long as they satisfy the Breitenlhoner-Freiedman stability bound.}.
We have the following general solution:
\begin{equation}\label{sol_nx}
 \hat{\phi} = \hat{\phi}_-\, z^{\Delta_-} + \hat{\phi}_+\, z^{\Delta_+} 
            = \hat{\phi}_-\, z^{4-\Delta} + \hat{\phi}_+\, z^\Delta \ .
\end{equation}
From the metric \eqref{AdS5} we have that $\sqrt{g}\sim 1/z^{5}$;
remembering that the action of the scalar action is quadratic in the field,
the solution $\hat{\phi}_-\, z^{4-\Delta}$ is always not-normalizable at the 
boundary whereas $\hat{\phi}_+\, z^\Delta$ can be both normalizable or not depending on the actual value of $\Delta$
\begin{equation}\label{nor_bou}
 \begin{split}
 &\int_{AdS_5, 0\leq z \leq \overline{z}} dx^5 \sqrt{g}\, |\hat{\phi}_-|^2 z^{2(4-\Delta)} 
  \sim \int_0^{\overline{z}} dz\  z^{2(4-\Delta)-5} = \infty \\
 &\int_{AdS_5, 0\leq z \leq \overline{z}} dx^5 \sqrt{g}\, |\hat{\phi}_+|^2 z^{2\Delta}
  \sim \int_0^{\overline{z}} dz\  z^{2\Delta-5} = \left\{\begin{array}{cl}
                                                       \infty \ \ &\text{for}\ \ 2\leq \Delta \leq 5/2\\
                                                       < \infty \ \ &\text{for}\ \ \Delta > 5/2 
                                                      \end{array}\right.\ ,
 \end{split}
\end{equation}
where $\overline{z}$ is some finite value of the $AdS_5$ radial coordinate $z$.
So, for $\Delta < 5/2$ we have two non-normalizable terms and, following \cite{Klebanov:1999tb}, we can choose between two
possibilities for the quantization on $AdS_5$ space; in other words, we can either consider $\phi_-$ to play the r\^{o}le 
of the source and interpret $\phi_+$ as the associated boundary operator or the opposite, namely $\phi_+$ the source and $\phi_-$ the
boundary operator.

So far we have assumed that the classical scalar profile does not depend on the coordinates $x^\mu$ which span the boundary.
Introducing this dependence, we extend the argument just performed and, in particular,
we obtain the following behavior for the solution of the bulk equation of motions
\begin{equation}
 \hat{\phi}(x^\mu,z) = \hat{\phi}_-(x^\mu) z^{4-\Delta} + \hat{\phi}_+(x^\mu) z^\Delta \ ,
\end{equation}
where $\Delta$ is the same as defined\footnote{This is not the most general solution;
a generalization can be obtained using the ansatz 
\begin{equation}
 z^\Delta \varphi(x^\mu)\ ,
\end{equation}
where the equation of motion imposes
\begin{equation}\label{genera}
 [\Delta(\Delta - 4) - m^2 ] \varphi + z^2 \partial_\mu \partial^\mu \varphi = 0 \ .
\end{equation}
In the main text we have asked that th e two terms in \eqref{genera} are solved separately.

 } in \eqref{Del_def} and
\begin{equation}\label{sol_x}
 \partial_\mu \partial^\mu \hat{\phi}_{\pm} (x^\mu) = 0 \ .
\end{equation}

\section{Boundary conditions and relation between bulk mass and conformal dimension of the dual operator}

As we have already observed, in the near-boundary analysis, the leading term in \eqref{sol_nx} and \eqref{sol_x}
is singular at $z=0$.
To impose the boundary conditions for $\hat{\phi}$ at $z=0$ we consider 
\begin{equation}\label{bou_con}
 \hat{\phi}(x^\mu,z) \overset{z\rightarrow 0}{\longrightarrow} z^{4-\Delta} \phi(x^\mu) \ .
\end{equation}
The function $\phi(x^\mu)$ is defined on the boundary and fixes the asymptotic boundary conditions for the
bulk field $\hat{\phi}$; furthermore, it is identified with the source of the dual operator ${\cal O}$ of the
conformal theory living at the boundary.

From the action of the massive scalar on $AdS_5$, \eqref{mass_sca_act}, we observe that
both the mass $m$ and the scalar field itself $\hat{\phi}$ are pure numbers.
Dimensional analysis on the boundary condition \eqref{bou_con} implies that the source function $\phi(x^\mu)$
possesses the dimension of $[\text{length}]^{\Delta - 4}$.
Eventually, from the source term of the boundary CFT \eqref{sou_ter} we see that the dimension of the operator
${\cal O}$ has to be therefore $\Delta$. We got a precise relation between the mass of the bulk field
and the scaling dimension of the corresponding operator.
Even though we are studying the detail of a simple case represented by a massive scalar,
similar arguments can be performed for any kind of bulk field and analogous relations between
their masses and the conformal dimension of the corresponding dual operator can be found.

As already mentioned, the two UV asymptotic terms of the scalar profile can have interchangeable roles: source and response%
\footnote{We refer to \cite{Klebanov:1999tb} to a complete analysis of this point.}.
The two choices correspond to two different boundary conformal models.
From the bulk viewpoint, the two choices correspond to two different ``quantizations''.
Indeed the two alternatives are related to the boundary condition choice, which represents the preliminary step
of quantizing the solutions on $AdS$.

\chapter{Clues for \texorpdfstring{$AdS$}{}/CFT Correspondence}
\label{clues}
\section{The decoupling limit}

In this section we describe the context which suggested Maldacena's conjecture.
The pivotal ingredient is a stack of D$3$-branes living in $10$-dimensional Minkowski space-time.
We are then working within the framework of Type IIB string theory.
In the full-fledged version of string theory, a stack of D$3$-branes is described as a source for
closed strings and represents a set of hyper-surfaces on which open strings can end.
The open strings describe the dynamics of the branes and the closed string sector describes the gravitational
interaction of the branes with the surrounding environment.

A second possibility is to describe the same stack of D-branes from the Type IIB supergravity viewpoint. 
Within the supergravity framework, the branes are solitonic solutions of the $10$-dimensional SUGRA equations of motion.

In order to appreciate how the two different points of view could suggest the $AdS$/CFT 
correspondence conjecture we have to
study a particular low-energy regime.
Let us focus on the low-energy Wilsonian effective description of Type IIB string theory in the presence of the D$3$-brane
stack. In other terms, we consider the effective theory of the massless modes obtained by integrating the 
effects of all the massive modes. More precisely, we move towards low energy maintaining fixed all the dimensionless
quantities (such as the string coupling constant $g_s$ and the number of branes $N$) and sending $\alpha'$ to zero.
As we have already seen, this implies that also the characteristic string length $l_s$ vanishes and the string tension
diverges.
Such effective description is associated to an action possessing the following schematic structure:
\begin{equation}
 S_{\text{eff}} = S_{\text{brane}} + S_{\text{bulk}} + S_{\text{interaction}}\ ,
\end{equation}
where $S_{\text{brane}}$ is the low-energy ${\cal N}=4$ SYM SU$(N)$ gauge theory accounting for the open-string massless modes,
$S_{\text{bulk}}$ describes the massless SUGRA multiplet in the $10$-dimensional bulk and 
$S_{\text{interaction}}$ describes the coupling of the 
stack of branes with the surrounding gravitational environment.

Let us remind the reader that $\kappa$, i.e. the square root of Newton's constant, is related to the string coupling $g_s$ and
Regge's slope parameter $\alpha'$ as follows:
\begin{equation}
 \kappa \propto g_s \alpha'^2\ .
\end{equation}
In the low-energy limit $\kappa$ vanishes and the gravitational interactions become negligible.
This has two consequences: $S_{\text{bulk}}$ reduces to an action describing free $10$-dimensional
propagation of the massless modes of the $10$-dimensional SUGRA multiplet, in other terms the gravitational
interaction terms can be forgotten; secondly, $S_{\text{interaction}}$ describing the coupling of the closed-string sector
with the stack of branes vanishes. Any term contained in $S_{\text{interaction}}$ is proportional
to some power of $\kappa$ because it describes how the stack of branes sources the closed string modes.

In the low-energy limit under consideration, the brane dynamics decouples from the free dynamics of string modes in the bulk;
the limit is commonly referred to as \emph{decoupling limit}.

Let us now turn the attention on the supergravity description of the same stack of D$3$-branes focusing on
the same low-energy decoupling limit.
In the SUGRA D-brane solution the metrics presents the following form \cite{Aharony:1999ti}
\begin{equation}\label{sugra_brane}
 ds^2 = f^{-1/2} (-dt^2 + \sum_{i=1}^3 dx_i^2) + f^{1/2} (d r^2 + r^2 d\Omega_5^2)\ ,
\end{equation}
where $r$ and $\Omega_5$ represent respectively the radius and the solid angle of the ``spherical''
coordinate in the space transverse to the branes, whereas $x_i$ span the space-like part of the D3's 
world-volume.
The function $f$ is
\begin{equation}\label{sugra_brane2}
 f = 1 + \frac{R^4}{r^4}\ , 
\end{equation}
where
\begin{equation}\label{sugra_brane3}
  R^4 = 4\pi g_s \alpha'^2 N \ .
\end{equation}

Since in the stringy picture we have concentrated on the massless modes, here we have to
study the low-energy excitations as well. 
Let us consider an observer placed at large distance from the stack, i.e. $r\gg 1$.
From the large-distance viewpoint, there are two kinds of low-energy modes:
the long-wave-length propagating in the $10$-dimensional space-time and the modes
living in the vicinity of the branes.
The latter type of low-energy excitations is due to red-shift effects seen from the observer at infinity.
Indeed, an object placed at radial distance $r_{obj}$ from the branes and having energy $E_{obj}$ in 
the proper frame, when observed from infinity it has the following red-shifted energy
\begin{equation}
 E' = \sqrt{g_{tt}(r_{obj})}\ E_{obj} = f^{-1/4}(r_{obj})\ E_{obj}\ ;
\end{equation}
when the object is close to the branes, $r_{obj}\rightarrow 0$, the red-shift factor $f^{-1/4}(r_{obj})$
vanishes.

On top of this, notice that if we consider the metric solution \eqref{sugra_brane} in the near-brane region,
we can trade $f$ with $R^4/r^4$ obtaining then:
\begin{equation}
 ds^2 \overset{r\ll R}{\longrightarrow} \frac{r^2}{R^2} (-dt^2 + \sum_{i=1}^3 dx_i^2) 
+ \frac{R^2}{r^2} dr^2 + R^2 d\Omega_5^2\ ;
\end{equation}
In the near-brane limit, the metric describes then the geometry of $AdS_5 \times S^5$ space-time.

The supergravity description of the D$3$-branes in the low-energy limit presents, from the large distance viewpoint,
two separate, i.e. decoupled, sectors: the long wave-length modes in the bulk and the near-brane excitations.
To become aware of their decoupling we can naively observe that the branes become ``transparent'' to the
long-wave bulk modes, their wave-length being much bigger than the characteristic near-brane region size $R$.
More precisely, one ought to consider the absorption cross section of bulk modes scattering on the branes;
for low energy this scattering is proportional to $\omega^3$ and therefore it vanishes for $\omega$ tending to zero.

The decoupling limit yielded two distinct and decoupled low-energy sectors both in the stringy and SUGRA descriptions
of the D$3$-branes. It is natural to identify the free propagation of the SUGRA closed string modes with
the low-energy bulk excitations of the supergravity solutions.
The next and essential step is about the possibility of identifying the near-branes SUGRA modes with the gauge
theory living on the branes accounting for open strings. This is indeed the inspiration to claim the $AdS$/CFT correspondence.

\subsection{Correspondence between symmetry groups}
\label{cor_group}

Anti-de Sitter space-times are maximally symmetric solutions of Einstein's equations with Minkowskian signature.
They present a negative value for the cosmological constant which is related to their constant radius of curvature.

Let us focus on the $5$-dimensional case.
The family of $AdS_5$ spaces corresponds to the space of solutions of the
quadratic equation
\begin{equation}\label{ads_qua}
 t^2 + z^2 - \sum_{i=1}^4 x_i^2 = R^2 \ ,
\end{equation}
where $R$ represents the radius of curvature.
In these terms, an $AdS_5$ space-time is interpreted as a quadratic surface of constant curvature
embedded in flat $6$-dimensional space-time with hyperbolic signature $(+,+,-,-,-,-)$.
From \eqref{ads_qua} it results manifestly that $AdS_5$ is invariant under symmetry transformations
belonging to O$(2,4)$ which constitutes its isometry group.

We turn the attention to the dual field theory and its conformal structure.
The base manifold is $4$-dimensional space-time with Minkowskian signature, 
whose conformal group is defined by the following set of infinitesimal transformations
\begin{equation}\label{conf_alg}
 \begin{array}{lccl}
  \delta x^\mu = a^\mu                         & \leftrightarrow & & \bm{P}^\mu \\
  \delta x^\mu = \omega^\mu_{\ \nu} x^\nu      & \leftrightarrow & & \bm{J}^\mu_{\ \nu} \\
  \delta x^\mu = \lambda x^\mu                 & \leftrightarrow & & \bm{D} \\
  \delta x^\mu = b^\mu x^2 - 2 x^\mu b\cdot x  & \leftrightarrow & & \bm{K}^\mu 
 \end{array} \ ,
\end{equation}
where $a,\omega,\lambda,b$ are the parameters and $\bm{P},\bm{J},\bm{D},\bm{K}$ are
the generators of the translations, rotations/Lorentz's transformations, dilatations and
special conformal transformations respectively.
The parameter counting gives
\begin{equation}
 4 + \frac{4(4-1)}{2} + 1 + 4 = 15 \ .
\end{equation}
Studying explicitly the algebra associated to the whole set of conformal generators \eqref{conf_alg},
it is possible to show that it is equivalent to the SO$(2,4)$ algebra.
In addition to the infinitesimal transformations \eqref{conf_alg}, there is also the discrete
conformal transformation
\begin{equation}\label{con_inv}
 x^\mu \rightarrow \frac{x^\mu}{x^2} \ .
\end{equation}
To appreciate that \eqref{con_inv} defines indeed a conformal transformations,
let us concentrate on the transformation of the metric element
\begin{equation}
 dx^\mu \rightarrow - 2\frac{x^\mu}{x^4} x^\nu dx_\nu + \frac{dx^\mu}{x^2} \ ,
\end{equation}
whose square is
\begin{equation}
 (dx)^2 \rightarrow \frac{1}{x^4} (dx)^2 \ ,
\end{equation}
representing actually a conformal transformation of the metric\footnote{
For a generic conformal (i.e. angle-preserving) transformation, the squared metric elements
transforms as
\begin{equation}\label{squa_ele}
 ds^2 \rightarrow \Lambda(x)\, ds^2 \ ,
\end{equation}
where the function $\Lambda$ is usually referred to as \emph{conformal factor}.
Notice that, being the angles defined by ratios of infinitesimal square elements
transforming as \eqref{squa_ele}, the factor $\Lambda$ introduced by a conformal
transformations both at the numerator at the denominator cancels leaving the angle invariant.}.

Considering also the discrete conformal inversion \eqref{con_inv} enhances the SO$(2,4)$ group
to full O$(2,4)$.
As a check, notice that the counting of the parameters matches, in fact for O$(2,4)$ we have
$6(6-1)/2=15$ independent generators.

The crucial point consists in conjecturing the correspondence of the two O$(2,4)$
groups that we have just found. 
This cannot be regarded as a proof of $AdS$/CFT correspondence, but is nevertheless a significant necessary clue.

\chapter{Meissner-Ochsenfeld Effect}
\label{MEI}
The Meissner-Ochsenfeld effect consists in the expulsion of the magnetic field from the interior of a superconductor.
The phenomenon can be described relying on the London equation \eqref{Londres}; specifically, let us consider
the curl of both members of \eqref{Londres},
\begin{equation}\label{curlo}
 \nabla \wedge \bm{j}_s = - \frac{n_s e^2}{m c} \bm{B}\ .
\end{equation}
Let us consider the Maxwell equation for the field $B$,
\begin{eqnarray}\label{Bequa}
 &\nabla \wedge \bm{B} &= \frac{4\pi}{c} \bm{j}\\
 &\nabla\cdot \bm{B}&=0 \label{Bequa2}
\end{eqnarray}
Taking again the curl of both sides of \eqref{curlo} and using \eqref{Bequa} and \eqref{Bequa2}, we obtain
\begin{equation}\label{BB}
 \nabla^2 \bm{B} = \frac{1}{\gamma^2} \bm{B}\ ,
\end{equation}
where we have defined
\begin{equation}
 \gamma = \left(\frac{mc^2}{4\pi n_s e^2}\right)\ .
\end{equation}
The quantity $\gamma$ is related to the penetration depth of the magnetic field inside the superconductor;
take a configuration depending only on one coordinate, say $x$, so that there is an interface at $x=0$ and
we have superconduction for $x>0$ and vacuum for $x<0$. The shape of the solution of
\eqref{BB} (which holds in the interior of the superconductor) is
\begin{equation}\label{penetration}
 e^{\pm\frac{x}{\gamma}}\ .
\end{equation}
The solution describing the profile within the superconductor is the exponentially suppressed one and it has to be matched to
the boundary value assumed by the magnetic field, i.e. $B(x=0)$.
It is then evident from \eqref{penetration} that $\gamma$ represents a characteristic penetration depth.

Note that the Meissner-Ochsenfeld effect is not simply a consequence of infinite DC conductivity.
Zero resistance alone would imply that the medium reacts to any attempt of magnetization with compensating currents loops
according to Faraday-Neumann-Lenz law. 
If the material is instead magnetized in a phase without perfect conductivity and later led thermodynamically to the
perfect-conductivity phase without altering the magnetization, we can have magnetization inside the medium and perfect conductivity together.
This contrasts the superconductor phenomenology where, in the superconducting phase, the Meissner-Ochsenfeld
occurs independently on the magnetic history of the system.
This fact emerges also from London theory: assuming just perfect conductivity $\bm{E} \propto \partial_t \bm{j}$ instead of the London equation \eqref{londra},
is not enough to derive the Meissner-Ochsenfeld effect; taking the curl of the perfect conductivity relation we obtain
\begin{equation}
 \partial_t \bm{B} \propto \partial_t (\nabla \wedge \bm{j})\ ,
\end{equation}
that, for stationary conditions, is automatically satisfied. 
As we have just said, perfect conductivity is compatible with constant but also
finite values of the magnetic field inside the material whereas superconductivity is not.

%
%
%
%
%

\chapter{Probe Approximation}
\label{probe}

As a first step into the detailed analysis of the system \eqref{laga} we
consider a particular simplifying limit,
the so called \emph{probe approximation}.
Instead of considering the full dynamics of the model, we regard the gauge field
$A$ and the scalar field $\psi$ as small perturbations
that leaves the background of the other fields unaffected.
In other terms, we consider classical solutions of the theory in which $A$ and
$\psi$ are disregarded and on these background solutions 
(which we hold fixed) we consider
the fluctuations of $A$ and $\psi$.
This particular choice is due to the fact that $A$ and $\psi$ are coupled.
Hence, we could not have chosen $A$ to be a ``small perturbation'' while
regarding $\psi$
as belonging to the background.

The probe approximation can be read as the large-charge limit, $q\gg 1$. 
Let us write the full Lagrangian density \eqref{laga} in terms of the rescaled
fields $\hat{\psi}=q\,\psi$ and $\hat{A}=qA$.
We consider the limit $q\rightarrow\infty$ while the hatted fields are kept
fixed.
The part of the Lagrangian density involving the rescaled fields
is\footnote{Here we
are considering the specific potential chosen in \eqref{potcho}.}:
\begin{equation}
 \hat{\cal L} = \frac{1}{2\kappa_4^2 q^2} \sqrt{-\text{det}\, g} \left[ 
-\frac{1}{4} \hat{F}^{ab}\hat{F}_{ab} - |\partial\hat{\psi} - \im  \hat{A}
\hat{\psi}|^2 - \frac{m^2}{L^2}|\hat{\psi}|^2 \right]\ .
\end{equation}
In the large $q$ limit it decouples from the metric and $B$ parts; these will be
then regarded as the Lagrangian density
for the background,
\begin{equation}\label{laga_back}
 {\cal L}_{\text{background}} = \frac{1}{2\kappa_4^2} \sqrt{-\text{det}\, g}
\left[ R + \frac{6}{L^2} - \frac{1}{4}Y^{ab}Y_{ab} \right]\ ,
\end{equation}
that admits Reissner-Nordstr\"om $AdS$ black hole solutions,
\begin{eqnarray}
 ds^2 &=& - g(r)\, dt^2 + \frac{r^2}{L^2} (dx^2 + dy^2) + \frac{dr^2}{g(r)}\\
 g(r) &=& \frac{r^2}{L^2}\left(1-\frac{r^3_H}{r^3}\right) + \frac{1}{4} \mu_B^2
\frac{r_H}{r}\left(1-\frac{r}{r_H}\right)\\
 B_t &=& \mu_B \left(1-\frac{r_H}{r}\right)\ .
\end{eqnarray}
The black hole solution is charged with respect to the (now only one) gauge
field $B$. 

\chapter{The near-horizon geometry of RN black hole at low temperature.}
\label{nearhor}
Let us study the near-horizon geometry of the Reisnner-Nordstr\"{o}m black hole solution in $d$ spatial dimensions (the holographic model
analyzed in the main text has $d=2$).
In the zero-temperature limit the horizon radius assumes the value
\begin{equation}
 r_H^2=\frac{1}{2d}\frac{(d-2)^2L^2\mu^2}{(d-1)}\, .
\end{equation}
In order to obtain the near-horizon behavior of the metric we expand the function $g(r)$ around $r_H$%
\footnote{The function $g(r)$, sometimes called \emph{blackening factor}, has been introduced in \eqref{metric}.}
\begin{equation}
g(r)\simeq g(r_H)+g^\prime(r_H)\tilde{r}+\frac{1}{2}g^{\prime\prime}(r_H)\tilde{r}^2\,,
\end{equation}
where $r=r_H+\tilde{r}$ with $\tilde{r}\rightarrow0$.
For the RN black hole we find
\begin{equation}\label{pappetta}
g(r_H)=0\,, \qquad g^\prime(r_H)\sim T=0\,,\qquad g^{\prime\prime}(r_H)=\frac{2d(d-1)}{L^2}\, .
\end{equation}
As a consequence, the near-horizon metric is
\begin{equation}
ds^2_{\mbox{\scriptsize{near horizon}}}\simeq -d(d-1)\frac{\tilde{r}^2}{L^2}dt^2+\frac{r_H^2}{L^2}d\vec{x}^2+\frac{L^2}{d(d-1)\tilde{r}^2}d\tilde{r}^2\,,
\end{equation}
from which we recognize the $AdS_2\times R^{d-1}$ form. 
In particular, the $AdS_2$ radius squared is 
\begin{equation}
 L_{(2)}^2=\frac{L^2}{d(d-1)} \ ,
\end{equation}
as stated in the main text.

\chapter{Green's Functions and Linear Response}
Let us define the \emph{retarded} Green function relating two operators labeled with $A$ and $B$
as follows
\begin{equation}\label{Green}
 G^{R}_{{\cal O}_A {\cal O}_B} (t_2 - t_1,\bm{x}_2 - \bm{x}_1)= -\im \ \theta(t_2 - t_1)\
<[{\cal O}_A(t_2,\bm{x}_1),{\cal O}_B(t_1,\bm{x}_1)]> 
\end{equation}
The Green function is essentially given by the expectation value of the commutator
between the two observables\footnote{
The corresponding Fourier transformed Green function (useful in the following) is:
\begin{equation}\label{GreenFou}
 G^{R}_{{\cal O}_A {\cal O}_B} (\omega,\bm{k})= -\im \int_0^t dt \int_V d^{d-1}x\
e^{\im (\omega t - \bm{k}\cdot\bm{x})}\ \theta(t)\
<[{\cal O}_A(t,\bm{x}),{\cal O}_B(0,0)]> 
\end{equation}
where the translation invariance has been used.}.
The $\theta(\Delta t)$ indicates the Heaviside step function and
it represents the key feature that makes the Green function retarded.

The retarded Green function \eqref{Green} expresses quantitatively how 
a perturbation proportional to ${\cal O}_B$ affects the expectation of ${\cal O}_A$.
Let us see this in detail.
We consider the following time-dependent perturbation to the Hamiltonian
\begin{equation}\label{pert}
 \delta H(t) = \int_V d^{d-1}x\ {\cal O}_B (t,\bm{x})\ \phi(t,\bm{x})\ ,
\end{equation}
where $\phi$ is a coefficient function and not an operator.
The expectation value of the observable ${\cal O}_A$ is
\begin{equation}\label{ExpH}
 <{\cal O}_A> (t,\bm{x}) =
\text{tr} \{ \, \rho \, {\cal O}_A(t,\bm{x}) \, \}\ ,
\end{equation}
where $\rho$ represents the density matrix.
As long as we consider the Heisenberg picture, the quantum states do not evolve and therefore
neither does the density matrix that is defined through them.
We now switch from the Heisenberg picture to the interaction picture.
We then express the time evolution of a state with the operator
\begin{equation}
 U(t) = e^{-\im \int_0^t dt' \delta H(t')} \ .
\end{equation}
We then have
\begin{equation}
 |\psi(t)> = U(t)|\psi(0)>\ .
\end{equation}
The expectation value of ${\cal O}_A$ becomes
\begin{equation}\label{ExpI}
 <{\cal O}_A> (t,\bm{x}) =
\text{tr} \{  U(t)  \, \rho \, U(t)^{-1} \, {\cal O}_A(t,\bm{x})\}\ .
\end{equation}
Here the time-dependence in ${\cal O}_A$ is due to the unperturbed Hamiltonian alone.
Using the cyclicity of the trace and expanding \eqref{ExpI} to the first order in the perturbation we obtain
\begin{equation}
 \begin{split}
 \delta <{\cal O}_A> (t,\bm{x}) =&\
\im\ \text{tr} \left\{  \rho \, \int_0^t dt' \int_V d^{d-1}x' \ [{\cal O}_B(t',\bm{x}'),{\cal O}_A(t,\bm{x})] \ \phi(t',\bm{x}') \right\} + ...\\
=&\ - \im\ \int_0^t dt' \int_V d^{d-1}x' \, <[{\cal O}_A(t,\bm{x}),{\cal O}_B(t',\bm{x}')]> \, \phi(t',\bm{x}') + ...\\
=&\ -\im\ \int_0^t dt' \int_V d^{d-1}x' \ G^{R}_{{\cal O}_A {\cal O}_B}(t-t',\bm{x}-\bm{x}') \ \phi(t',\bm{x}') + ...
 \end{split}
\end{equation}
The linear response of the mean value of the operator ${\cal O}_A$ to the perturbation \eqref{pert}
is accounted for by the Green function \eqref{Green}. Assuming translation invariance and taking the Fourier transform, we obtain
\begin{equation}\label{FouLin}
 \delta <{\cal O}_A> (\omega,\bm{k}) = \tilde{G}^{R}_{{\cal O}_A {\cal O}_B}(\omega,\bm{k}) \ \phi(\omega,\bm{k}) + ...
\end{equation}

\section{Causality and analyticity properties}
\label{causa}

The retarded function is causal by definition. 
This is manifest noticing the presence of the Heaviside step function in \eqref{Green}.
Let us consider the inverse Fourier transform of the Green function \eqref{GreenFou},
\begin{equation}
 G^{R}_{{\cal O}_A {\cal O}_B}(t,\bm{x}) = \int \frac{d\omega}{2\pi}\, e^{-\im\omega t} 
\, G^{R}_{{\cal O}_A {\cal O}_B}(\omega,\bm{x})
\end{equation}
Since for $t<0$ we want the Green function to be zero (this is another way to state causality)
the function inside the integral must not have poles in the upper complex $\omega$-plane.
Said otherwise, it is there analytic.
We are of course assuming that the integrand vanishes fast enough as $|\omega|\rightarrow \infty$
so that we can evaluate the integral with the residues method.
The Kramers-Kronig equations relating the imaginary and real parts of the retarded Green function 
descend directly from this analyticity condition and then from causality.

%

\chapter{Onsager's Reciprocity Relation}
\label{Onsa}
In this appendix we follow the lines described in \cite{Herzog:2009xv}.
The entries of the conductivity matrix 
\begin{equation}\label{calJ}
 {\cal J}(\omega) = \hat{\sigma}(\omega)\ {\cal E}(\omega),
\end{equation}
are proportional to the corresponding two-point retarded correlators,
\begin{equation}\label{condgre}
 \hat{\sigma}_{IJ}(\omega,\bm{k}) =\, \frac{1}{\im\, \omega}\, \tilde{G}^R_{IJ} (\omega,\bm{k}).
\end{equation}
Note that \eqref{calJ} is the generalization of the static Ohm law to the case presenting harmonic time dependence
and it represents a rewriting of \eqref{FouLin}\footnote{We
have set the momentum $\bm{k}$ to zero; moreover we identified $\delta<{\cal O}_A>$ with $\cal J$ and $\phi$ with the electromagnetic potential.
The factor $\im\, \omega$ in \eqref{condgre} is due to the fact that for $\bm{k}=0$ the electric field is the time derivative of the potential.}.

Onsager's reciprocity relation implies that a system possessing time-reversal invariant equilibrium states
presents a symmetric conductivity matrix for zero-momentum external perturbations, namely
\begin{equation}\label{symcon}
 \hat{\sigma}_{IJ}(\omega,\bm{0}) = \hat{\sigma}_{JI}(\omega,\bm{0})\ .
\end{equation}

To derive Onsager's formula in full generality, let us consider the possibility of having a non-vanishing external magnetic field $\bm{B}$ which
is mapped to $-\bm{B}$ by the time-reversal operation.
In the presence of a finite magnetic field, the assumption of time-reversal invariance of the equilibrium states generalizes to
\begin{equation}
 \Theta |\bm{B}> = |-\bm{B}> ,
\end{equation}
where $\Theta$ is the time-reversal operator.

Let us consider two observables (i.e. Hermitian operators) $\phi$ and $\psi$ that behave as follows
\begin{eqnarray}
 \Theta \phi \Theta &=& \eta_\phi\, \phi^\dagger = \eta_\phi\, \phi\\
 \Theta \psi \Theta &=& \eta_\psi\, \psi^\dagger = \eta_\psi\, \psi,
\end{eqnarray}
where the $\eta$'s are either $+$ or $-$ one. In coordinate space, we have that the correlator between
the two operators in the equilibrium state $|\bm{B}>$ is given by
\begin{equation}\label{greons}
 G^R_{\phi\psi}(t,\bm{x};\bm{B}) = <\bm{B}| [\phi(t,\bm{x}),\psi(0,\bm{0})]|\bm{B}> .
\end{equation}
Let us consider the time-reversal transformation of \eqref{greons}:
\begin{equation}
 \begin{split}
 <\bm{B}| [\phi(t,\bm{x}),\psi(0,\bm{0})]|\bm{B}> &= <-\bm{B}|\Theta
[\phi(t,\bm{x}),\psi(0,\bm{0})] \Theta|-\bm{B}>^*\\
                                                  &= \eta_\phi \eta_\psi
<-\bm{B}| [\phi(-t,\bm{x}),\psi(0,\bm{0})] |-\bm{B}>^*\\
                                                  &= \eta_\phi \eta_\psi
<-\bm{B}| [\psi(0,\bm{0}),\phi(-t,\bm{x})] |-\bm{B}>\\
                                                  &= \eta_\phi \eta_\psi
<-\bm{B}| [\psi(t,-\bm{x}),\phi(0,\bm{0})] |-\bm{B}>
 \end{split}
\end{equation}
In the last passage we have used the assumption of translation invariance.
Therefore, if we consider the Fourier component corresponding to $k^\mu =
(\omega,\bm{0})$, we have
\begin{equation}
 \tilde{G}^R_{\phi\psi}(\omega,\bm{0};\bm{B}) = \eta_\phi \eta_\psi\, 
\tilde{G}^R_{\psi\phi}(\omega,\bm{0};-\bm{B}) \ .
\end{equation}
In the case of zero external magnetic field it reduces to
\begin{equation}
 \tilde{G}^R_{\phi\psi}(\omega,\bm{0};\bm{0}) = \eta_\phi \eta_\psi\, 
\tilde{G}^R_{\psi\phi}(\omega,\bm{0};\bm{0}) \ ,
\end{equation}
which is known as \emph{Onsager's reciprocity relation}.
In \eqref{calJ} $\cal J$ is an array of vector currents. Any current $J_I^\mu$
is an odd operator under $\Theta$, then $\eta_I=-1$ for all values of $I$.
This leads to
\begin{equation}
 \tilde{G}^R_{IJ}(\omega,\bm{0};\bm{0}) = \tilde{G}^R_{JI}(\omega,\bm{0};\bm{0})
\ ,
\end{equation}
Looking at \eqref{condgre}, we have that \eqref{symcon} is proved\footnote{In
the case $\bm{B}\neq0$ it simply generalizes to
\begin{equation}
\hat{\sigma}_{IJ}(\omega,\bm{0};\bm{B}) =
\hat{\sigma}_{JI}(\omega,\bm{0};-\bm{B}) \ ,
\end{equation}}.

\chapter{Josephson Effect}
\label{joseph}
In this appendix we take inspiration from similar arguments described in \cite{2009arXiv0908.1761P}.
In a superconducting state, quantum coherence is macroscopic.
Let us describe the superconducting state with a single macroscopic wave-function:
\begin{equation}
 \Psi = \Psi(\bm{r}) = |\Psi(\bm{r})|\, e^{\im \phi(\bm{r})} \ .
\end{equation}
Loosely speaking, it describes the wave-function of the Cooper pairs condensate.

By definition a Josephson junction is constituted by two superconductors divided by a thin insulating layer.
In other terms, we have two weakly coupled distinct superconductors which we label $1$ and $2$.
The coupling between the two superconductors is due to the fact that, being the insulating layer very thin 
there is a non-null overlap between the wave-functions describing the two superconductors. 
Moreover, the fact that a wave-function of, say, superconductor $1$ is non-vanishing also in a region beyond the insulating layer, so 
in the volume of superconductor $2$, allows for quantum tunneling of Cooper pairs of species $1$. The converse is analogously true.

Let us describe the whole Josephson junction as a single macroscopic quantum state,
\begin{equation}
 |\Psi\rangle = |\Psi_1 \rangle + |\Psi_2 \rangle \ ,
\end{equation}
each terms refer respectively to the superconductors $1$ and $2$.
The density of elementary carriers in the condensate (i.e. the Cooper pairs) of each superconductor is given by
\begin{equation}\label{orma}
 n_i = |\Psi_i|^2 = \langle \Psi_i |\Psi_i\rangle
\end{equation}
The time evolution of the junctions is described by the Schr\"{o}dinger equation
\begin{equation}
 \im \hbar \partial_t |\Psi\rangle = H |\Psi\rangle \ ,
\end{equation}
where the Hamiltonian has the following shape
\begin{equation}
 H=H_1+H_2+H_{\text{int}} \ .
\end{equation}
Let us account for the weak coupling between the two superconductors assuming an interaction term of the following form
\begin{equation}
 H_{\text{int}} = -\frac{\alpha}{2} \left(|\Psi_1\rangle\langle \Psi_2| + |\Psi_2\rangle\langle \Psi_1 | \right)
\end{equation}

If we project the Schr\"{o}dinger equation on the two states $\langle \Psi_i|$ we obtain the following system of coupled  equations
\begin{eqnarray}\label{ems}
 & \im \hbar \partial_t \Psi_1 = E_1 \Psi_1 -\frac{\alpha}{2} \Psi_2 &\\
 & \im \hbar \partial_t \Psi_2 = E_2 \Psi_2 -\frac{\alpha}{2} \Psi_1 & \label{ems2}
\end{eqnarray}
Let us rewrite the wave-functions in the following fashion
\begin{equation}
 \Psi_i = \sqrt{n_i} e^{\im \phi_i}
\end{equation}
Taking then the real and imaginary parts of the system \eqref{ems} and \eqref{ems2} we can write it in terms of the densities and phases,
\begin{eqnarray}
 & \partial_t n_1 = -\partial_t n_2 &= \frac{\alpha}{\hbar} \sqrt{n_1 n_2} \sin(\phi_1-\phi_2)\\
 & \partial_t (\phi_2-\phi_1) & = \frac{1}{\hbar} (E_1-E_2) + \frac{\alpha}{2\hbar} \frac{n_1-n_2}{\sqrt{n_1n_2}} \cos(\phi_1-\phi_2)
\end{eqnarray}
From the continuity equation, accounting for the conservation of the number of carriers, we have that the current through the junction
is equal to the time variation of the densities (with the appropriate signs depending on how we define positive the direction of the flow)
\begin{equation}
 j = \partial_t n_1 = -\partial_t n_2 = j_c \sin(\phi_1-\phi_2)\ ,
\end{equation}
where $j_c= \alpha \sqrt{n_1 n_2} /\hbar$. It is to be underlined that the current is not proportional to the energy difference
between the two superconductors. We can have neat flow also in the case $E_1=E_2$, that is when no external voltage unbalances the
junction. 

If we specialize to the case in which the Copper pair density is equal at the two sides of the junction,
namely $n_1=n_2=n$, the phase equation becomes
\begin{equation}\label{faso}
 \partial_t(\phi_2-\phi_1) = \frac{1}{\hbar} (E_1-E_2) \ .
\end{equation}
The phase difference does not evolve if there is no energy difference between the two states, i.e. when the 
junction is balanced. Given the normalization \eqref{orma}, the energy eigenvalues $E_i$ represent
the energy of an elementary carrier, therefore the energy of a Cooper pair. 
As a consequence, the difference $E_1-E_2$ is given by the potential energy lost or gained by a Cooper pair 
traversing the junction, i.e. $E_1-E_2 = 2 e V$ where $e$ is the charge of the single electron.
We can rewrite equation \ref{faso}
\begin{equation}
 \partial_t (\phi_2 - \phi_1) = \frac{2eV}{\hbar}
\end{equation}
For further details see, for instance, \cite{Jose} and \cite{2009arXiv0908.1761P}.

\chapter{Temperature Gradient, Heat Flow, Electric Fields and the Metric}
\label{QET}
\section{Vector fluctuations of the metric and temperature gradient}

This appendix in inspired by similar arguments in \cite{Herzog:2008he}.

The temperature is identified with the inverse period of the compact imaginary time.
Indicating the time with $t$ and the temperature with $T$ we have that 
$ \text{Im}(t) \in \left[0,\frac{1}{T}\right) $
with $g_{tt}=-1$ (the original metric for our boundary theory is Minkowski).
We define a new rescaled time coordinate $\tilde{t}= T t$ so that
$ \text{Im}(\tilde{t}) \in [0,1) $
and the temperature dependence is ``transfered'' to the $\tilde{t}\tilde{t}$ metric component,
\begin{equation}
 g_{\tilde{t}\tilde{t}} = -\frac{1}{T^2}\ .
\end{equation}
The mixed space/time components of the metric are null, $g_{i\tilde{t}}=0$ (where $i$ represents
the spatial index associated to the coordinate $x^i$).

Let us consider a small variation in the temperature, i.e. $\delta T(x)$, that can be space-dependent; the corresponding 
variation of $g_{\tilde{t}\tilde{t}}$ is
\begin{equation}
 \delta g_{\tilde{t}\tilde{t}} = 2 \frac{\delta T}{T^3}\ .
\end{equation}
As usual, we consider harmonic time dependence but this time in the rescaled time coordinate (and correspondingly rescaled frequency)
$ e^{\im \tilde{\omega}\tilde{t}}$.

We can perform a ``gauge'' transformation, i.e. a diffeomorphism generated by the vector parameter
\begin{equation}\label{vecgen}
 \xi_{\tilde{t}} =  \im \frac{\delta T}{\tilde{\omega} T^3}\ , \ \ \ \ \
 \xi_x = 0 
\end{equation}
The metric is correspondingly transformed
\begin{equation}
 g_{ab} \rightarrow g_{ab} + \partial_a \xi_b + \partial_b \xi_a + \xi_c \partial_c g_{ab}\ . 
\end{equation}
Specifically, the effect of the transformation generated by \eqref{vecgen} on $\delta g_{\tilde{t}\tilde{t}}$ is
\begin{equation}
 \begin{split}
 \delta g_{\tilde{t}\tilde{t}} \rightarrow& \delta g_{\tilde{t}\tilde{t}} + 2 \partial_{\tilde{t}} \xi_{\tilde{t}}\\
 =& \delta g_{\tilde{t}\tilde{t}} - 2 \frac{\delta T}{T^3} = 0
 \end{split}
\end{equation}
where, in the first passage, we have used $\partial_{\tilde{t}} g_{\tilde{t}\tilde{t}}=  0 $.
The re parametrization affects the mixed $i \tilde{t}$ components as well, actually we have
\begin{equation}
 \delta g_{i\tilde{t}} =  \partial_i \xi_{\tilde{t}} = \im \frac{\partial_i \delta T}{\tilde{\omega}T^3}\ ,
\end{equation}
where we have to remember that initially both $g_{i\tilde{t}}$ and $\delta g_{i\tilde{t}}$ were null.
Scaling back from $\tilde{t}$ to $t$ we eventually get
\begin{equation}\label{tempmet}
 \delta g_{i t} = \im \frac{\partial_i \delta T}{\omega T}\ ,
\end{equation}
where it is manifest that a vector fluctuation of the metric is associated to the gradient of the temperature fluctuation.

\section{Heat flow and electrical fields}

The variation of the vector potential due to a reparametrization is
\begin{equation}\label{gauvara}
 \delta_g A_\mu = A_\nu \xi^\nu_{\ ,\mu} + A_{\mu,\nu} \xi^\nu
\end{equation}
Since we start with Minkowski metric, the covariant derivatives in \eqref{gauvara} are normal partial derivatives.
We consider the reparametrization transformation induced by the vector field 
\begin{equation}\label{repanosca}
 \xi_{t} = \im \frac{\delta T}{\omega T} \ , \ \ \
 \xi_x = 0 
\end{equation}
This is the same transformation as in \eqref{vecgen} but without the temporal rescaling.
The reparameterization variation of the gauge potential due to \eqref{repanosca} is given by
\begin{equation}
 \delta_g A_i = -A_t \partial_i \xi_t = - \im \mu \frac{\partial_i \delta T}{\omega T}\ .
\end{equation}
The total variation of $A$ is given by two contributions:
as just seen, one is induced by the temperature fluctuation through the metric, the second is instead related to a proper, external electrical field,
\begin{equation}
 \begin{split}
  \delta A_\mu &= \delta_g A_\mu + \frac{E_\mu}{\im \omega} \\
                    &= - \mu \delta g_{i t}  + \frac{E_\mu}{\im \omega} \\
 \end{split}
\end{equation}
In the last passage we used \eqref{tempmet}.
Reading it the other way around, we have that the electrical field is related to the fluctuation of the bulk fields in the following way
\begin{equation}
 E_\mu = \im \omega \left(\delta A_\mu + \mu\, \delta g_{i t}\right)\ .
\end{equation}
Eventually the variation of the Hamiltonian is
\begin{equation}
 \delta {\cal H} = \int d^d x \left( T^{\mu\nu} \delta g_{\mu\nu} + J^\mu \delta A_\mu \right)\ .
\end{equation}
Expressing it in terms of the temperature gradient fluctuation (see \eqref{tempmet}) and the electric field we obtain
\begin{equation}
 \delta {\cal H} = \int d^d x \left[ \left(T^{i t} - \mu J^i \right) \frac{\partial_i \delta T}{\im \omega T} + J^\mu \frac{E_\mu}{\im \omega} \right]\ ,
\end{equation}
from which it arises naturally that the response to a temperature gradient fluctuation, i.e. the \emph{heat flow}, is given by
\begin{equation}
 Q^i = T^{it} - \mu J^i\ .
\end{equation}

\chapter{Holographic Renormalization of our Model}
\label{appe}
The covariant divergence of a vector $V$ is:
\begin{equation}\label{veto}
 \nabla_m V^m = \partial_m V^m + \Gamma^m_{\ ma}V^a\ .
\end{equation}
The connection symbol with the first two indexes saturated, i.e. $\Gamma^m_{\
ma}$, admits the following compact expression
\begin{equation}\label{logacov}
 \begin{split}
 \Gamma^m_{\ ma} &= \frac{1}{2} g^{mc} \left\{\frac{\partial g_{ac}}{\partial
x^m} + \frac{\partial g_{cm}}{\partial x^a} - \frac{\partial g_{ma}}{\partial
x^c} \right\} 
 =\frac{1}{2} g^{mc} \left\{ \frac{\partial g_{cm}}{\partial x^a} \right\} 
 =\frac{1}{2} \text{tr} \left(\hat{g}^{-1} \partial_a \hat{g} \right) \\
 &=\frac{1}{2} \text{tr} \ \partial_a \ln \hat{g} 
 =\frac{1}{2} \partial_a \text{tr} \ln \hat{g} 
 =\frac{1}{2} \partial_a \ln \text{det} \hat{g} 
 =\partial_a \ln \sqrt{g} \ ,
 \end{split}
\end{equation}
where $\hat{g}$ represents the metric in matrix notation, while $g$ represents
its determinant. 
From the passages in \eqref{logacov} we obtained
\begin{equation}
 \Gamma^m_{\ ma} = \frac{1}{\sqrt{g}} \partial_a \sqrt{g}\ .
\end{equation}
Putting this compact expression into \eqref{veto} we get
\begin{equation}
\nabla_m V^m = \partial_m V^m +\frac{1}{\sqrt{g}} (\partial_m \sqrt{g})\, V^m 
             = \frac{1}{\sqrt{g}} \partial_m (\sqrt{g}\, V^m) \ .
\end{equation}

Let us study in detail the asymptotic behavior of the extrinsic curvature
counter-term \eqref{ctK},
\begin{equation}\label{cgK}
 S_K = \int d^4x\, \sqrt{-\tilde{g}}\ \left(2 K - \frac{4}{L}\right)  \ .
\end{equation}
Considering \eqref{boumet} and the asymptotic behavior of the fields given in
\eqref{boufie} and \eqref{bouder}, we obtain
\begin{equation}\label{fullme}
 \begin{split}
 \sqrt{-\tilde{g}} &= g(r)^{1/2}\ e^{-\chi(r)/2}\ r^2 + \frac{e^{\chi(r)/2}}{2
g(r)^{1/2}} g_{tx}(r,t)^2 \\
                    &\overset{r_{\infty}\rightarrow\infty}{\longrightarrow}\ 
                    \left(\frac{r^3}{L}-\frac{\epsilon L^3}{4}\right)
+\frac{1}{2}\left(\frac{L}{r} + \frac{1}{4}\frac{\epsilon L^5}{r^4} \right)
g_{tx}^2 + ... \\                   
                    &\sim \frac{r^3}{L} -\frac{\epsilon L^3}{4} + \frac{L}{2}
r^3 g_{tx}^{(0)} g_{tx}^{(0)} + L\, g_{tx}^{(0)} g_{tx}^{(1)} 
+ \frac{\epsilon L^5}{8} g_{tx}^{(0)} g_{tx}^{(0)} + ... \\
 \end{split}
\end{equation}
We want to analyze the large $r$ behavior of the extrinsic curvature
\eqref{exspa}.
Let us consider the factors that appear there singularly; the square root of the
determinant of the full metric is
\begin{equation}\label{unosufullme}
 \begin{split}
 \sqrt{-g} &= e^{-\chi(r)/2}\ r^2 + \frac{e^{\chi(r)/2}}{2 g(r)} g_{tx}(r,t)^2
\\
           &\overset{r_{\infty}\rightarrow\infty}{\longrightarrow}\ 
            r^2 + \frac{1}{2}\frac{1}{\frac{r^2}{L^2}-\frac{\epsilon L^2}{2r}}
g_{tx}^2 + ... \ ,\\
           & \sim r^2 + \frac{L^2}{2} r^2 g_{tx}^{(0)} g_{tx}^{(0)} +
\frac{L^2}{r} g_{tx}^{(0)} g_{tx}^{(1)} 
+ \frac{\epsilon L^6}{4 r} g_{tx}^{(0)} g_{tx}^{(0)} + ...  \ ,
 \end{split}
\end{equation}
and its inverse is
\begin{equation}
 \begin{split}
 \frac{1}{\sqrt{-g}} \overset{r_{\infty}\rightarrow\infty}{\longrightarrow}\    
                                    & \frac{1}{r^2} - \frac{L^2}{2 r^2} g_{tx}^{(0)}
g_{tx}^{(0)} - \frac{L^2}{r^5} g_{tx}^{(0)} g_{tx}^{(1)}
- \frac{\epsilon L^6}{4 r^5} g_{tx}^{(0)} g_{tx}^{(0)} + ...                                     
 \end{split}
\end{equation}
The leading behavior of the factor $1/\sqrt{g_{rr}}$ is given by%
\footnote{Attention not to confuse the $g$ in \eqref{fullme} and
\eqref{unosufullme} which represents the full metric with
the $g$ in \eqref{grr} that represents the function $g(r)$ appearing in the
metric ansatz.}
\begin{equation}\label{grr}
 \frac{1}{\sqrt{g_{rr}}} = \sqrt{g(r)}
\overset{r_{\infty}\rightarrow\infty}{\longrightarrow}\ 
                           \frac{r}{L} - \frac{\epsilon L^3}{4 r^2}  + ... \ ,
\end{equation}
Putting these results in \eqref{exspa} we obtain (up to quadratic contributions
in the fluctuations and discarding subleading terms)
\begin{equation}
  \begin{split}
 & K = \frac{1}{\sqrt{-g}}\,\partial_r \left(\frac{\sqrt{-g}}{\sqrt{g_{rr}}}
\right) \overset{r_{\infty}\rightarrow\infty}{\longrightarrow}\ 
 \frac{3}{L} -\frac{\epsilon}{2}\frac{L^5}{r^3}g_{tx}^{(0)} g_{tx}^{(0)} -
\frac{3L}{r^3} g_{tx}^{(0)} g_{tx}^{(1)}
 \end{split}    
\end{equation}
%
We eventually have the following leading terms (for the part of the counter-term
which is quadratic in the metric fluctuations)
\begin{equation}\label{compe}
 \begin{split}
 S_K^{(2)}  &\sim  \int d^3x\, \sqrt{- \tilde{g}}\ \left(2K - \frac{4}{L}
\right) 
\sim \int d^3x\  \left( r^3 g_{tx}^{(0)} g_{tx}^{(0)} -\frac{3}{2}\epsilon L^4
g_{tx}^{(0)} g_{tx}^{(0)} - 4 r^3 g_{tx}^{(0)} g_{tx}^{(1)}\right)\ .
 \end{split}
\end{equation}
The terms in the action involving the metric fluctuations are
\begin{equation}\label{ugo}
 \begin{split}
  - e^{\chi/2} g_{tx} g'_{tx} &\sim  - \left(r^2 g_{tx}^{(0)} + \frac{1}{r}
g_{tx}^{(1)}\right) \left(2 r g_{tx}^{(0)} - \frac{1}{r^2} g_{tx}^{(1)}\right)\\
                              &\sim  - 2 r^3 g_{tx}^{(0)} g_{tx}^{(0)} - 2
g_{tx}^{(0)} g_{tx}^{(1)} + g_{tx}^{(0)} g_{tx}^{(1)} + \frac{1}{r^3}
g_{tx}^{(1)} g_{tx}^{(1)}\\
                              &\sim  - 2 r^3 g_{tx}^{(0)} g_{tx}^{(0)} -
g_{tx}^{(0)} g_{tx}^{(1)}+ ...
 \end{split}
\end{equation}
and
\begin{equation}\label{duna}
 \begin{split}
  \frac{e^{\chi/2}}{2} \frac{g'}{g} g_{tx}^2 &\sim \frac{1}{2}
\frac{\frac{2r}{L^2} + \frac{\epsilon L^2}{2 r^2}}
{\frac{r^2}{L^2} - \frac{\epsilon L^2}{2r}} \left( r^2 g_{tx}^{(0)} +
\frac{1}{r} g_{tx}^{(1)} \right)^2\\
                &\sim r^3 g_{tx}^{(0)}g_{tx}^{(0)} + \frac{3}{4}\epsilon L^4
g_{tx}^{(0)}g_{tx}^{(0)} + 2 g_{tx}^{(0)} g_{tx}^{(1)} 
 \end{split}
\end{equation}
Putting together the results obtained in \eqref{compe}, \eqref{ugo} and \eqref{duna}, we obtain the terms 
containing the metric fluctuations that appear in the renormalized action \eqref{resuquad}.

\chapter{Thermal Conductivity}
\label{therm}
From \eqref{SA}, \eqref{SB} and \eqref{Qt} we have
\begin{equation}\label{QQ}
 Q = \frac{\delta S_{quad}}{\delta g^{(0)}_{tx}} - \mu \frac{\delta S_{quad}}{\delta A^{(0)}_{x}} - \delta\mu \frac{\delta S_{quad}}{\delta B^{(0)}_{x}}
\end{equation}
Besides, from the conductivity matrix \eqref{matrix} using \eqref{gra} we obtain
\begin{equation}\label{kappaT}
 Q = \left.\kappa T \left(- \frac{\nabla_x T}{T}\right) \right|_{E^A_x=E^B_x=0}
   = \kappa T \left(\im \omega g_{tx}^{(0)}\right) 
\end{equation}
Again using \eqref{gra}, the $E_x^A=E_x^B=0$ condition implies the following relations
\begin{eqnarray}\label{zerocampiA}
 E_x^A &=& 0 \ \Rightarrow \ A_x^{(0)} = - \mu g_{tx}^{(0)}\\
 E_x^B &=& 0 \ \Rightarrow \ B_x^{(0)} = - \delta\mu g_{tx}^{(0)}
 \label{zerocampiB}
\end{eqnarray}
Let us compute explicitly the terms in \eqref{QQ} using all the just mentioned relations
\begin{equation}
\frac{\delta S_{quad}}{\delta g^{(0)}_{tx}} = -3\, g_{tx}^{(1)} - \epsilon\, g_{tx}^{(0)}
                                            = - \rho A_x^{(0)} - \delta\rho B_x^{(0)} - \epsilon\, g_{tx}^{(0)}
                                            = \left(\mu\, \rho +\delta\mu \delta\rho - \epsilon\right) g_{tx}^{(0)}
\end{equation}
\begin{equation}
 \begin{split}
 - \mu \, \frac{\delta S_{quad}}{\delta A^{(0)}_{x}} 
 &=- \mu \left( \frac{1}{2} A_x^{(1)} + \frac{ \im \omega \sigma_A}{2} A_x^{(0)} + \frac{ \im \omega \gamma}{2} B_x^{(0)} - g_{tx}^{(0)} \rho \right)\\
 &= \left(\mu^2 \, \im\, \omega \sigma_A  + \mu\, \delta\mu\, \im\, \omega \gamma  + \mu \rho \right)    g_{tx}^{(0)}
  \end{split}
\end{equation}
\begin{equation}
 \begin{split}
 - \delta \mu \, \frac{\delta S_{quad}}{\delta B^{(0)}_{x}} 
 &=- \delta \mu \left( \frac{ \im \omega \gamma}{2} A_x^{(0)} + \frac{1}{2} B_x^{(1)}  + \frac{ \im \omega \sigma_B}{2} B_x^{(0)} - g_{tx}^{(0)} \delta\rho \right)\\
 &=  \left(\mu\, \delta \mu\, \im \omega \gamma  + \delta\mu^2 \, \im \omega \sigma_B + \delta \mu  \delta\rho \right) g_{tx}^{(0)}
  \end{split}
\end{equation}
Substituting back into \eqref{QQ} we obtain
\begin{equation}
 \begin{split}
 Q &=\left(2\mu\, \rho + 2 \delta\mu \delta\rho - \epsilon
+ \mu^2 \, \im\, \omega \sigma_A  + 2\mu\, \delta\mu\, \im\, \omega \gamma  
 + \delta\mu^2 \, \im \omega \sigma_B  \right) g_{tx}^{(0)}\ ,
 \end{split}
\end{equation}
which eventually, comparing with \eqref{kappaT}, leads to
\begin{equation}
 \kappa T 
 = \frac{\im}{\omega}\left(-2\mu\, \rho - 2 \delta\mu \delta\rho + \epsilon \right)
+ \mu^2 \,  \sigma_A  + 2\mu\, \delta\mu\,  \gamma  
 + \delta\mu^2 \,  \sigma_B  
\end{equation}
coinciding with \eqref{kappa} up to the term in the pressure (see the comments below \eqref{kappa} in the main text).

\section{Simple observation on the sign of the thermo-electric conductivity}
\label{simple}

We look at a simple medium whose transport properties are due to the presence of carriers;
in this medium a temperature gradient can generate an electric transport.
The purpose is to show that the sign of the thermo-electric effect can be either positive or negative already in a simple model.
Consider a cylinder of material having unitary section; within the material there are electrically charged carriers that can flow and interact.
They are characterized by their volume density $n$, their mean free path $\lambda$ and the mean time between consecutive interactions $\tau$.
Suppose that in two neighboring slices of the material there is a difference in temperature; as the quantities
characterizing the carriers are temperature-dependent, the two slices at different values of the temperature are unbalanced.
Let us study the net flux of charges through the section of the cylinder separating the two slices.
The two slices $1$ and $2$ are respectively characterized by $T_i,n_i,\lambda_i$ and $\tau_i$ where $i=1,2$%
\footnote{Where $T_i = T(x_i)$ and similarly for the other quantities.}.
The fluxes of carriers $1\rightarrow 2$ and $2\rightarrow 1$ through $\Sigma$ in the lapse of time $dt$ are respectively given by
\begin{equation}
 \begin{split}
 &d\Gamma_{1\rightarrow 2} = \frac{n_1 \lambda_1}{2} \frac{dt}{\tau_1}\\
 &d\Gamma_{1\leftarrow 2}  = \frac{n_2 \lambda_2}{2} \frac{dt}{\tau_2}
 \end{split}
\end{equation}
where the factor $1/2$ is due to the fact that only half of the carriers move towards $\Sigma$.
The total flux through $\Sigma$ is
\begin{equation}
 d\Gamma_{\text{tot}} = \frac{1}{2} \left(\frac{n_1 \lambda_1}{\tau_1} - \frac{n_2 \lambda_2}{\tau_2} \right)\ dt
\sim \frac{1}{2} \left(- \frac{n}{\tau} \frac{d\lambda}{dx}dx - \frac{\lambda}{\tau} \frac{d n}{dx}dx 
 + \frac{n \lambda}{\tau^2} \frac{d\tau}{dx}dx \right)_{x=x_1}\ dt
\end{equation}
where 
\begin{equation}
 X_2 \sim X_1 + \left.\frac{dX}{dx}\right|_{x=x_1} dx \ \ \ \ \text{with}\ \ \ \ X= n, \lambda, \tau \ .
\end{equation}
Since $dx$ is of the order of the mean free path we can consider $dx \sim \lambda$ up to corrections of higher order, so
\begin{equation}\label{fluss}
 \frac{d\Gamma_{\text{tot}}}{dt} \sim \frac{\lambda}{2} \left(- \frac{n}{\tau} \frac{d\lambda}{dx} - \frac{\lambda}{\tau} \frac{d n}{dx}
 + \frac{n \lambda}{\tau^2} \frac{d\tau}{dx} \right)_{x=x_1}\ .
\end{equation}
From \eqref{fluss} it is comprehensible that the sign of the charged carriers flux is sensitive to how the various quantities $n, \lambda$ and $\tau$
vary as a consequence of the thermal gradient along the rod.

Notice that it is not redundant to retain both the mean free path $\lambda$ and the characteristic time between two interactions $\tau$,
indeed they can have different behaviors with respect to the temperature and, for instance, the composition or disorder of the system.
Consider a simple example in which the cross-section scattering of the carriers on impurities is independent of the temperature and the velocity of the carriers:
we can change the temperature and the impurity density so that $\tau$ remains constant and, at the same time, $\lambda$ results instead affected.

\@openrighttrue\makeatother


\chapter*{Brief Conclusion}
\addcontentsline{toc}{chapter}{Brief Conclusion}
The general picture which hopefully this thesis has helped to convey,
is that the string formalism offers an extremely fertile framework for studying field theory beyond the perturbative regime.
Such possibility proves significant for the theoretical investigations in general
and, maybe, it will play a crucial role in  exposing new generations of string-inspired models to experimental tests.

In the first part of the thesis we described the computation of stringy instanton effects in $4$-dimensional ${\cal N}=2$ SU$(N)$ gauge theory with matter in the symmetric
representation. The main conclusion there is that stringy effects give non-vanishing contributions both in the conformal SU$(2)$ case
and in the general non-conformal SU$(N>2)$ models.
In the former case, the stringy corrections to the effective prepotential suggest a very interesting structure
which leads one to conjecture a close compact form for the resummed stringy contributions at all orders in the topological charge.
This point has to be further investigated as it could signal a still unknown duality of the model.
It is to be noted that our model represents the first instance of stringy instanton effects appearing directly 
(i.e. without the need of any compactification procedure)
in a $4$-dimensional theory.

The second part of the thesis concerns the detailed analysis of a holographic system with two Abelian currents.
The presence of a scalar undergoing condensation allows us to study a model of superfluid or superconductor transitions
at strong coupling.
The emergence of superconductivity has been checked precisely by means of the study of the DC transport properties of the system.
We have characterized the phase diagram of the system observing that it suggests the absence of any Chandrasekar-Clogston bound at strong-coupling
and also the absence of a non-homogeneous LOFF-like phase.
The system, in its normal phase, generalizes at strong-coupling Mott's ``two-current'' model and shows interesting spintronic properties.
One of the main features of the mixed ``electric-spin''  characteristics of the system concerns the occurrence of ``spin-superconductivity'';
this phenomenon occurs even though the Cooper-like condensate is  neutral with respect to the effective magnetic field.
The model offers suggestive behaviors readable in terms of carrier-like transport and possesses various viable generalizations, 
among which, the finite-momentum analysis and the introduction of impurities or lattice-like features.

\chapter*{Acknowledgements}
\addcontentsline{toc}{chapter}{Acknowledgements}
I would like to thank my Ph.D. advisor Prof. Alberto Lerda not only for his help in the preparation of this thesis, but especially
for his guiding role throughout the last three years. As the Ph.D. is the phase in which a student is supposed to learn from his mentor the job of the researcher,
I would like to resemble him in his calm, prompt and aware attitude towards the wide range of theoretical physics.

I want to express my gratitude also toward Dr. Aldo Cotrone that introduced me to the holographic topics and accompained me
like a second advisor. In the last year I asked him a huge amount of questions and he had always the patience of answering;
his helpfulness and insightful answers have been essential to me.
I take the chance of promising him that I will keep on seeking his advice and my questions will not end together with my doctorate!

A special ``thank you'' goes to Dr. Francesco Bigazzi for his precious collaboration and his support in seeking a job which has yielded me the possibility
of spending my first postdoc in Brussels. I am also very grateful to Prof. Riccardo Argurio who offered me this concrete chance and
for his important comments and corrections to the thesis.
I want to thank Prof. Nick Evans for his corrections to the draft of my thesis and for his interest and encouragement.
I thank also the members of my Ph.D. committee, Prof. Matteo Bertolini, Prof. Marco Billo and Prof. Silvia Penati for the
attention tributed to my work and for the very stimulating discussion session I had in occasion of the thesis defence.
I am indebted also to Prof. Carlo Maria Becchi and Prof Massimo D'Elia for their teachings, support and encouragement
since my undergrad years.

I thank Dr. Parsa Ghorbani who has been my fellow in the stringy instanton project; with him I have shared many stimulating and intense discussions
in front of a blackboard. Our paths will maybe detach, but I hope to encounter him again for blackboard debates in the future.

A particular ackonowledgment goes to Dr. Davide Forcella and Dr. Alberto Mariotti: firstly I have known them because of their paper 
\cite{Amariti:2010hw} which I liked very much and had stimulated me to get seriously involved with the holographic panorama, secondly because I spent
a very nice period in Brussels with them. I am looking forward to collaborating with them in the near future and have further great time together.

A thanks goes to Dr. Natalia Pinzani and Dr. Domenico seminara for the valuable collaboration about the unbalanced superconductor project.

I would like to thank my father, Girogio Musso, and my friend Katherine Deck for helping me revising the draft.

On the personal side, the biggest thank you goes to my parents. 
I would like to thank them for a specific thing, the fact that they center on me their deepest hopes will always be an inexhaustible 
source of courage to me. A special thanks goes to my grandmother because thinking to her makes the perspective of getting old a dreamlike one.
I would like to thank Claudio for much fun in my childhood. 

A big thanks goes to all of my friends. I want to single out the closest fellows of this thesis-writing experience:
Anita Laurina Polonia Terry Esmeralda, Cesare Papiani, Gibbone Sexy and Lucca (senza tazzina proletaria). 
If the future of humanity were a dystopia as in Sci-Fi stories, living in an underground bunker in their company 
will be an enjoyable condition, somehow similar to write a thesis in our subterranean office!

I beg pardon for reducing the remainder of the acknowledgements to a list, it absolutely does not make justice to the individuals, but allows me to mention 
the numerous people to which I am grateful. I start with my office mates and then the ``rest of the world'':
A great ``thank you'' goes (alphabetically) to Camerello Annual Report, Diogo Bunda, Giulio, Graziella,  
Marcogallo, Marcogall'Ines, Mario, Nicoletta Stockfish,  
Notaio Pochi Parametri, Roberto Pellegri, Rosco Siffredi, Roverto, Stefano XXVIB,
the two new Andrea's, Voera, Wikimur.
Eventually, a big thank you to: Alessandrioli, Cavallaro, Claudia, CS libri, Dar\'io, Dopocos\'i, Fragolina, 
Francesco Final Countdown, Giulia Bobbola, Giulia Mezzas, Hamdy, Irene Adler Santandrea, Johanna, John, Emiu, Evelina, Fabioli, Fragolina, Livio Tesi, Kat, Lara, 
Maria Pilar, Martina, Neda, Parthenope's staff, Pi\'upagato, Sebastian, Seriet\'a Giovane, Tamu, Test di Roscah, Torre di Pisa, Veronica, Victoria, Zappia, the 5 shrimps and Mia.
All this names could sound imaginary but fortunally to me they correspond to real people, a real orangutan, a real dog and 5 real crustaceans. 
I chersih to have shared the time in Torino (and Paris and Corf\'u) with them!

\bibliographystyle{utphys}
\bibliography{ESTI}

\end{document}